\documentclass[]{aa}

\usepackage{graphics}
\usepackage{times}  

\begin{document}

\title{The Hamburg/ESO R-process Enhanced Star survey (HERES) 
\thanks{Based on observations collected at the European Southern Observatory, Paranal, Chile (Proposal Number 68.B-0320).} 
\thanks{Tables 1 and 2 are only available in electronic form at the CDS via anonymous ftp to cdsarc.u-strsbg.fr (130.79.125.5) or via http://cdsweb.u-strasbg.fr/Abstract.html}
} 

\subtitle{II. Spectroscopic analysis of the survey sample}

\author{P. S. Barklem\inst{1} \and N. Christlieb\inst{2} \and T. C. Beers\inst{3} \and V. Hill \inst{4} \and M. S. Bessell\inst{5} \and J. Holmberg\inst{6,7,8} \and B. Marsteller\inst{3} \and S. Rossi\inst{9} \and F.-J. Zickgraf\inst{2} \and D. Reimers\inst{2}}
 
\offprints{P. S. Barklem,
\email{barklem@astro.uu.se}}

\institute{Department of Astronomy and Space Physics, Uppsala University, Box 515, S 751-20 Uppsala, Sweden \and
Hamburger Sternwarte, Universit\"at Hamburg, Gojenbergsweg 112, 21029 Hamburg, Germany \and
Department of Physics and Astronomy and JINA: Joint Institute for Nuclear Astrophysics, Michigan State University, East Lansing, MI 48824 USA  \and
GEPI, Observatoire de Paris \-Meudon, F-92125 Meudon Cedex, France \and
Research School of Astronomy and Astrophysics, Australian National University, Cotter Rd, Weston, ACT 2611, Australia \and
Tuorla Observatory, V\"ais\"al\"antie 20, FI-21500 Piikki\"o, Finland \and
Astronomical Observatory, NBIfAFG, Juliane Meries Vej 30, 2100 Copenhagen, Denmark \and
Nordic Optical Telescope Scientific Association, Apartado 474, ES-38\,700 Santa Cruz de La Palma, Spain \and
Departamento de Astronomia, IAG, Universidade de S{\~a}o Paulo, Rua do Mat{\~a}o 1226, 05508-900 S{\~a}o Paulo - SP, Brazil
}

\date{Received 2 March 2005 / Accepted 2 May 2005}

\abstract{We present the results of analysis of ``snapshot'' spectra of 253 metal-poor halo stars $-3.8\leq \mathrm{[Fe/H]} \leq -1.5$ obtained in the HERES survey.    The snapshot spectra have been obtained with VLT/UVES and have typically $S/N\sim 54$ per pixel (ranging from 17 to 308), $R\sim20000$, $\lambda =$ 3760--4980~\AA.  This sample represents the major part of the complete HERES sample of 373 stars; however, the CH strong content of the sample is not dealt with here. 

The spectra are analysed using an automated line profile analysis method based on the Spectroscopy Made Easy (SME) codes of Valenti \& Piskunov.
Elemental abundances of moderate precision (absolute rms\ errors of order 0.25~dex, relative rms\ errors of order 0.15~dex) have been obtained for 22 elements, C, Mg, Al, Ca, Sc, Ti, V, Cr, Mn, Fe, Co, Ni, Zn, Sr, Y, Zr, Ba, La, Ce, Nd, Sm, and Eu, where detectable. Of these elements, 14 are usually detectable at the 3$\sigma$ confidence level for our typical spectra.  The remainder can be detected in the least metal-poor stars of the sample, spectra with higher than average $S/N$, or when the abundance is enhanced.    

Among the sample of 253 stars, disregarding four previously known comparison stars, we find 8 r-II stars and 35 r-I stars.  The r-II stars, including the two previously known examples CS~22892-052 and CS~31082-001, are centred on a metallicity of $\mathrm{[Fe/H]} = -2.81$, with a very small scatter, on the order of 0.16~dex. The r-I stars are found across practically the entire metallicity range of our sample.  We also find three stars with strong enhancements of Eu which are s-process rich.  A significant number of new very metal-poor stars are confirmed: 49 stars with $\mathrm{[Fe/H]}<-3$ and 181 stars with $-3<\mathrm{[Fe/H]}<-2$.  We find one star with $\mathrm{[Fe/H]}<-3.5$.

We find the scatter in the abundance ratios of Mg, Ca, Sc, Ti, Cr, Fe, Co, and Ni, with respect to Fe and Mg, to be similar to the estimated relative errors and thus the cosmic scatter to be small, perhaps even non-existent.  The elements C, Sr, Y, Ba and Eu, and perhaps Zr, show scatter at $\mathrm{[Fe/H]} \la -2.5$ significantly larger than can be explained from the errors in the analysis, implying scatter which is cosmic in origin.  Significant scatter is observed in abundance ratios between light and heavy neutron-capture elements at low metallicity and low levels of r-process enrichment.
    
}

\maketitle

\section{Introduction}

The Hamburg/ESO R-process Enhanced Star (HERES) survey has been described in the preceding paper in this series (Christlieb et~al.~\cite{heres1}; hereafter Paper I).  HERES is mostly based on confirmed metal-poor stars from the Hamburg/ESO survey (HES; Wisotzki et~al.~\cite{wisotzki00}). In HERES, ``snapshot'' spectra of 373 very metal-poor stars, here meaning with $\mathrm{[Fe/H]} \leq -1.5$ \footnote{$\mathrm{[A/B]} = \log ( N_\mathrm{A} / N_\mathrm{B} )_\star - \log (N_\mathrm{A} / N_\mathrm{B})_\mathrm{\sun}$ where $N_\mathrm{X}$ are number densities.} as judged from medium resolution spectra, have been obtained with VLT2-UVES, with the main goal of finding stars enhanced in the r-process elements through detection of strong Eu II lines.  Though the snapshot spectra are of what would generally be considered low quality for abundance analysis (typically $S/N\sim 54$, $R\sim20000$, $\lambda =$ 3760--4980~\AA), they contain a wealth of information and abundances may be derived for a significant number of elements with moderate precision (absolute rms\ errors of order 0.25~dex, relative errors of order 0.15~dex).  Modern surveys of metal-poor stars, such as the ESO ``First Stars'' Large programme (Cayrel et~al.~\cite{cayrel04}, Hill~et~al.~\cite{hill02} and references therein), now obtain significantly better quality spectra for of the order of 70 stars, yet just a decade ago spectra of similar quality to our snapshot spectra were typical for studies of very metal-poor stars (e.g. McWilliam et~al.~\cite{mcw95a, mcw95b}).  While high precision (better than 0.1 dex) is expected to be needed to discern detailed patterns in abundance distributions which might serve as diagnostics of early nucleosynthesis (e.g. Karlsson \& Gustafsson~\cite{karlsson01}), the large number of stars observed in the HERES survey offers the possibility to investigate more general trends in metal-poor star abundances in a previously unexplored statistical regime.  In particular, the scatter in abundance distributions may provide insights into mixing and the diversity of supernovae at early epochs.  The study of Norris~et~al.~(\cite{norris01}), which investigated such scatter, drawing from different surveys in the literature, had of order 70 stars in this metallicity regime.  This project provides a homogeneously analysed sample of several hundred stars. 

In this paper we analyse a total of 253 of the spectra using an automated spectrum analysis technique based on the Spectroscopy Made Easy (SME) codes by Valenti \& Piskunov~(\cite{sme}).  In Sect.~\ref{sect:obs} the sample, observations, and choice of initial stellar parameters will be described.  In Sect.~\ref{sect:automated_analysis} the automated analysis technique and codes are described, including the line list, detection classification and error analysis, and the method is tested through comparisons with previously studied stars and tests of the robustness of the method.  The results are presented in Sect.~\ref{sect:results}, including the new interesting stars found in the survey.  In Sect.~\ref{sect:discussion}, we discuss general trends in the results and other interesting features to emerge.  Finally, in Sect.~\ref{sect:conclusions} we present our conclusions and discuss possible uses for the data set.

\section{Observations and Sample}
\label{sect:obs}

\begin{figure}
\begin{center}
\resizebox{\hsize}{!}{\rotatebox{0}{\includegraphics{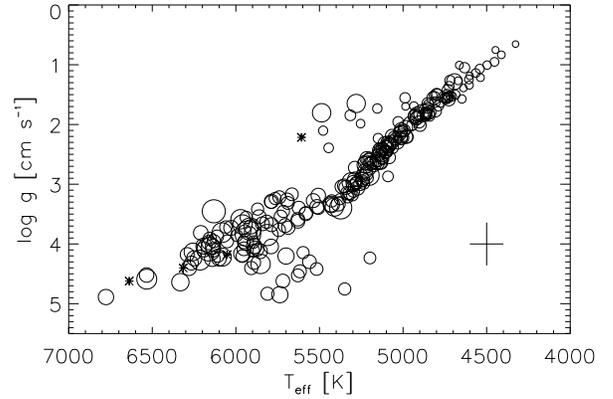}}}
\end{center}
\caption{The $T_\mathrm{eff}$ versus $\log g$ Kiel diagram for the sample.  The diameter of the circle for each star is proportional to [C/Fe]; cases where C is undetected are shown with an asterisk.  The mean error bar is shown in the bottom right.}
\label{fig:kiel}
\end{figure}

\begin{figure}
\begin{center}
\resizebox{\hsize}{!}{\rotatebox{0}{\includegraphics{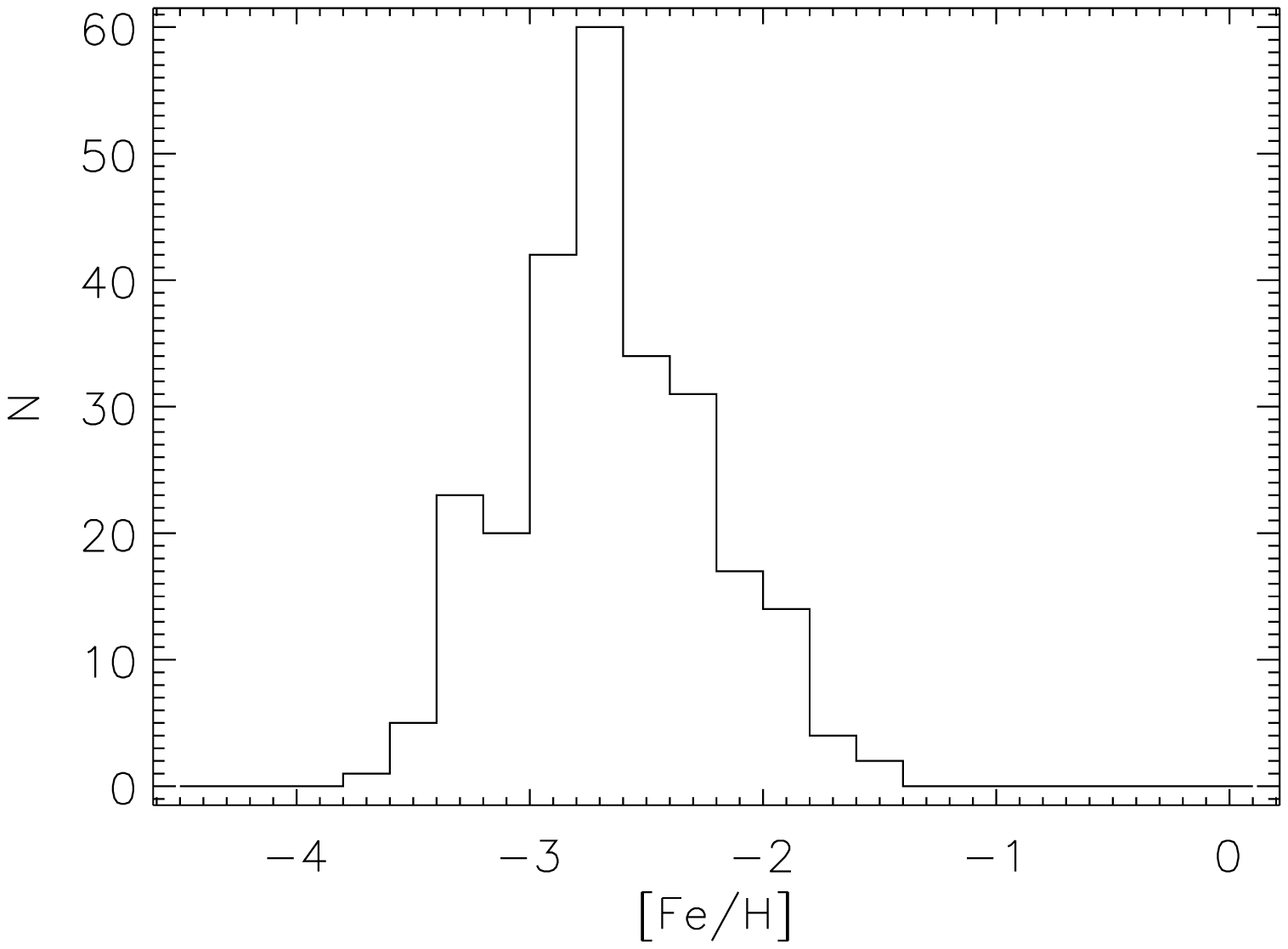}}}
\end{center}
\caption{The distribution of analysed stars in [Fe/H].  The [Fe/H] values are from the final analysis of the spectra.}
\label{fig:fe_dist}
\end{figure}

The target selection for HERES and observational details have been described in Sect.~3 of Paper I.  The spectra were obtained with the ESO-VLT2 and UVES, and cover a wavelength range of 3760--4980~\AA, and have an average signal-to-noise ratio of $S/N \sim 54$ per pixel over the spectral range, though some spectra have $S/N$ as low as 17 and as high as 308.  A 2\arcsec\ slit is employed giving a minimum resolving power of $R\approx 20000$, though typically the resolving power is seeing limited and thus slightly better.  As mentioned in Paper~I, the pipeline-reduced spectra are employed, corrected to the stellar rest frame. The sample of analysed stars is given in the electronic Table~\ref{tab:sample} \footnote{Table~\ref{tab:sample} in its entirety is available only electronically} with coordinates and barycentric radial velocities.  The final stellar parameters of the sample are presented in Table~\ref{tab:abunds},\footnote{Table~\ref{tab:abunds} in its entirety is available only electronically} together with the abundances for convenience, and are plotted in Fig.~\ref{fig:kiel}.  The [Fe/H] distribution for the sample from the final analysis is shown in Fig.~\ref{fig:fe_dist}.  Though observations of adequate quality have been obtained for 373 stars, not all are analysed in this paper.  Spectra showing strong molecular carbon features cannot be analysed by our current method, and thus 72 such spectra from the survey are not included in this sample for this reason. These spectra are, however, of great interest and it is planned that they will be analysed separately in future.  Further, some stars which turned out to be too Fe-rich ($\mathrm{[Fe/H]}>-1.5$) or too cool for our analysis method ($<4200$~K), or are suspected to be spectroscopic binaries or rotators, were also removed from the sample. For a small number of stars, we have not yet obtained adequate photometry and they are also removed from the sample.  Thus, the final sample of stars analysed in this work presented in Tables~\ref{tab:sample} and~\ref{tab:abunds} contains 253 stars.

\begin{table*}
\begin{center}
\caption{The sample.  The entire table is available only electronically.  Coordinates of the HES stars (prefix HE) have been derived from the Digitized Sky Survey~I and are accurate to 1\arcsec; the coordinates of the HK survey stars (prefix CS) are from identifications of the sources in the 2MASS All Sky Release.  Barycentric radial velocities $v_\mathrm{rad}$ were measured from the snapshot spectra and are accurate to a few km~s$^{-1}$. }
\label{tab:sample}
\begin{tabular}{crrr}
\hline
Star & $\alpha(2000.0)$ & $\delta(2000.0)$ & $v_\mathrm{rad}$ \\
     &                  &                  &  [km s$^{-1}$]   \\
\hline
 CS~22175-007  &  02 17 26.6 &	 $-$09 00 45  &  $	  -18.3   $ \\	 
 CS~22186-023  &  04 19 45.5 &	 $-$36 51 35  &  $	   51.9   $ \\	 
 CS~22186-025  &  04 24 32.8 &	 $-$37 09 02  &  $	 -122.6   $ \\	 
 CS~22886-042  &  22 20 25.8 &	 $-$10 23 20  &  $	 -220.6   $ \\	 
 CS~22892-052  &  22 17 01.6 &	 $-$16 39 27  &  $	   14.5   $ \\	 
 CS~22945-028  &  23 31 13.5 &	 $-$66 29 57  &  $	  388.9   $ \\	 
 CS~22957-013  &  23 55 49.0 &	 $-$05 22 52  &  $	 -213.5   $ \\	 
 CS~22958-083  &  02 15 42.7 &	 $-$53 59 56  &  $	  175.6   $ \\	 
 CS~22960-010  &  22 08 25.3 &	 $-$44 53 56  &  $	   49.2   $ \\	 
 CS~29491-069  &  22 31 02.1 &	 $-$32 38 36  &  $	 -375.5   $ \\	 
   ~$\vdots$   &    $\vdots$ &	 $ \vdots$    &   	$\vdots   $ \\ 
 HE~2338-1618  &  23 40 36.2 &	 $-$16 01 28  &  $	 -219.4   $ \\	 
 HE~2345-1919  &  23 47 55.4 &	 $-$19 02 37  &  $	  119.4   $ \\	 
 HE~2347-1254  &  23 50 09.8 &	 $-$12 37 50  &  $	  -43.5   $ \\	 
 HE~2347-1334  &  23 50 26.9 &	 $-$13 17 39  &  $	  -54.8   $ \\	 
 HE~2347-1448  &  23 49 58.3 &	 $-$14 32 16  &  $	 -165.1   $ \\	 
\hline
\end{tabular}
\end{center}
\end{table*}

\begin{table*}
\tabcolsep 0.5mm
\begin{center}
\caption{Derived stellar atmosphere parameters and elemental abundances for the sample.  The entire table is available only electronically.  This portion of the table is given as a guide to its form.  For each star we report the average signal-to-noise ratio per pixel for the whole observed spectrum, the stellar parameters with their respective relative and absolute rms\ error estimates,  $\sigma^\mathrm{rel}$ and $\sigma^\mathrm{abs}$, then for each element X the abundance $\log \epsilon_\mathrm{X}$ and its relative and absolute rms\ error estimates $\sigma_{\log\epsilon_\mathrm{X}}$, [X/Fe] and its relative and absolute rms\ error estimates $\sigma_\mathrm{[X/Fe]}$.  N gives the number of features of the element used in each star, noting that this varies from star to star as certain features may be automatically rejected if the star has a strong G band, if the feature is near a Balmer line in warmer stars (see Sect.~\ref{sect:lines}), or if they are determined to be affected by a cosmic ray hit or bad pixel. N3 then gives the number of those features classified as 3$\sigma$ detections.  In the case of CH bands we simply classify detections and non-detections with 1 and 0 respectively.  In the case of a non-detection, this is signified by an abundance of $-9.99$ and error of $0.00$.}
\label{tab:abunds}
\scriptsize
\begin{tabular}{ccccccccccccccccrccccc}
\hline
Star        & $S/N_\mathrm{av}$ & $T_\mathrm{eff}$ & $\sigma_{T_\mathrm{eff}}$  & $\log g$ & $\sigma_{\log g}^\mathrm{rel}$ & $\sigma_{\log g}^\mathrm{abs}$ & [Fe/H] & $\sigma_\mathrm{[Fe/H]}^\mathrm{rel}$ & $\sigma_\mathrm{[Fe/H]}^\mathrm{abs}$ & $\xi$ & $\sigma_\xi^\mathrm{rel}$ & $\sigma_\xi^\mathrm{abs}$ &  $\log\epsilon_\mathrm{C}$ & $\sigma_{\log\epsilon_\mathrm{C}}^\mathrm{rel}$ & $\sigma_{\log\epsilon_\mathrm{C}}^\mathrm{abs}$  & [C/Fe] & $\sigma_\mathrm{[C/Fe]}^\mathrm{rel}$ & $\sigma_\mathrm{[C/Fe]}^\mathrm{abs}$ & N & N3 & \ldots \\    
 & & [K] & & [cm s$^{-2}$] & & & & & & [km s$^{-1}$] & & &  &  & & & & & & & \ldots \\    
\hline 
  CS~22175-007      &	   31 &	5108 &	100 & 2.46&  0.26 & 0.36 & $-2.81$ &  0.13 & 0.18&  1.67&  0.14&  0.24 &     5.77 & 0.19 & 0.26 & $   0.19$ &  0.18&  0.27&   1 & 1 & $\ldots$ \\	
  CS~22186-023      &	   55 &	5066 &	100 & 2.19&  0.24 & 0.34 & $-2.72$ &  0.13 & 0.18&  1.58&  0.11&  0.21 &     5.97 & 0.19 & 0.25 & $   0.30$ &  0.18&  0.27&   1 & 1 & $\ldots$ \\	
  CS~22186-025      &	   32 &	4985 &	100 & 1.70&  0.26 & 0.36 & $-2.87$ &  0.14 & 0.19&  2.14&  0.13&  0.23 &     4.83 & 0.20 & 0.27 & $  -0.68$ &  0.19&  0.29&   1 & 1 & $\ldots$ \\	
  CS~22886-042      &	   39 &	4881 &	100 & 1.85&  0.25 & 0.35 & $-2.68$ &  0.12 & 0.18&  1.84&  0.12&  0.22 &     5.72 & 0.19 & 0.26 & $   0.01$ &  0.19&  0.28&   1 & 1 & $\ldots$ \\	
  CS~22892-052      &	   46 &	4884 &	100 & 1.81&  0.26 & 0.36 & $-2.95$ &  0.14 & 0.19&  1.67&  0.12&  0.22 &     6.44 & 0.19 & 0.26 & $   1.00$ &  0.18&  0.28&   1 & 1 & $\ldots$ \\	
  CS~22945-028      &	   31 &	5126 &	100 & 2.55&  0.26 & 0.36 & $-2.66$ &  0.13 & 0.18&  1.53&  0.14&  0.24 &     5.93 & 0.19 & 0.26 & $   0.21$ &  0.18&  0.27&   1 & 1 & $\ldots$ \\	
  CS~22957-013      &	   35 &	4904 &	100 & 1.96&  0.25 & 0.35 & $-2.64$ &  0.14 & 0.19&  1.79&  0.12&  0.22 &     5.85 & 0.19 & 0.26 & $   0.10$ &  0.17&  0.27&   1 & 1 & $\ldots$ \\	
  CS~22958-083      &	   32 &	5101 &	100 & 2.40&  0.26 & 0.36 & $-2.79$ &  0.12 & 0.18&  1.50&  0.13&  0.23 &     6.24 & 0.20 & 0.26 & $   0.64$ &  0.18&  0.27&   1 & 1 & $\ldots$ \\	
  CS~22960-010      &	   35 &	5737 &	100 & 4.85&  0.25 & 0.35 & $-2.65$ &  0.12 & 0.17&  1.53&  0.17&  0.27 &     6.56 & 0.17 & 0.24 & $   0.82$ &  0.16&  0.25&   1 & 1 & $\ldots$ \\	
  CS~29491-069      &	   57 &	5103 &	100 & 2.45&  0.24 & 0.34 & $-2.81$ &  0.13 & 0.18&  1.54&  0.12&  0.22 &     5.76 & 0.19 & 0.25 & $   0.18$ &  0.17&  0.27&   1 & 1 & $\ldots$ \\	
$\vdots$  &$\vdots$  &$\vdots$  &$\vdots$  &$\vdots$  &$\vdots$  & $\vdots$ & $\vdots$ & $\vdots$ & $\vdots$ & $\vdots$ & $\vdots$ & $\vdots$ & $\vdots$ & $\vdots$ & $\vdots$ &  $\vdots$ & $\vdots$& $\vdots$ &  $\vdots$& $\vdots$ &$\ldots$ \\ 
  HE~2338-1618      &	   40 &	5515 &	100 & 3.38&  0.27 & 0.37 & $-2.65$ &  0.12 & 0.18&  1.43&  0.16&  0.26 &     6.22 & 0.18 & 0.25 & $   0.47$ &  0.18&  0.27&   1 & 1 & $\ldots$ \\	
  HE~2345-1919      &	   44 &	5617 &	100 & 4.46&  0.25 & 0.35 & $-2.46$ &  0.12 & 0.18&  1.47&  0.16&  0.26 &     6.17 & 0.19 & 0.24 & $   0.24$ &  0.17&  0.26&   1 & 1 & $\ldots$ \\	
  HE~2347-1254      &	   80 &	6132 &	100 & 3.95&  0.24 & 0.34 & $-1.83$ &  0.13 & 0.18&  1.67&  0.12&  0.22 &     6.83 & 0.15 & 0.22 & $   0.27$ &  0.16&  0.26&   1 & 1 & $\ldots$ \\	
  HE~2347-1334      &	   81 &	4453 &	100 & 0.95&  0.24 & 0.34 & $-2.55$ &  0.12 & 0.19&  2.38&  0.11&  0.21 &     5.33 & 0.21 & 0.26 & $  -0.50$ &  0.18&  0.27&   1 & 1 & $\ldots$ \\	
  HE~2347-1448      &	   43 &	6162 &	100 & 3.98&  0.27 & 0.37 & $-2.31$ &  0.12 & 0.18&  0.84&  0.17&  0.27 &     6.58 & 0.18 & 0.24 & $   0.50$ &  0.19&  0.27&   1 & 1 & $\ldots$ \\	
\hline
\end{tabular}
\end{center}
\end{table*}

\subsection{Photometry and Effective Temperatures}
\label{subsect:photometry}

To begin our analysis we require as accurate as possible estimates of the effective temperatures of HERES targets. This is particularly the case since, as noted below, the snapshot spectra we obtain are not generally of sufficiently high quality to derive precise spectroscopic temperature estimates. Hence, over the course of the past few years, we have obtained broadband $BVR_CI_C$ (where the subscript ``C'' indicates the Cousins system) observations for as many HERES targets as possible, using the ESO/Danish 1.5m telescope on La Silla and the DFOSC instrument. The observing and reduction techniques for this data are described in Beers et~al. (2005, in preparation). We also make use of near-infrared $JHK$ photometry from the 2MASS Point Source Catalog (Cutri et~al. \cite{cutri03}). The complete set of available photometry for stars in the HERES sample will be published in Beers et~al. (2005, in preparation), a compilation of photometry for more than 1500 metal-poor stars and horizontal-branch candidates.  The adopted reddening is taken from the maps of Schlegel et~al. (\cite{schlegel98}), adjusted to account for distance.  Note, all but the brightest giants of our sample lie outside the reddening layer and thus for most stars the full reddening is applied.  The reddening data will also be presented in Beers et~al.  

Temperature estimates are obtained on the scale of Alonso et~al. (\cite{alonso99}). Note that, while the majority of the colour calibrations of Alonso et~al. require measurements on the Johnson system, the near-IR colours $J-H$ and $J-K$ are on the TCS system (the photometric system at the 1.54~m Carlos Sanchez telescope; Arribas \& Martinez-Roger \cite{arribas87}). Thus we first have to apply several transformations of our observed colours. We follow the prescription described by Sivarani et~al. (\cite{sivarani04}). In performing our transformations, and for carrying out the Alonso et~al. estimates of effective temperature, we made use of a program kindly provided to us by T. Sivarani.   In order to arrive at a final ``best estimate'' of the effective temperatures for our HERES pilot sample stars, we take a straight average of the derived estimates from different colour criteria, after trimming off the highest and lowest estimates for each star. The estimated effective temperatures, for the colours we had available, are provided in Table~\ref{tab:abunds}.  Although errors in the determinations of temperature for individual colours arising from photometric errors can range from 50-100 K (for the optical colours), and up to several hundred K for the near-IR colours (owing to the generally larger errors in the 2MASS photometry), we conservatively estimate that our final determination of effective temperature has an absolute rms\ error of 100 K.  We note that our photometry is incomplete, i.e.\ we do not have measurements of all colours for every star, and this incompleteness, together with the trimming in the averaging procedure, means there is an error introduced due to differences in the temperature scales from different colour criteria.  Since the incompleteness is effectively random, this should lead only to a slight increase in random scatter.  This has been considered in our error estimates.  The error due to uncertain reddening (see discussion below) was also considered.  We plan to obtain improved photometric estimates of $T_{\rm eff}$ for stars of particular interest on the HERES program (for instance, those noted to be r- and/or s-process enhanced) by measuring more precise $JHK$ photometry in the near future. 

\begin{figure}
\begin{center}
\resizebox{\hsize}{!}{\rotatebox{0}{\includegraphics{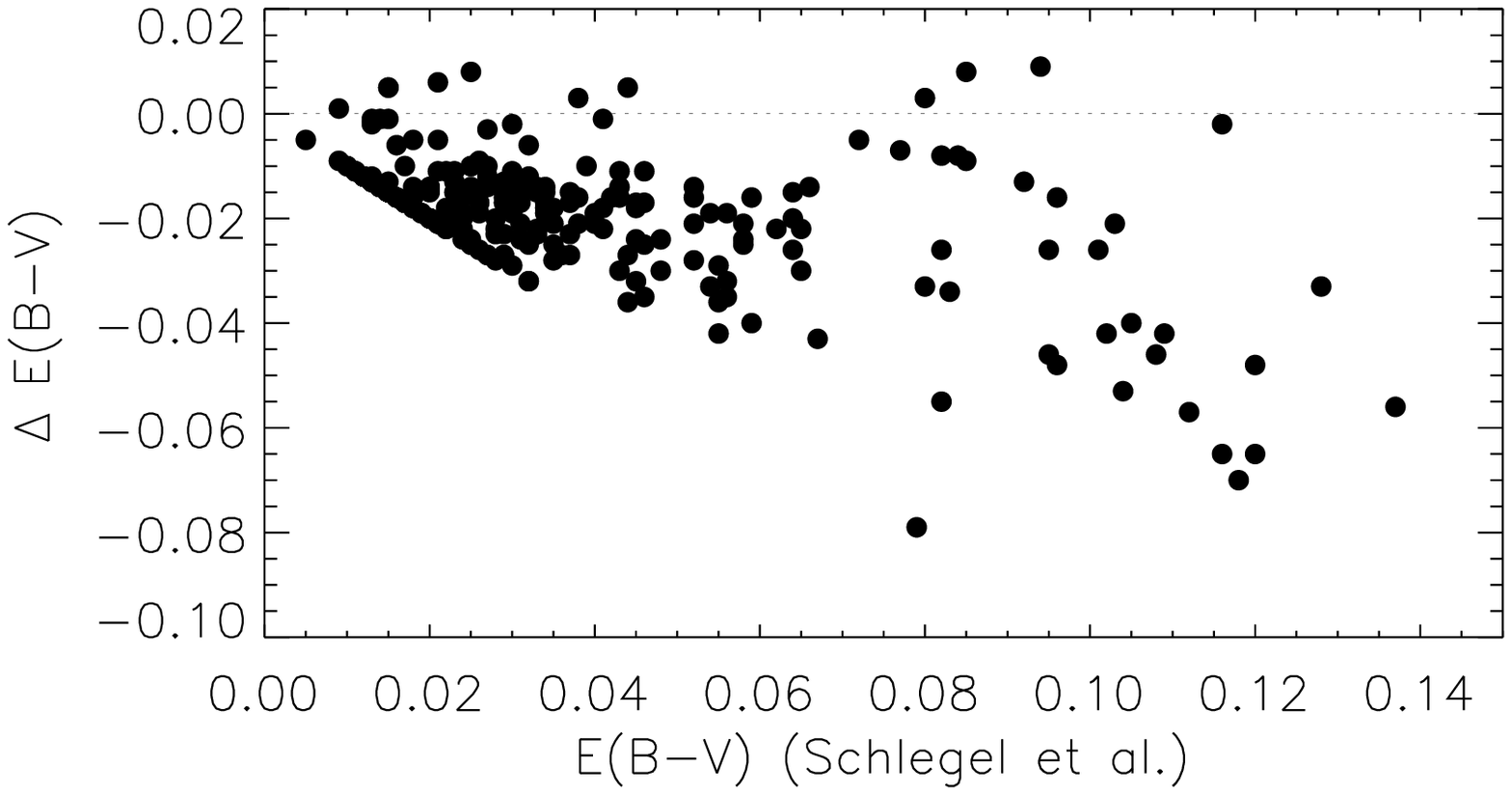}}}
\resizebox{\hsize}{!}{\rotatebox{0}{\includegraphics{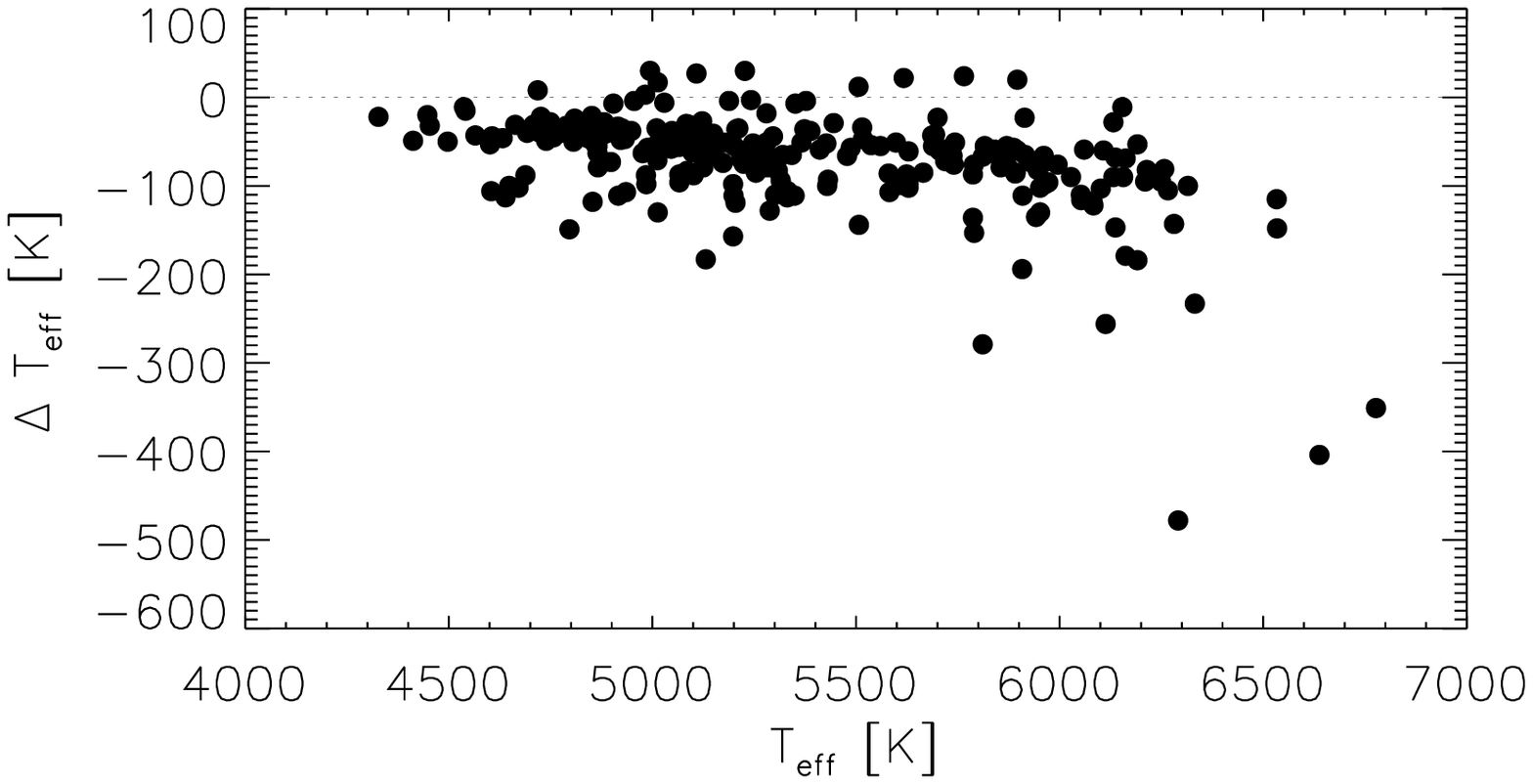}}}
\end{center}
\caption{Change in $E(B-V)$ (upper panel) and $T_\mathrm{eff}$ (lower panel) if we adopt the reddening maps of Burstein \& Heiles~(\cite{burstein82}) instead of those of Schlegel~et~al.~(\cite{schlegel98}).  The reddening corrections are usually reduced and thus the temperatures become somewhat cooler.}
\label{fig:reddening}
\end{figure}

As will be discussed in Sect.~\ref{sect:comparison}, our temperatures are found to be systematically warmer than those found in the literature for the stars where comparisons can be made.  This discrepancy was traced primarily to the use of different reddening maps.  In this work, the maps of Schlegel~et~al.~(\cite{schlegel98}) have been employed, while in the past most workers have used the maps of Burstein \& Heiles~(\cite{burstein82}).  We computed reddenings and effective temperatures using both maps and find systematic differences as seen in Fig.~\ref{fig:reddening}, the results using the maps of Burstein \& Heiles being cooler than those found using Schlegel~et~al.\ by $70\pm58$~K.  We have chosen to use the Schlegel~et~al.\ estimates, which have superior spatial resolution and are thought to have a better determined zero point. We note, however, that Arce \&~Goodman~(\cite{arce99}) have pointed out that the Schlegel~et~al.\ maps may overestimate the reddening values when $E(B-V)$ exceeds about 0.15 mag, though none of the stars considered here exceed that value (see also Beers~et~al. \cite{beers02} for further discussion).

It should be noted that while the HERES target selection aims at a cutoff of $B-V>0.5$, a number of warmer stars which turned out to be bluer than our targeted cutoff have also been included due to incorrect estimated $B-V$ colours from the HES prism plates.

\begin{figure}
\begin{center}
\resizebox{8cm}{!}{\rotatebox{0}{\includegraphics{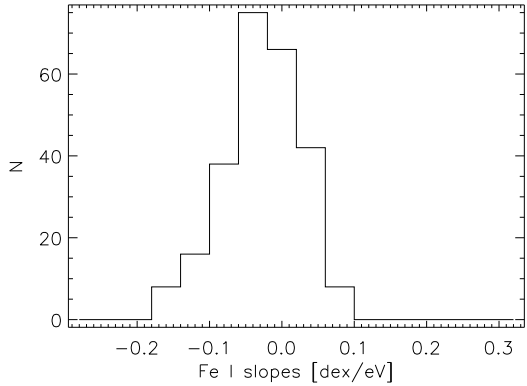}}}
\end{center}
\caption{Histogram of the slopes of fits to the trends of Fe I line abundances with excitation for the sample.  }
\label{fig:equilibria_hist}
\end{figure}

As will be described later in Sect.~\ref{sect:procedure}, we compute individual abundances for all Fe lines and thus can compare our adopted temperature scale with that implied by excitation equilibrium.  Given the lines available to us, the spectroscopic data are not of sufficient quality to determine precise excitation temperatures, but one can make a comparison of the average temperature scale.  Figure~\ref{fig:equilibria_hist} plots the histogram of slopes of the best fit to the Fe I line abundances against excitation.  On average the distribution is skewed towards negative slopes, indicating that our temperature scale is on average warmer than would be obtained from excitation equilibrium.  Such trends have been noted in previous works on metal-poor giants and dwarfs, for example by Norris et~al.~(\cite{norris01}) and Cohen et~al.~(\cite{cohen04}).

\subsection{Other Atmospheric Parameters}
\label{subsect:other_params} 

\begin{figure}
\begin{center}
\resizebox{\hsize}{!}{\rotatebox{0}{\includegraphics{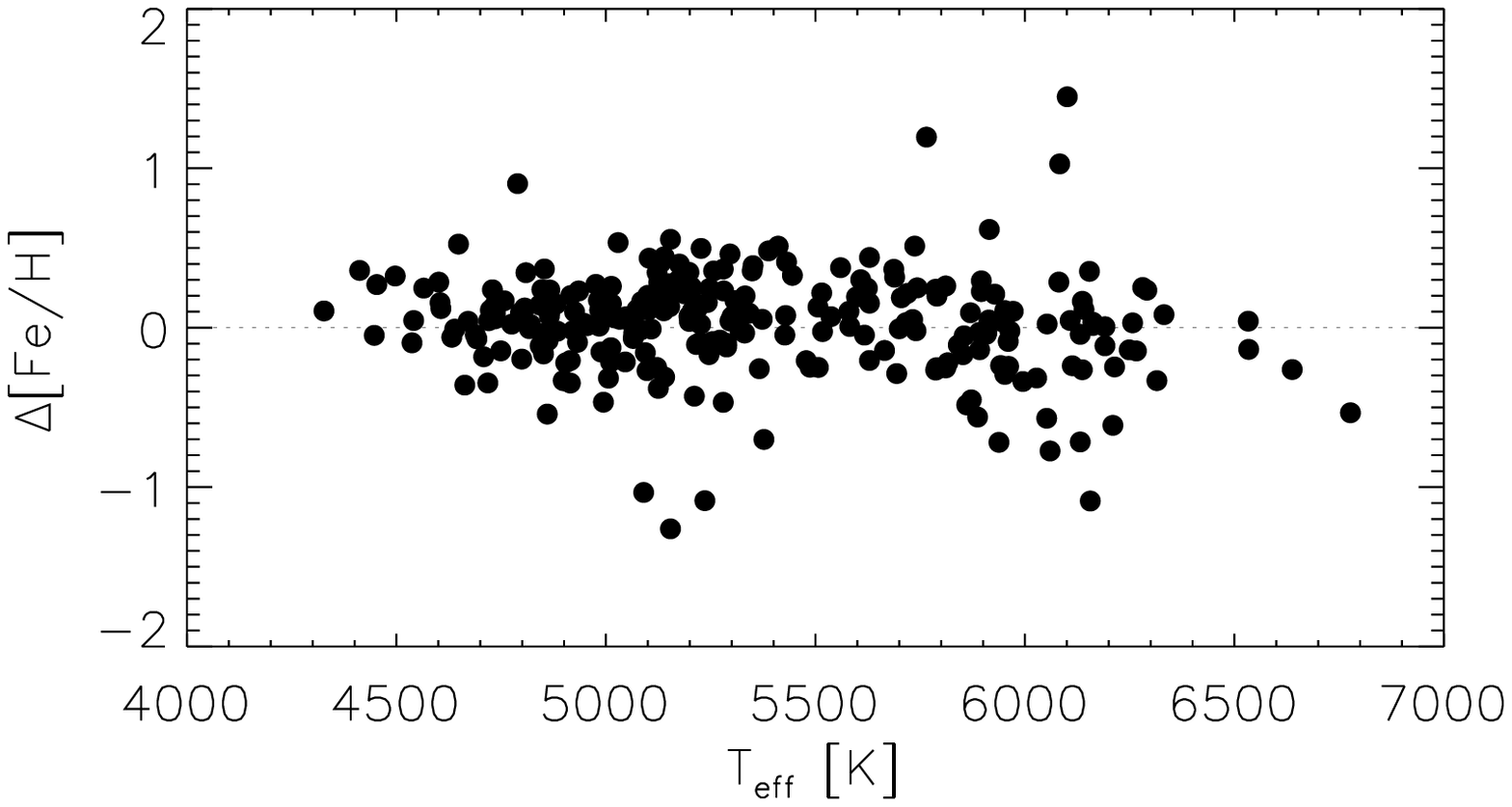}}}
\resizebox{\hsize}{!}{\rotatebox{0}{\includegraphics{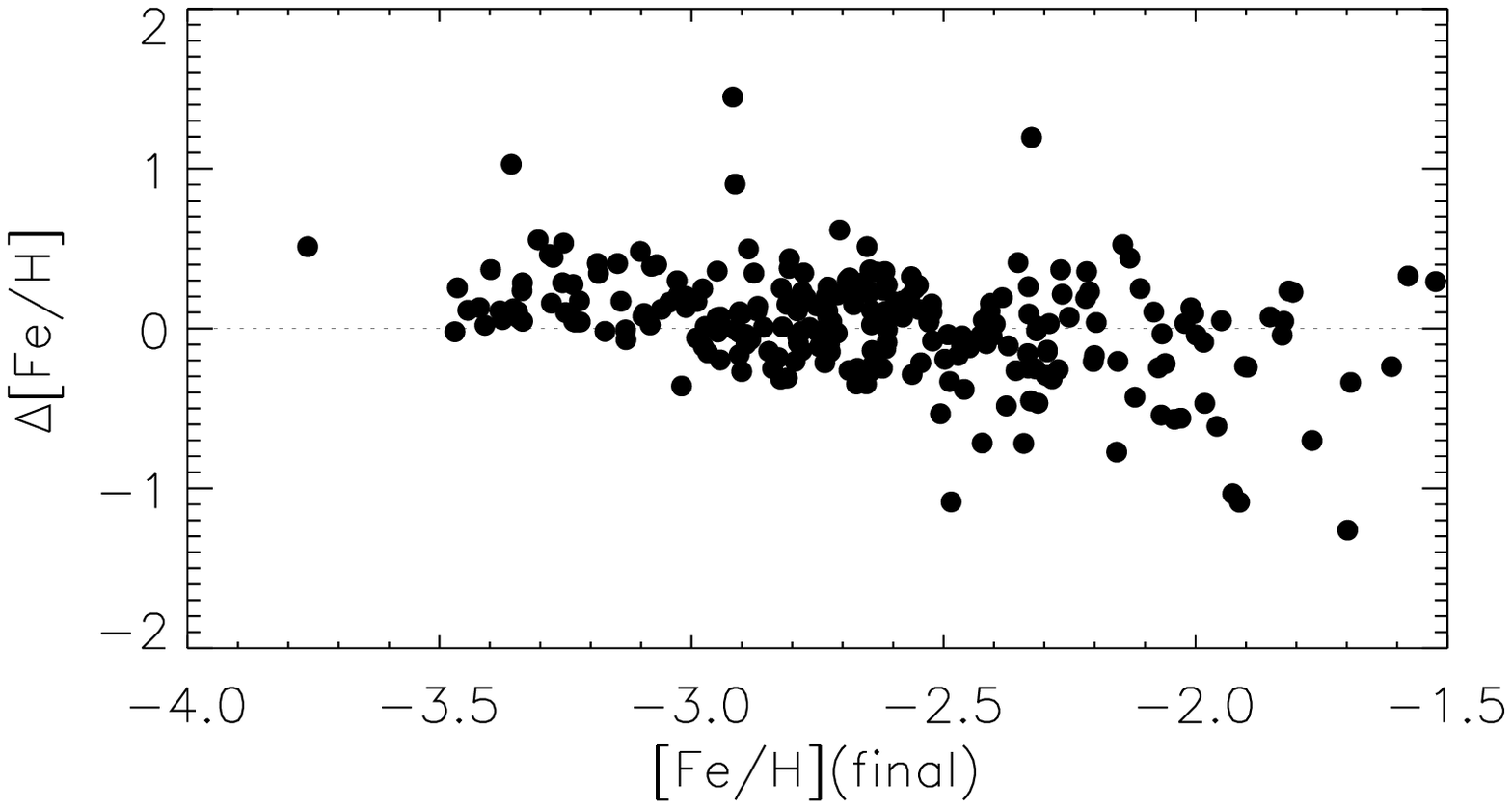}}}
\end{center}
\caption{Difference between the final [Fe/H] as determined from the spectrum and the initial guess determined from medium-resolution follow-up spectroscopy, where $\Delta$[Fe/H]$ = $[Fe/H](initial)$ - $[Fe/H](final), plotted against $T_\mathrm{eff}$ and [Fe/H](final).}
\label{fig:met_diff}
\end{figure}

Estimates of the metallicity, [Fe/H], which serve as initial guesses for our automated analysis, are derived from the medium-resolution spectra obtained during the course of the large-scale campaign to identify suitable targets for the HERES program (see Paper~I), in combination with available $B-V$ and $J-K$ photometry. Estimates of [Fe/H] were obtained from the calibration of the Ca~II K-line index, KP, along with the $B-V$ colour, described by Beers et~al.~(\cite{beers99}). For most stars this calibration should be accurate to on the order of 0.2~dex, given spectra and photometry of suitable quality. In addition, we carried out a new calibration, making use of available $J-K$ colours, and a newly-trained artificial neural network procedure (see Snider et~al.~\cite{snider01}), to provide an alternative initial estimate of [Fe/H]. This latter step was taken because the HERES targets include objects that are known to be carbon-enhanced; the strong CH G-bands in these objects could possibly perturb the $B-V$ colours used in the Beers et~al. calibration.  We take the average of the two determinations as our initial guess for [Fe/H].  The differences between the adopted initial guess and the final [Fe/H] values from the automated analysis of the high-resolution spectra are plotted in Fig.~\ref{fig:met_diff}, and application of robust methods (see, e.g., Beers, Flynn, \& Gebhardt \cite{beers90}) to the set of differences in the initial and final metallicity determinations yield estimates of the mean offset and standard deviation of +0.04~dex and 0.27~dex, respectively.  The mean estimated relative error for the high-resolution abundance determinations of [Fe/H] is 0.12~dex (see Sect.~\ref{subsect:detect_errors} and Table~\ref{tab:abunds}), indicating that the medium-resolution estimates have relative errors of about 0.24~dex.  Note that the [Fe/H] estimates for the worst outliers were often based on rather noisy medium-resolution spectra. 

\begin{figure}
\begin{center}
\resizebox{\hsize}{!}{\rotatebox{0}{\includegraphics{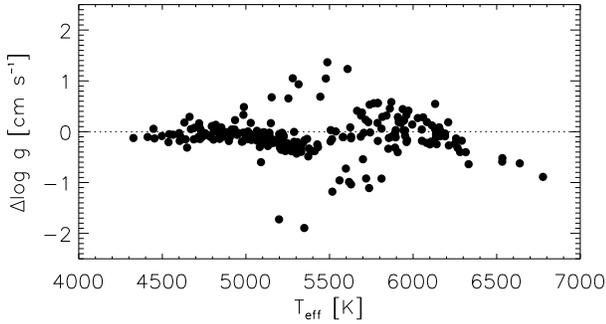}}}
\end{center}
\caption{Difference between the final $\log g$ as determined from the spectrum and the initial guess as described in the text, where $\Delta \log g = \log g \mathrm{(initial)} - \log g \mathrm{(final)}$.}
\label{fig:logg_diff}
\end{figure}

An initial estimate of the surface gravity, $\log g$, is also required, which is refined in the automated analysis.  For stars on the subgiant and giant-branch stages of evolution, it is well known that $\log g$ correlates very well with effective temperature.  Hence, in order to obtain a first-pass estimate of surface gravity we used the reported temperatures and surface gravities by Honda et~al.~(\cite{honda04}) to derive the regression relation:
\begin{equation}
\log g = -9.301 + 2.273  \; T_{\rm eff} / 1000    
\label{eqn:init_logg}
\end{equation}
Note that the Honda et~al. program considers stars over a very similar range of metallicities and temperatures as the HERES program.  
For warmer stars, this fit leads to overly high initial estimates, so we limited the initial guess to a maximum value of $\log g=4$.  That is, if $T_{\rm eff} > 5850$~K an initial guess of $\log g=4$ is adopted, which was found to be adequate.  Figure~\ref{fig:logg_diff} shows the differences between the adopted initial estimate and the determination obtained from automated analysis of the snapshot spectra.

\begin{figure}
\begin{center}
\resizebox{\hsize}{!}{\rotatebox{0}{\includegraphics{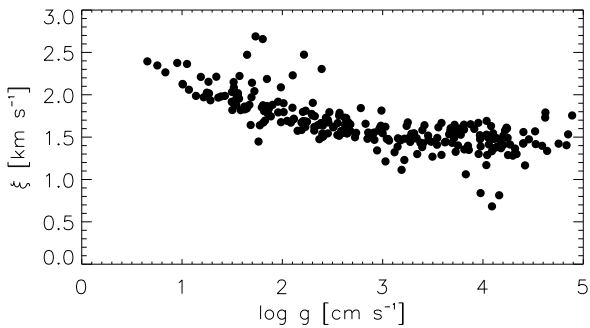}}}
\end{center}
\caption{Plot of final microturbulence values $\xi$ as derived from the spectrum against gravity.  Note that the stars lying higher than the general trend are the red horizontal branch stars (see Fig.~\ref{fig:kiel}).}
\label{fig:micro}
\end{figure}

As will be described in the next section, a 1D analysis is employed and turbulent line broadening is modelled via the classical microturbulence $\xi$ and macroturbulence $v_\mathrm{macro}$ parameters.  These parameters are derived from the spectrum.  Initial guesses of $\xi=1.8$~km~s$^{-1}$ and $v_\mathrm{macro}=1.5$~km~s$^{-1}$ were employed.  The final values of $\xi$ derived from the spectrum are given in Table~\ref{tab:abunds} and plotted in Fig.~\ref{fig:micro}, showing the usual correlation with gravity. Note that the stars lying above the main trend, around $\log g \sim 2$ are the red horizontal branch stars (see Fig.~\ref{fig:kiel}).   

\section{Automated Spectrum Analysis}
\label{sect:automated_analysis}

\subsection{Analysis code}
\label{subsect:program}

Software for automated analysis of the spectrum has been developed, based on the Spectroscopy Made Easy (SME) package by Valenti \&\ Piskunov~(\cite{sme}). SME consists of three components: a spectrum synthesis component written in C++, a parameter optimisation component written in IDL, and a user interface written in IDL.  In this work we make use of the first two components.  Our developed software is written in IDL, and essentially provides an alternative interface to the parameter optimisation component, which in turn calls the spectrum synthesis component.  We also made some minor adaptations and improvements to the SME codes, the most important of which are discussed below.  The new interface software takes our input data and provides it to the parameter optimisation component without any need for user interaction.  The input data in this case consist of a pipeline-reduced count spectrum with corresponding measurement error spectrum (corrected to the stellar rest frame), initial guess stellar atmosphere parameters and abundances (see Sect.~\ref{sect:procedure}), and a list of spectral features with corresponding atomic or molecular data (see Sect.~\ref{sect:lines}).  The main tasks of the new interface software are, in a completely automated fashion, to extract relevant spectral regions appropriate for the line list, to identify continuum points and normalise the spectra relative to the continuum, to make small adjustments (within the error of the wavelength calibration) to line central wavelengths such that they match the observed line centres, and to reject lines polluted by artifacts such as cosmic ray hits.  The continuum points are identified by an automated procedure where for each spectral region required for the analysis, the continuum is determined for a wide region, including the desired spectral region and 7~\AA\ on each side by iteratively fitting a low order polynomial to the count spectrum and discarding points more than one standard deviation below the fitted line in the subsequent iteration, thereby converging to the continuum high points.  Cosmic ray hits are identified by being significantly above this fit.  

Given the input data in the correct form, the SME parameter optimisation code can then solve for any desired model parameters.  The parameter optimisation code uses the Marquardt algorithm (Marquardt~\cite{marquardt63}, Press~et~al.~\cite{numrec}) to obtain estimates of the parameters, through minimising the $\chi^2$ statistic comparing model and observed spectra.  

The spectrum synthesis assumes LTE and a 1D plane-parallel model of the atmosphere, where turbulence is modelled through the classical microturbulence and macroturbulence parameters.  Note that in recent versions of SME, molecular line formation is supported (see Valenti et~al.~\cite{valenti98}), and for speed the radiative transfer is solved using a Feautrier scheme (e.g.~Mihalas~\cite{mihalas}) not a Runge-Kutta scheme (cf. Valenti \&\ Piskunov~\cite{sme}).  The original spectrum synthesis code did not account for scattering in the source function, i.e.\ continuous scattering is treated as absorption and thus $S_\nu = B_\nu$.  Rayleigh scattering by neutral hydrogen can be an important contributor to continuous opacity particularly in the ultraviolet for metal-poor stars, especially giants, and thus it is important to treat continuous scattering correctly in the source function (e.g. Griffin~et~al.~\cite{griffin82}) . It should be computed as $S_\nu = (1-\epsilon_\nu) J_\nu + \epsilon_\nu B_\nu$ where $\epsilon_\nu$ is the photon thermalisation probability given by $\kappa_\nu/(\kappa_\nu+\sigma_\nu)$ where $\kappa_\nu$ is the absorption coefficient (including line absorption) and $\sigma_\nu$ is the continuum scattering coefficient.  This has been implemented into the SME spectrum synthesis code, where the mean intensity $J_\nu$ has been solved using the perturbation technique of Cannon~(\cite{cannon71}).  Further minor improvements included correction of Thorium partition functions (see Paper~I), 
and the ability to handle data for line broadening by neutral hydrogen collisions from the Anstee, Barklem \& O'Mara theory (see Sect.~\ref{sect:lines}).

A grid of 1D, plane-parallel, LTE models covering the stellar atmosphere parameter space of the stars of interest, namely low metallicity F, G, and K stars from the main sequence to the giant stage, was computed using the 1997 MARCS code (Asplund~et~al.~\cite{marcs_asp}, Gustafsson~et~al.~\cite{marcs_gust}).  All models use scaled solar abundances with the exception of the alpha-elements, which are enhanced by 0.4 dex.  Throughout this paper solar abundances means those of Grevesse \&\ Sauval~(\cite{grevesse98}), except C which is taken from Allende Prieto et~al. (\cite{allende02}).  Microturbulence in the line blanketing of 2 km s$^{-1}$ was adopted as most of our stars are giants (see Fig.~\ref{fig:micro}).  This should not have a great effect in metal-poor stars in any case.  This model grid was incorporated into SME, allowing SME to solve for atmospheric parameters, where specific models are obtained by interpolation.

In Sect.~\ref{sect:procedure} we will describe how we apply this code. First we describe the adopted line list.

\subsection{Line list}

\label{sect:lines}

A list of spectral lines and corresponding atomic and molecular data, suitable for our sample of stars, observational data, and analysis technique is required.  For our automated line profile analysis technique, spectral windows where the model and observed spectrum are to be compared need also to be defined. 

The lines and spectral windows used are listed in Tables~\ref{tab:lines_ch1}--\ref{tab:lines_heavy} in Appendix~\ref{app:linelist}, along with comments on each element.   Note that in order to avoid the cores of strong lines, any wavelength points in the chosen spectral windows where the observed flux is less than 0.5 of the continuum are automatically rejected.  The line list and windows were, for the most part, built by selecting suitable lines from previous studies, particularly McWilliam~et~al.~(\cite{mcw95a,mcw95b}), Norris et~al.~(\cite{norris96}), and Sneden~et~al.~(\cite{sneden96}).  As our technique is automated, our aim is to build a line list which requires either no adjustment from star to star, or possibly a small amount of adjustment that can be automated based on criteria that are known in advance of the analysis, such as stellar temperature.  We must therefore be more selective than in interactive analyses. One must keep in mind, however, the need to have as many lines as possible to provide better statistics, which is particularly important for low-quality spectra. Ideally we would like to choose lines which {\it for all stars} of the type we are examining are unblended, or which are partially blended in such a way that they can be used if an appropriate comparison window is chosen that neglects the blended part of the line (usually one wing of the line).   

Our first step for selecting the lines was to apply all candidate lines to an automated abundance analysis of four stars viewed to be extreme cases of possible blending for metal-poor stars.  We used HD~20, moderately r-process enhanced, which is one of the most metal-rich of our stars, non-carbon-enhanced; CS~31082-001, an r-process enhanced, non-carbon-enhanced star; CS~22892-052 an r-process enhanced, carbon-enhanced star; and HE~0338-3945, an r- and s-process enhanced, carbon-enhanced star.   Based on examination of the four spectra we chose appropriate spectral windows for each line, rejecting lines where a suitable window for all four stars could not be easily chosen.  We then performed an abundance analysis of each star and rejected lines which were not well fit by the derived abundance with respect to the majority of lines of that element, indicating most likely an unresolved blend, or possibly poor atomic data.  Special attention was paid to lines that were blended in the carbon-enhanced stars, but not in the carbon-normal stars.  We chose to only make use of these lines for the carbon-normal stars, as described below.  These lines are marked with an asterisk in the line list.  To err on the side of caution, we also used some other very carbon-enhanced stars in the sample, which we deemed too carbon-rich for our present method, to indicate other lines which could be blended by carbon features.  Finally, the line list and windows were refined based on testing using the complete sample.

We empirically determined that the lines noted above to be blended in the carbon-enhanced stars, could be safely applied in stars where the G~band has a depth not greater than 0.6 of the continuum flux.  Further, in the warmer stars some chosen lines lie in the wings of high Balmer lines.  Thus, we empirically derived a scheme for application of the line list to a given star.  The list is applied to all stars with the following adjustments:
\begin{itemize}
\item if the maximum depth of the G~band of CH has a depth greater than 0.6 of the continuum flux, the lines seen or suspected to be blended in carbon-enhanced stars (marked with asterisks in the tables) are removed from the list;
\item lines close to Balmer lines are removed depending on the stellar temperature.
\end{itemize}
These adjustments are fully automated.

While the employed spectral lines were chosen from the above mentioned previous studies, we compiled our own atomic data for these lines, though often guided by these works.  In particular, we attempted to update the data.  The adopted oscillator strengths are discussed for each element in Appendix~\ref{app:linelist}.  The VALD database\footnote{http://www.astro.univie.ac.at/\~{}vald/} (Kupka~et~al.~\cite{vald}) and NIST Atomic Spectra Database\footnote{http://physics.nist.gov/cgi-bin/AtData/main\_asd}  (see also Wiese et~al.~\cite{wiese69}, Wiese \& Martin~\cite{wiese80}, Martin et~al.~\cite{martin88}, Fuhr et~al.~\cite{fuhr88}) were used extensively, particularly for wavelengths and excitation energies.  Radiative broadening, Stark broadening and collisional broadening by neutral hydrogen are included, though the data have not been presented in the tables; they may be obtained from the authors on request.  Radiative broadening is taken in all cases from VALD, which is supplied from the calculations by Kurucz~(\cite{kurucz}).  Collisional broadening by neutral hydrogen is, where possible, described by the Anstee, Barklem and O'Mara theory (e.g. Anstee \& O'Mara~\cite{anstee91}, Barklem et~al.~\cite{barklem00} and references therein).  In the absence of such calculations the data are taken from VALD which again originates from Kurucz~(\cite{kurucz}) using the van der Waals theory (Uns\"old~\cite{unsold55}) with a detailed calculation of the long range interaction constant $C_6$.  In the absence of data from both of these sources the classical van der Waals theory is used, where $C_6$ is estimated by the usual Uns\"old approximation.  In both the last two cases we apply an enhancement of a factor of~2.  Stark broadening is taken from VALD where available, which often originates from Kurucz~(\cite{kurucz}).  We note that since the lines are typically weak, collisional mechanisms are generally not important, except in a few lines which attain significant strength (e.g. the strongest lines of Mg, Sr and Ba).

Where possible we have considered the hyperfine structure of spectral lines.  Hyperfine structure is typically seen in elements with odd atomic numbers.  Though Sr and Ba have even atomic numbers, in the solar system they have non-negligible abundances of isotopes with odd atomic masses, and thus hyperfine structure must be considered in these elements also.  Isotopic mixtures and splitting have been considered in Sr, Ba and Eu.  In all other elements the dominating natural isotope is assumed (or equivalently in the case of even atomic numbers that isotopes with even atomic masses dominate and they have negligible isotope shift).  Details of adopted hyperfine structure and isotopic ratios are discussed for each element in Appendix~\ref{app:linelist} where it is considered.  Hyperfine structure and isotopic splitting have been omitted from the tables for compactness.  They may be obtained on request from the authors.

\subsection{Procedure}
\label{sect:procedure}

We now describe the procedure used for analysis of the spectra.  The goal is to determine the required stellar atmosphere parameters and abundances from the spectra in the best possible way.  In Sect.~\ref{subsect:program} a code was described which can analyse a given spectrum to obtain best estimates of chosen model parameters.  While the code can in principal analyse the entire set of spectral lines and obtain all desired stellar parameters and abundances simultaneously in one step, this is not done, since for cases with many free model parameters such a procedure is susceptible to errors if strong dual dependencies of the model spectra on parameters exists.  Thus, our procedure, like traditional analyses, separates the determination of stellar parameters using appropriate spectral features from the determination of abundances as we now describe.    

First, the $T_\mathrm{eff}$ from Sect.~\ref{subsect:photometry} is adopted and held fixed throughout.  We experimented with refining $T_\mathrm{eff}$ from the spectra (essentially excitation equilibrium of Fe I), but found that the spectra are not of sufficiently high quality given the spectral features available to us.  Thus we need to solve for the atmosphere parameters surface gravity $\log g$, microturbulence $\xi$, macroturbulence $v_\mathrm{macro}$, and metallicity [Fe/H].  Once these parameters are known we then solve for the individual abundances.   All stars are assumed to be slow rotators; we adopt $v \sin i = 1$~km s$^{-1}$ for all stars, and this has no effect on the final abundances.  Since the resolving power of observations can vary from star to star, we fix the instrumental broadening at a value $R=50000$, significantly higher than expected and allow the derived macroturbulence to compensate for the difference.  Thus our derived $v_\mathrm{macro}$ includes both the macroturbulence and a degree of the instrumental profile.

Our basic procedure is broken into three steps.  In the first two steps we employ only the Fe and Ti lines of our list.  In the first step initial estimates of $T_\mathrm{eff}$, $\log g$, [Fe/H], $\xi$ and $v_\mathrm{macro}$ as described in Sect.~\ref{subsect:photometry} and \ref{subsect:other_params} are adopted.  The initial guess Ti abundance is chosen to be a scaled solar value based on the [Fe/H] initial guess. We then solve for the Fe and Ti abundances, $\log \epsilon_\mathrm{Fe}$ \footnote{We define the abundance parameter for element X in the standard notation $\log \epsilon_\mathrm{X} = \log(N_\mathrm{X}/N_\mathrm{H}) + 12$ where $N_\mathrm{X}$ are number densities.} and $\log \epsilon_\mathrm{Ti}$ to obtain better initial estimates of these parameters before solving for the remaining stellar parameters. 

The second step repeats the first, but now $\log g$, $\xi$ and $v_\mathrm{macro}$ are also free parameters. The minimisation to solve for $\log g$ is in essence equivalent to performing an ionisation equilibrium procedure for Fe and Ti.  The resultant parameters ($\log g$, $\xi$, $v_\mathrm{macro}$, [Fe/H], $\log \epsilon_\mathrm{Ti}$) are then adopted and held fixed for stage three, the analysis of lines of elements other than Fe and Ti in the list, where the only free parameters are the elemental abundances.  All the remaining elements are considered separately, as this is most computationally efficient.  Solar abundances scaled by [Fe/H] are adopted as initial guesses, which was found usually to be satisfactory; on rare occasions where convergence was not achieved this could be solved by searching for an initial guess which resulted in convergence.  Note that in all steps, the single abundance which gives the best global fit to the spectrum is derived.  It is also worth noting that we adopt scaled solar abundances for all elements which we do not analyse, except O which we assume enhanced by $+0.4$~dex; these abundances enter the molecular equilibrium and continuous opacity calculations. 

An additional stage is performed for checking purposes.  In this stage we analyse the Fe and Ti line list keeping all parameters fixed to those derived in the above stages except the abundance which may now vary from line to line (in SME this can be achieved in practice by fixing the abundance and solving for $\log gf$ for each line, and observing the change from the $\log gf$-value in the line list).  This permits us to check for trends with excitation and ionisation stage, which might warn of deficiencies in the adopted model.  

Three further steps are then performed which are required for the error analysis, the details of which will be described below in Sect.~\ref{subsect:detect_errors} and in Appendix~\ref{app:errors}.  However, for the error analysis we will require partial derivatives describing the dependence of each abundance on $T_\mathrm{eff}$, $\log g$ and $\xi$.  Thus we solve for all abundances with fixed stellar parameters except that $T_\mathrm{eff}$, $\log g$ and $\xi$ are offset by 100~K, 0.3~dex and 0.2~km~s$^{-1}$ in each step respectively.  We note that these last three steps increase computing time significantly (they comprise more than 1/2 of the computing time).  However, we view the error analysis as important enough to justify this increase.

It is worthwhile to make some comment on computing times.  With the large number of molecular lines, isotopic and hyperfine components, our line list results in approximately 900 individual line components to be included in the spectrum synthesis, which is where the vast majority of computation is used.  Model convergence typically requires approximately 3.5$N$+13 model spectrum calculations, where $N$ is the number of free parameters (Valenti \&\ Piskunov~\cite{sme}).  This means for the main abundance step (step 3) we require of the order of 15 model spectrum calculations of the complete spectrum.  This is repeated three additional times for the error calculations.  With a modern workstation typical computation times per star are of the order of 2--3 hours.

\subsection{Elemental detections and errors}
\label{subsect:detect_errors}

A first important step before assigning an error to a derived abundance is to determine if there is in fact a reliable detection.  A Gaussian is fit to each line and the approximate equivalent width $W$ is determined.  Following Norris~et~al.~(\cite{norris01}), the error in $W$ is then computed as $\sigma_W = \lambda\sqrt{n}/(R[S/N])$ where $R$ is the resolving power, $n$ the number of pixels integrated to obtain $W$, and $S/N$ the signal to noise ratio per pixel.  In all cases we adopted $R=20000$ and adopted the full-width of the fitted Gaussian measured in pixels for $n$.  By then computing $m= W/\sigma_W$ we classify a $m\sigma$ detection of the line. For each element we count the number of lines detected at the $3\sigma$ level or better.  If the element has at least one line detected at the $3\sigma$ level we assign the abundance as a detection.  Otherwise the element is regarded as undetected, and the derived abundance discarded.  Due to the possibility of weak blending, even in high quality spectra, it was further required that a line must be deeper than 10\% of the continuum flux for a valid detection.  Due to decreased likelihood of blending in the redder parts of the spectrum, we relaxed this condition to 5\% for lines redder than 4500 \AA.  A similar detection scheme was used for CH bands, though $W$ and $n$ for the considered band must be estimated by a different technique. 

As one of the main goals of this work is to look at abundance scatter at low metallicities, and given the relatively low quality of the data, it is vital to have error estimates which are as accurate as possible to distinguish real scatter from that due to uncertainties.  Thus, we have put significant effort into obtaining realistic error estimates.

Usually in analyses of stellar spectra, systematic errors are expected to dominate.  This is not necessarily the case here.  For elements with a small number of spectral features, the low $S/N$ means that random measurement errors can be significant.  However, for elements with a significant number of spectral features, e.g. Fe, measurement errors are in fact quite small, and systematic errors will dominate.

Our method for calculating errors in abundances and abundance ratios is detailed in Appendix~\ref{app:errors}.  First, we should emphasise that we make a distinction between {\em absolute} and {\em relative} errors.  In our discussion we will refer to the uncertainty in the absolute abundance or abundance ratio as the absolute error, and the uncertainty in the relative abundance between stars, that is from star to star, as the relative error.  Appendix~\ref{app:errors} provides a formalism for calculation of both absolute and relative errors; however, throughout this paper, unless otherwise stated we discuss the relative error.  This is of most interest as we are interested in distributions of abundances.  The main difference between absolute errors and relative errors is that some systematic sources of error are presumed to cancel in the latter, in particular, errors in oscillator strengths and {\em some part} of the errors incurred from modelling uncertainties such as LTE and 1D.  We note however, that as our sample covers such a large stellar parameter space, the cancellation of errors from the modelling will be only partial.

The error estimates include contributions from uncertainties in the observations, atomic and molecular data, continuum fitting, stellar parameters, model atmospheres and inherent assumptions in the spectrum modelling such as LTE and 1D.  We have included as many sources of error as possible, however, some of these uncertainties are difficult to estimate quantitatively, particularly those related to the model atmospheres and modelling assumptions.  Our abundances and error estimates are provided with the explicit understanding that they are based on traditional 1D LTE models and spectrum synthesis.   Corrections for 3D and non-LTE effects often should be applied by the user.  For example, it is known that the resonance line of Al is subject to strong deviations from LTE (e.g. Baum\"uller \&\ Gehren~\cite{baumueller97}).  While corrections for cases such as Al where only a single line is employed are relatively straight forward, we point out that for other elements one needs to consider the correction to the best fit to all spectral features used here.  

\subsection{Tests of the automated method}

Before presenting the results for the sample, we present the results of some tests on our automated method.  In particular, we compare our results where possible with the literature, and perform test calculations to assess the robustness of the automated method at low $S/N$.

\subsubsection{Comparison stars}
\label{sect:comparison}

Five stars, namely HD~20, HD~221170, CS~22186-025, CS~22892-052 and CS~31082-001, which have been studied in detail with better quality observational material by others, were included in our pilot sample.  Four of these were included in the sample for the specific purpose of use as comparison stars, while CS~22186-025 has been recently observed as part of the ESO First Stars programme (Cayrel~et~al.~\cite{cayrel04}). We now present a comparison of our results with some of the literature studies.  For CS~22186-025, CS~22892-052, and CS~31082-001, we compare with the studies of Cayrel~et~al., Sneden~et~al.~(\cite{sneden03}), and Hill~et~al.~(\cite{hill02}), respectively.  For HD 20 and HD 221170 there are a number of different studies which have included these stars.  For the purpose of clarity, for the abundances of these two stars we chose to concentrate our comparison on just one study, that of Burris~et~al.~(\cite{burris00}), which includes both these stars and has reasonable overlap with our work in terms of the elements analysed. 

First, in Table~\ref{tab:params_comparison} we compare our stellar parameters for these stars with those used in the literature.  Our temperatures are seen to be systematically warmer than those found in the literature for the comparison stars by of order 100~K.  As mentioned in Sect.~\ref{subsect:photometry}, and has been noted by other authors for individual stars, (e.g. Hill~et~al.~\cite{hill02}, Sneden et~al.~\cite{sneden03}), the reddening maps of Schlegel~et~al.~(\cite{schlegel98}) give larger reddenings than those of Burstein \& Heiles (\cite{burstein82}).  As shown in Fig.~\ref{fig:reddening} this leads to systematic differences in $T_\mathrm{eff}$, of around 100~K.  In Table~\ref{tab:params_comparison} we also quote our temperatures using the lower reddenings of Burstein \& Heiles.  These temperatures agree well with the literature within error (of order $\pm 100$~K).   The remaining parameters agree reasonably well, given the typical absolute error bars of $\sigma_{\log g} \sim 0.35$~dex, $\sigma_\mathrm{[Fe/H]}\sim 0.10$~dex, and $\sigma_\xi \sim 0.15$~km~s$^{-1}$.  Considering the effect of the systematic offset in $T_\mathrm{eff}$ we see, in line with expectations, that it usually leads to slightly higher $\log g$ values (+100~K $\rightarrow$ +0.2 dex), and slightly higher metallicities (+100~K $\rightarrow$ +0.1 dex).

\begin{table*}
\begin{center}
\caption{Comparison of our final stellar parameters with those adopted in the literature for the comparison stars.  $T_\mathrm{eff}$ derived with the reddening maps of Burstein \& Heiles (\cite{burstein82}) instead of Schlegel~et~al.~(\cite{schlegel98}) are also given for comparison.}
\label{tab:params_comparison}
\begin{tabular}{lccccccccl}
\hline
Star & \multicolumn{4}{c}{This Work} & \multicolumn{4}{c}{Literature} & Reference/note \\ 
     & $T_\mathrm{eff}$ & $\log g$ & [Fe/H] & $\xi$ &$T_\mathrm{eff}$ & $\log g$ & [Fe/H] & $\xi$ & \\
     & [K] & [cm s$^{-2}$] & & [km~s$^{-1}$] & [K] & [cm s$^{-2}$] & & [km~s$^{-1}$] & \\  
\hline
HD 20       & 5445 & 2.39 & $-$1.58 & 2.30 & 5351 & & &                 & Alonso et~al.~(\cite{alonso99}) \\
         &&&&                              & 5475 & 2.8 & $-$1.4 & 2.0  & Burris et~al.~(\cite{burris00})\\
         &&&&                              & 5375 & 2.41& $-$1.4 & 1.90 & Fulbright \& Johnson (\cite{fulbright03}) \\
            & 5416 & & & & & & & & Using Burstein \& Heiles (\cite{burstein82}) \\
HD 221170   & 4648 & 1.57 & $-$2.14 & 2.22 & 4410 & & &                 & Alonso et~al.~(\cite{alonso99}) \\
         &&&&                              & 4425 & 1.0 & $-$2.0 & 1.5  & Burris et~al.~(\cite{burris00})\\
         &&&&                              & 4500 & 0.9 & $-$2.1 & 2.75 & Fulbright (\cite{fulbright00}) \\
         &&&&                              & 4475 & 1.15& $-$2.2 & 2.40 & Fulbright \& Johnson (\cite{fulbright03}) \\
            & 4548 & & & & & & & & Using Burstein \& Heiles (\cite{burstein82}) \\
CS~22186-025 & 4985 & 1.70 & $-$2.87 & 2.14 & 4900 & 1.5 & $-$3.0 & 2.0  & Cayrel et~al.~(\cite{cayrel04}) \\
            & 4887 & & & & & & & & Using Burstein \& Heiles (\cite{burstein82}) \\
CS~22892-052 & 4884 & 1.81 & $-$2.95 & 1.67 & 4800 & 1.5 & $-$3.1 & 1.95 & Sneden et~al.~(\cite{sneden03}) \\
            & 4843 & & & & & & & & Using Burstein \& Heiles (\cite{burstein82}) \\
CS~31082-001 & 4922 & 1.90 & $-$2.78 & 1.88 & 4825 & 1.5 & $-$2.9 & 1.8  & Hill et~al.~(\cite{hill02}) \\
            & 4874 & & & & & & & & Using Burstein \& Heiles (\cite{burstein82}) \\
\hline
\end{tabular}
\end{center}
\end{table*}

\begin{figure*}
\begin{center}
\begin{tabular}{cc}
\resizebox{80mm}{!}{\rotatebox{0}{\includegraphics{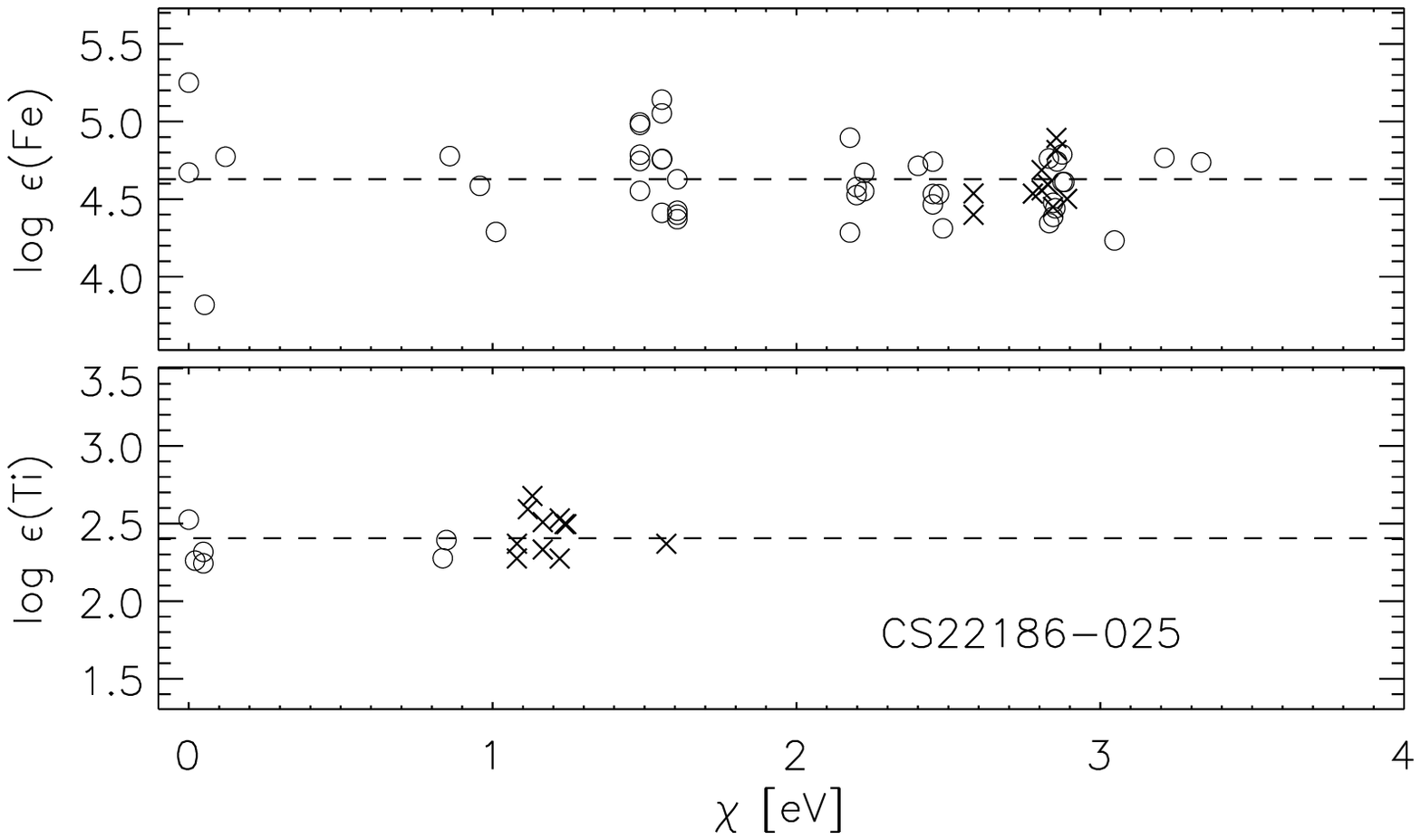}}} &
\resizebox{80mm}{!}{\rotatebox{0}{\includegraphics{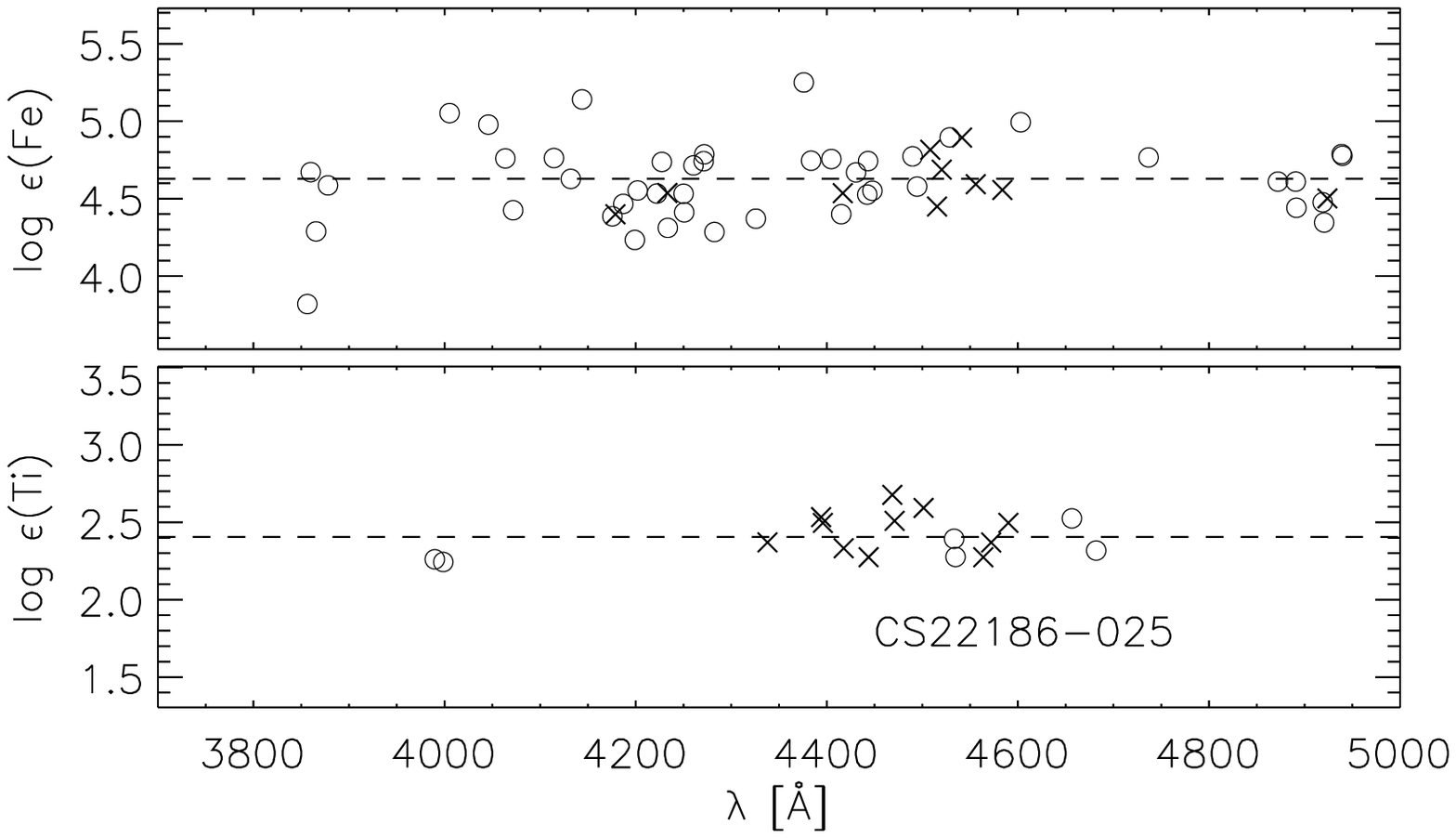}}} \\
\resizebox{80mm}{!}{\rotatebox{0}{\includegraphics{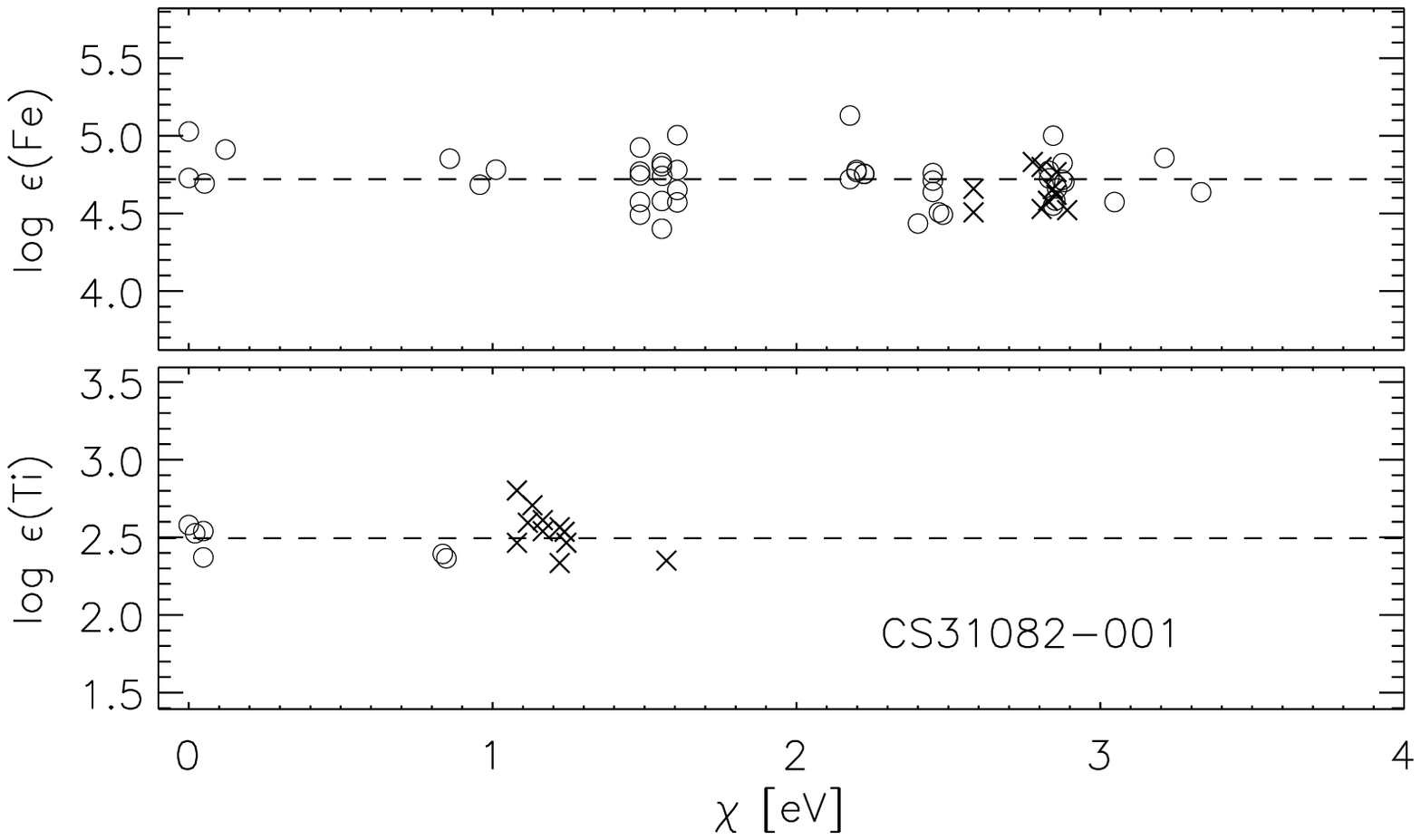}}} &
\resizebox{80mm}{!}{\rotatebox{0}{\includegraphics{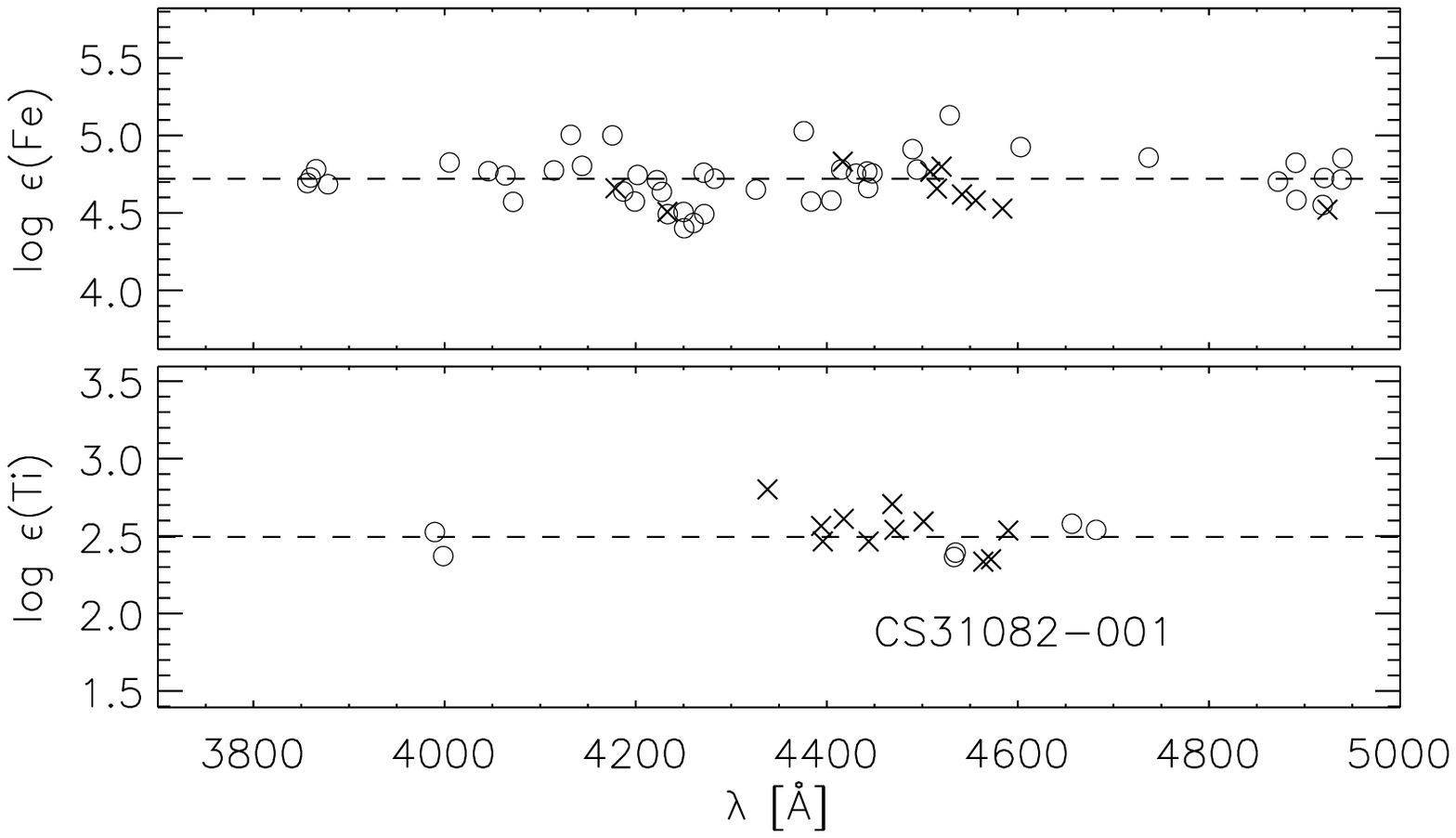}}} \\
\end{tabular}
\end{center}
\caption{Plots showing trends of abundances found from individual Fe and Ti lines with excitation potential $\chi$ (left column) and wavelength $\lambda$ (right column) for two comparison stars, CS~22186-025 $S/N\sim 32$ and CS~31082-001 $S/N\sim 90$.  Circles and crosses indicate lines from the neutral and singly ionised species respectively.  The dashed line in each case shows the determined abundance from the simultaneous best fit to all lines.}
\label{fig:equilibria}
\end{figure*}

To demonstrate the quality of the stellar parameter solutions, Fig.~\ref{fig:equilibria} plots abundances derived for individual Fe and Ti lines (the checking step of the procedure) against lower level excitation potential and wavelength of the line.  Lines of the neutral and singly ionised species are distinguished in the plots, and thus it is clearly seen that the derived $\log g$ values satisfy ionisation equilibrium for both Fe and Ti.  The plots against excitation potential, particularly for Fe, demonstrate that our spectra are not of sufficiently high quality to specify $T_\mathrm{eff}$ with the required precision as seen from the large scatter in the plot for CS~22186-025 which has $S/N\sim 32$.  The plots show our temperatures are reasonably consistent with excitation equilibrium (noting of course that one may question the validity of 1D LTE excitation equilibria); however, as discussed in Sect.~\ref{subsect:photometry} we find that on average excitation equilibria would give a cooler temperature scale.   Finally, the plots against line wavelength indicate no significant systematic effects in the bluer lines due to blending or continuum placement.  We also checked for significant trends with line strength and found none.  Similar results were obtained for all stars in the sample.

\begin{figure*}
\begin{center}
\begin{tabular}{cc}
\resizebox{80mm}{!}{\rotatebox{0}{\includegraphics{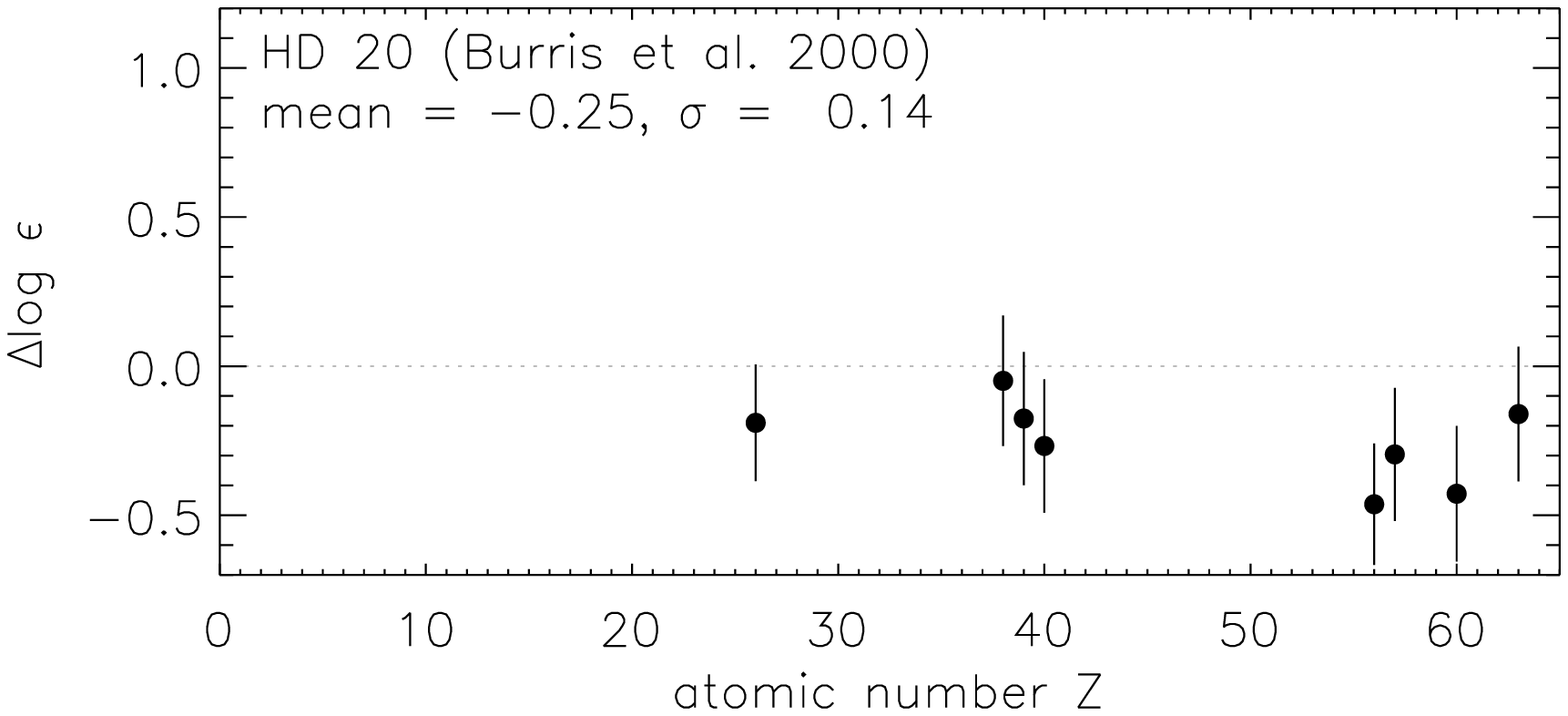}}} &
\resizebox{80mm}{!}{\rotatebox{0}{\includegraphics{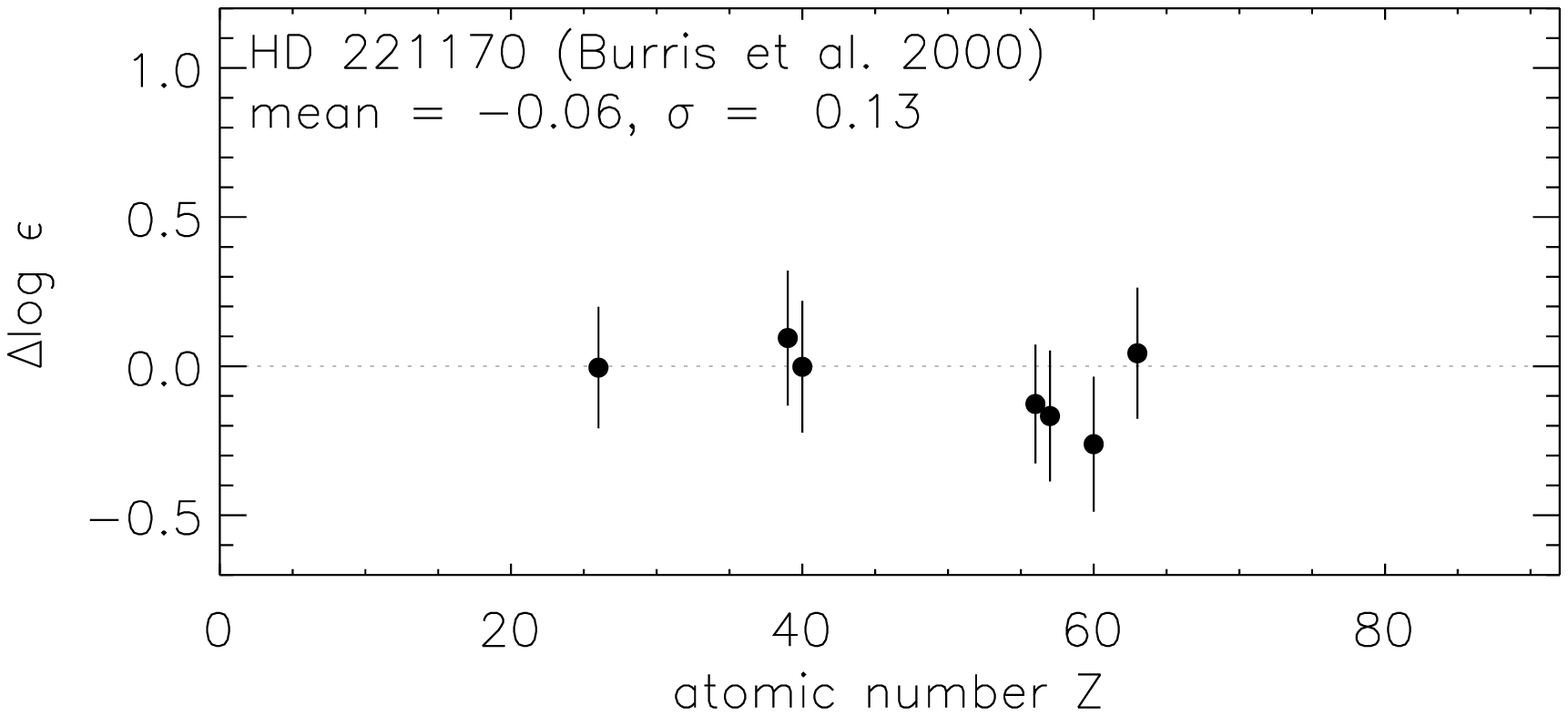}}}  \\
\resizebox{80mm}{!}{\rotatebox{0}{\includegraphics{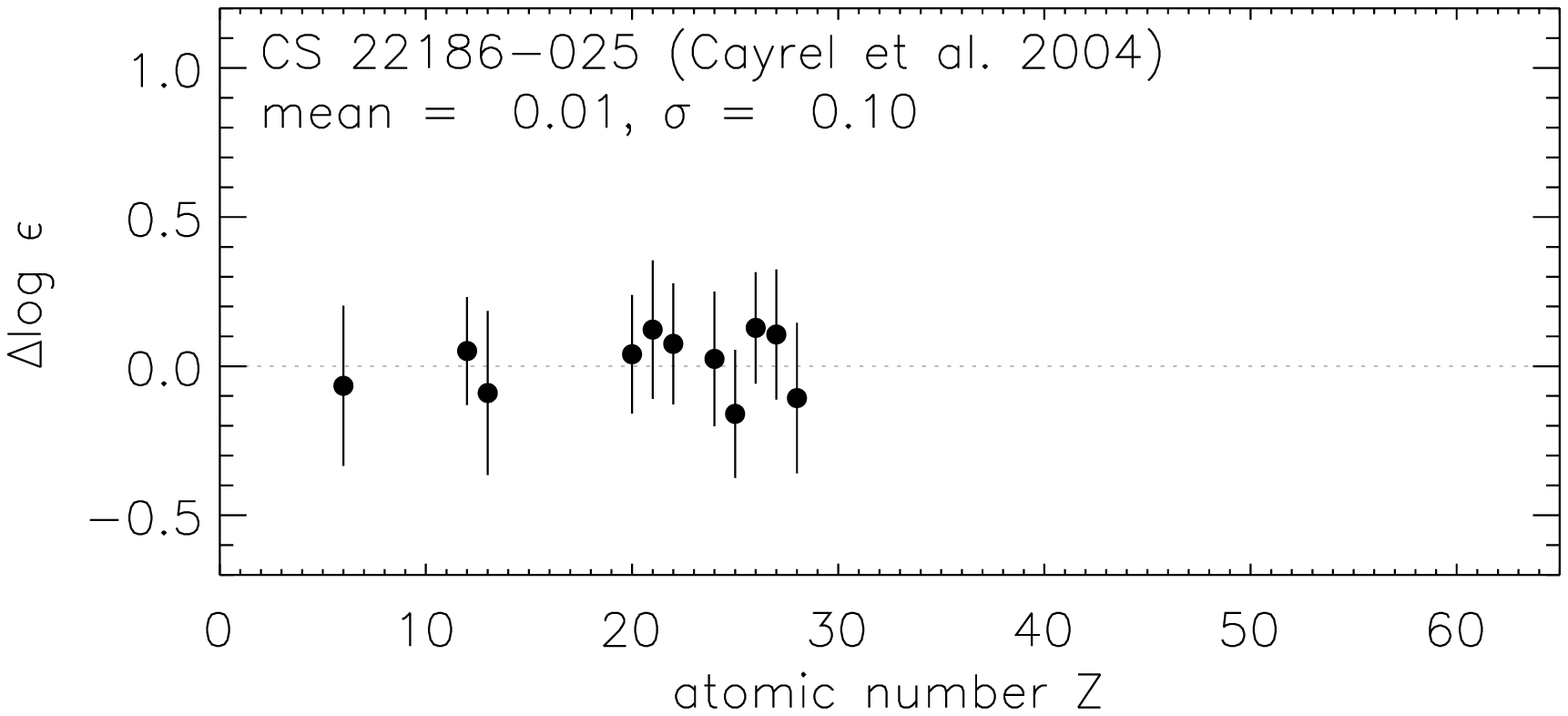}}} & 
\resizebox{80mm}{!}{\rotatebox{0}{\includegraphics{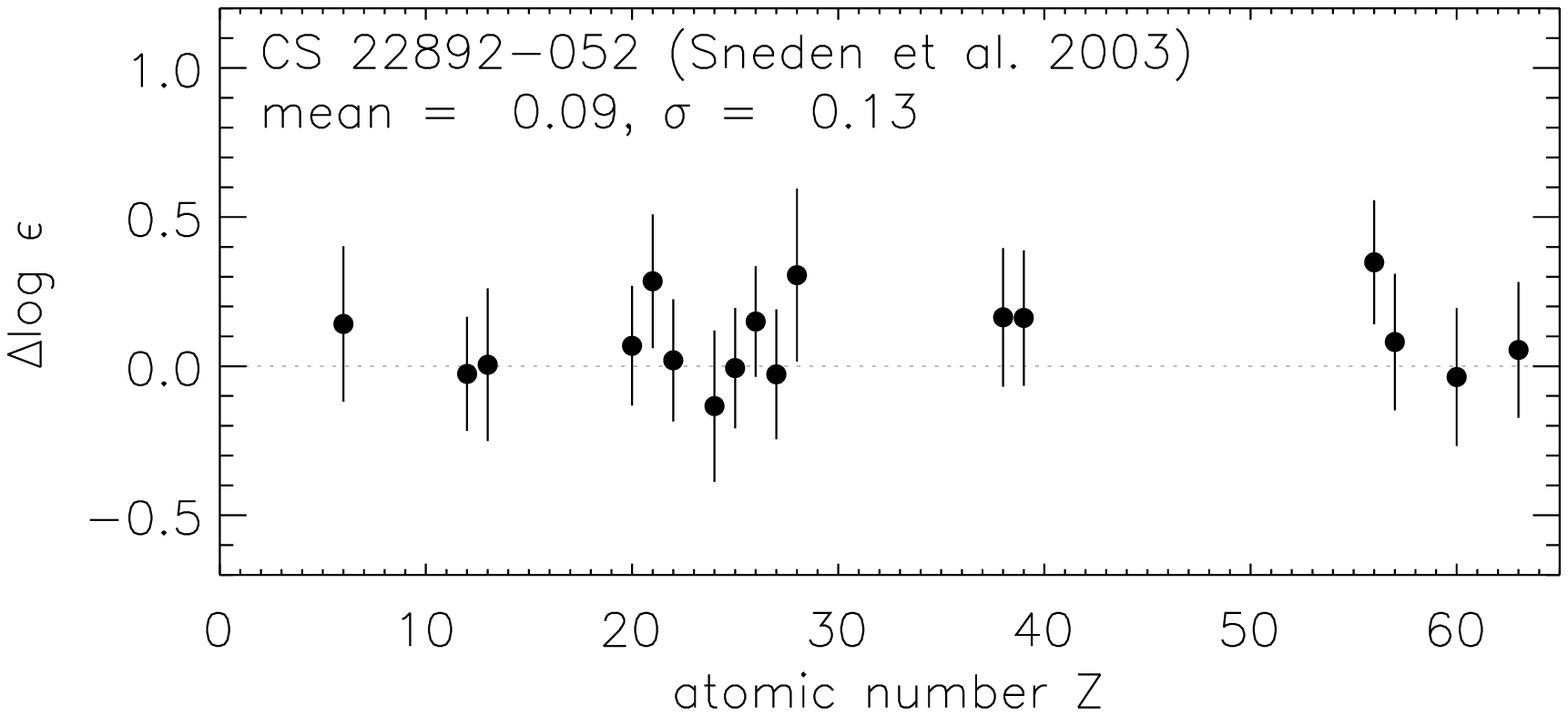}}} \\
\resizebox{80mm}{!}{\rotatebox{0}{\includegraphics{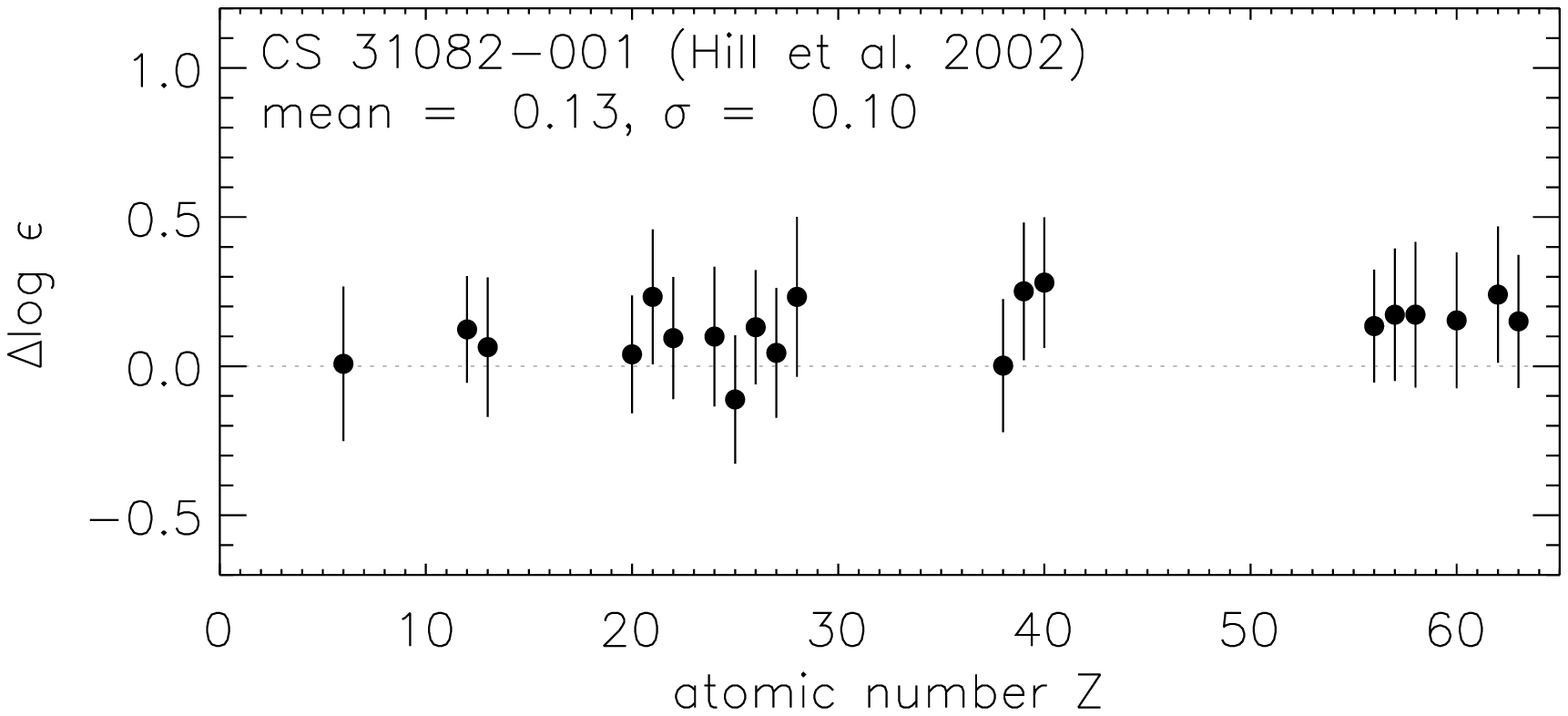}}} & \\
\end{tabular}
\end{center}
\caption{Comparison of abundances for stars previously studied in the literature.  Differences are in the sense of $\Delta \log \epsilon = \log \epsilon_\mathrm{this\; work} - \log \epsilon_\mathrm{literature}$.  Error bars indicate the estimated absolute error in our result.  The mean difference and standard deviation $\sigma$ are reported.
}
\label{fig:compare_other}
\end{figure*}

\begin{figure*}
\begin{center}
\begin{tabular}{cc}
\resizebox{80mm}{!}{\rotatebox{0}{\includegraphics{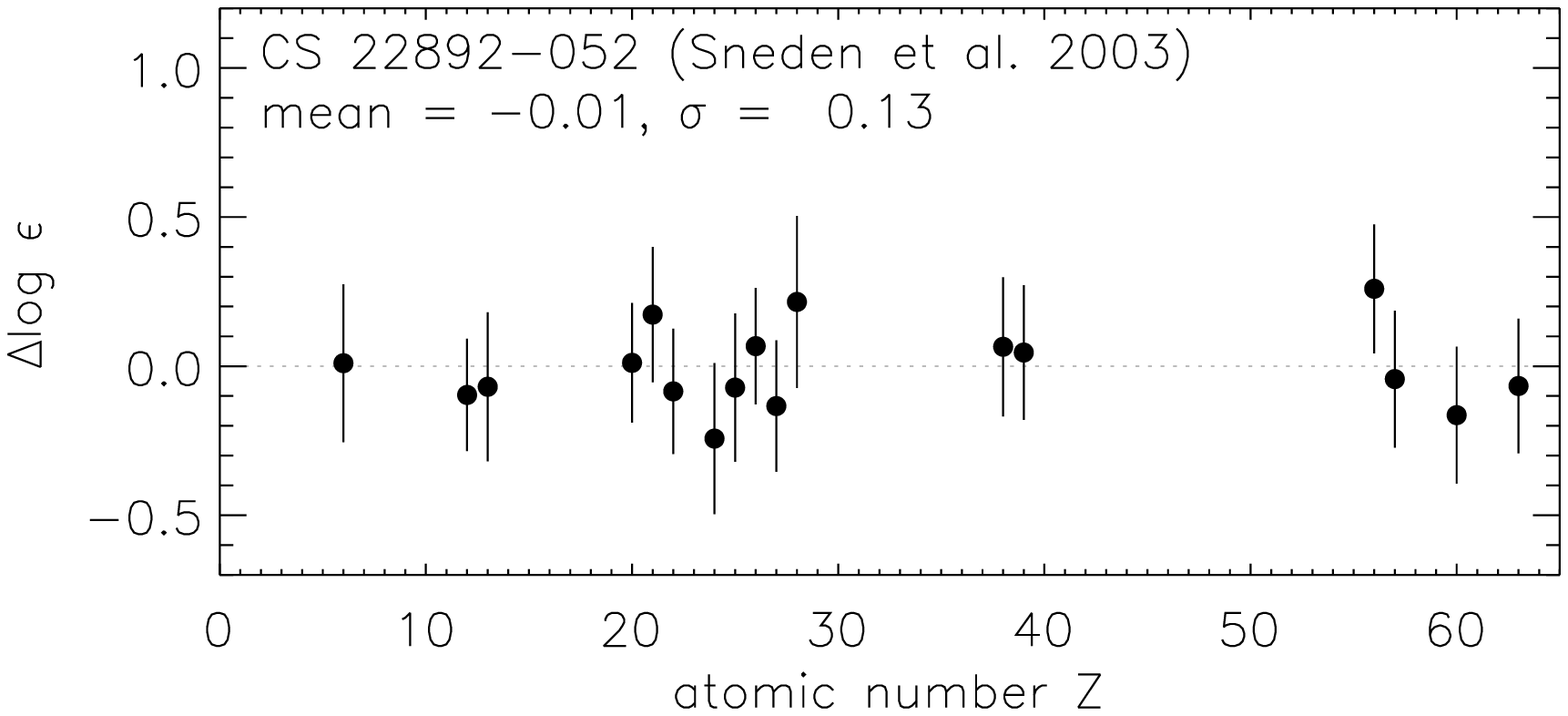}}} &
\resizebox{80mm}{!}{\rotatebox{0}{\includegraphics{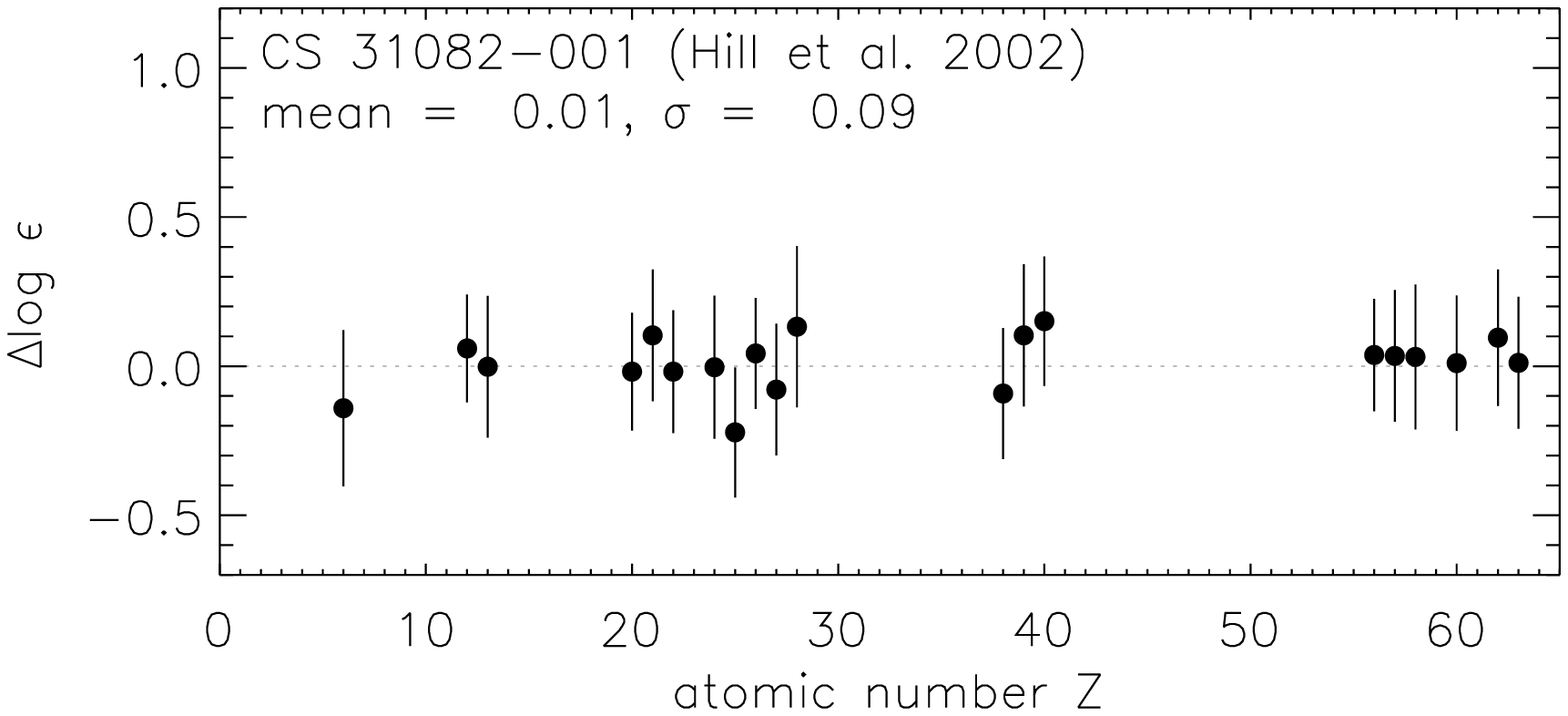}}} \\
\end{tabular}
\end{center}
\caption{Comparison of abundances as in Fig.~\ref{fig:compare_other}, where we have adopted the same temperature as the literature study.  
}
\label{fig:compare_other_wparams}
\end{figure*}

Figure~\ref{fig:compare_other} compares abundances for our comparison stars with results from the literature.  The results show generally reasonable agreement within error, thus supporting not only our method for obtaining abundances, but our error estimates as well.  For CS~22892-052 and CS~31082-001 there is a systematic offset which is a result of our slightly warmer temperatures.  This is clearly demonstrated by Fig.~\ref{fig:compare_other_wparams}, where we compare results for CS~22892-052 and CS~31082-001 where we have reanalysed the spectra adopting the temperatures used in the relevant literature work.  The remaining small differences are likely attributable to use of different lines and atomic data, particularly oscillator strengths and hyperfine structure.

\begin{figure}
\begin{center}
\resizebox{80mm}{!}{\rotatebox{0}{\includegraphics{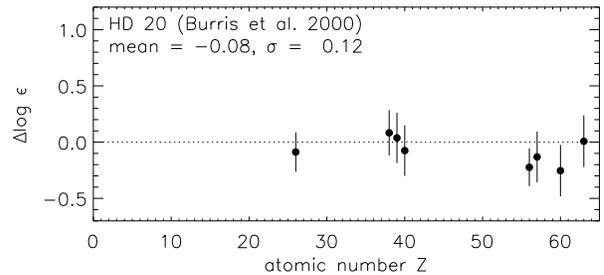}}}
\end{center}
\caption{Comparison of abundances as in Fig.~\ref{fig:compare_other}, for HD~20 where we have adopted the same temperature, gravity and microturbulence as Burris~et~al.~(\cite{burris00}).  
}
\label{fig:compare_other_wparamsHD20}
\end{figure}

The differences for HD~20 and HD~221170 are not improved by adopting temperatures used by Burris~et~al.~(\cite{burris00}), which originate from Pilachowski et~al.~(\cite{pilachowski96}).  For HD~20, our temperature is already in reasonable agreement with that of Burris~et~al., yet we obtain systematically lower abundances by of the order of 0.3~dex.  We reanalysed the spectrum adopting the stellar parameters of  Burris~et~al., and the new comparison shown in Fig.~\ref{fig:compare_other_wparamsHD20} shows a significant improvement. Thus, the main reasons for the differences are our significantly higher microturbulence and lower gravity.  We found no evidence of a significant trend of Fe abundance with line strength or ionisation stage to indicate our microturbulence or gravity solutions are in error.  In spite of the fact that our stellar parameters for HD~221170 differ significantly from those adopted in Burris et~al., the abundances are in reasonable agreement, probably due to compensation of the $T_\mathrm{eff}$, $\log g$ and microturbulence differences.  Note that this star has a large colour excess $E(B-V)\approx 0.12$ in our sample, and a large discrepancy with the Burstein \&\ Heiles maps (see Fig.~\ref{fig:reddening}).

While these systematic offsets, of order 0.1--0.2~dex, are present and attributed to differences in stellar parameters, perhaps the most important result is the {\it scatter} in the difference, as this indicates the accuracy with which the abundance pattern in a given star is reproduced.  We find for all comparisons scatter of 0.1 to 0.15~dex.  Thus we conclude that the relative abundance patterns are accurate to of order 0.15 dex, within our quoted error bars.

\subsubsection{Robustness at low $S/N$}
\label{subsect:snr}

Since most of our comparison stars are  relatively bright, the spectra are usually of higher $S/N$ than is typical for this project.  The lowest quality comparison star spectrum is that of CS~22186-025, which has on average $S/N\approx 32$ per pixel.  As some of our spectra are of even lower quality, it is important to ascertain what effect $S/N$ has on our results.  It should be pointed out that, provided our detection classification method and error estimates are reasonable, the pertinent issues are: (a) what is the minimum precision in abundances that is useful, and (b) what $S/N$ is necessary to achieve it.  

 To investigate this we took the spectrum of CS~31082-001, which has $S/N \approx 90$, and degraded it to lower $S/N$ values by multiplying with appropriate Poisson noise and rescaling so that count numbers are consistent with the noise level.  These spectra were then run through the entire automated spectrum analysis process completely independently from the original spectrum.    The differences in derived atmosphere parameters (noting the model [Fe/H] is determined by the Fe abundance) and abundances for four degraded spectra compared to the original spectrum are given in Table~\ref{tab:sn_comparison}, along with the relative error bar for that derived parameter.  First, we note that almost every difference reported was less than our computed error bar for the degraded spectrum, giving confidence in our estimated errors.  Secondly, the differences are generally reasonable for most elements, less than 0.1~dex.  However, it is immediately obvious that for most elements there is a systematic trend towards slight underestimation of the abundances.  We attribute this systematic trend to systematically lower continuum placement, which is to be expected, since at low $S/N$ it is impossible to distinguish weak lines from continuum.  Therefore, there should be a natural trend towards the continuum being placed too low, and thus underestimation of abundances, as the weak lines make the continuum appear slightly lower.  This problem could perhaps be circumvented if one {\it a priori} defined continuum regions from higher quality spectra to be used for normalising the spectra; however such a procedure would be dependent on the similarity of the spectra.  As noted, the error in the abundance at this $S/N$ is much larger in any case. 
 
Based on these results, we conclude that our method seems to be reasonably robust even at $S/N$ as low as 15 for elements with a large number of lines, such as Fe and Ti.  However, as one expects, detections for a number of elements cannot be made at such low $S/N$ for stars of this metallicity.  Elements with a small number of lines which are still detectable at low $S/N$ at this metallicity, for example Mg, Ca and Ni, are less robust, but at $S/N>15$ the induced error usually does not appear to exceed of order 0.1 dex.  

\begin{table*}
\begin{center}
\caption{Comparison of results for CS~31082-001 using spectra with degraded $S/N$.  The difference from our results using the original spectrum $S/N\approx 90$ in Table~\ref{tab:abunds} is quoted, with the relative error in the derived abundance for the degraded $S/N$ in parentheses. A blank means a non-detection (below 3$\sigma$) at this $S/N$ for this star. }
\label{tab:sn_comparison}
\scriptsize
\begin{tabular}{lrrrr}
\hline
Parameter  & \multicolumn{4}{c}{$\Delta$ Parameter} \\
           & $S/N = 30$ & 20 & 15 & 10 \\
\hline
$\log g$                      & $+0.01$ (0.26)& $+0.05$ (0.27)& $-0.08$ (0.26)& $ 0.00$ (0.34) \\
$\xi$                         & $-0.04$ (0.18)& $-0.07$ (0.18)& $-0.08$ (0.18)& $-0.16$ (0.20) \\
$\log\epsilon(\mathrm{C}) $   & $-0.02$ (0.20)& $-0.10$ (0.20)& $-0.07$ (0.20)& $-0.19$ (0.22) \\
$\log\epsilon(\mathrm{Mg})$   & $-0.04$ (0.12)& $-0.06$ (0.12)& $-0.02$ (0.12)& $-0.02$ (0.13) \\
$\log\epsilon(\mathrm{Al})$   & $-0.19$ (0.19)& $-0.14$ (0.30)& ---           & ---            \\
$\log\epsilon(\mathrm{Ca})$   & $-0.02$ (0.12)& $-0.04$ (0.12)& $-0.06$ (0.11)& $-0.06$ (0.14) \\
$\log\epsilon(\mathrm{Sc})$   & $-0.02$ (0.17)& $ 0.00$ (0.18)& $-0.09$ (0.19)& ---            \\
$\log\epsilon(\mathrm{Ti})$   & $-0.03$ (0.16)& $-0.02$ (0.16)& $-0.04$ (0.16)& ---            \\
$\log\epsilon(\mathrm{V}) $   & ---           & ---           & ---           & ---            \\
$\log\epsilon(\mathrm{Cr})$   & $-0.06$ (0.17)& $-0.09$ (0.19)& ---           & ---            \\
$\log\epsilon(\mathrm{Mn})$   & $-0.07$ (0.16)& $-0.11$ (0.18)& $-0.03$ (0.19)& ---            \\
$\log\epsilon(\mathrm{Fe})$   & $-0.02$ (0.13)& $ 0.00$ (0.12)& $-0.02$ (0.13)& $ 0.00$ (0.15) \\
$\log\epsilon(\mathrm{Co})$   & $-0.05$ (0.16)& ---           & ---           & ---            \\
$\log\epsilon(\mathrm{Ni})$   & $-0.08$ (0.20)& $-0.09$ (0.23)& $-0.21$ (0.25)& ---            \\
$\log\epsilon(\mathrm{Zn})$   & ---           & ---           & ---           & ---            \\
$\log\epsilon(\mathrm{Sr})$   & $+0.01$ (0.15)& $+0.01$ (0.17)& $ 0.00$ (0.19)& $-0.07$ (0.23) \\
$\log\epsilon(\mathrm{Y}) $   & $-0.04$ (0.18)& $-0.03$ (0.19)& ---           & ---            \\
$\log\epsilon(\mathrm{Zr})$   & $-0.04$ (0.18)& ---           & ---           & ---            \\
$\log\epsilon(\mathrm{Ba})$   & $-0.04$ (0.15)& $-0.03$ (0.16)& $+0.05$ (0.16)& $+0.01$ (0.21) \\
$\log\epsilon(\mathrm{La})$   & $-0.02$ (0.18)& ---           & ---           & ---            \\
$\log\epsilon(\mathrm{Ce})$   & ---           & ---           & ---           & ---            \\
$\log\epsilon(\mathrm{Nd})$   & $-0.02$ (0.19)& ---           & ---           & ---            \\
$\log\epsilon(\mathrm{Sm})$   & ---           & ---           & ---           & ---            \\
$\log\epsilon(\mathrm{Eu})$   & $-0.01$ (0.18)& $-0.01$ (0.19)& $-0.15$ (0.19)& ---            \\
\hline
\end{tabular}
\end{center}
\end{table*}

\subsubsection{Example fits to spectra}

\begin{figure*}
\begin{center}
\begin{tabular}{cc}
\resizebox{8.5cm}{!}{\rotatebox{0}{\includegraphics{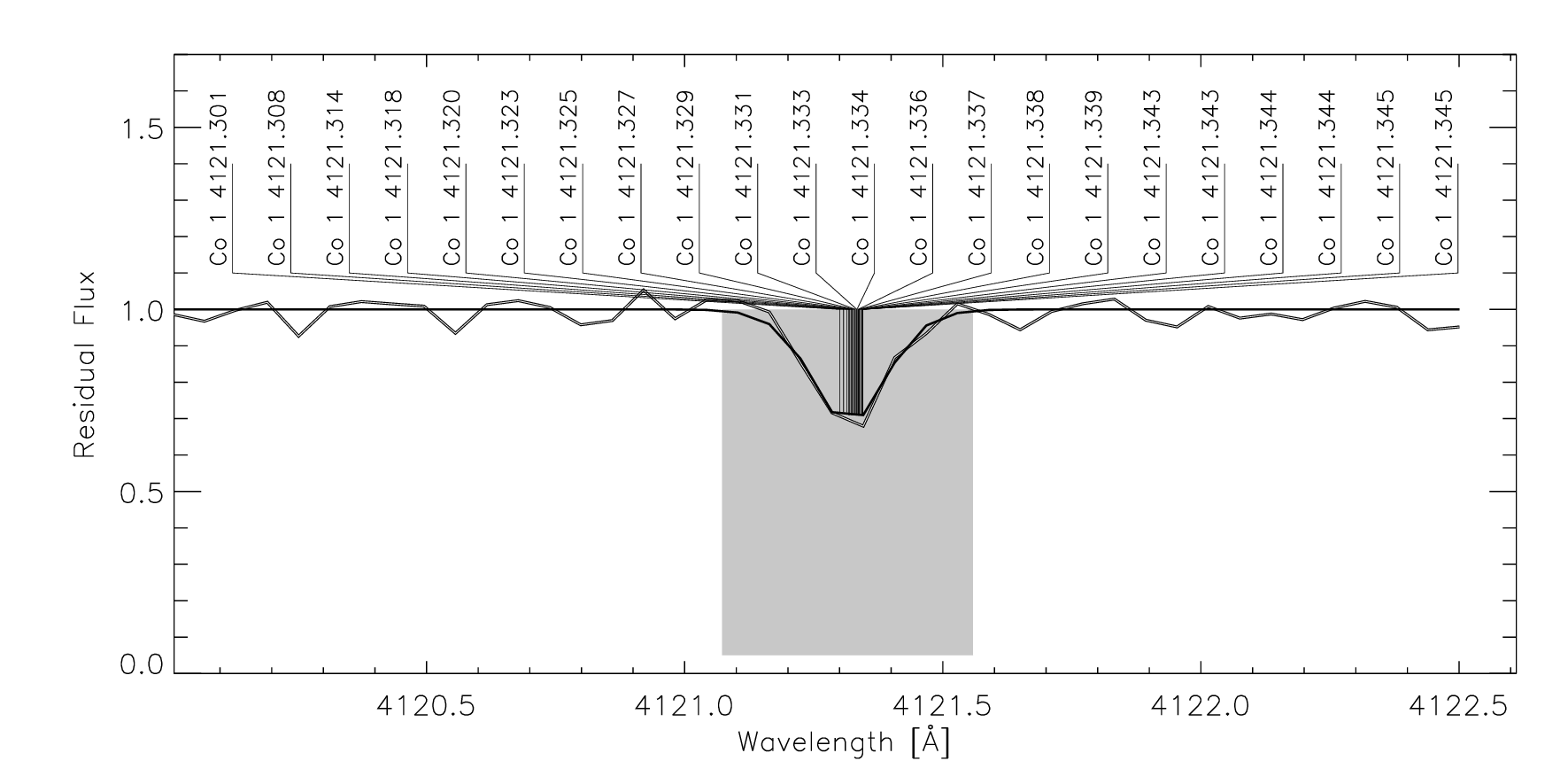}}} &
\resizebox{8.5cm}{!}{\rotatebox{0}{\includegraphics{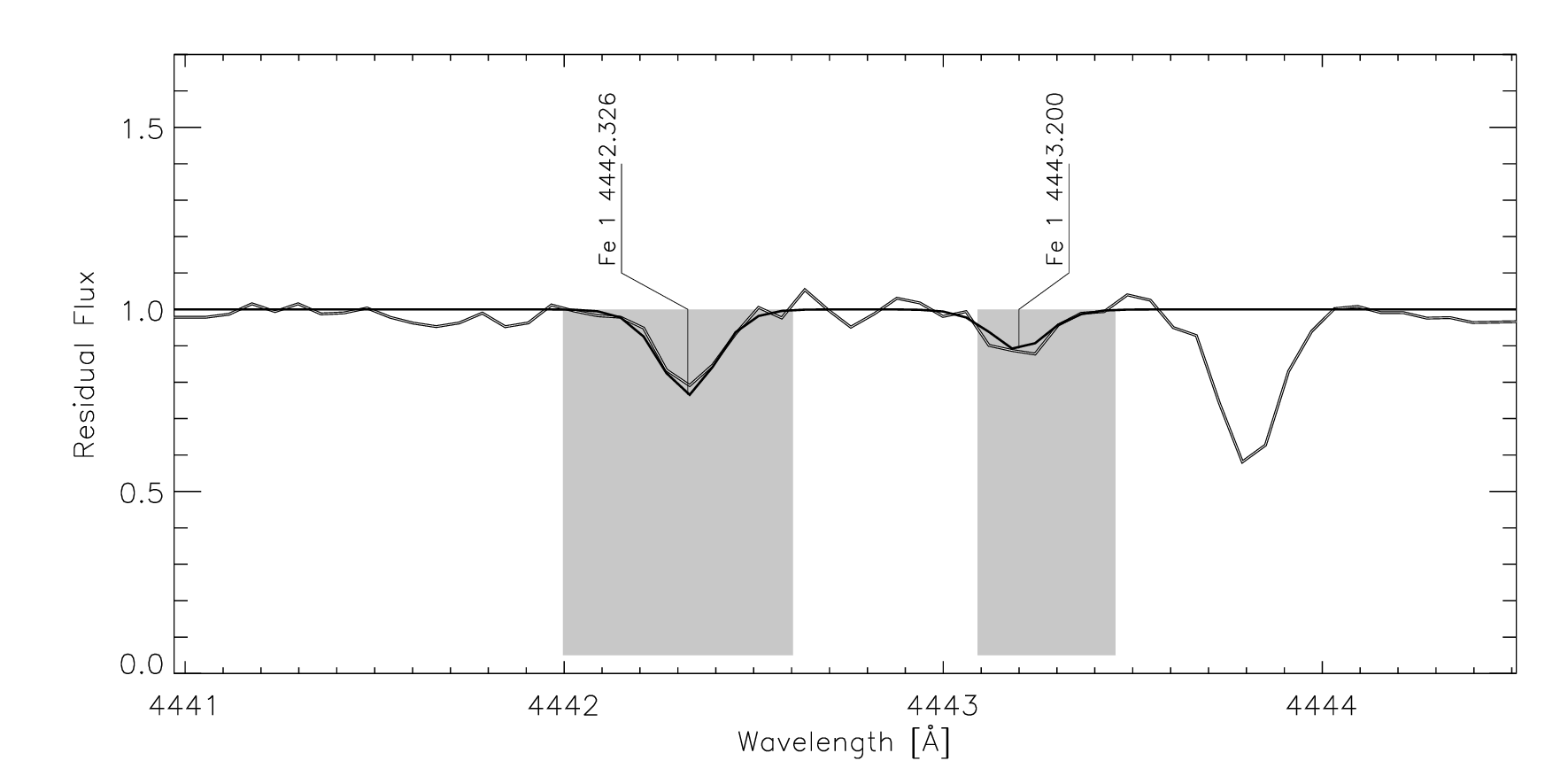}}} \\
\multicolumn{2}{c}{\resizebox{\hsize}{!}{\rotatebox{0}{\includegraphics{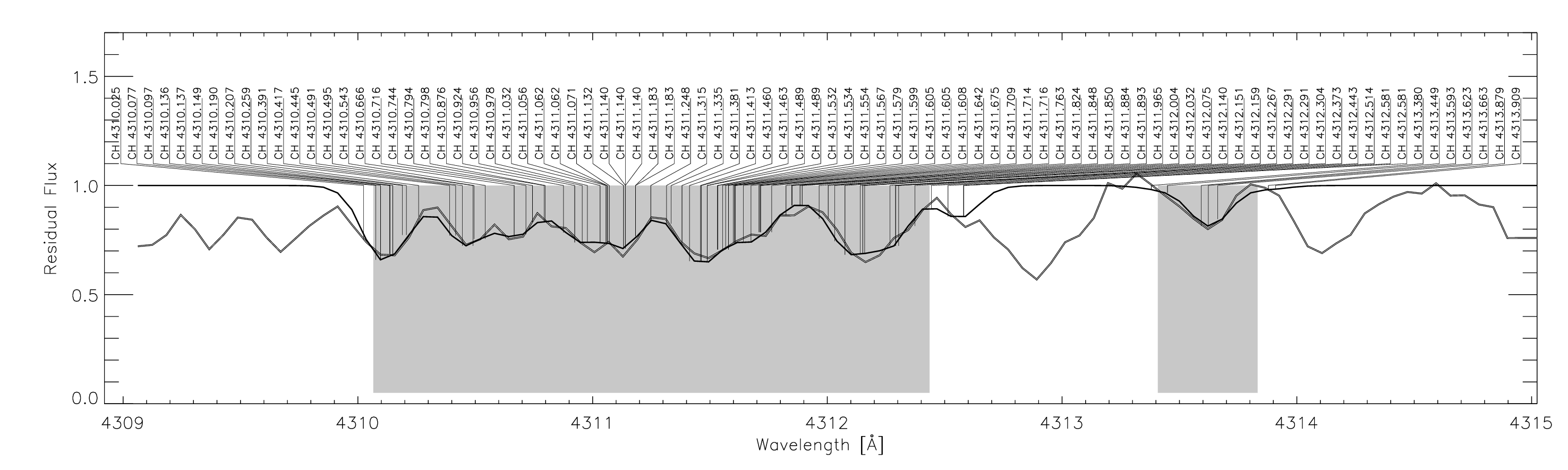}}}}
\end{tabular}
\end{center}
\caption{Examples of the spectra and fits for HE~1127-1143 (an r-process rich star), which has average $S/N\sim49$, $T_\mathrm{eff}=5224$~K, $\log g\sim2.6$, $\mathrm{[Fe/H]}\sim-2.7$, and is thus reasonably typical for the sample.  The double line is the observed spectrum and the single line the fitted model spectrum (noting that it is not only the plotted observations which are being fit, but all considered lines of the element).  The shaded regions are the comparison windows used for fitting.}
\label{fig:example_spectra}
\end{figure*}

Examples of portions of the spectra, and fits to the data obtained, are shown in Fig.~\ref{fig:example_spectra} for an r-process enhanced star HE~1127-1143.  The example was chosen to be representative of the sample; it is a giant star with $S/N\sim49$, typical for the sample.  

\section{Results}
\label{sect:results}

The derived stellar parameters and elemental abundances for the sample are presented in Table~\ref{tab:abunds}.\footnote{Table~\ref{tab:abunds} in its entirety is available only electronically}  Below, we note and discuss objects of particular interest found in the sample.   In the next section we will discuss the more general behaviour of the abundances and implications for understanding the chemical evolution of the Galactic halo. 

It is worth emphasising that some care must be taken in interpreting the results presented here involving elements which are not always, or even usually, detected in our spectra, which may lead to selection effects.  Fourteen elements, C, Mg, Al, Ca, Sc, Ti, Cr, Mn, Fe, Co, Ni, Sr, Y and Ba, are almost always detectable at the 3$\sigma$ level in the spectra.  Another eight elements are analysed, V, Zn, Zr, La, Ce, Nd, Sm and Eu, which can usually only be detected in the spectra of the least metal-poor stars of our sample, spectra with higher than usual $S/N$, if the abundance is enhanced, or a combination of these factors.  Thus, particular care must be taken in interpreting results involving these eight elements as the data are incomplete.
      
\subsection{Objects of Interest}
\label{sect:new_objects}

The complete sample consists of 253 stars, of which four were comparison stars already known to be r-process enhanced metal-poor stars.  Among the remainder we have identified a number of interesting objects, which we will now summarise.  For the discussion in this subsection (\S~\ref{sect:new_objects}) we consider only the 249 other stars as ``the sample''.  Though CS~22186-025 has been recently observed by Cayrel et~al.~(\cite{cayrel04}) it is included in our statistics as it had not been observed at the time of selection.

First, this work adds significantly to the number of confirmed very metal-poor stars.  The [Fe/H] distribution for the sample is shown in Fig.~\ref{fig:fe_dist}.  The sample contains 49 stars with $\mathrm{[Fe/H]}<-3$, and 181 stars with $-3<\mathrm{[Fe/H]}<-2$.   Only 12 stars are currently known with $\mathrm{[Fe/H]} < -3.5$ (Beers \& Christlieb~\cite{beers05}) and we find just one new star in this regime, HE~1300+0157 with $\mathrm{[Fe/H]}=-3.76\pm0.18$.  This star is at the base of the giant branch ($T_\mathrm{eff}\sim 5400$~K, $\log g \sim 3.4$) and is carbon enhanced with $\mathrm{[C/Fe]} \sim 1.2$. 

\begin{figure}
\begin{center}
\resizebox{\hsize}{!}{\rotatebox{0}{\includegraphics{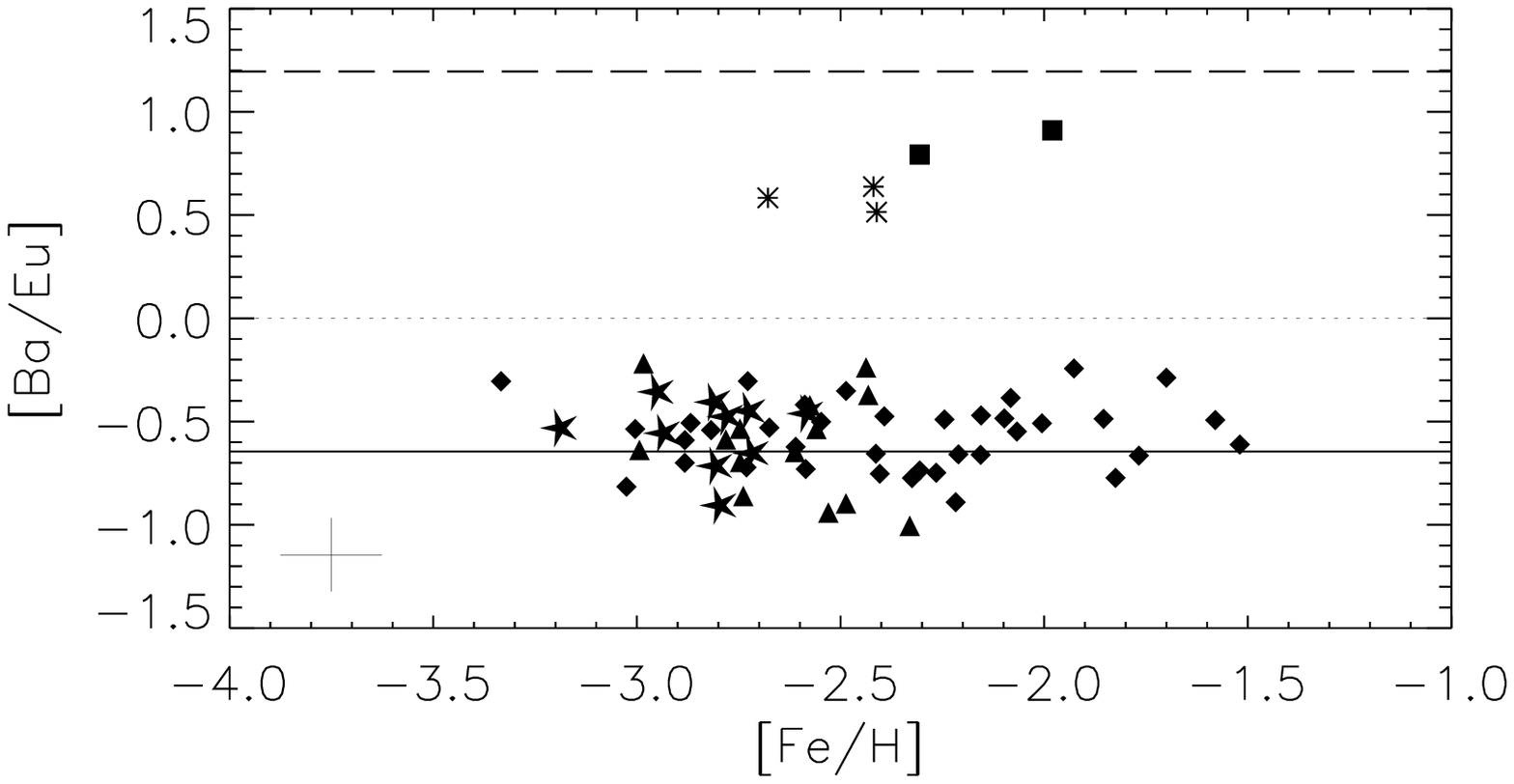}}}
\resizebox{\hsize}{!}{\rotatebox{0}{\includegraphics{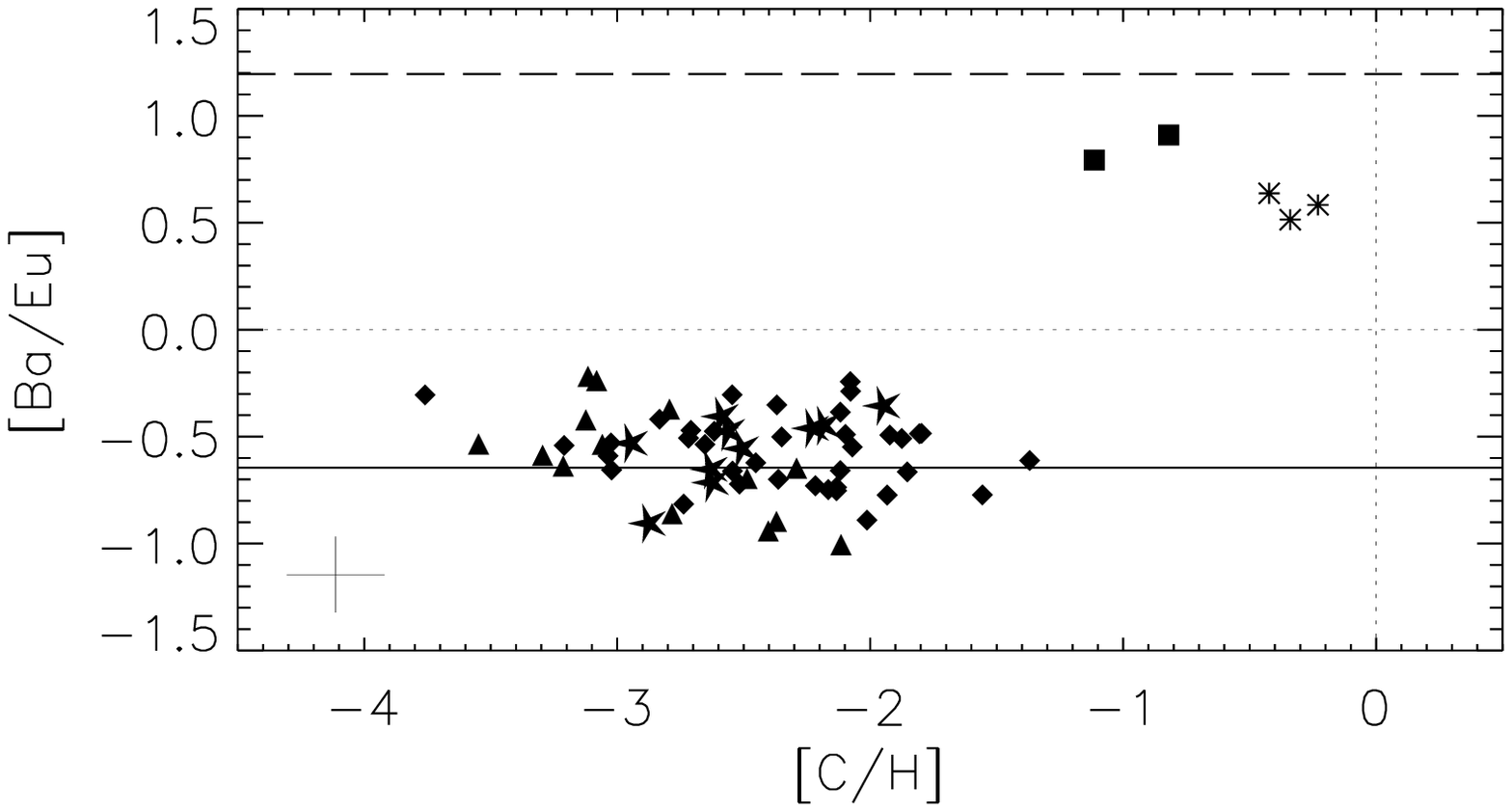}}}
\end{center}
\caption{A plot of [Ba/Eu] vs [Fe/H] (upper panel) and [C/H] (lower panel).  The horizontal full line shows the pure solar r-process value, computed from the solar r-process fractions of Arlandini et~al.~(\cite{arlandini99}), and the dashed line the solar s-process value.   The pure r-II stars are shown as stars, the r-I stars as diamonds, the pure r-process stars without excess r-process elements as triangles, the s-II stars as asterisks, and the two remaining s-process-rich stars as squares.  The average relative error bar is shown in the bottom left.
}
\label{fig:Ba_Eu}
\end{figure}

\begin{figure}
\begin{center}
\resizebox{\hsize}{!}{\rotatebox{0}{\includegraphics{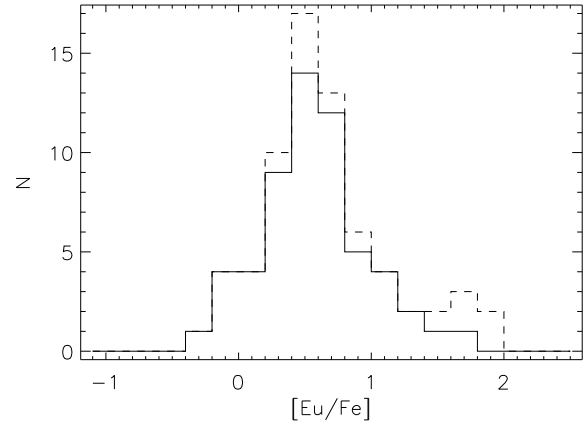}}}
\end{center}
\caption{Histogram of [Eu/Fe] values.  The dashed line includes all stars where Eu was detected, including the four comparison stars.  The full line includes the pure r-process stars only (i.e.\ $\mathrm{[Ba/Eu]}<0$ noting Ba must be detected; for a small number of stars where Eu was detected, Ba was not detected due to cosmic ray pollution of the Ba II resonance line) and excludes the comparison stars. }
\label{fig:Eu_Fe_histogram}
\end{figure}

The main goal of the HERES survey is to identify r-process enhanced stars, to address questions about the r-process.  We were able to detect both Eu and Ba at the $3\sigma$ level in 62 stars in the sample.  Of these, 57 are judged to be ``pure'' r-process stars, as they have $\mathrm{[Ba/Eu]}<0$, see Fig.~\ref{fig:Ba_Eu}.   In Paper~I we made the distinction between r-I and r-II stars, $0.3 \leq \mathrm{[Eu/Fe]} \leq 1.0$ and $\mathrm{[Eu/Fe]} > 1.0$ respectively (and $\mathrm{[Ba/Eu]}<0$ in both cases), on the basis that previous work had suggested a bimodal distribution of [Eu/Fe] with a lack of stars found in the range between 1.0 and 1.5.  A histogram of {\em detected} [Eu/Fe], is plotted in Fig.~\ref{fig:Eu_Fe_histogram}; however, due to the significant incompleteness of the Eu abundances and the detection bias towards high [Eu/Fe], only the high [Eu/Fe] side of this distribution is reliable, and thus this plot gives little information about the cosmic distribution.  We will return to the question of the distribution of r-process enhancement in Sect.~\ref{subsect:heavyneutron}.  Following this classification system, of these 57 stars, eight are r-II stars ($\mathrm{[Eu/Fe]} > 1.0$) which are listed in Table~\ref{tab:interesting} (two of which were announced in Paper~I), and 35 are r-I stars ($0.3 \leq \mathrm{[Eu/Fe]} \leq 1.0$), while 14 do not have r-process enhancement ($\mathrm{[Eu/Fe]} < 0.3$).  This implies a frequency of around 3\% and $>14$\% for r-II and r-I stars respectively among metal-poor stars ($\mathrm{[Fe/H]}<-1.5$).  It must be borne in mind that these frequencies are dependent on detection in our spectra, and thus are certainly lower limits, particularly for r-I stars.  For example, at the lowest metallicities, an r-I star would often not be detectable with our typical spectra.

\begin{table*}
\tabcolsep 1.5mm
\begin{center}
\caption{New r-II and s-II objects.  Quoted error estimates are relative errors.  N3 is defined in the caption to table~\ref{tab:abunds}.}
\label{tab:interesting}
\begin{tabular}{crrrcrcrcrcr}
\hline
star    &    $T_\mathrm{eff}$ & $\log g$ &  [Fe/H] & N3  & [C/Fe] &  N3  & [Ba/Fe]  & N3 &  [Eu/Fe]  & N3 & [Ba/Eu] \\
        &    [K]  & [cm s$^{-2}$]&         &    &        &     &          &   &           &   &         \\
\hline
&&&&&&&&&&&\\
\multicolumn{12}{c}{\underline{r-II stars}} \\
&&&&&&&&&&&\\
CS~29491-069 & 5103 &  2.45 & $	 -2.81\pm  0.13$ & 46 & $   0.18 \pm 0.17$ &  1 & $   0.34 \pm 0.19$ &	1  & $	1.06\pm	 0.15$ &  2 & $	 -0.71\pm  0.17$ \\
CS~29497-004 & 5013 &  2.23 & $	 -2.81\pm  0.13$ & 42 & $   0.22 \pm 0.18$ &  1 & $   1.21 \pm 0.15$ &	1  & $	1.62\pm	 0.15$ &  4 & $	 -0.41\pm  0.17$ \\
HE~0430-4901 & 5296 &  3.12 & $	 -2.72\pm  0.12$ & 46 & $   0.09 \pm 0.18$ &  1 & $   0.50 \pm 0.20$ &	1  & $	1.16\pm	 0.17$ &  3 & $	 -0.65\pm  0.17$ \\
HE~0432-0923 & 5131 &  2.64 & $	 -3.19\pm  0.13$ & 41 & $   0.24 \pm 0.17$ &  1 & $   0.72 \pm 0.18$ &	1  & $	1.25\pm	 0.15$ &  2 & $	 -0.53\pm  0.16$ \\
HE~1127-1143 & 5224 &  2.64 & $	 -2.73\pm  0.14$ & 46 & $   0.54 \pm 0.17$ &  1 & $   0.63 \pm 0.19$ &	1  & $	1.08\pm	 0.15$ &  3 & $	 -0.45\pm  0.18$ \\
HE~1219-0312 & 5140 &  2.40 & $	 -2.81\pm  0.12$ & 44 & $  -0.08 \pm 0.19$ &  1 & $   0.51 \pm 0.25$ &	1  & $	1.41\pm	 0.17$ &  3 & $	 -0.91\pm  0.21$ \\
HE~2224+0143 & 5198 &  2.66 & $	 -2.58\pm  0.12$ & 54 & $   0.35 \pm 0.17$ &  1 & $   0.59 \pm 0.18$ &	1  & $	1.05\pm	 0.15$ &  4 & $	 -0.46\pm  0.16$ \\
HE~2327-5642 & 5048 &  2.22 & $	 -2.95\pm  0.12$ & 50 & $   0.43 \pm 0.19$ &  1 & $   0.66 \pm 0.19$ &	1  & $	1.22\pm	 0.17$ &  4 & $	 -0.56\pm  0.17$ \\
&&&&&&&&&&&\\
\multicolumn{12}{c}{\underline{s-II stars}} \\
&&&&&&&&&&&\\
HE~0131-3953 & 5928 &  3.83 & $	 -2.71\pm  0.11$ & 22 & $   2.45 \pm 0.18$ &  1 & $   2.20 \pm 0.16$ &	2  & $	1.62\pm	 0.19$ &  2 & $	  0.58\pm  0.18$ \\
HE~0338-3945 & 6162 &  4.09 & $	 -2.41\pm  0.13$ & 23 & $   2.07 \pm 0.16$ &  1 & $   2.41 \pm 0.15$ &	2  & $	1.89\pm	 0.16$ &  2 & $	  0.51\pm  0.16$ \\
HE~1105+0027 & 6132 &  3.45 & $	 -2.42\pm  0.13$ & 34 & $   2.00 \pm 0.17$ &  1 & $   2.45 \pm 0.16$ &	2  & $	1.81\pm	 0.16$ &  4 & $	  0.64\pm  0.19$ \\
\hline
\end{tabular}
\end{center}
\end{table*}

\begin{figure*}
\tabcolsep 0mm
\begin{center}
\begin{tabular}{cccc}
\resizebox{58mm}{!}{\rotatebox{0}{\includegraphics{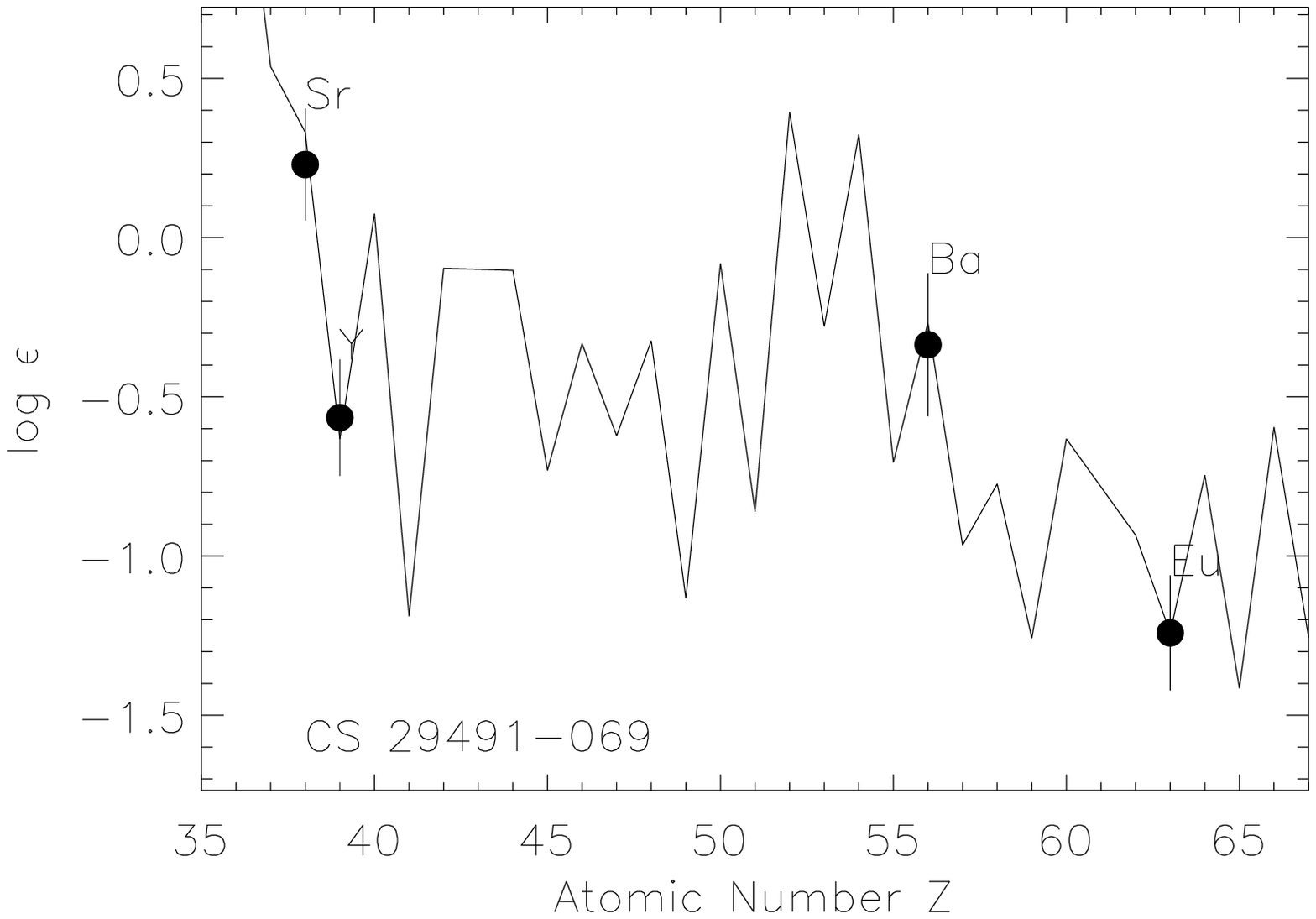}}} &
\resizebox{58mm}{!}{\rotatebox{0}{\includegraphics{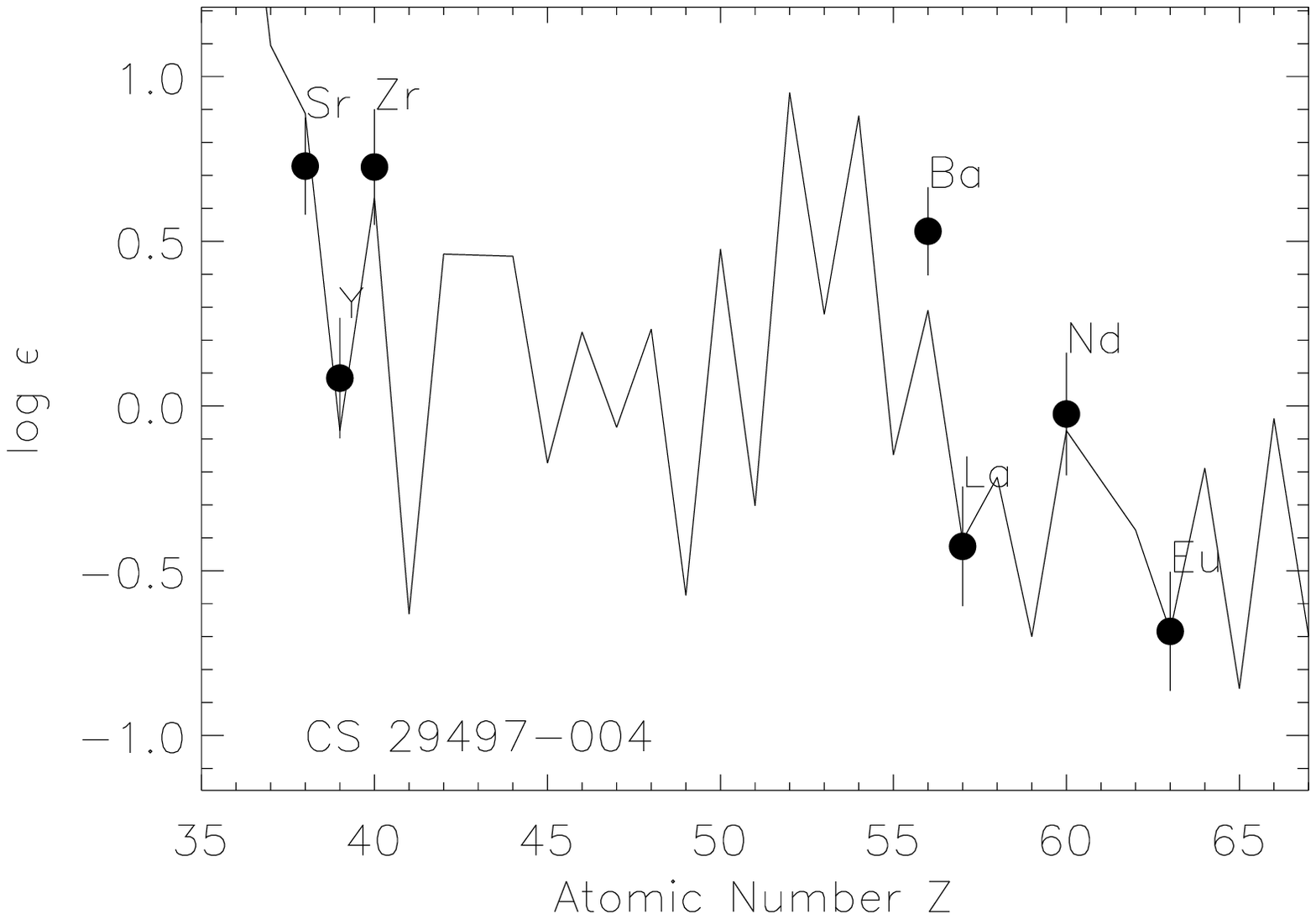}}} &
\resizebox{58mm}{!}{\rotatebox{0}{\includegraphics{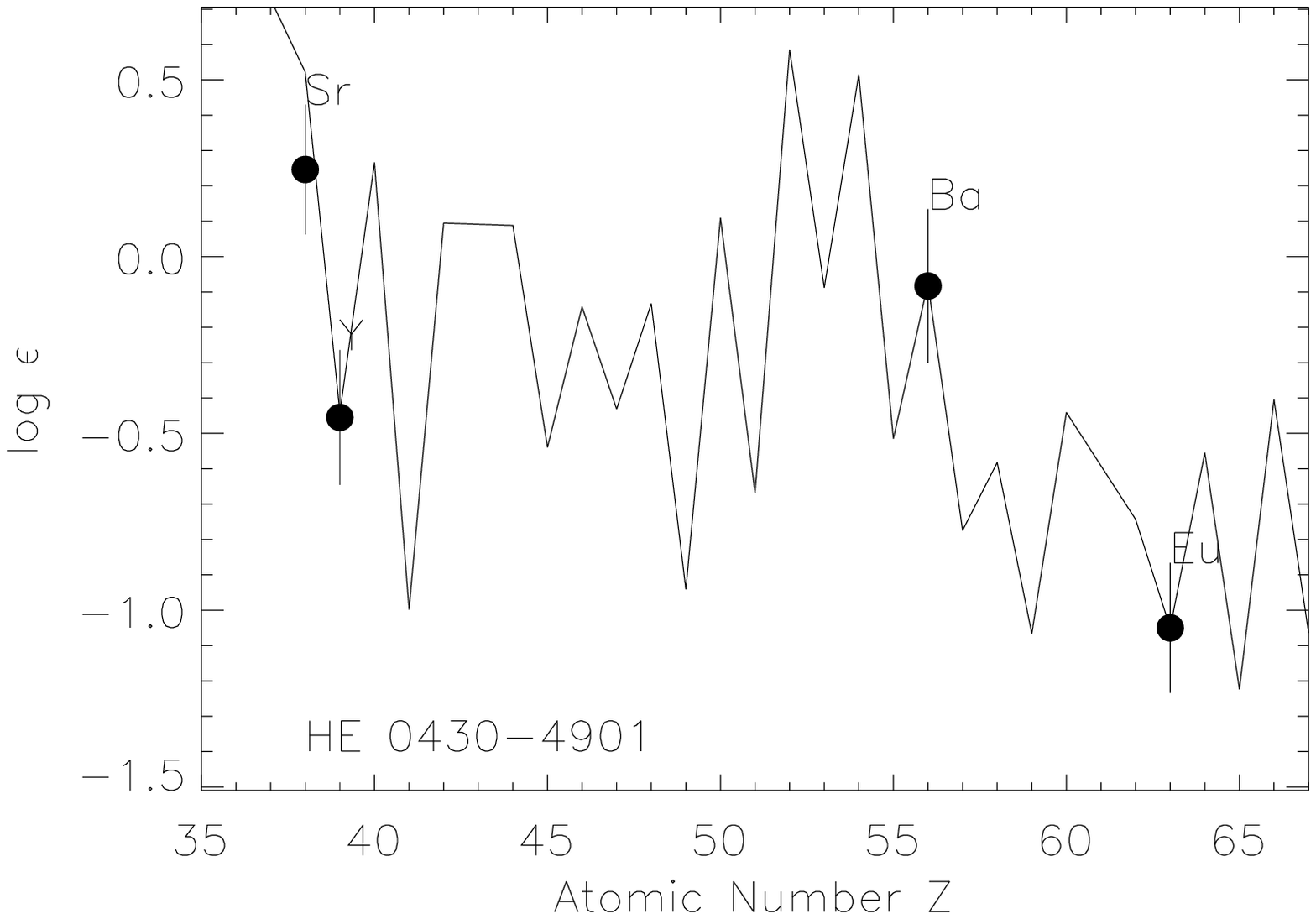}}} \\
\resizebox{58mm}{!}{\rotatebox{0}{\includegraphics{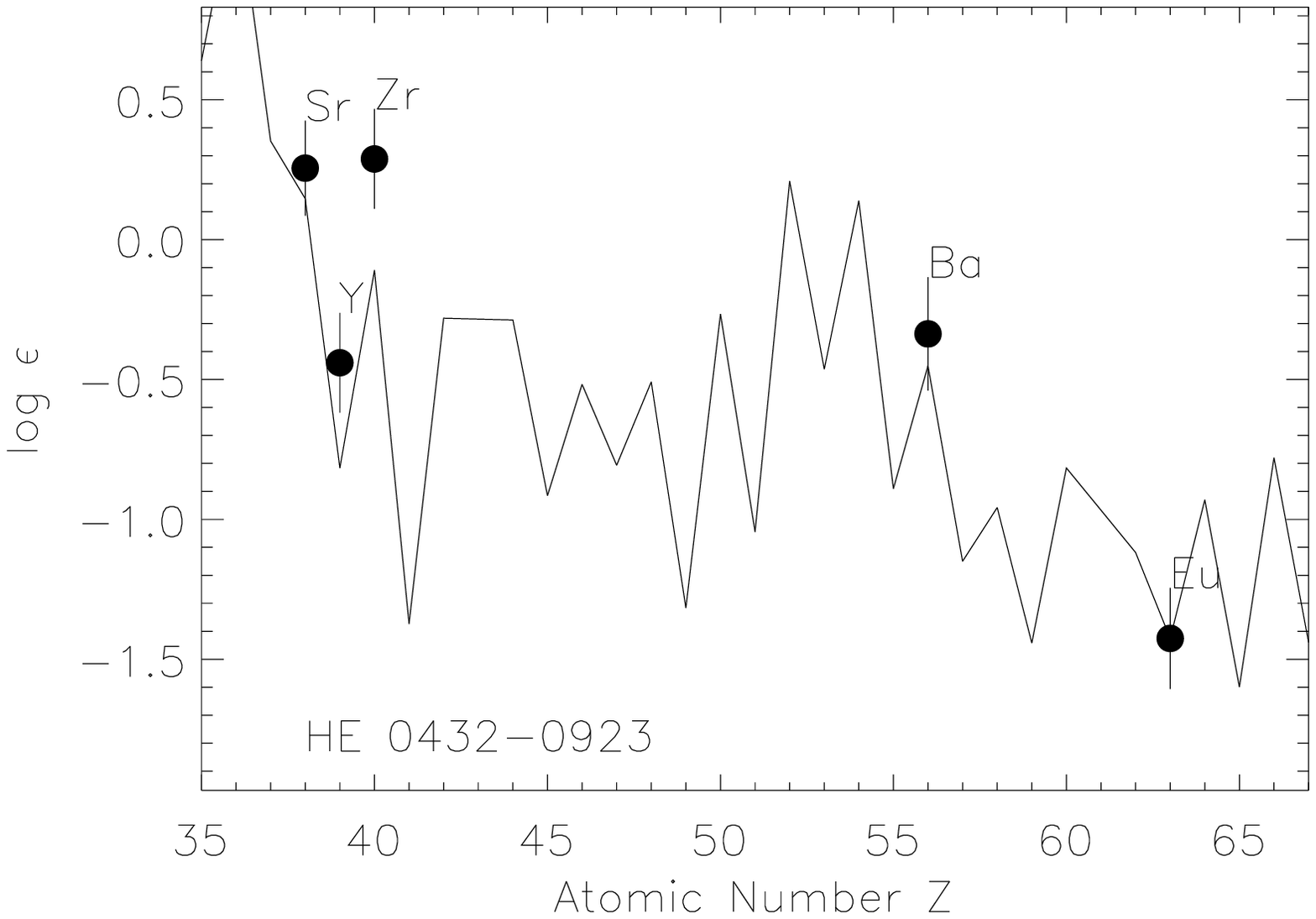}}} &
\resizebox{58mm}{!}{\rotatebox{0}{\includegraphics{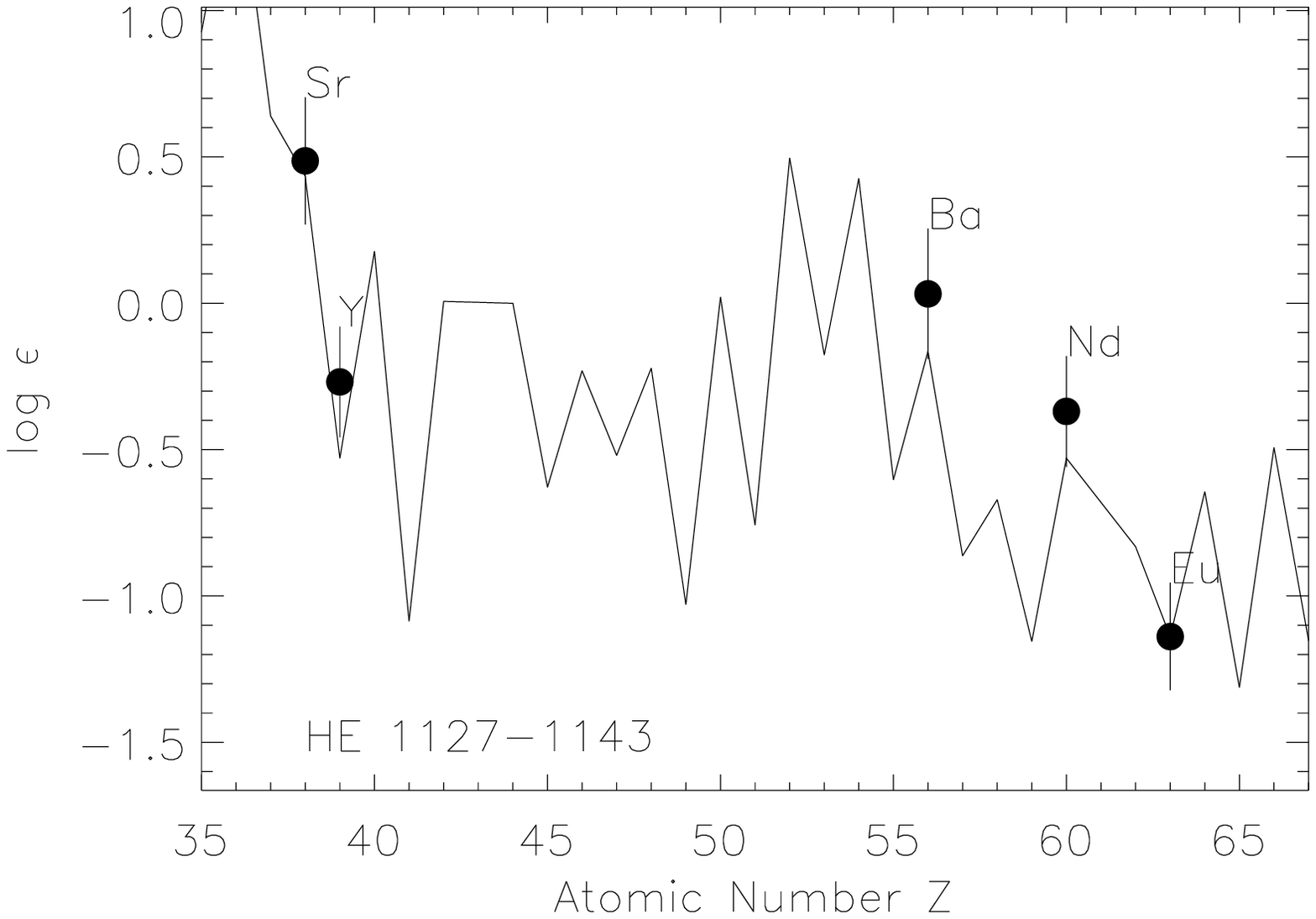}}} &
\resizebox{58mm}{!}{\rotatebox{0}{\includegraphics{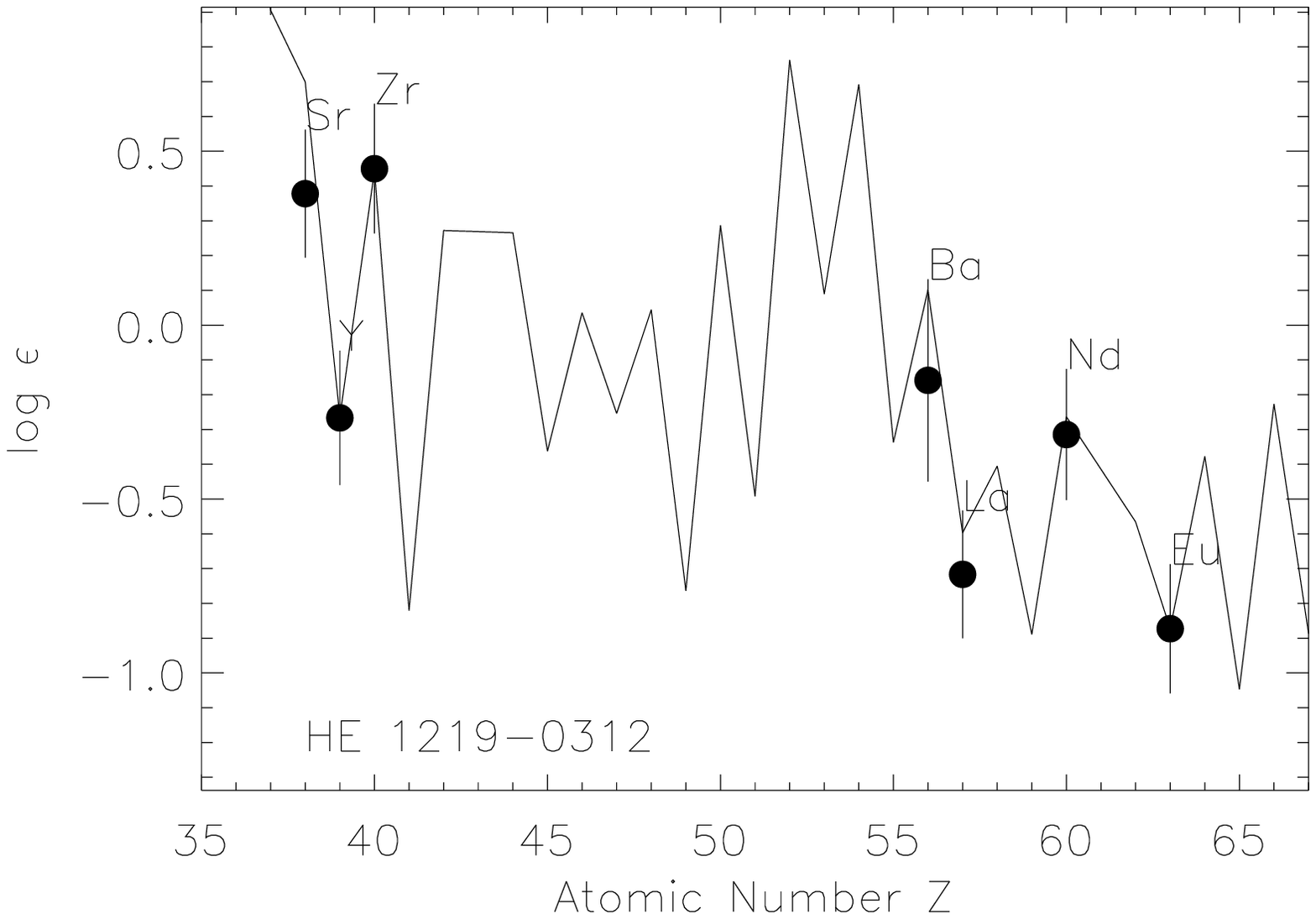}}} \\
\resizebox{58mm}{!}{\rotatebox{0}{\includegraphics{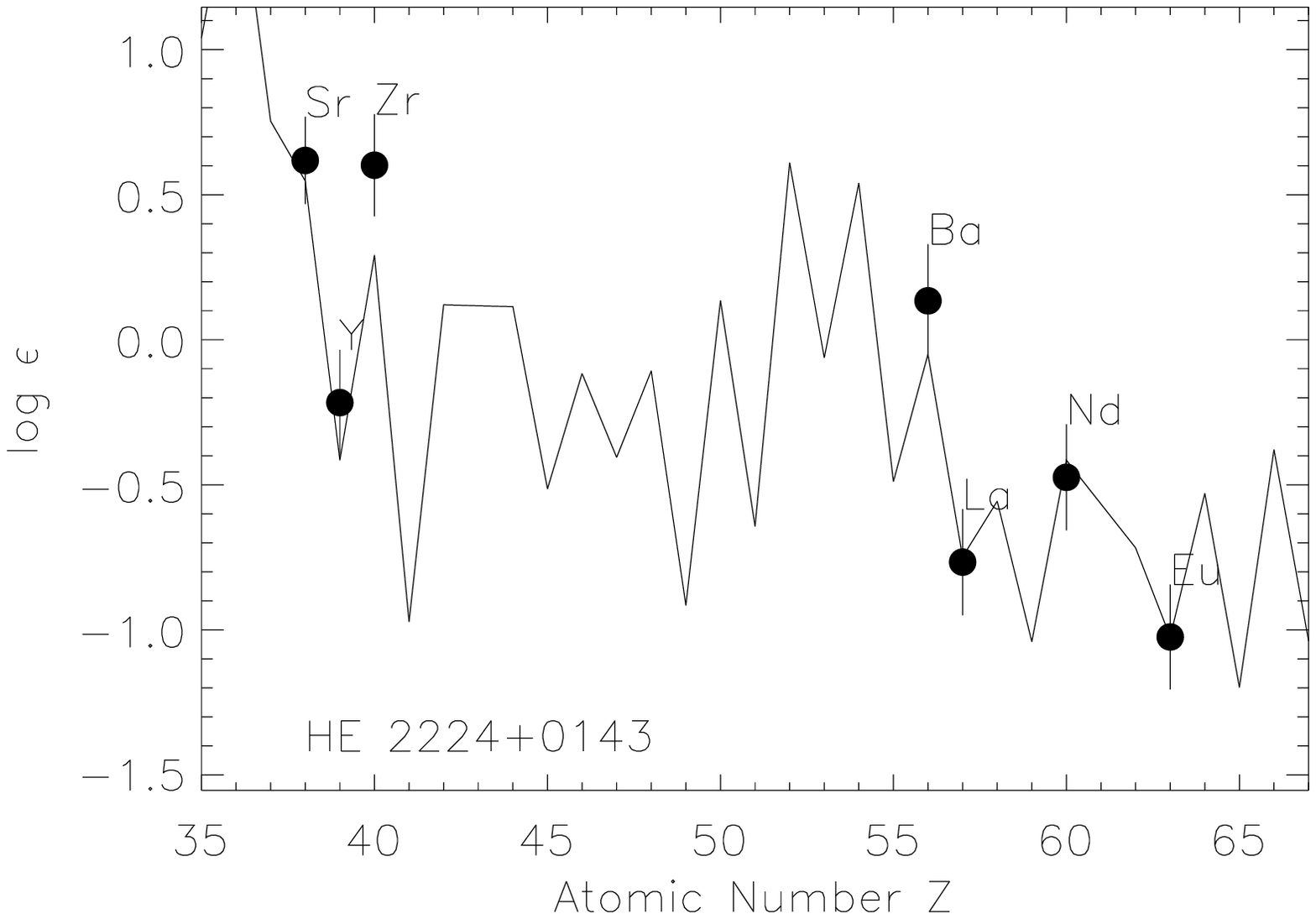}}} &
\resizebox{58mm}{!}{\rotatebox{0}{\includegraphics{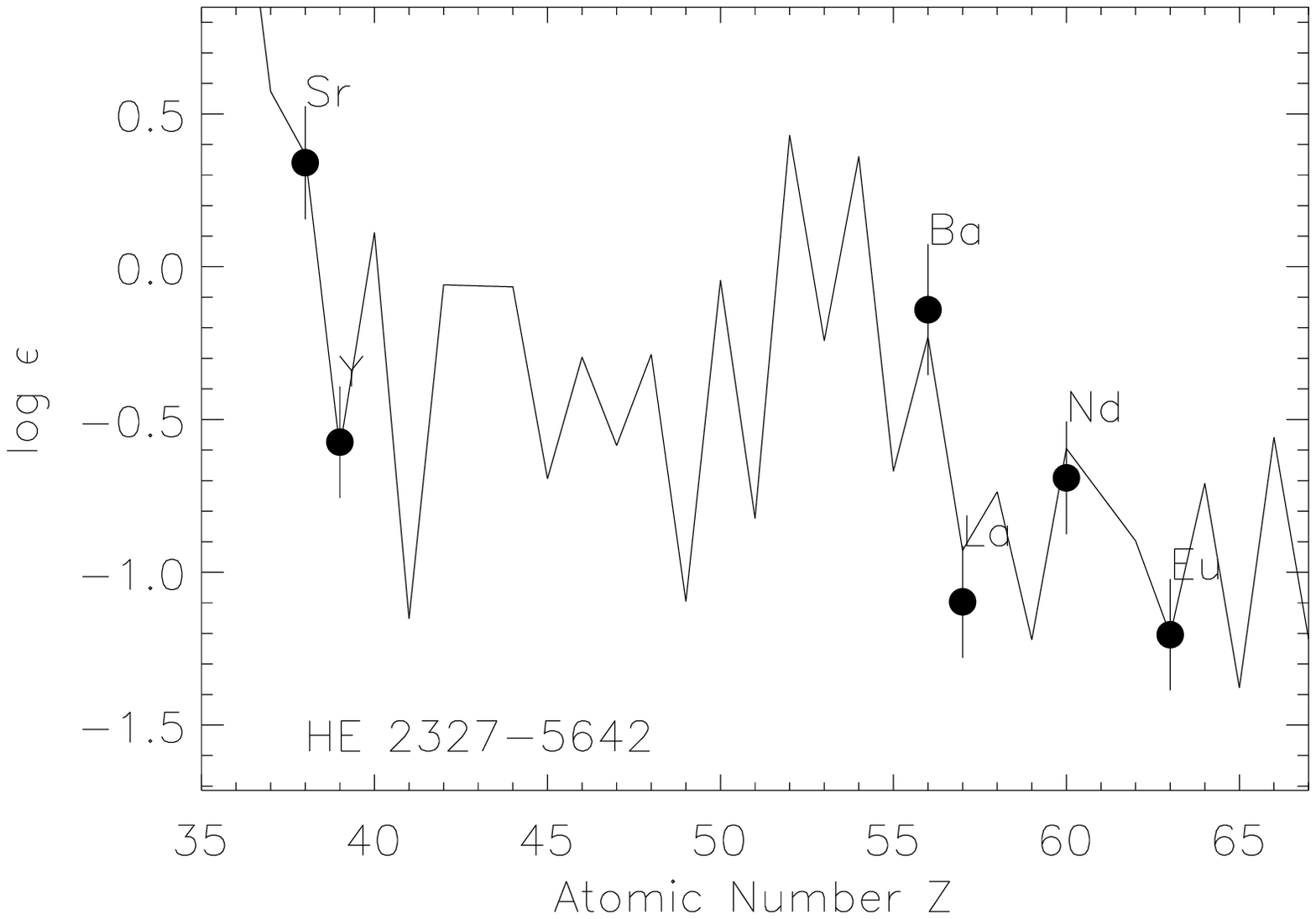}}} &
 \\
\end{tabular}
\end{center}
\caption{Abundance patterns for neutron-capture elements in the r-II stars of the sample.  The full line shows the scaled solar system r-process abundances based on the r-process fractions from Arlandini et~al.~(\cite{arlandini99}), meteoritic abundances from Grevesse \& Sauval~(\cite{grevesse98}), scaled to match our observed Eu abundance. Plotted error estimates are relative errors. }
\label{fig:abund_patterns_rII}
\end{figure*}

The abundance patterns for the neutron-capture elements for the eight r-II stars are shown in Fig.~\ref{fig:abund_patterns_rII}, along with the scaled solar system r-process abundances, demonstrating that these elements follow this pattern in these stars.  In fact, all 57 pure r-process stars, as judged from Ba/Eu, follow the pattern quite well for Ba and heavier elements.  The lighter neutron-capture elements, Sr, Y and Zr, tend to deviate from the solar-r-process pattern in stars of lower r-process enrichment, which will be discussed in Sect.~\ref{subsect:lightneutron}.  The scaled solar system s- and r-process abundances used throughout this paper are based on the s-process fractions from the stellar model of Arlandini et~al.~(\cite{arlandini99}), the assumption that the remainder of neutron-capture elements are produced by the r-process, and meteoritic abundances from Grevesse \& Sauval~(\cite{grevesse98}).  We chose the Arlandini et~al.~(\cite{arlandini99}) fractions over those of Burris~et~al.~(\cite{burris00}) for our comparisons, since we found consistently better agreement with Y abundances in the pure r-process stars.  The abundance patterns are otherwise very similar for the other neutron-capture elements considered here.

The eight r-II stars are all giants with $T_\mathrm{eff}\sim 5100$~K, and have a quite narrow range in metallicity, $-3.2 \leq \mathrm{[Fe/H]} \leq -2.6$.  The r-II stars, including the two previously known examples CS~22892-052 and CS~31082-001, are centred on a metallicity of $\mathrm{[Fe/H]} = -2.81$, with a very small scatter, on the order of 0.16~dex.   Note, we find one r-II star with $\mathrm{[Fe/H]}<-3$ among 49 stars in our sample, a frequency of 2\%, and 7 r-II stars among 118 stars in the range $-3<\mathrm{[Fe/H]}<-2.5$, a frequency of 6\%.  This perhaps indicates that r-II stars are more rare at $\mathrm{[Fe/H]}<-3$; further study in a larger sample with better spectra would be desirable to resolve this question more definitively.  All eight r-II stars have quite normal C/Fe abundances, the largest being $\mathrm{[C/Fe]} \sim 0.5$, noting of course that the present sample is biased towards CH weak stars.  The low C abundances mean that these stars are all candidates for studies of the actinide elements, particularly Th and U, in such stars.  CS~22892-052 remains a unique object as the only r-II star known with $\mathrm{[C/Fe]}\sim 1$. 

The majority of the r-I stars are also giants, however, four such stars are unevolved (i.e.\ not yet on the giant branch $T_\mathrm{eff} > 5500$~K, $\log g > 3$), namely HE~0341-4024, HE~0534-4615, HE~0538-4515 and HE~2301-4024.  As seen in Fig.~\ref{fig:abunds} the \mbox{r-I} stars span the range $-3.4 \leq \mathrm{[Fe/H]} \leq -1.5$, practically the entire metallicity range of our sample.  We note that the four unevolved stars all have $\mathrm{[Fe/H]} \geq -2.1$, but that there is a selection effect at work here; the Eu~II lines observed will be weaker in unevolved stars than in evolved stars with the same abundances, and thus our survey preferentially detects Eu in giant stars.  Our survey is also biased towards cool giants as described in Paper~I.

Five stars in which both Eu and Ba were detected are found to be s-process rich as judged from their having $\mathrm{[Ba/Eu]}>0.5$. \footnote{As described in appendix~\ref{app:linelist}, we have assumed r-process isotopic composition of Ba in all stars which may lead to underestimation of the Ba abundance and thus Ba/Eu in stars with significant s-process contributions.  Similarly, we assume no $^{13}$C which may lead to errors in the C abundances for such stars.}   All five are carbon-enhanced, and three are strongly Eu enhanced with $\mathrm{[Eu/Fe]}>1.0$.  The three stars with $\mathrm{[Ba/Eu]}>0.5$ and $\mathrm{[Eu/Fe]}>1.0$, are listed in Table~\ref{tab:interesting}.  A number of similar stars are known; e.g.\ Hill~et~al.~(\cite{hill00}), Aoki~et~al.~(\cite{aoki02b}) and Cohen~et~al.~(\cite{cohen03}).  We shall refer to these stars as ``s-II'' stars, though we note that whether the neutron-capture elements in such stars are produced predominantly by the s-process or by both the s- and r-processes is a matter of current debate (e.g.\ Johnson \& Bolte~\cite{johnson04}).  We note that these three stars all have large C enhancements $\mathrm{[C/Fe]}>1.5$, and are unevolved.  It must be borne in mind, however, that since this sample is limited to the stars with spectra only weakly polluted by CH features, we were only able to analyse the warmest strongly C enhanced stars.  The complete HERES sample contains 72 CH strong stars which were not analysed here, among which there will be a large number of s-process rich stars.  As we will discuss in the next section, a further such s-II star is suspected based on other abundances, but is not confirmed as Eu is detected at below the 3$\sigma$ level.  

\begin{figure*}
\tabcolsep 0mm
\begin{center}
\begin{tabular}{ccc}
\resizebox{58mm}{!}{\rotatebox{0}{\includegraphics{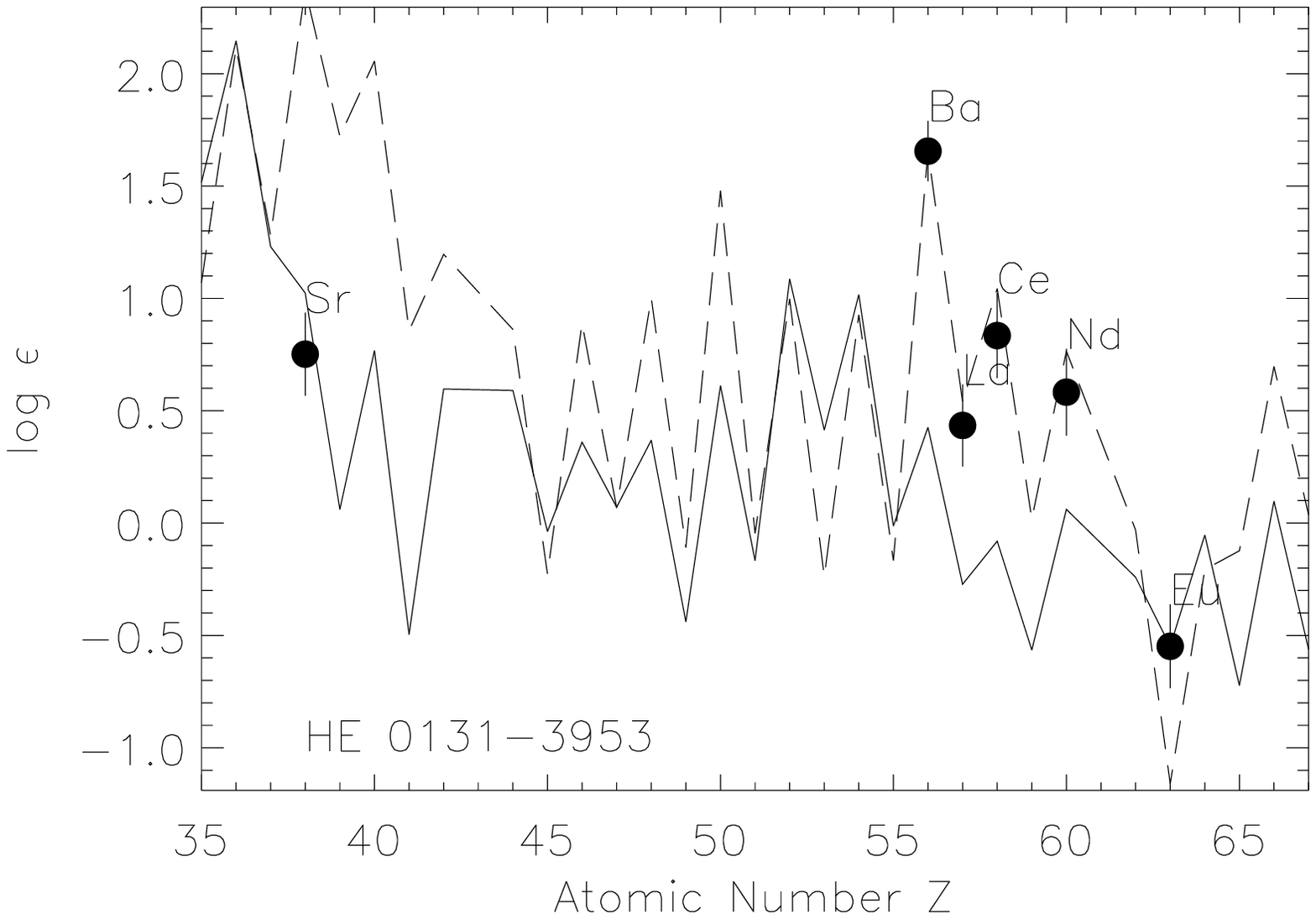}}} &
\resizebox{58mm}{!}{\rotatebox{0}{\includegraphics{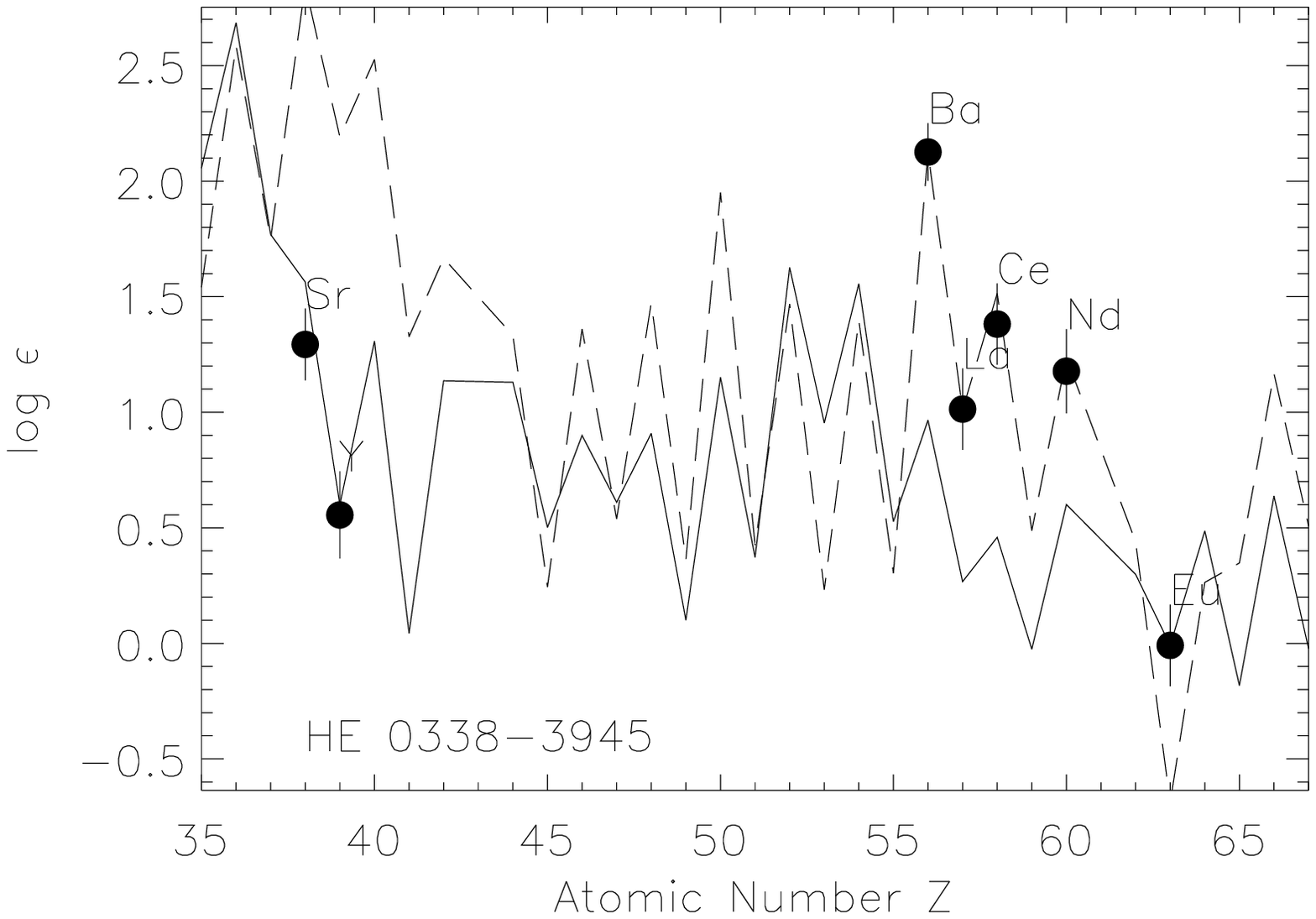}}} &
\resizebox{58mm}{!}{\rotatebox{0}{\includegraphics{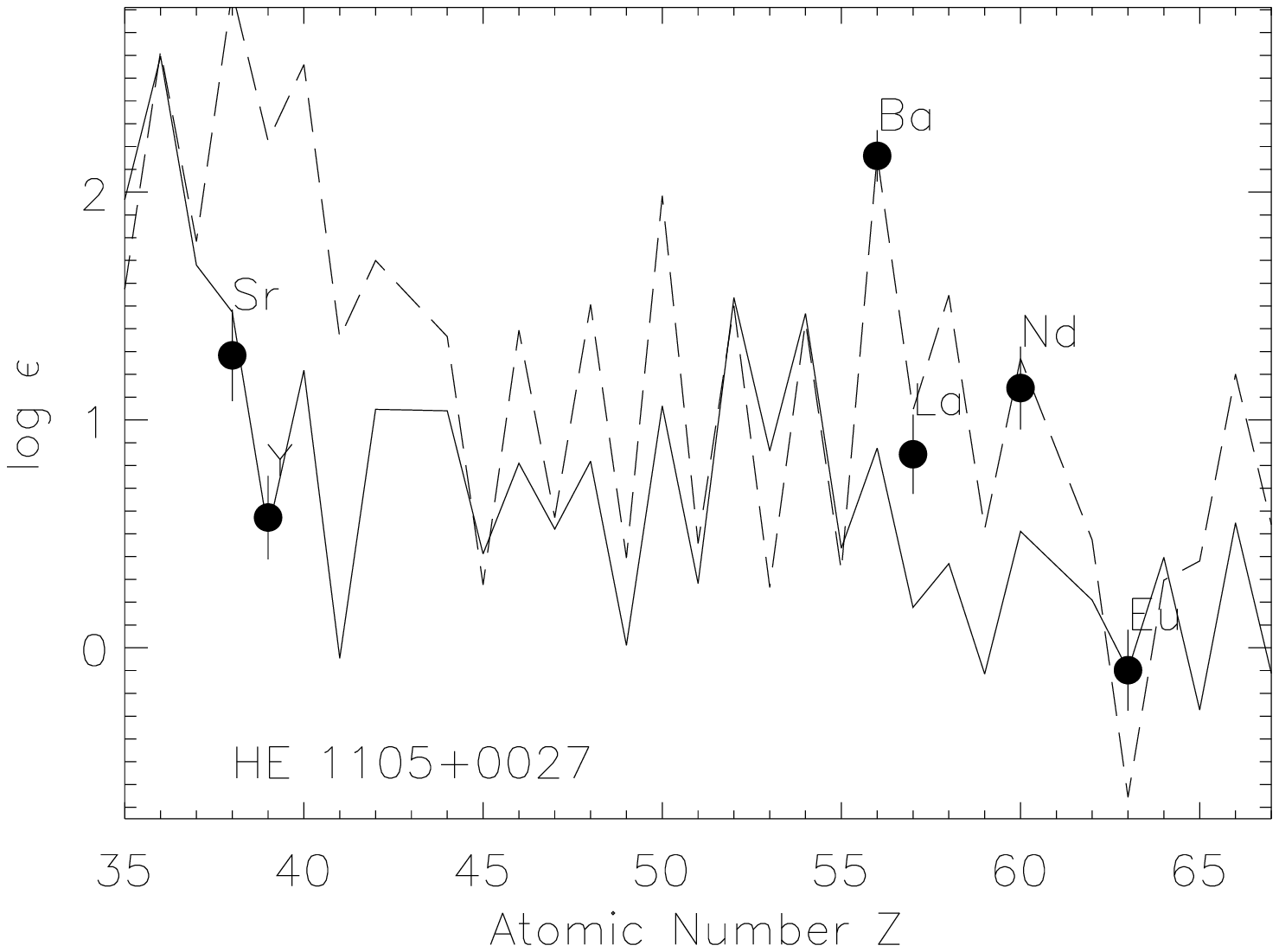}}} \\
\end{tabular}
\end{center}
\caption{Abundance patterns for neutron-capture elements in the s-process rich stars of the sample with strong Eu enhancement, $\mathrm{[Eu/Fe]}> 1.5$, the s-II stars.  The full line shows the scaled solar system r-process abundances based on the r-process fractions from Arlandini et~al.~(\cite{arlandini99}), meteoritic abundances from Grevesse \& Sauval~(\cite{grevesse98}), scaled to match our observed Eu abundance.   The dashed line shows the equivalent scaled solar system s-process abundances, scaled to match our observed Ba abundance.  Plotted error estimates are relative errors. }
\label{fig:abund_patterns_rs}
\end{figure*}

The remaining two s-process rich stars with $\mathrm{[Ba/Eu]}>0.0$ yet more normal Eu enhancement ($\mathrm{[Eu/Fe]}<0.6$) have abundance patterns of neutron-capture elements which are reasonably well matched by a scaled solar s-process abundance pattern.   Figure~\ref{fig:abund_patterns_rs} shows the abundance patterns for the three s-II stars.  In these cases the Ba, La, Ce and Nd abundances can be well reproduced by a scaled solar-system s-process abundance pattern.  The Sr and Y abundances, however, are much lower.  We notice, however, that these abundances are consistent with the scaled solar system r-process abundance pattern normalised to Eu.  On the other hand, the s-process is expected to be skewed towards heavier elements in metal-poor environments (e.g.\ Busso et~al.~\cite{busso99}).  A summary of suggested possible production scenarios for the s-process rich metal-poor stars has been recently given by Johnson \& Bolte~(\cite{johnson04}).  As stated by those authors, to distinguish different scenarios more strongly, measurements of as many heavy elements as possible are needed, and studies of these stars using better observational material will be the subject of future work as part of the HERES survey.

\section{Discussion}
\label{sect:discussion}

\subsection{Abundance Trends and Scatter with Metallicity}

\begin{figure*}
\tabcolsep 0mm
\renewcommand{\arraystretch}{3.0}
\begin{center}
\begin{tabular}{ccc}
\resizebox{60mm}{!}{\rotatebox{0}{\includegraphics{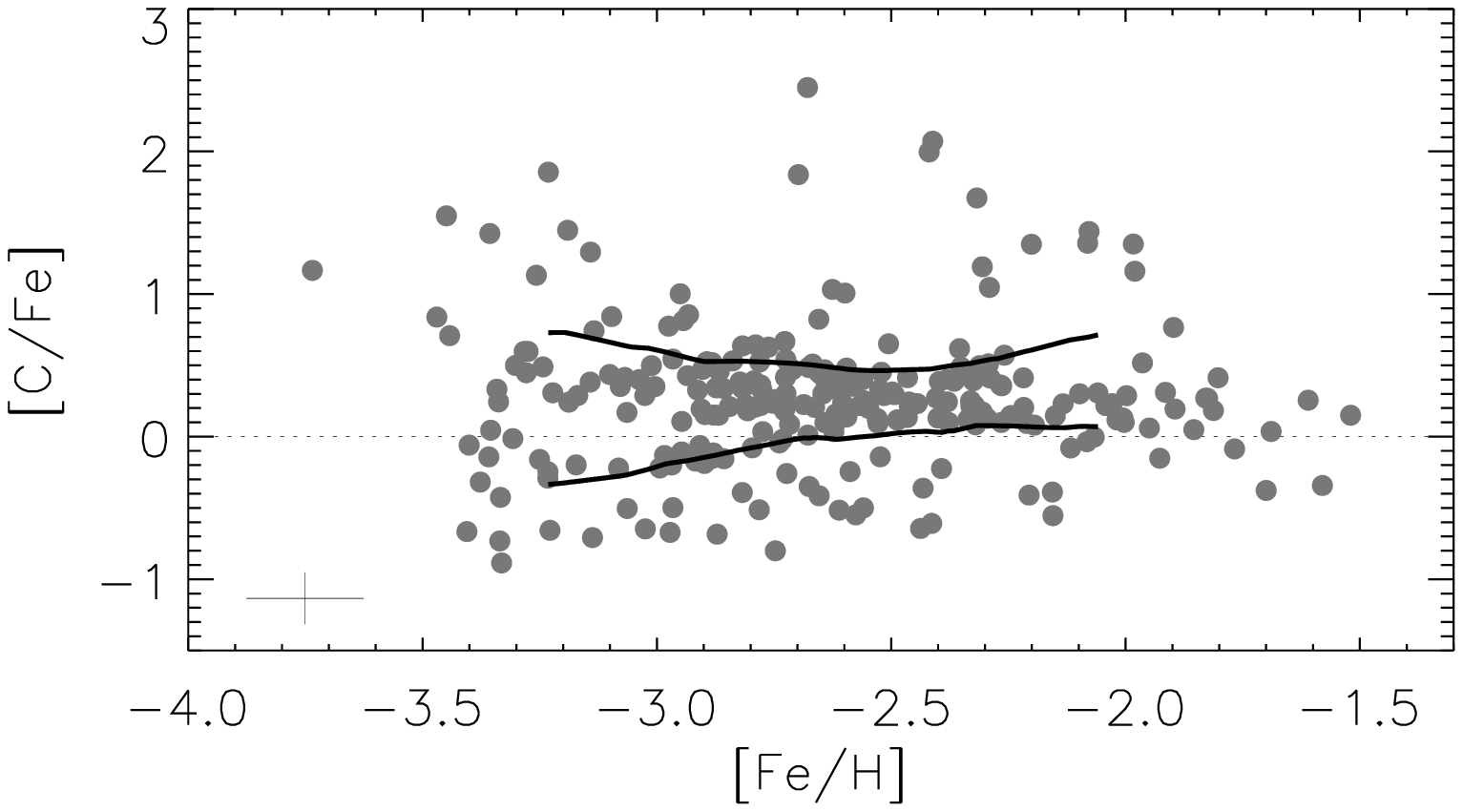}}}  &
\resizebox{60mm}{!}{\rotatebox{0}{\includegraphics{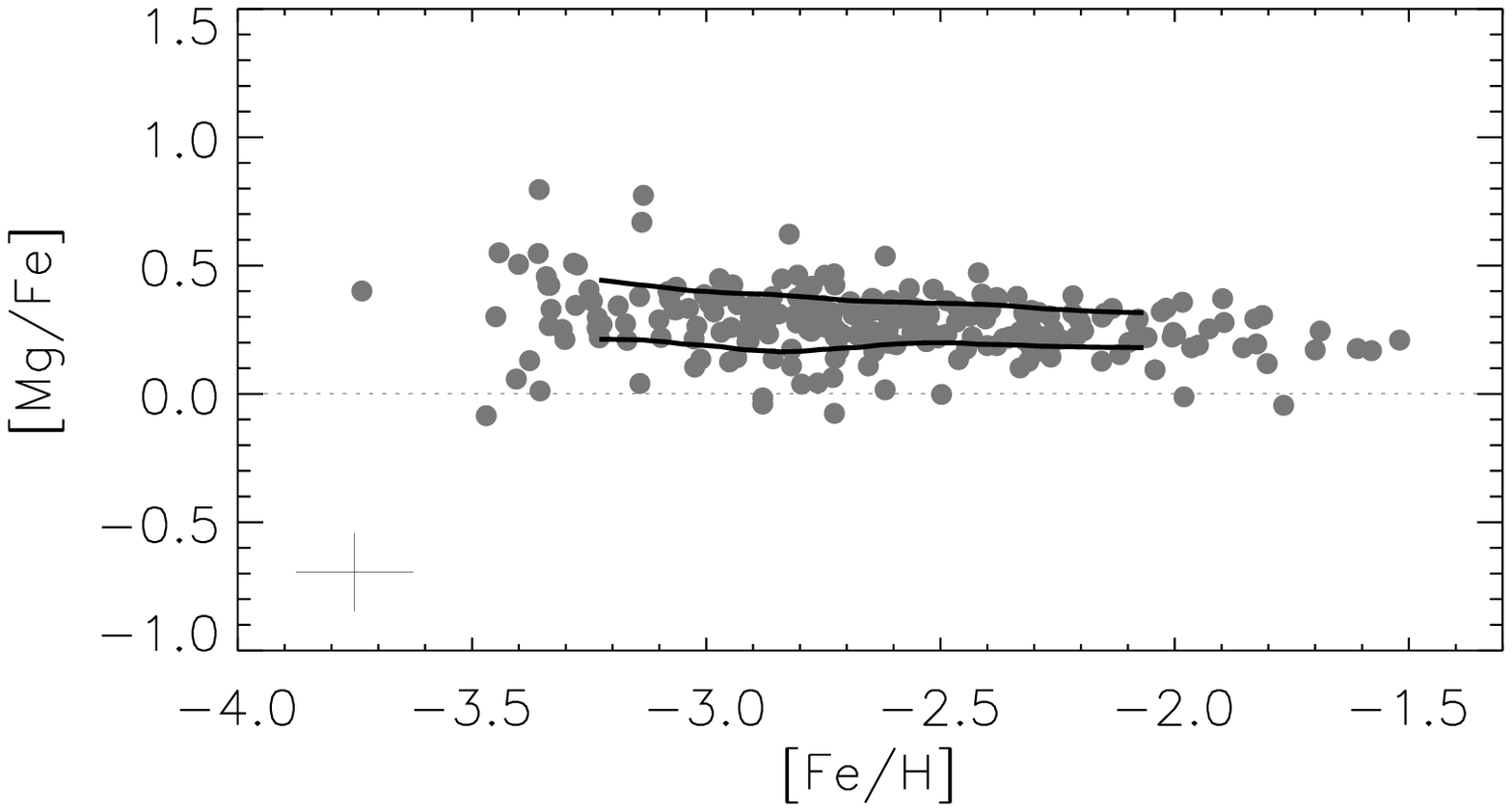}}} &
\resizebox{60mm}{!}{\rotatebox{0}{\includegraphics{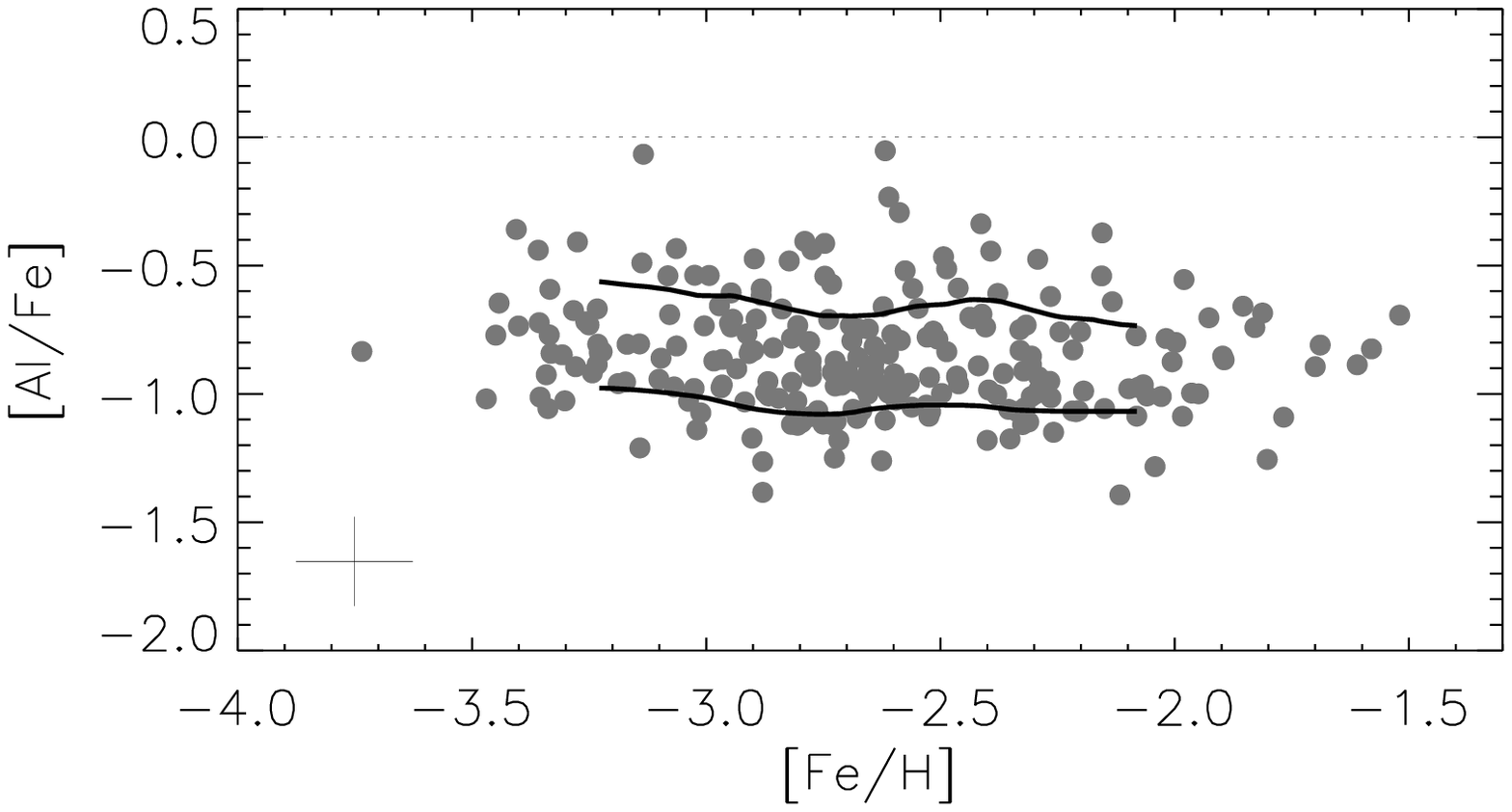}}} \\
\resizebox{60mm}{!}{\rotatebox{0}{\includegraphics{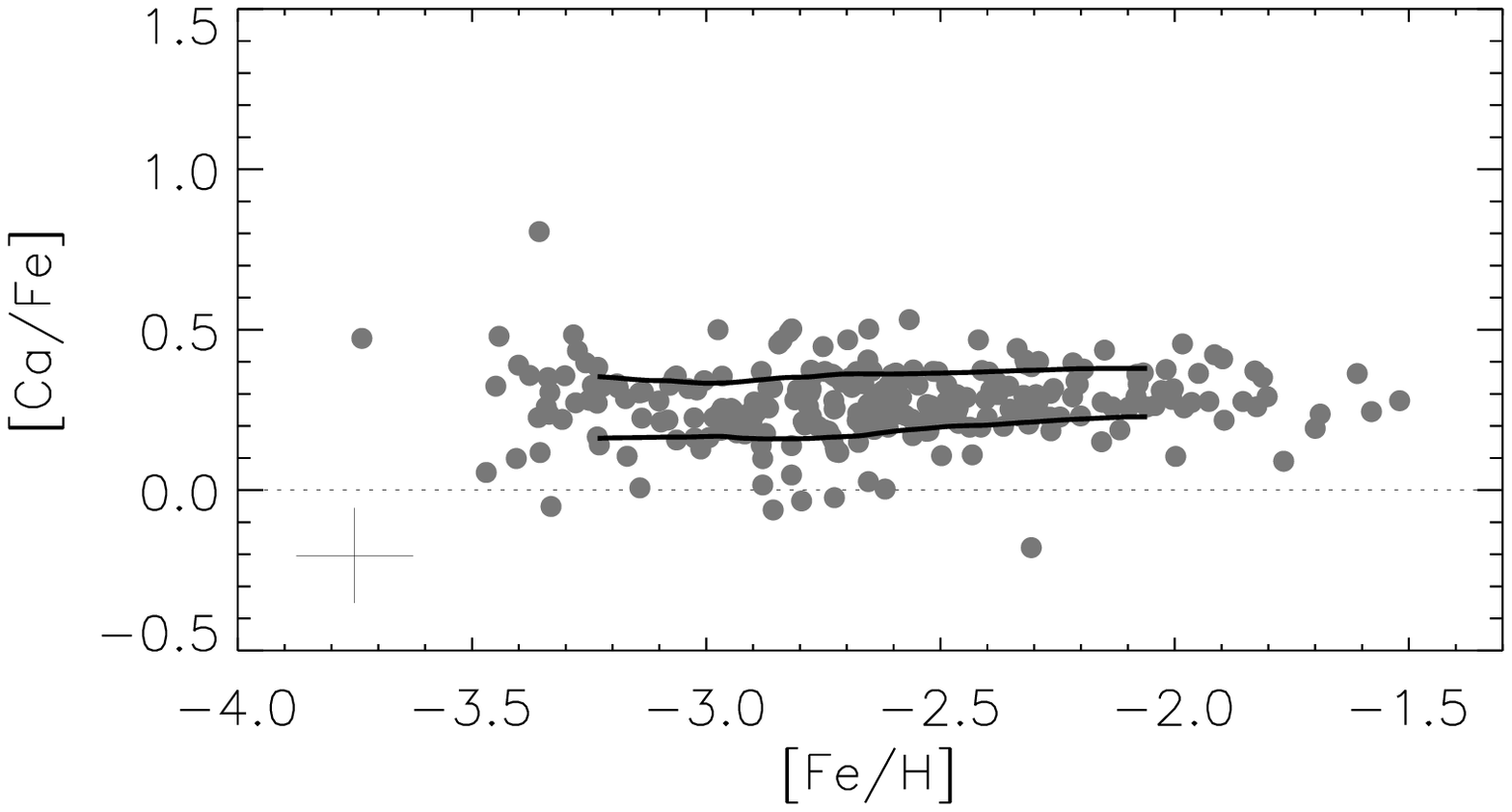}}} &
\resizebox{60mm}{!}{\rotatebox{0}{\includegraphics{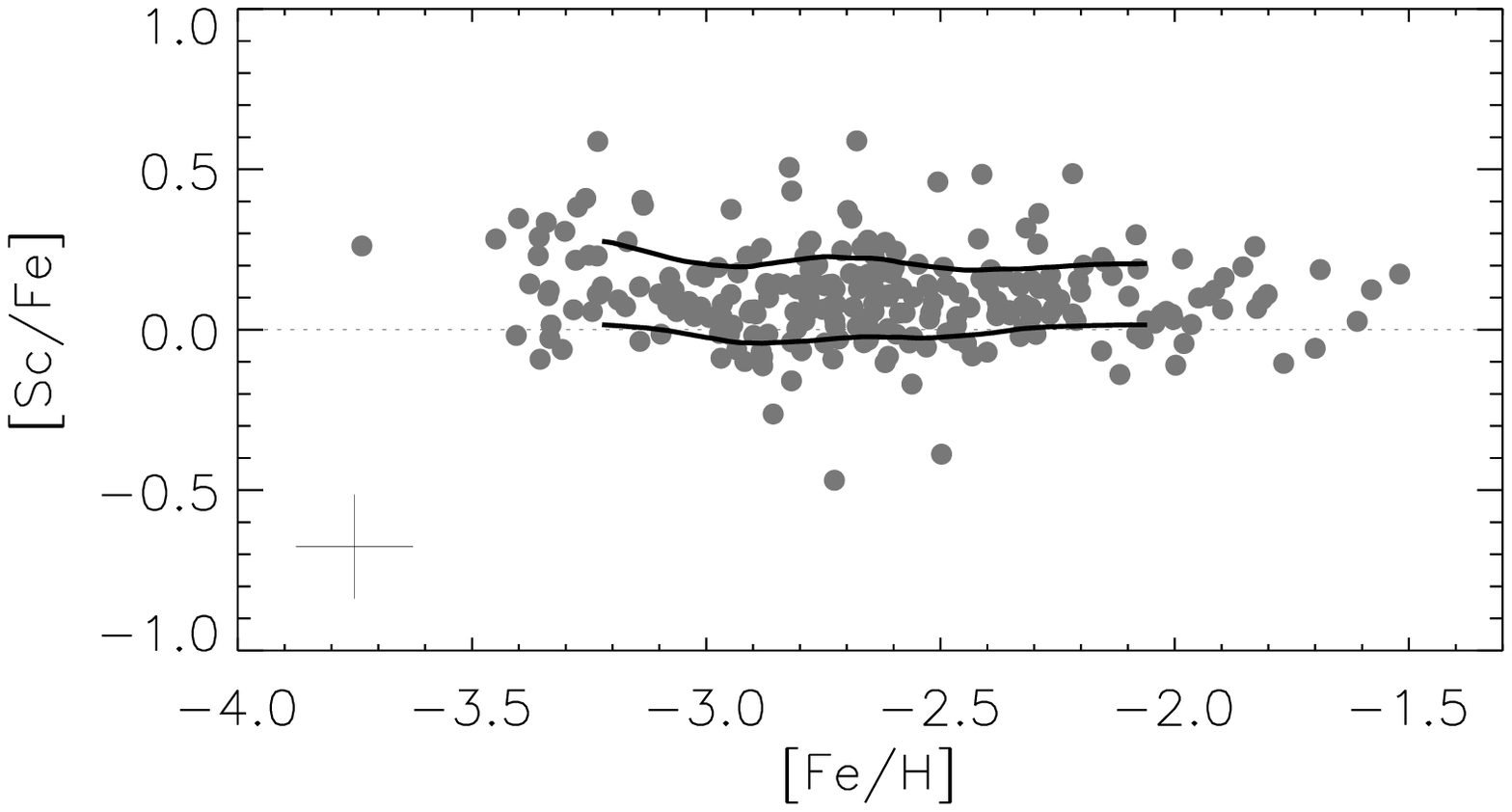}}} &
\resizebox{60mm}{!}{\rotatebox{0}{\includegraphics{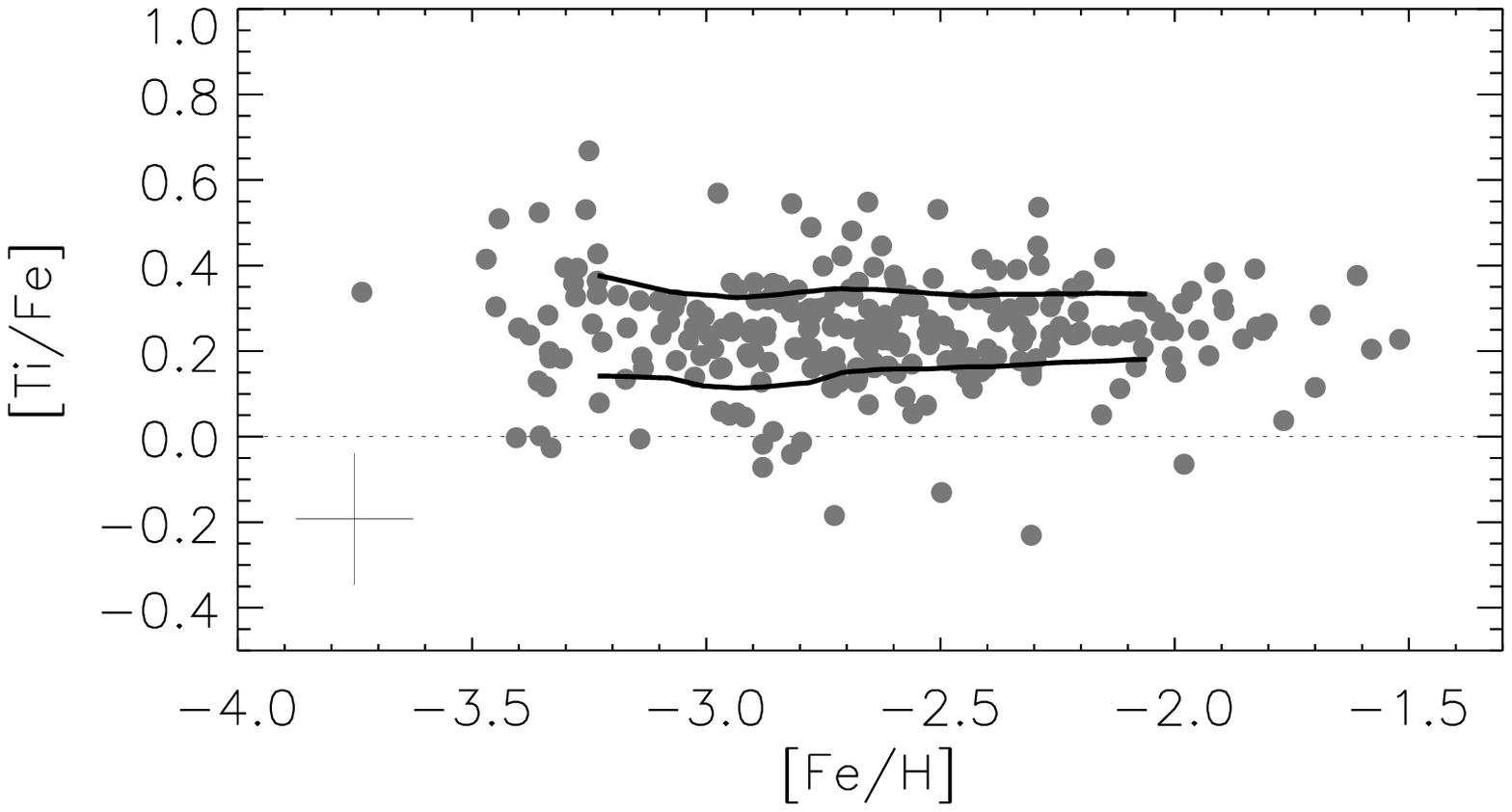}}} \\
\resizebox{60mm}{!}{\rotatebox{0}{\includegraphics{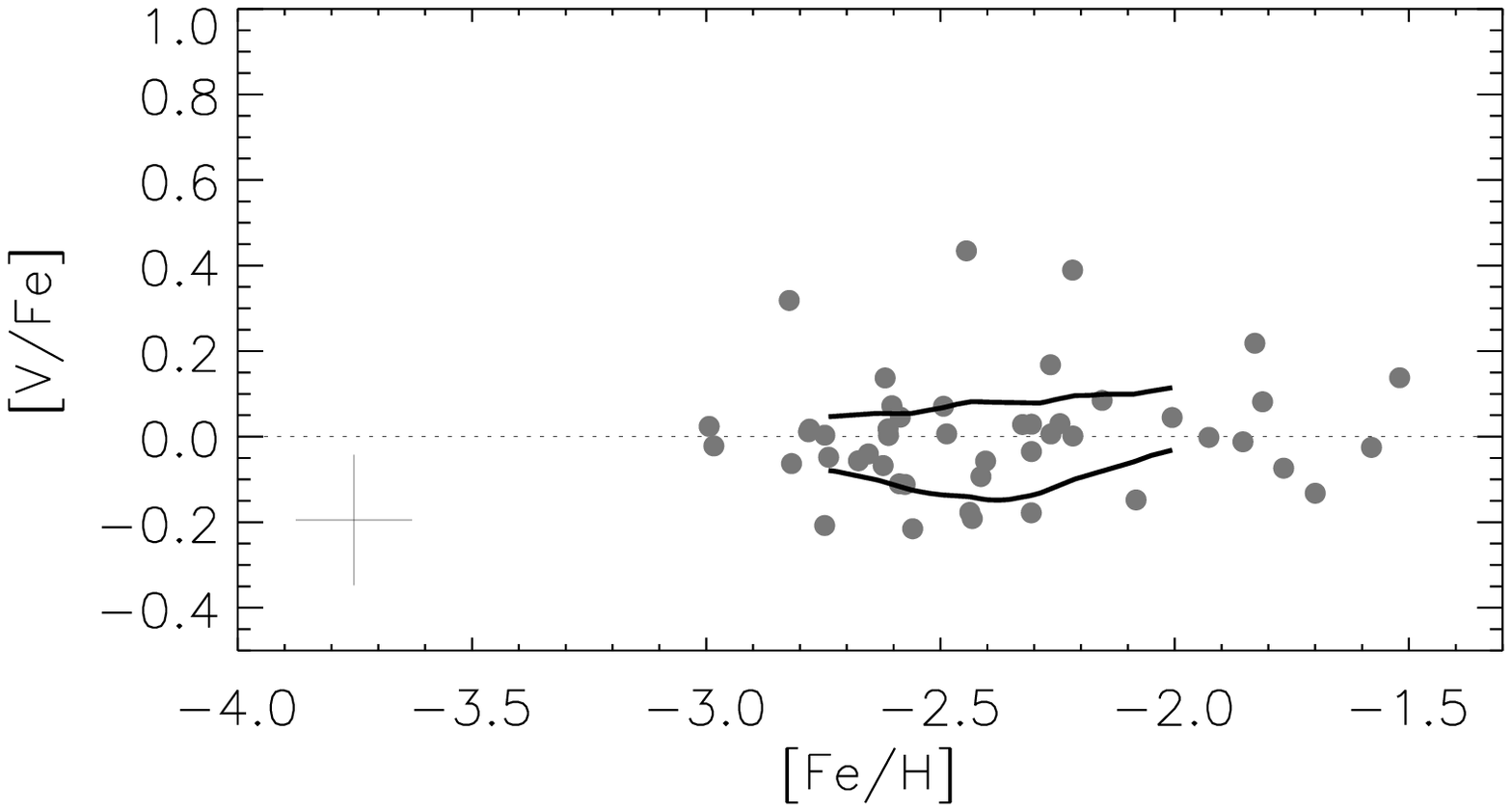}}}  &
\resizebox{60mm}{!}{\rotatebox{0}{\includegraphics{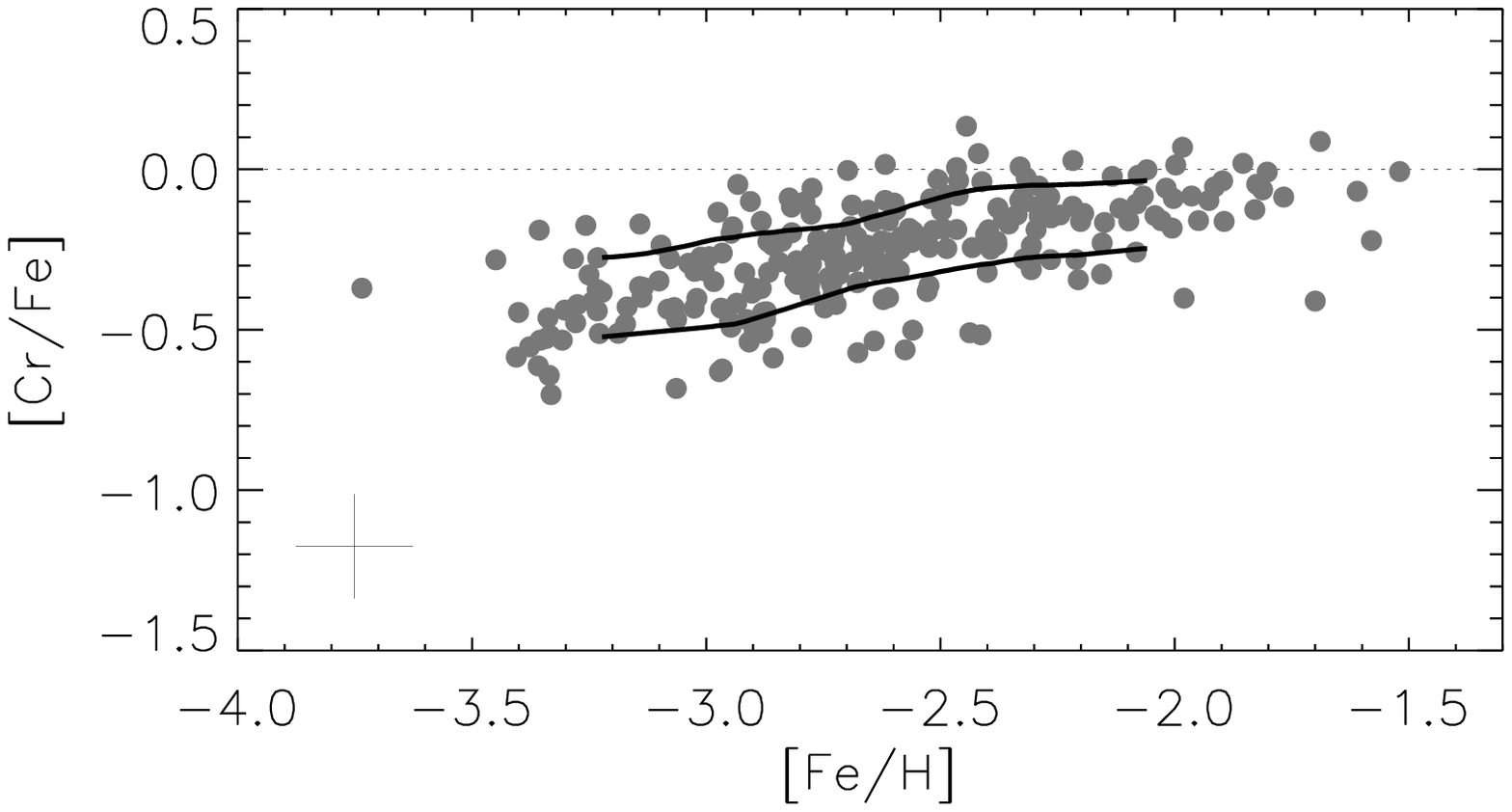}}} &
\resizebox{60mm}{!}{\rotatebox{0}{\includegraphics{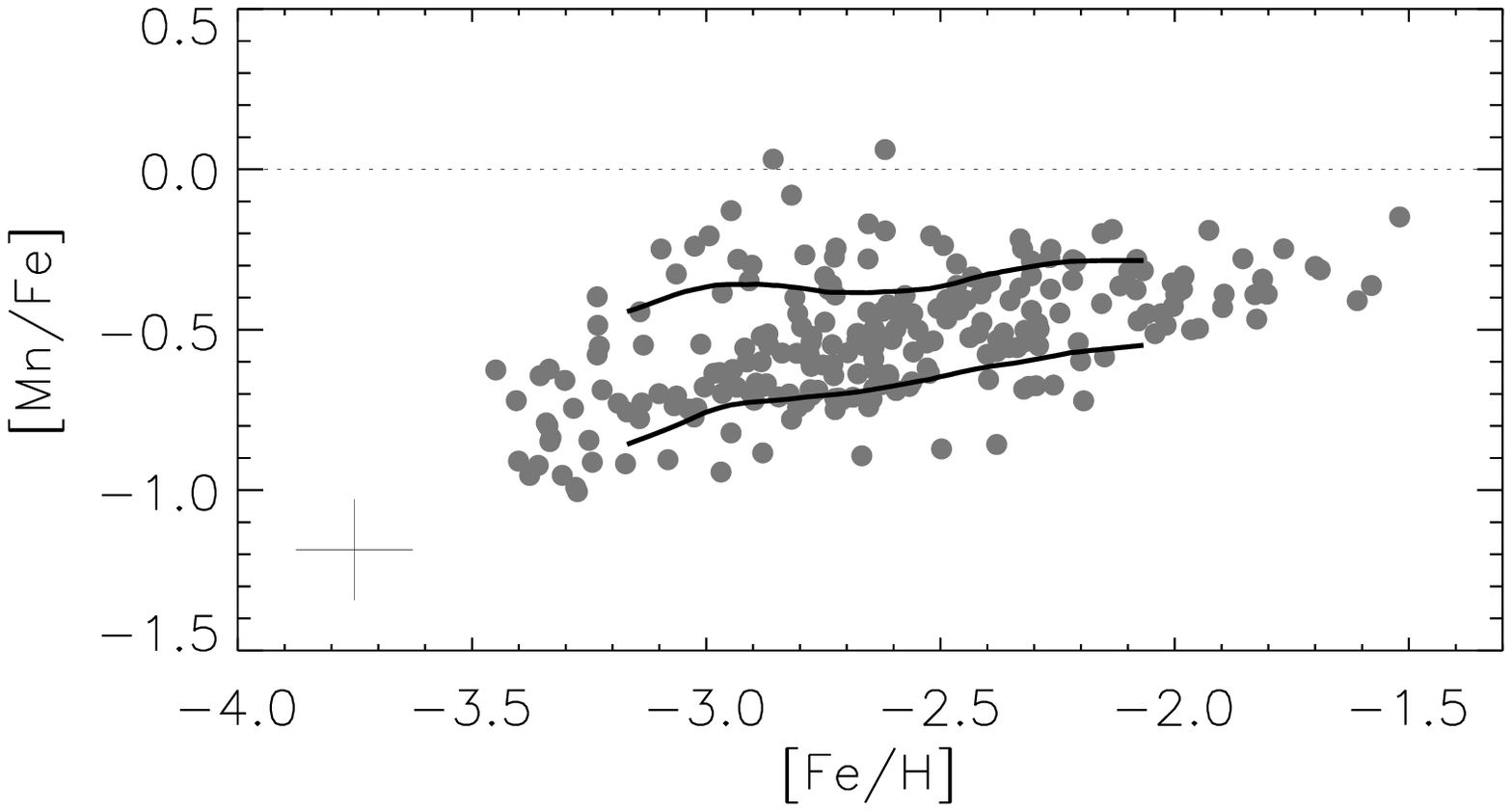}}} \\
\resizebox{60mm}{!}{\rotatebox{0}{\includegraphics{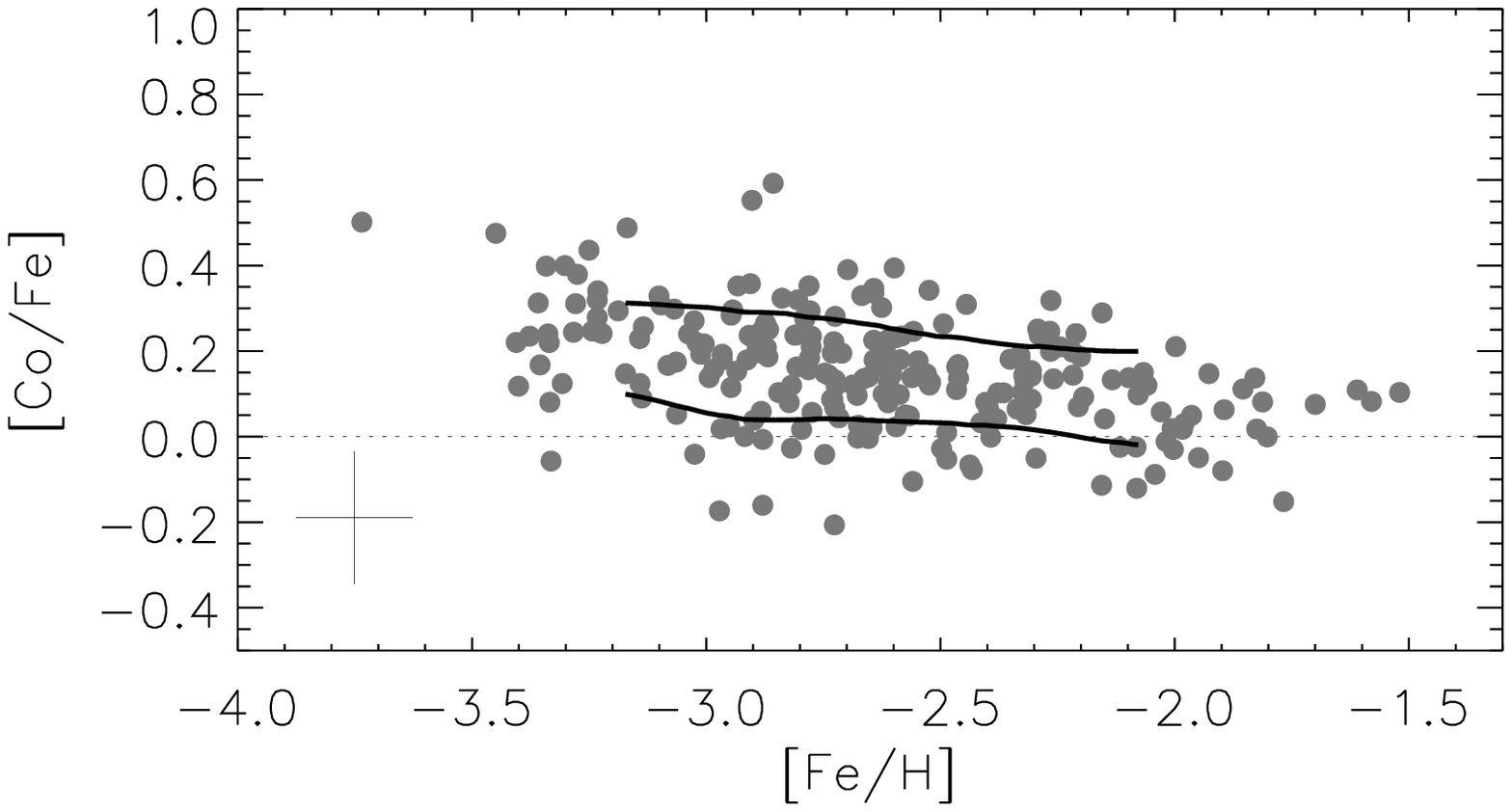}}} &
\resizebox{60mm}{!}{\rotatebox{0}{\includegraphics{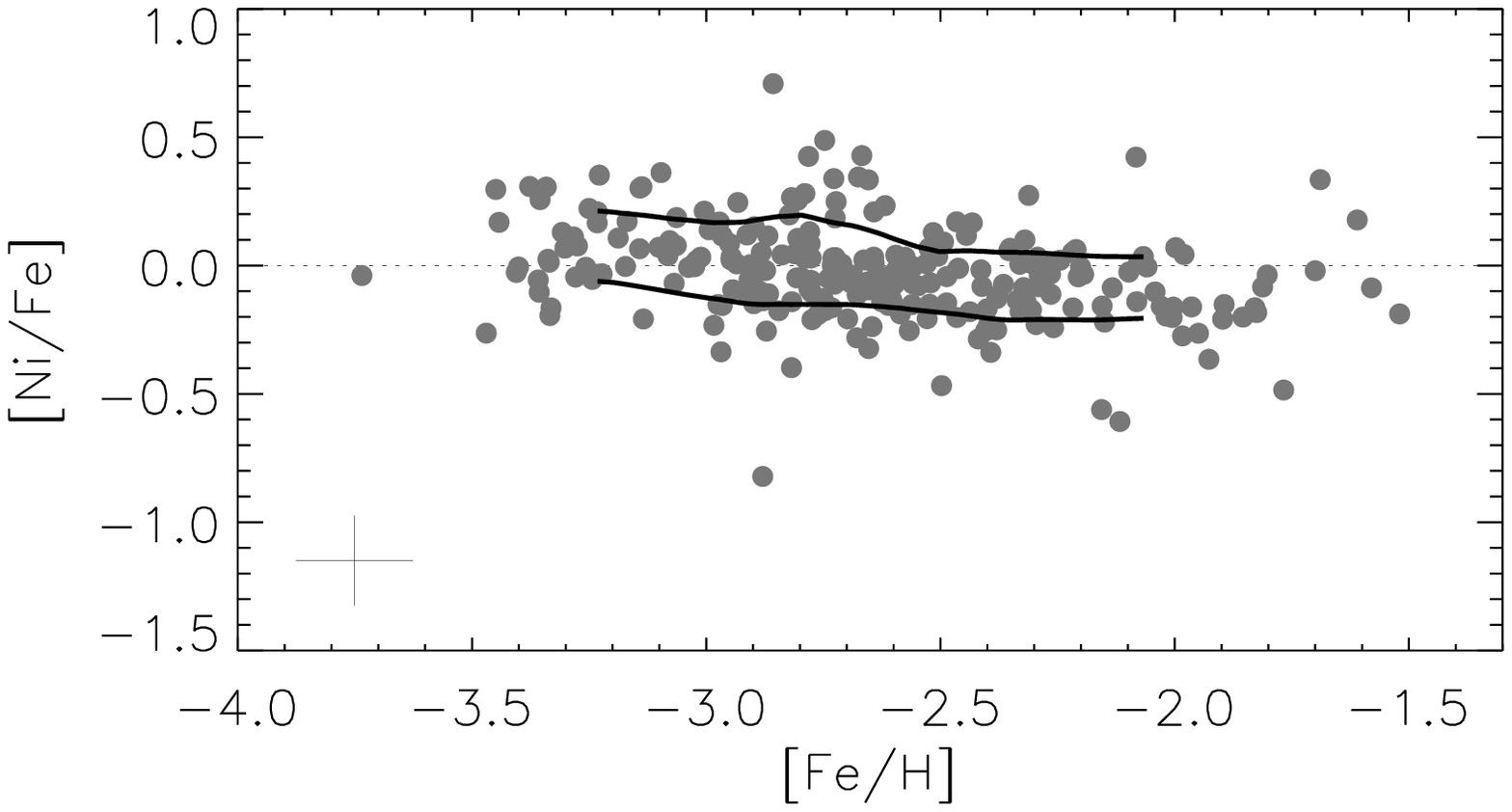}}} &
\resizebox{60mm}{!}{\rotatebox{0}{\includegraphics{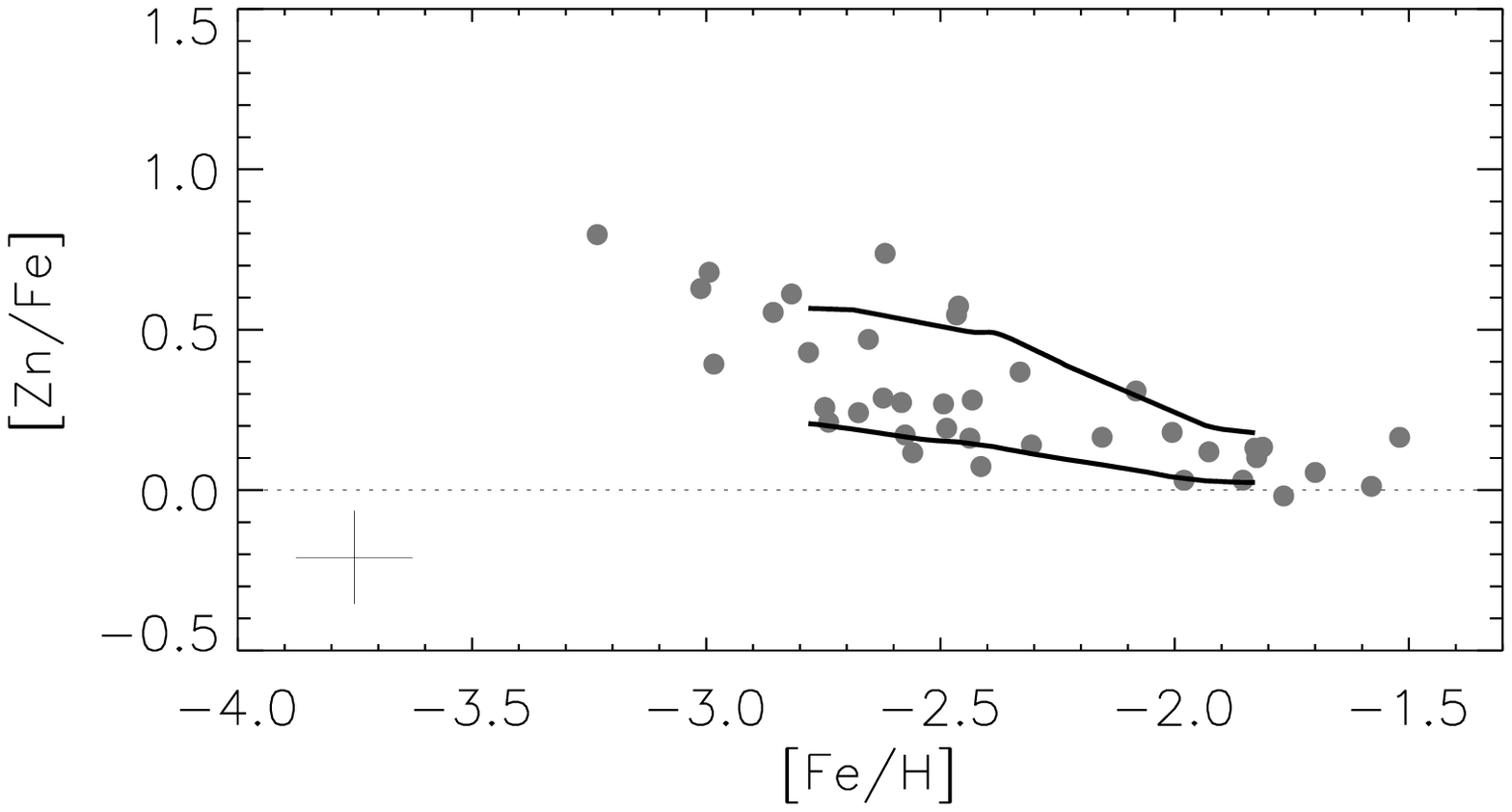}}} \\
\resizebox{60mm}{!}{\rotatebox{0}{\includegraphics{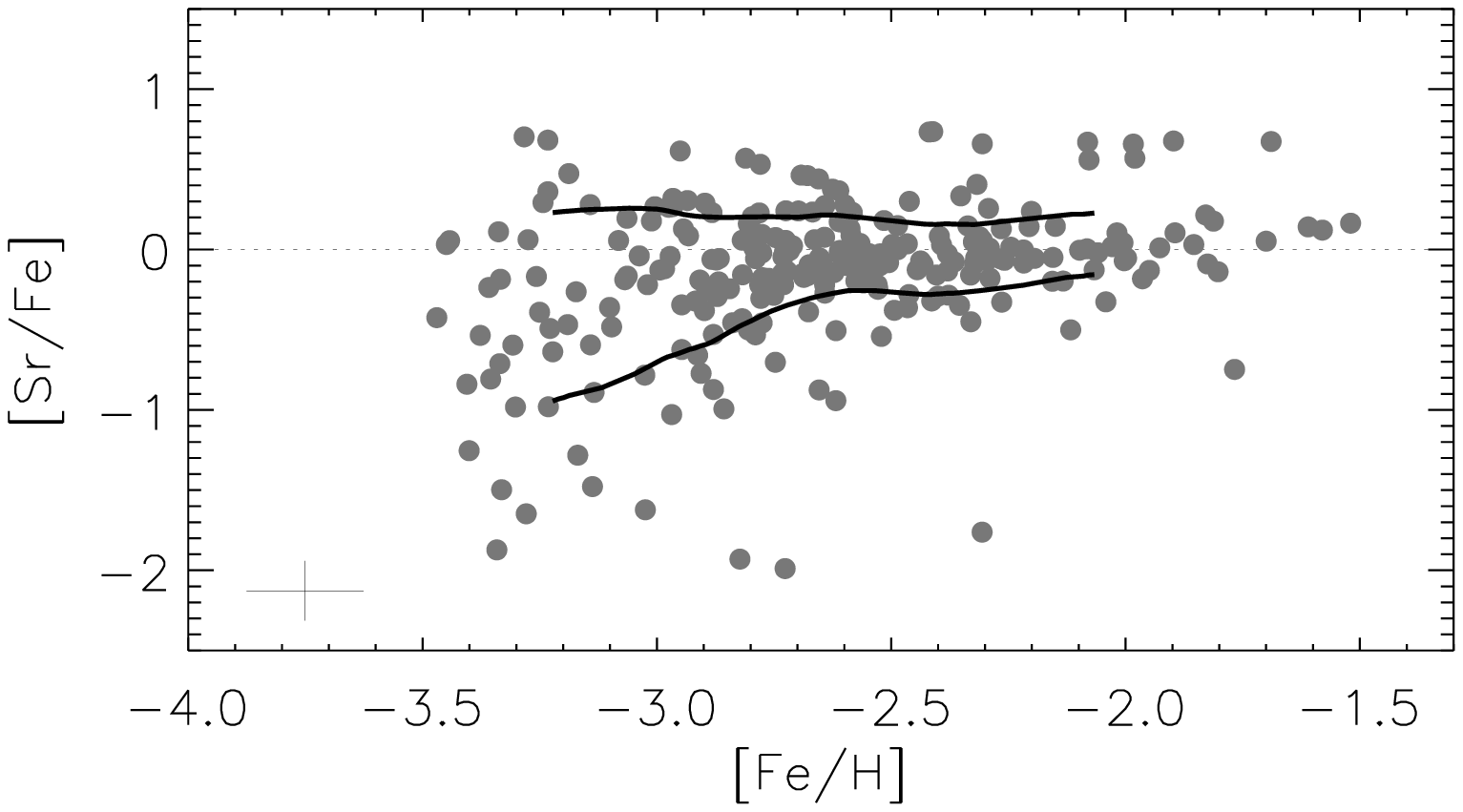}}} &
\resizebox{60mm}{!}{\rotatebox{0}{\includegraphics{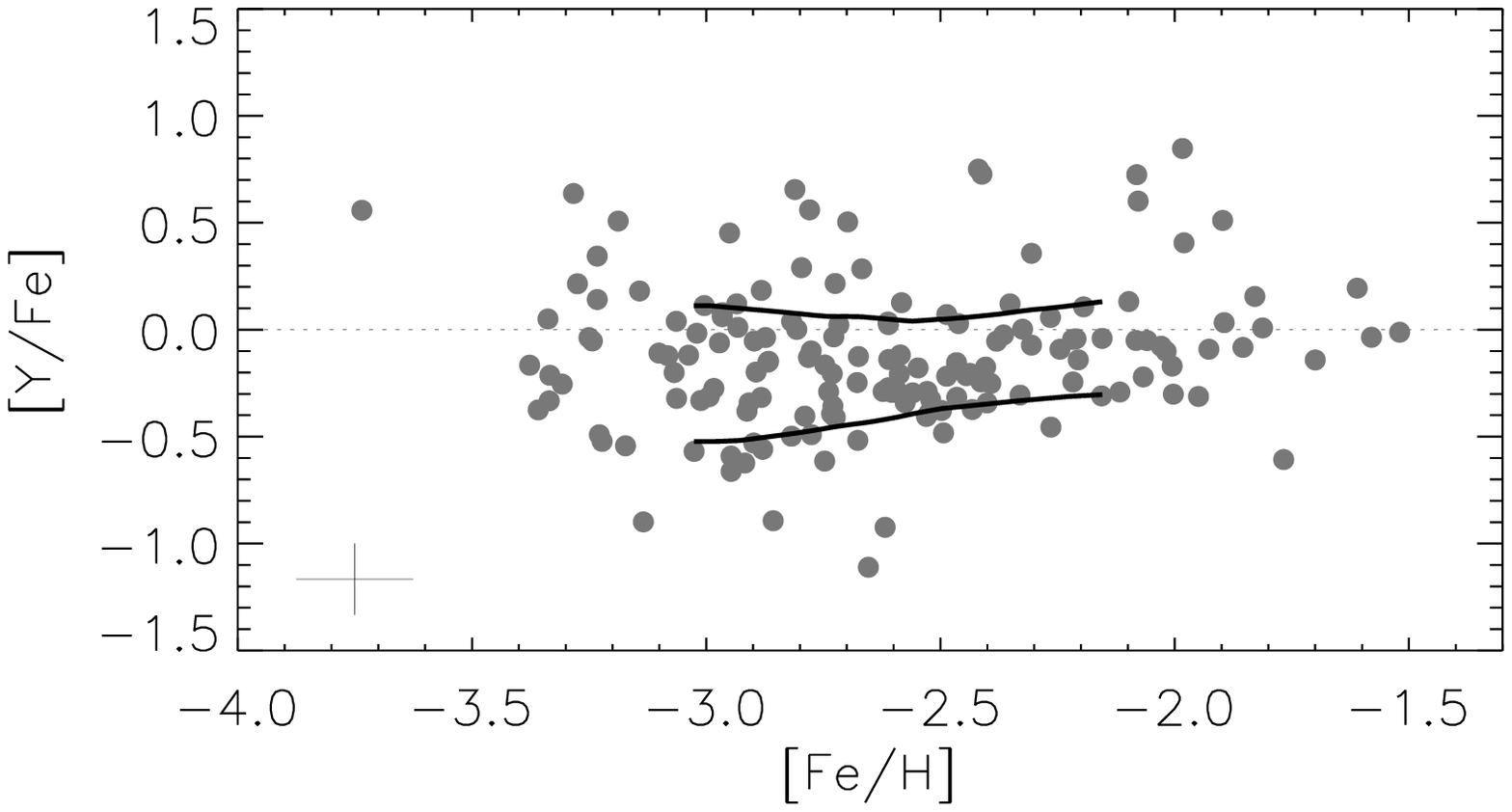}}}  &
\resizebox{60mm}{!}{\rotatebox{0}{\includegraphics{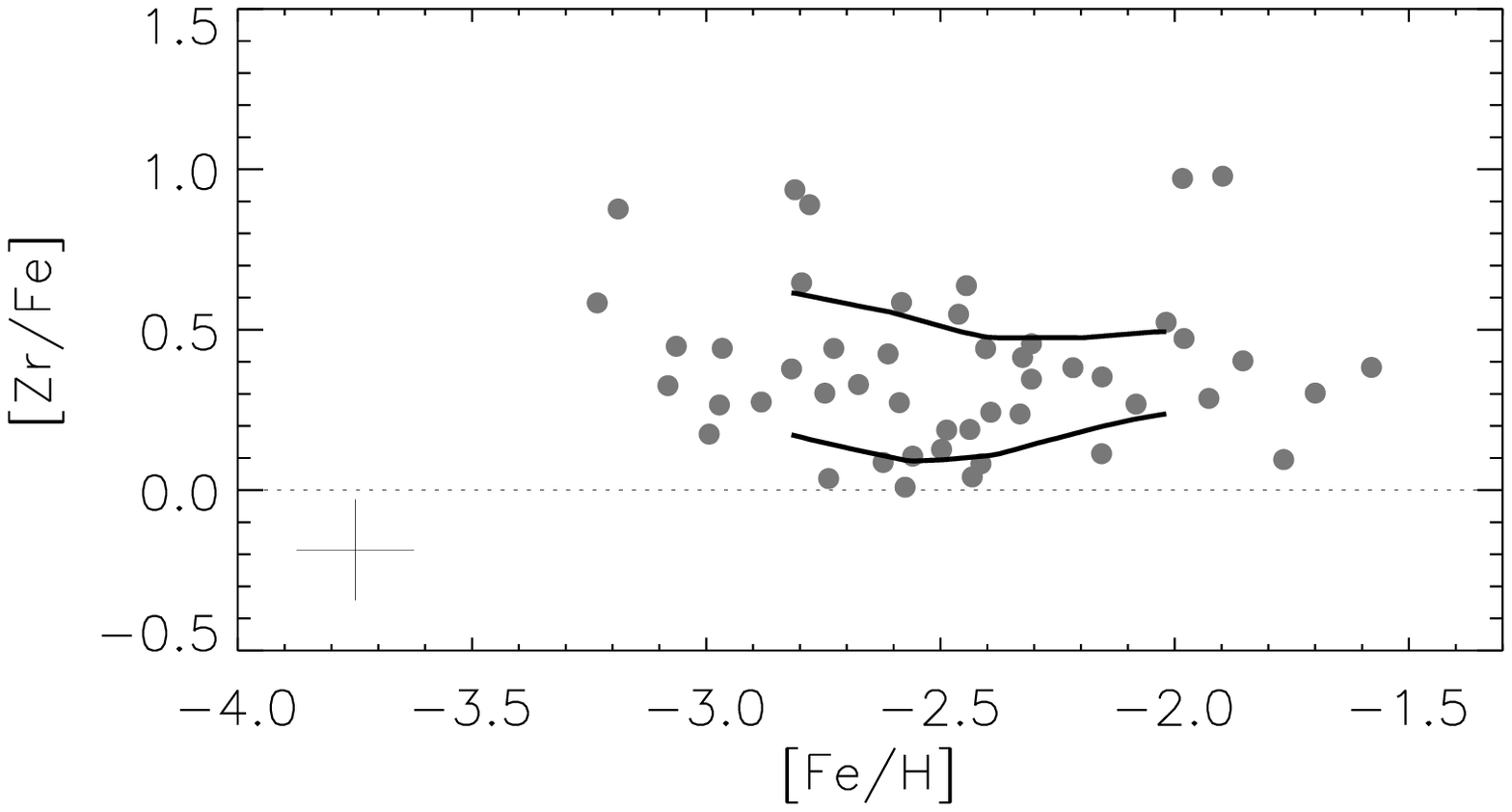}}} \\
\resizebox{60mm}{!}{\rotatebox{0}{\includegraphics{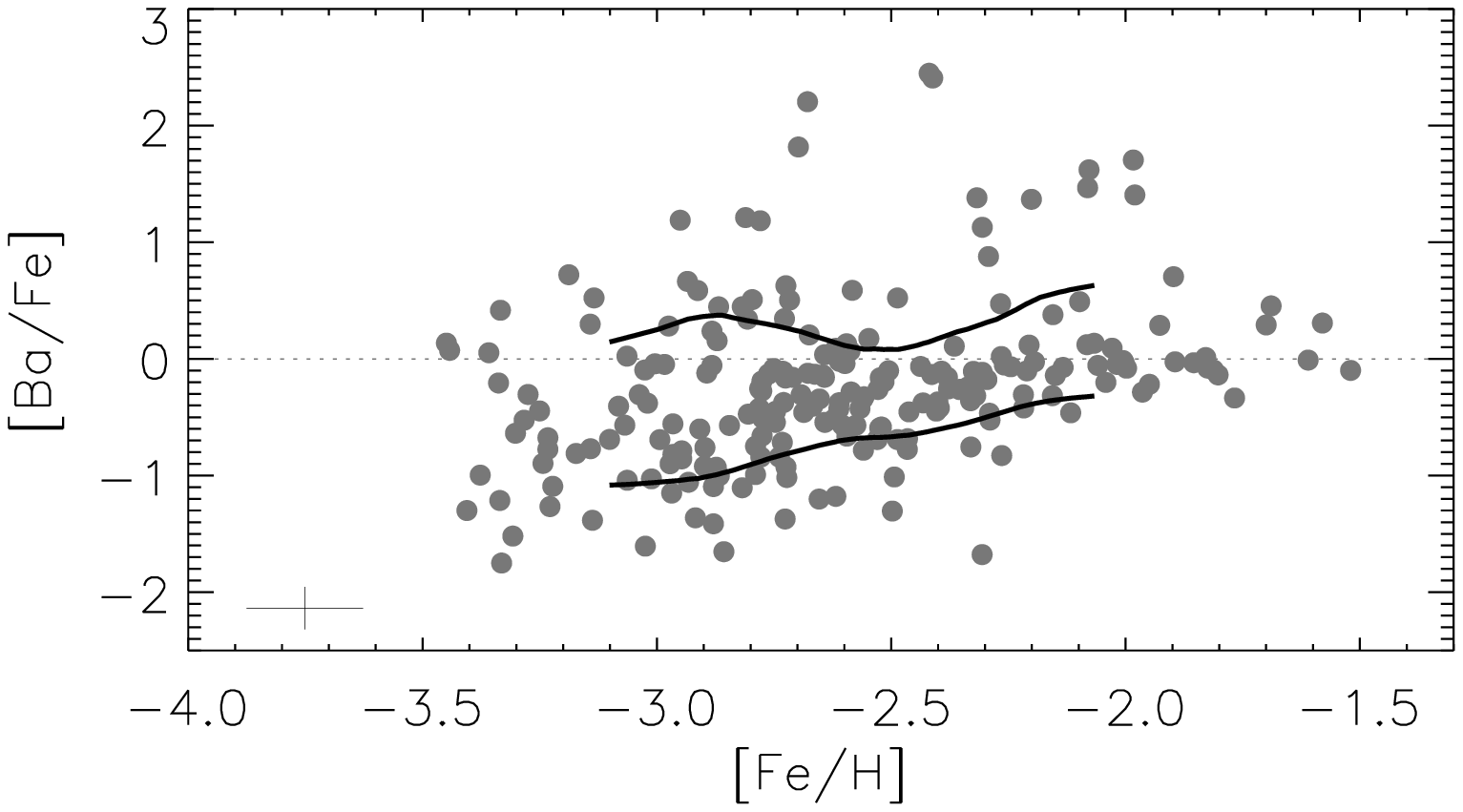}}} &
\resizebox{60mm}{!}{\rotatebox{0}{\includegraphics{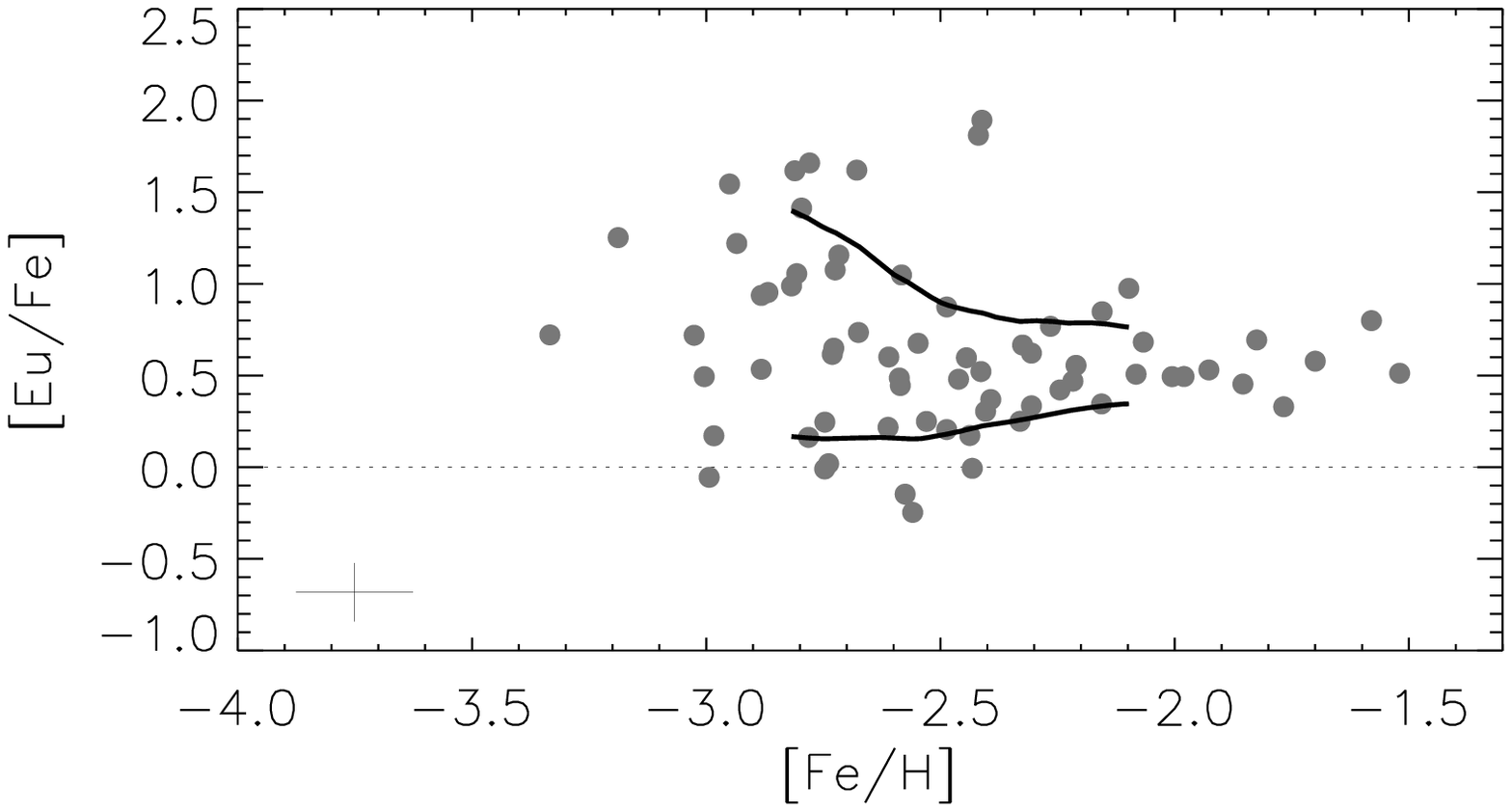}}} \\
\end{tabular}
\end{center}
\caption{Abundances ratios [X/Fe] plotted against [Fe/H] for elements with significant numbers of detections.  Full lines show estimated $1\sigma$ scatter.  The average relative error bars are shown in the bottom left.  Note differing scales on the $y$-axes.}
\label{fig:abunds}
\end{figure*}

\begin{figure*}
\tabcolsep 0mm
\renewcommand{\arraystretch}{3.0}
\begin{center}
\begin{tabular}{ccc}
\resizebox{60mm}{!}{\rotatebox{0}{\includegraphics{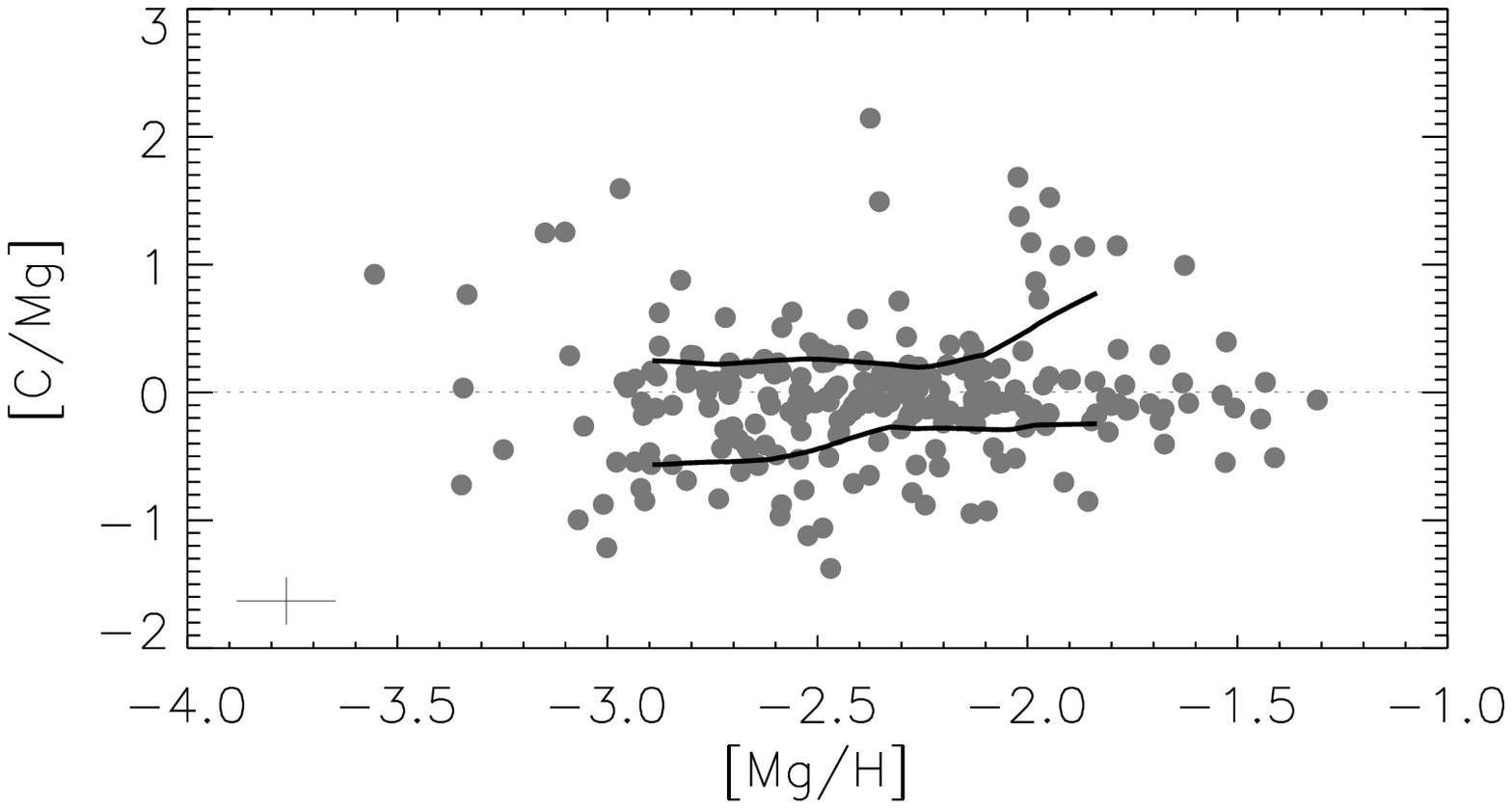}}} &
\resizebox{60mm}{!}{\rotatebox{0}{\includegraphics{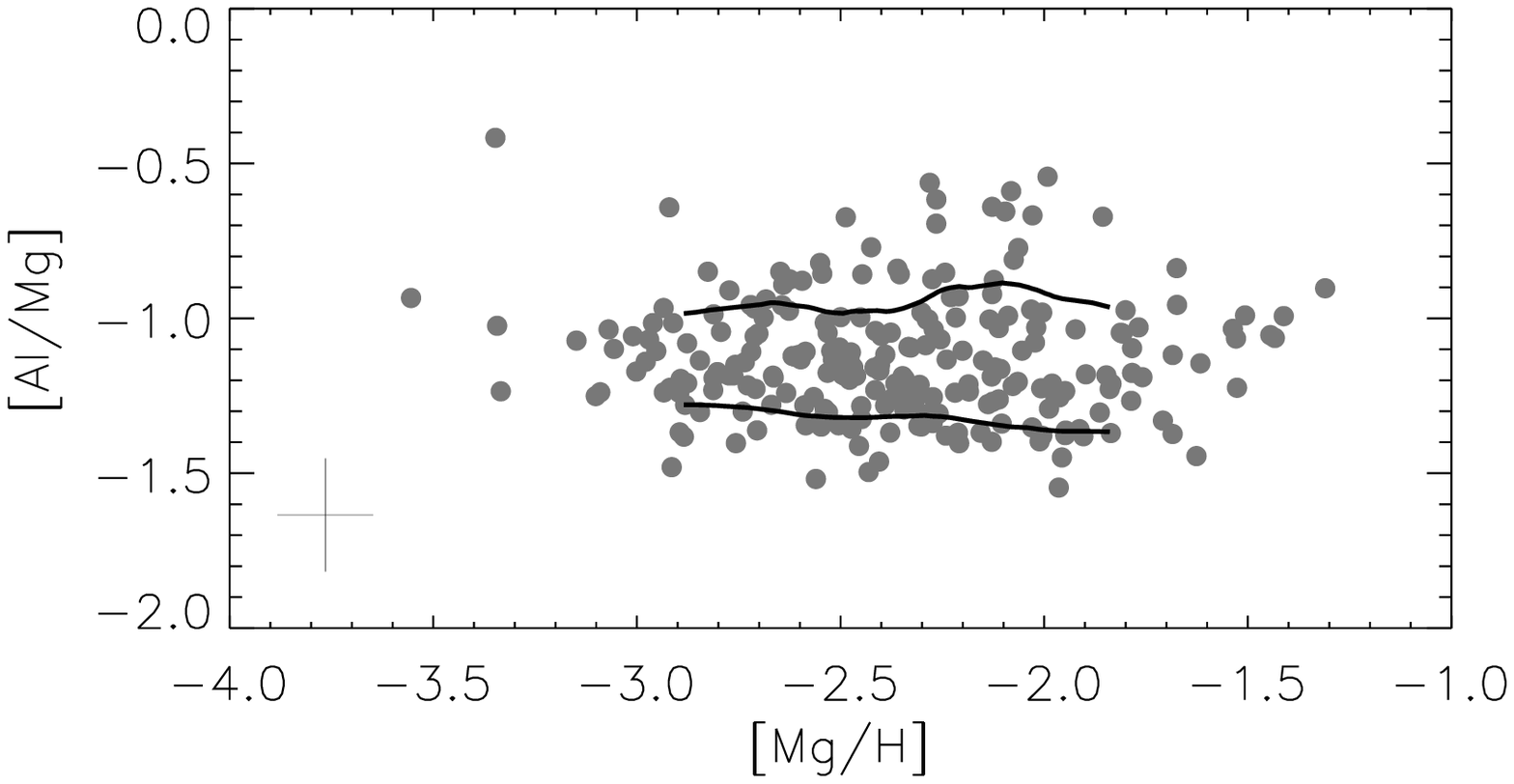}}} &
\resizebox{60mm}{!}{\rotatebox{0}{\includegraphics{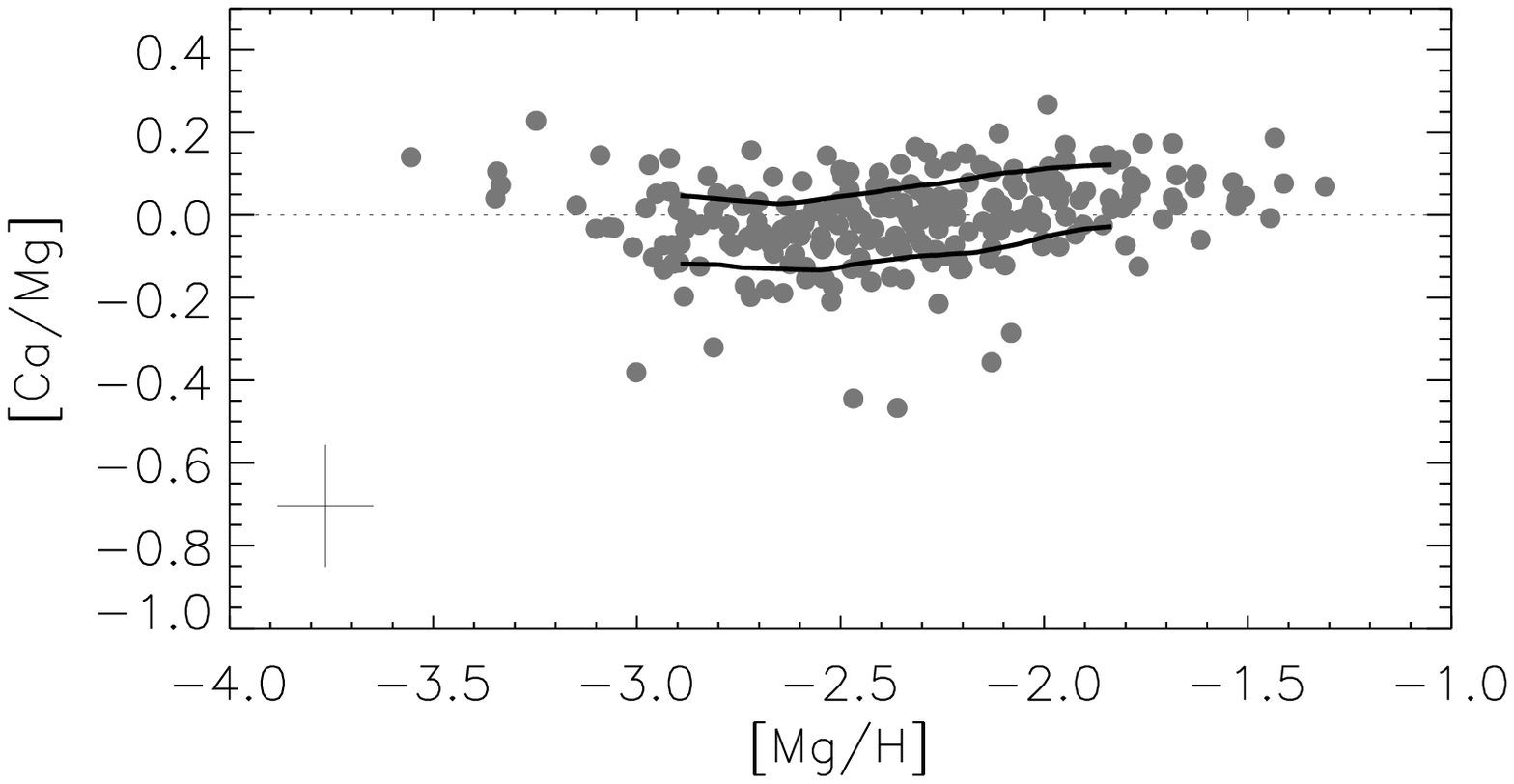}}} \\
\resizebox{60mm}{!}{\rotatebox{0}{\includegraphics{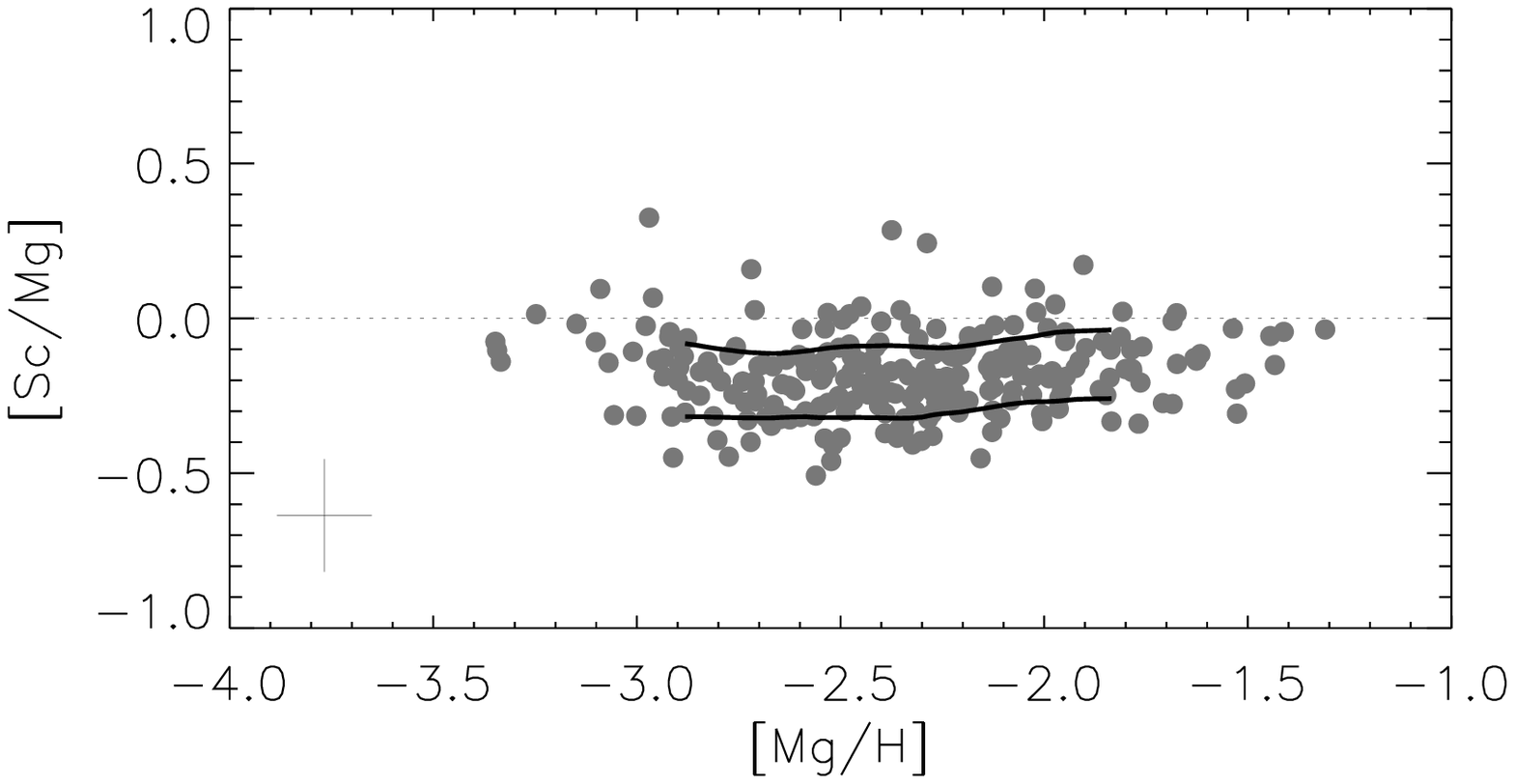}}} &
\resizebox{60mm}{!}{\rotatebox{0}{\includegraphics{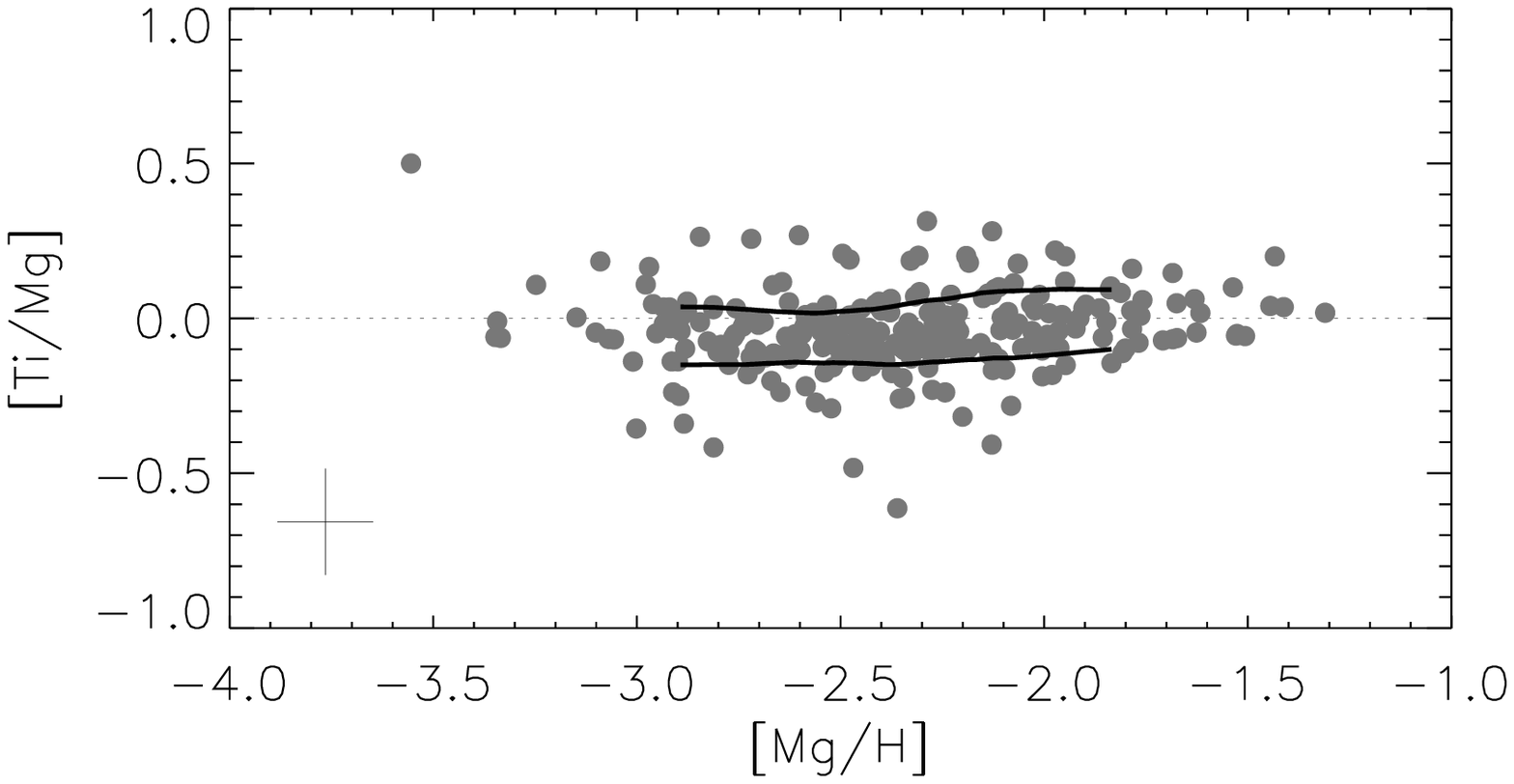}}} &
\resizebox{60mm}{!}{\rotatebox{0}{\includegraphics{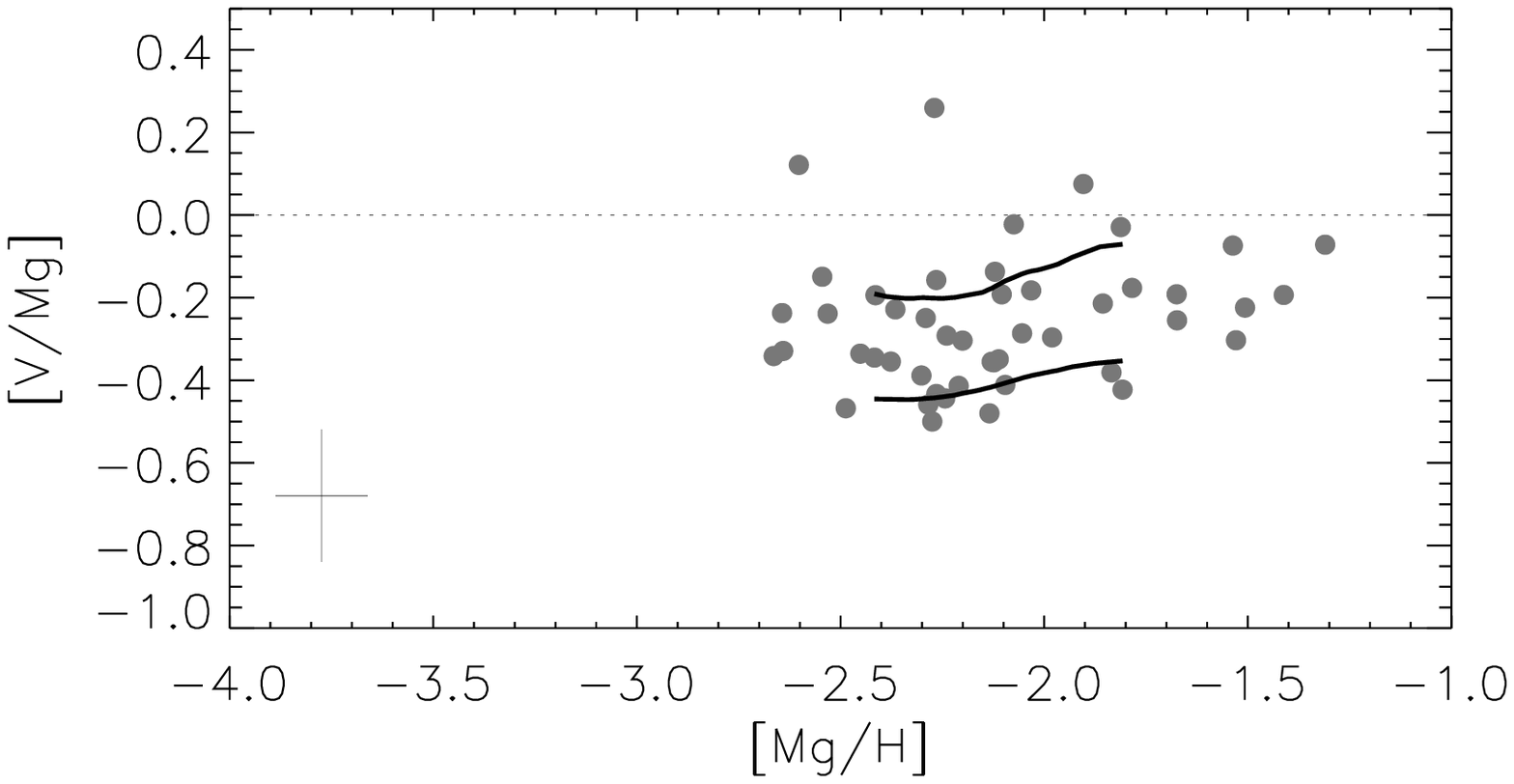}}}  \\
\resizebox{60mm}{!}{\rotatebox{0}{\includegraphics{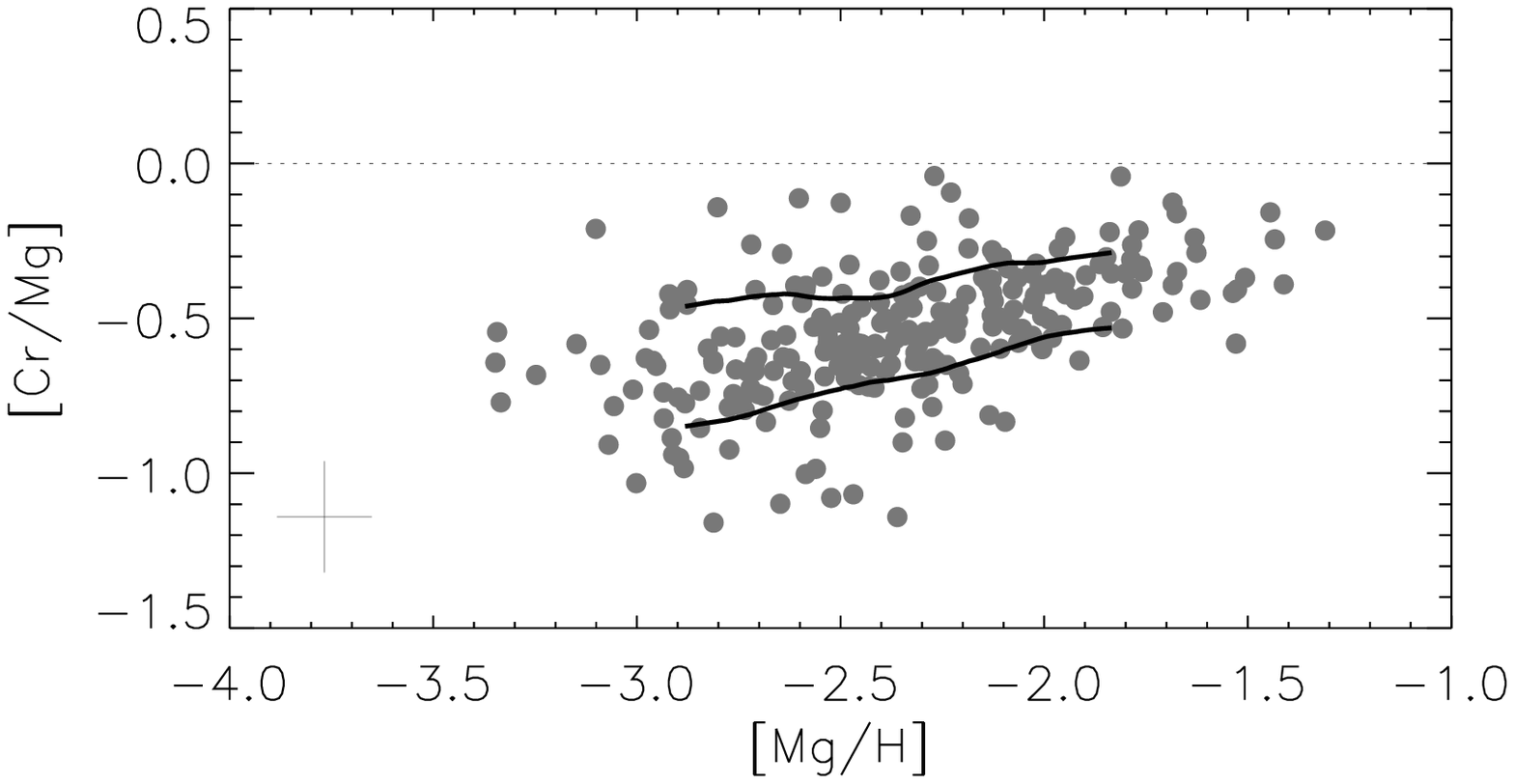}}} &
\resizebox{60mm}{!}{\rotatebox{0}{\includegraphics{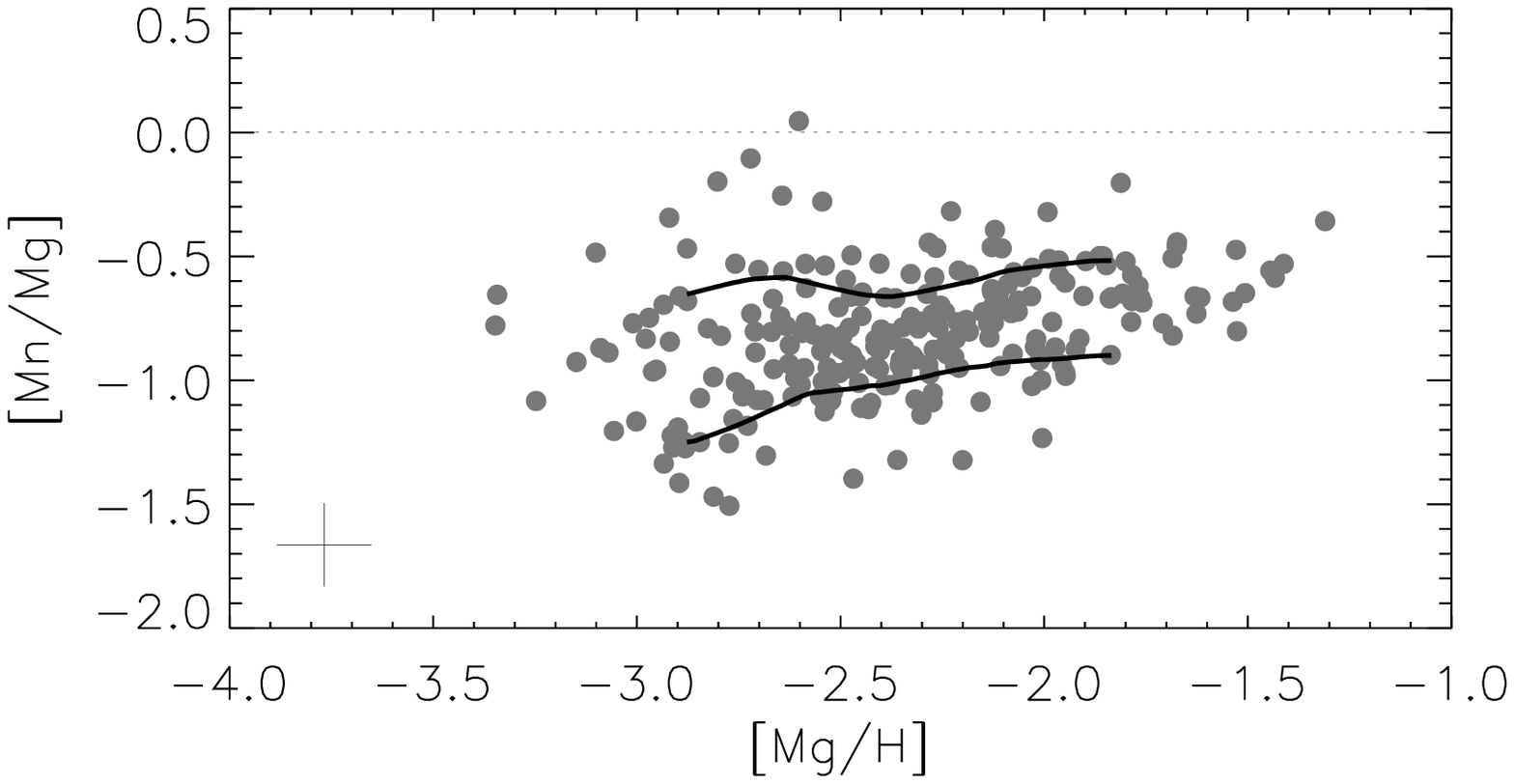}}} &
\resizebox{60mm}{!}{\rotatebox{0}{\includegraphics{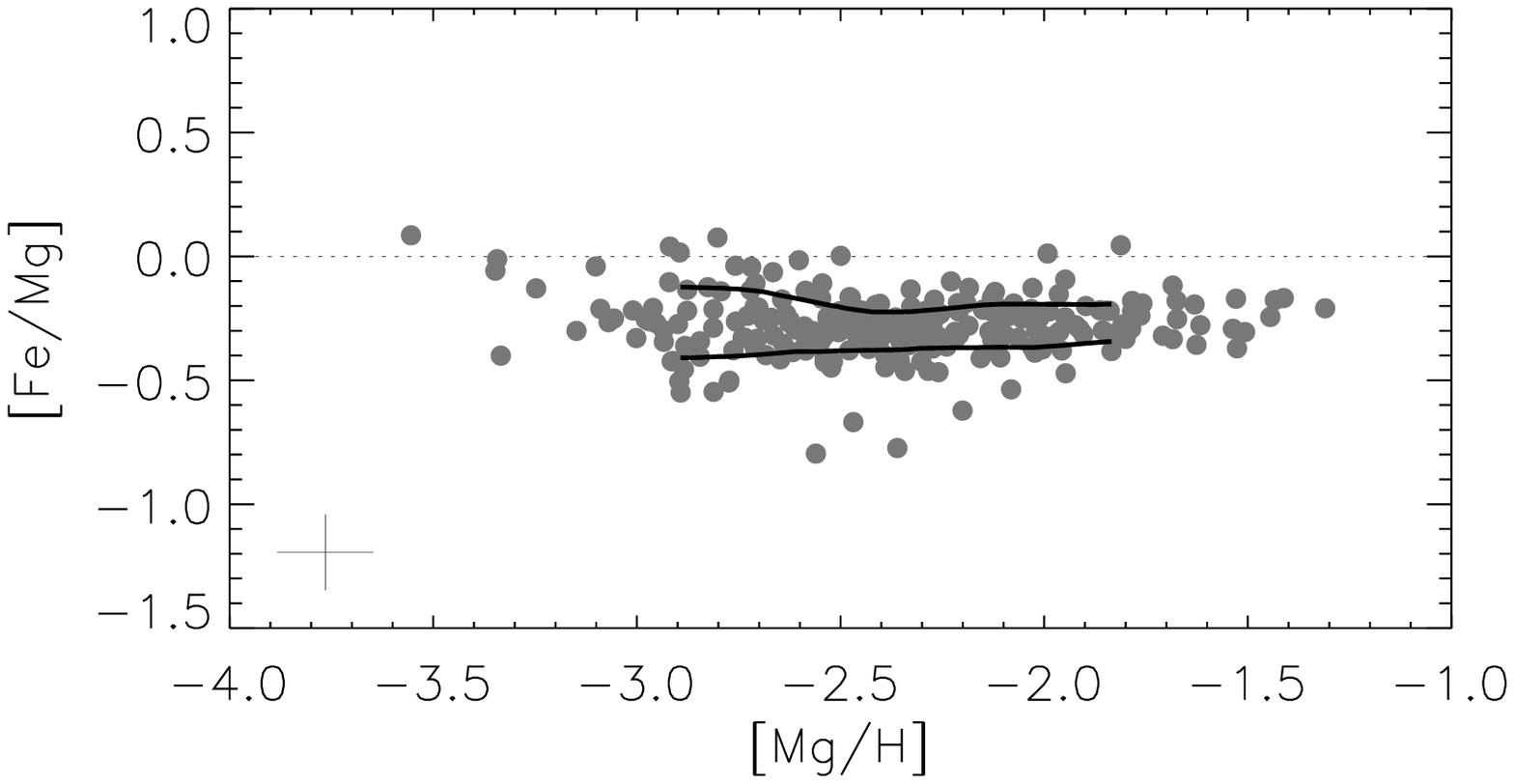}}} \\
\resizebox{60mm}{!}{\rotatebox{0}{\includegraphics{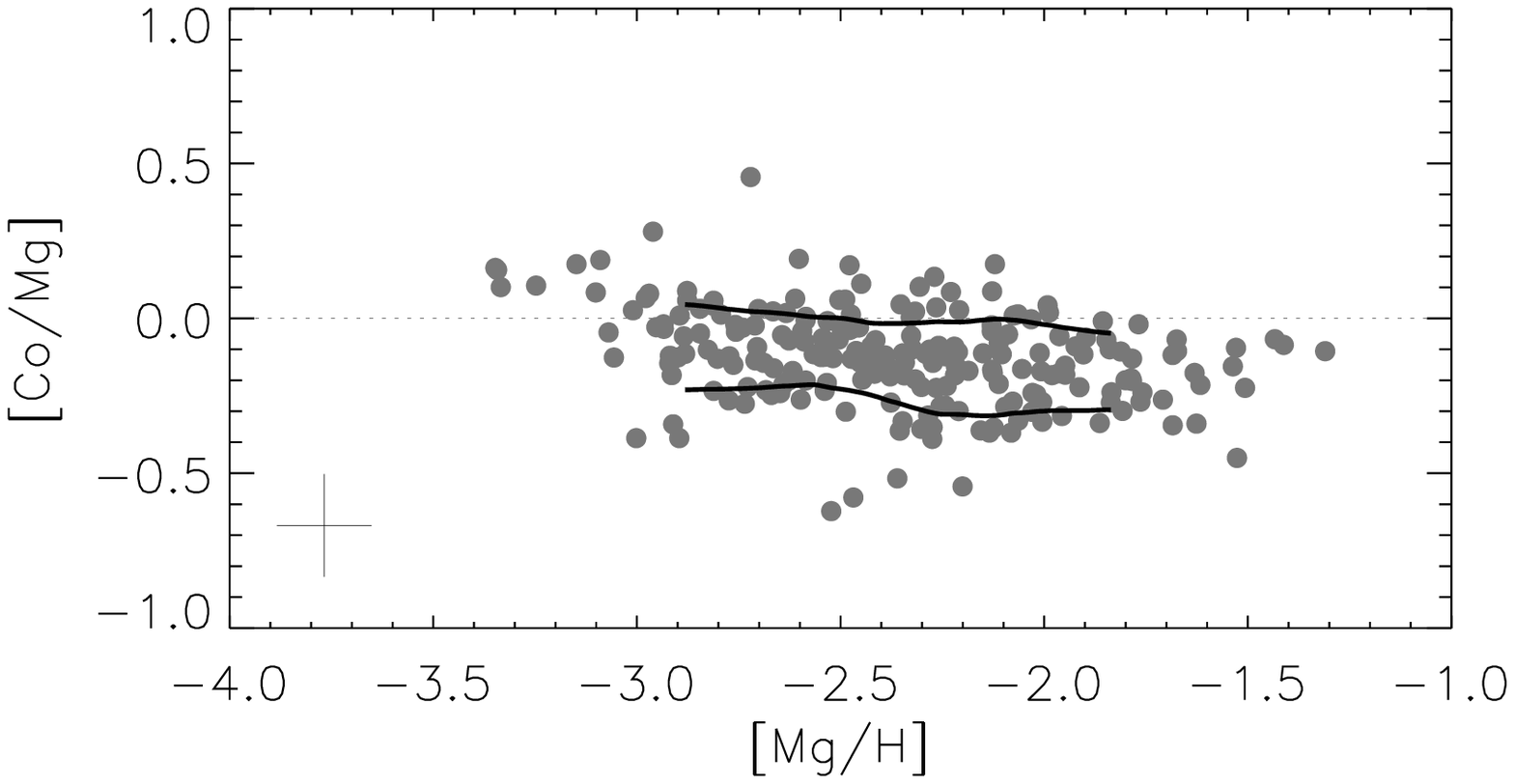}}} &
\resizebox{60mm}{!}{\rotatebox{0}{\includegraphics{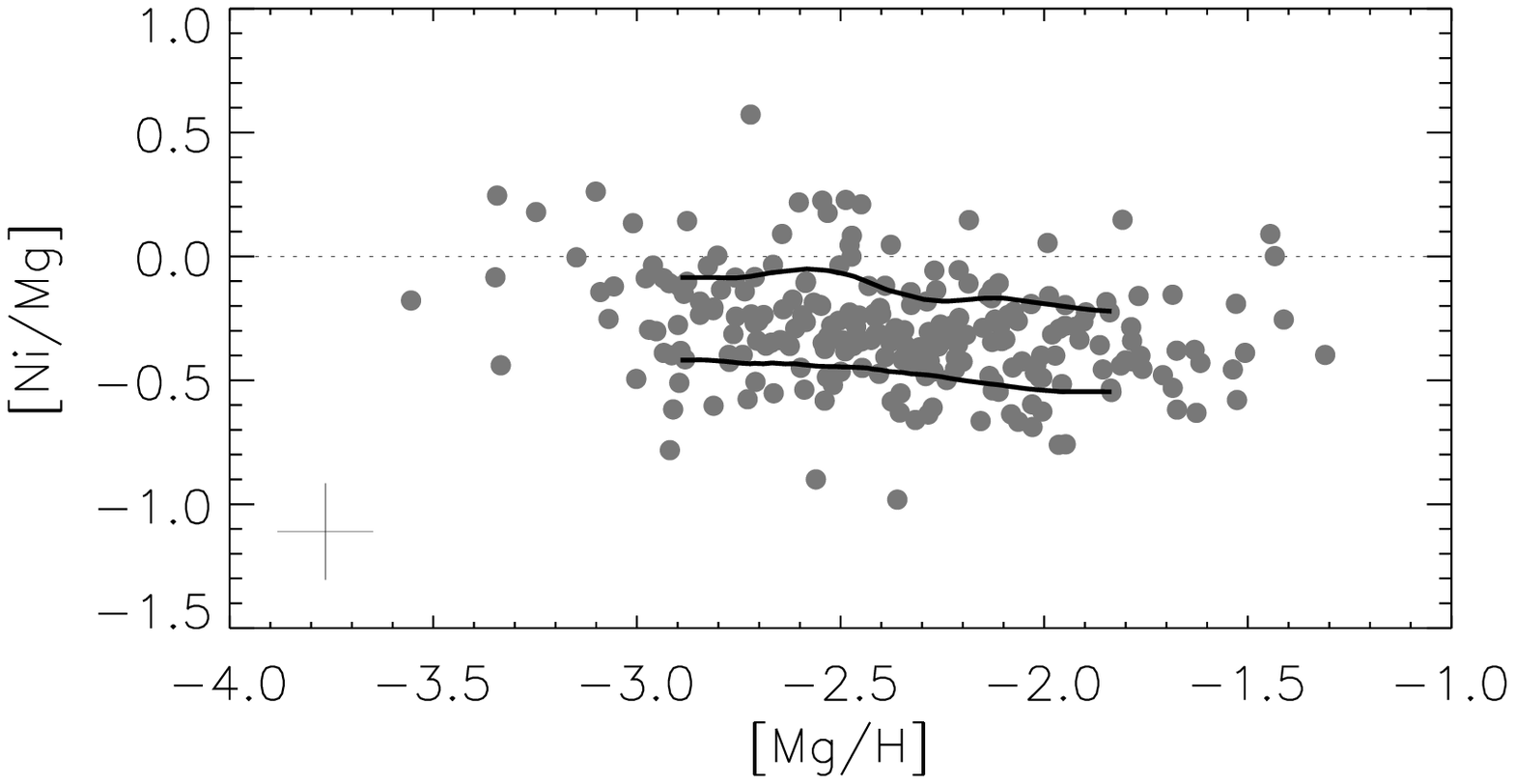}}} &
\resizebox{60mm}{!}{\rotatebox{0}{\includegraphics{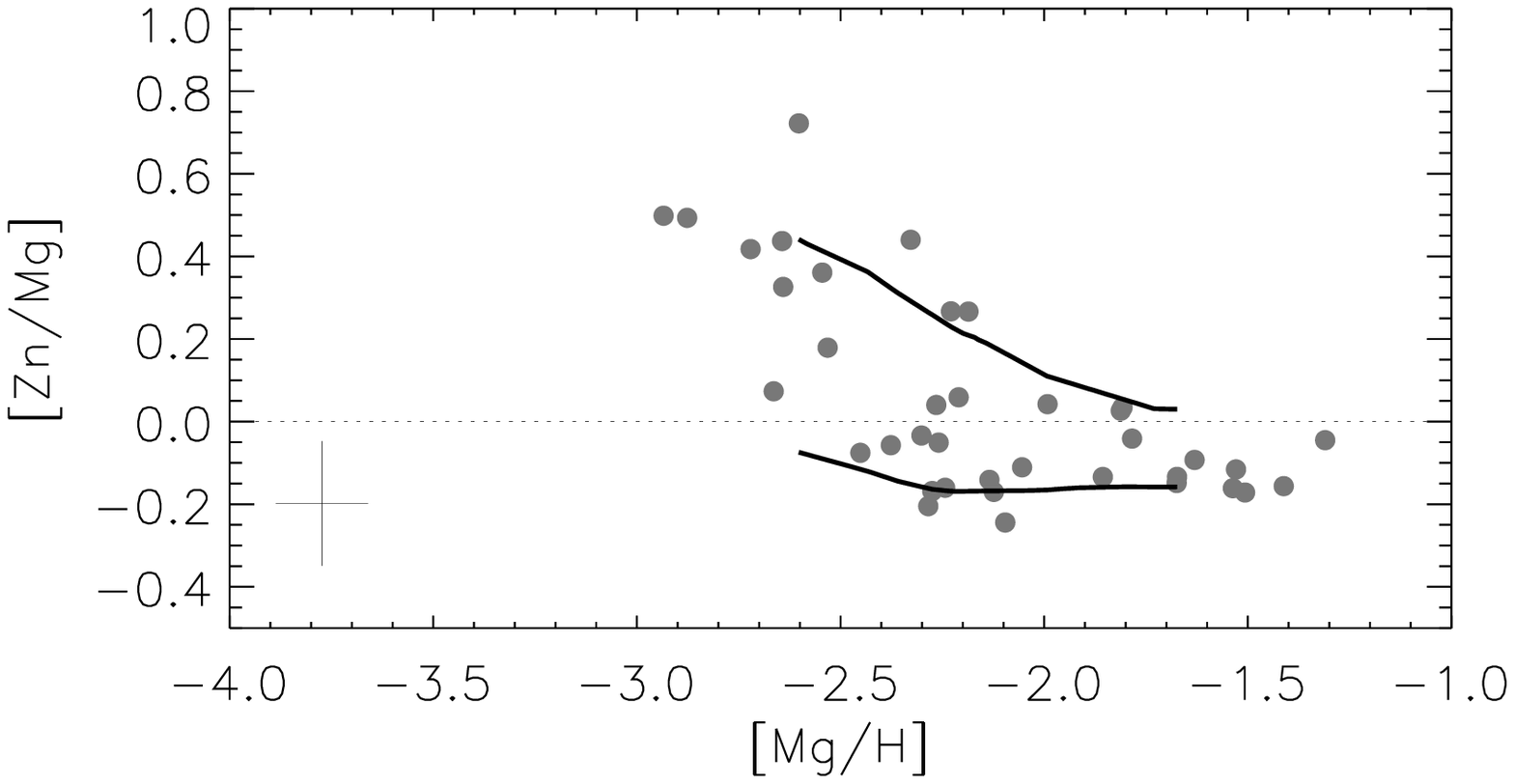}}} \\
\resizebox{60mm}{!}{\rotatebox{0}{\includegraphics{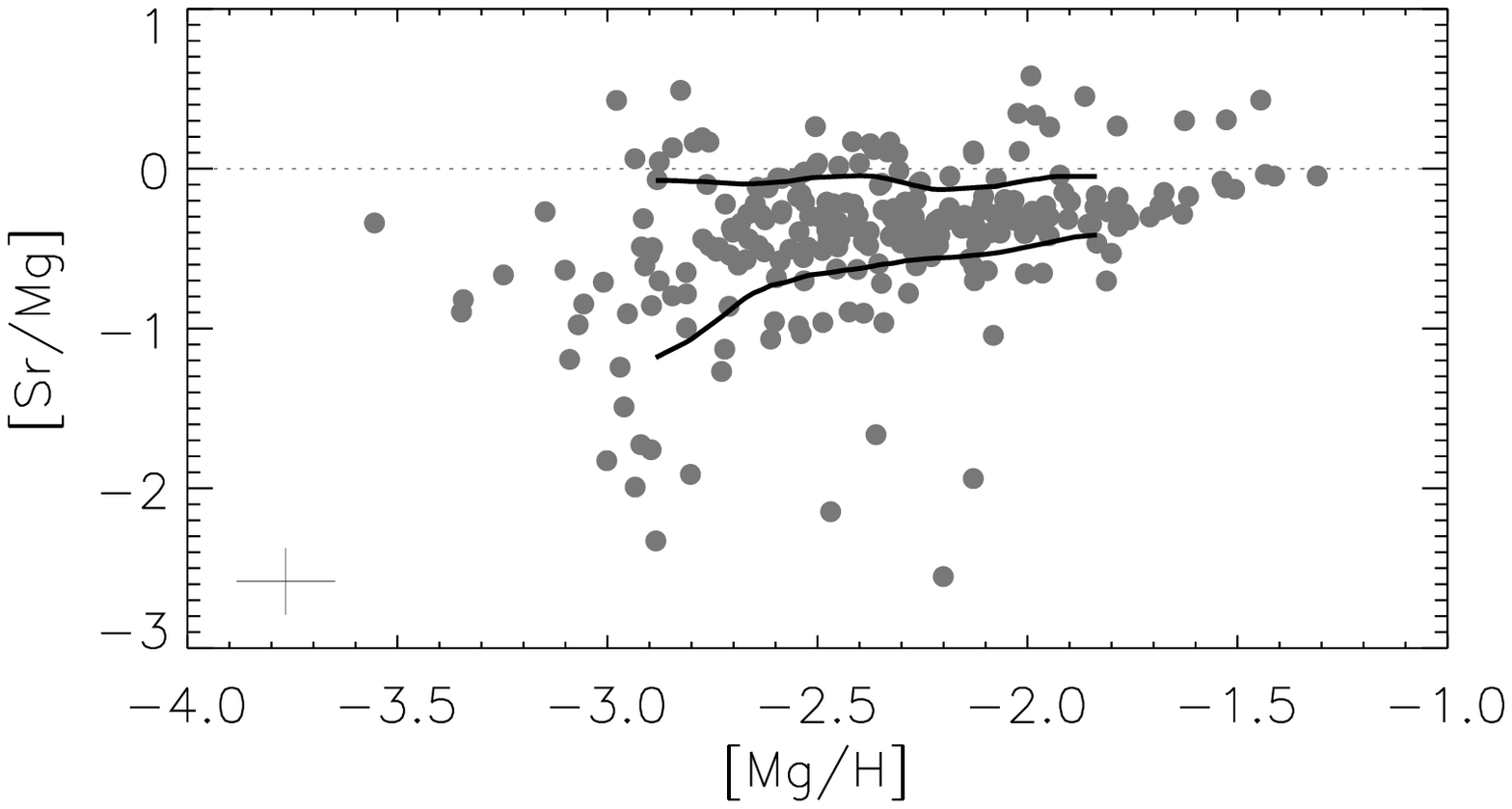}}} &
\resizebox{60mm}{!}{\rotatebox{0}{\includegraphics{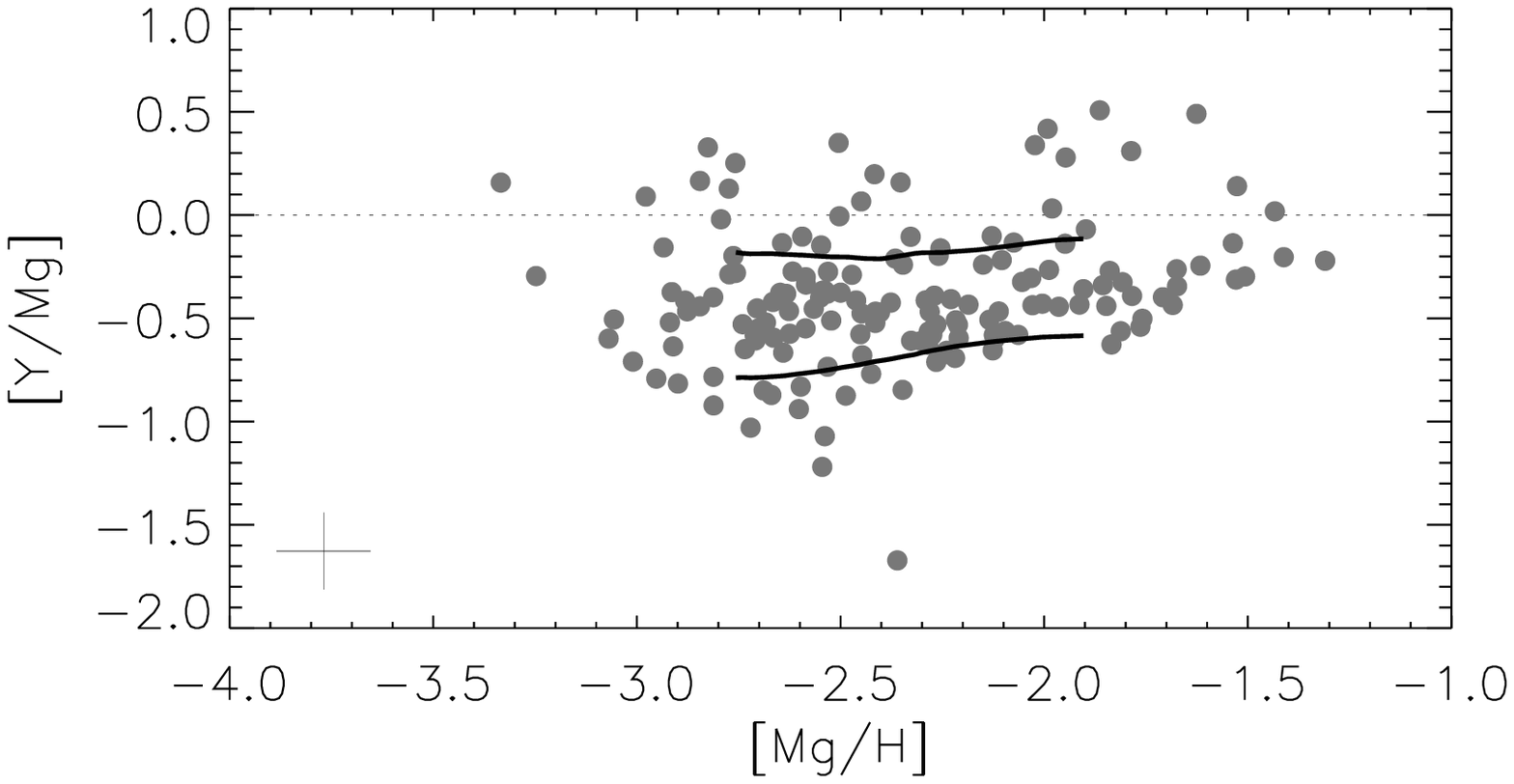}}}  &
\resizebox{60mm}{!}{\rotatebox{0}{\includegraphics{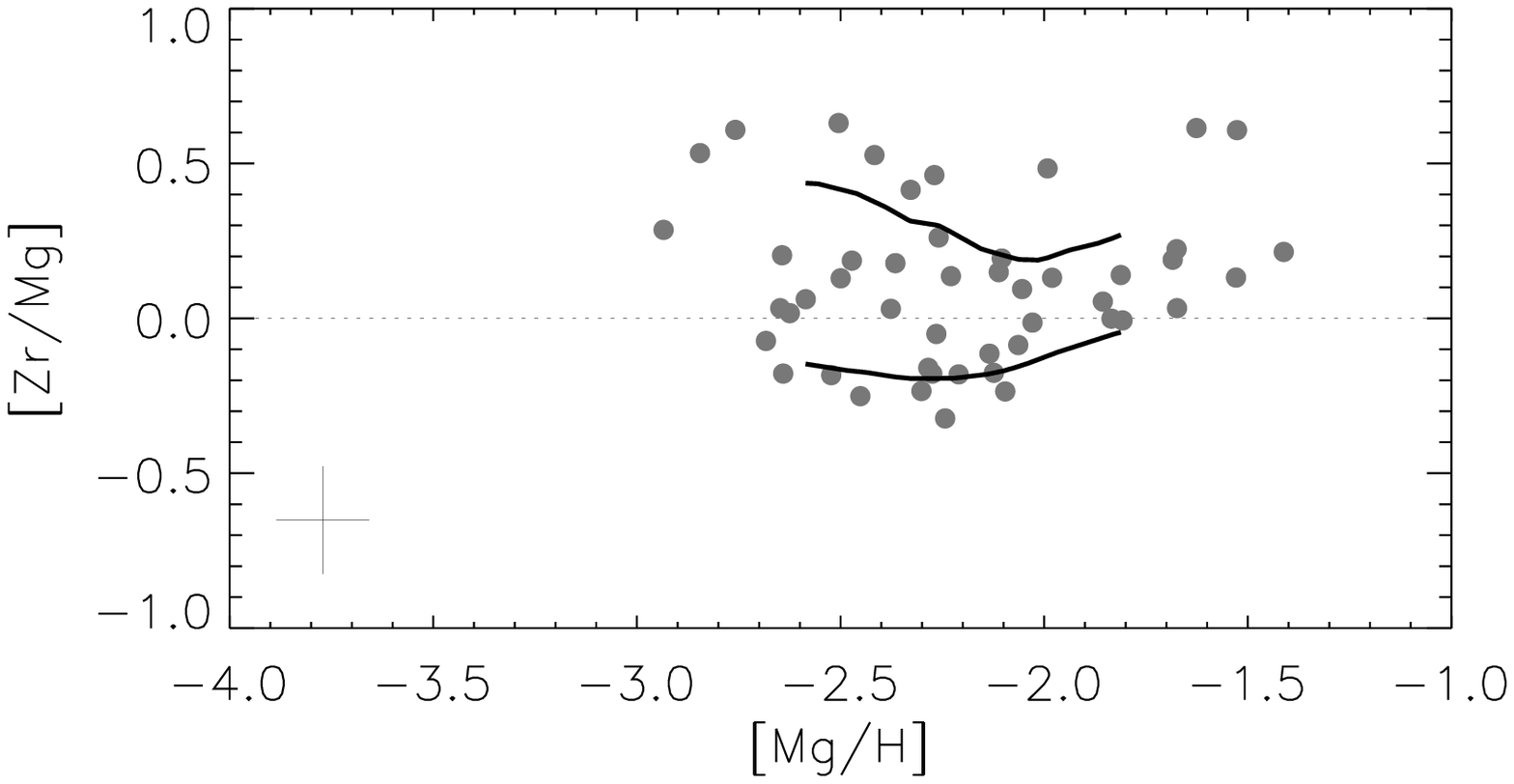}}} \\
\resizebox{60mm}{!}{\rotatebox{0}{\includegraphics{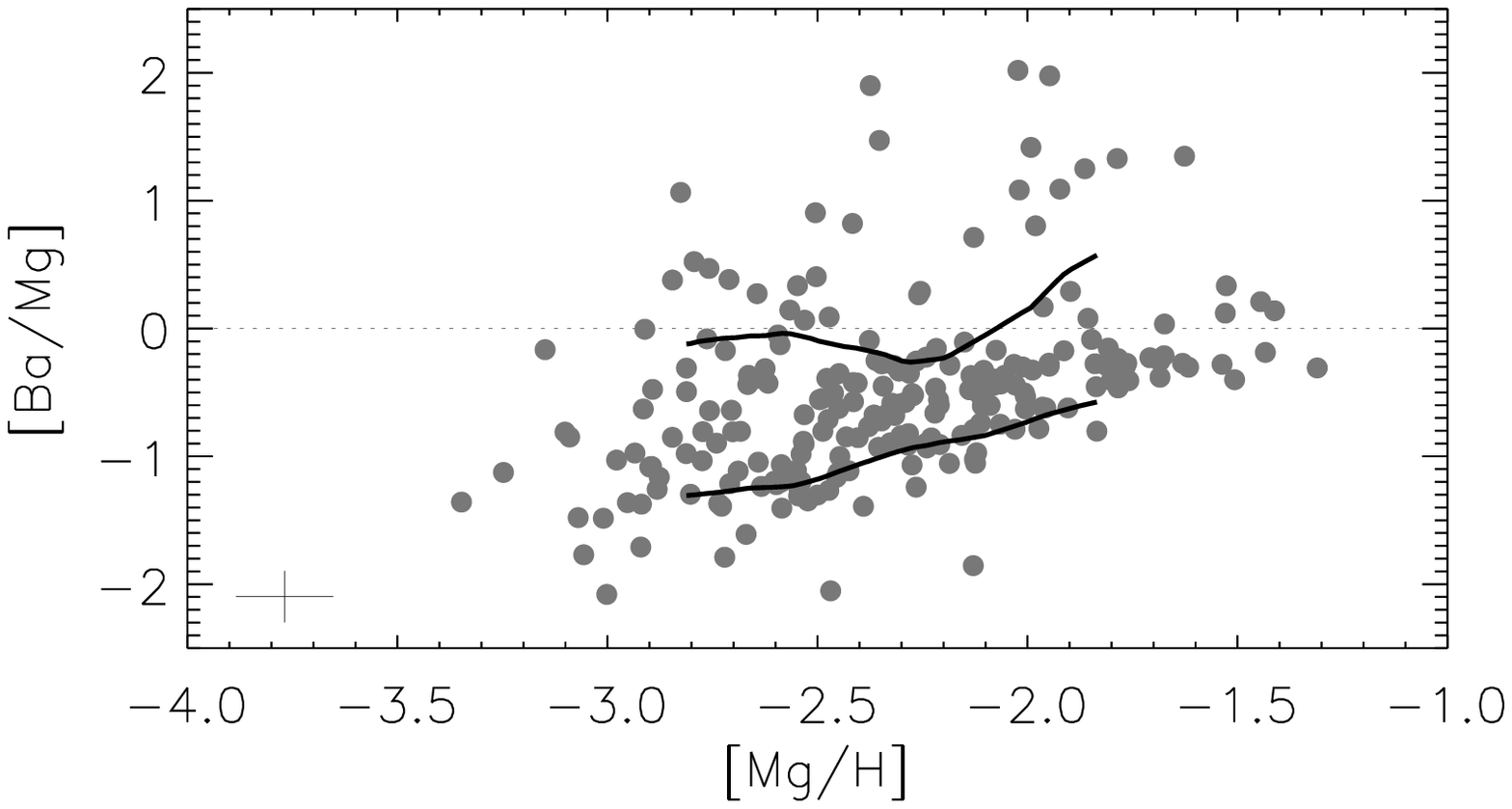}}} &
\resizebox{60mm}{!}{\rotatebox{0}{\includegraphics{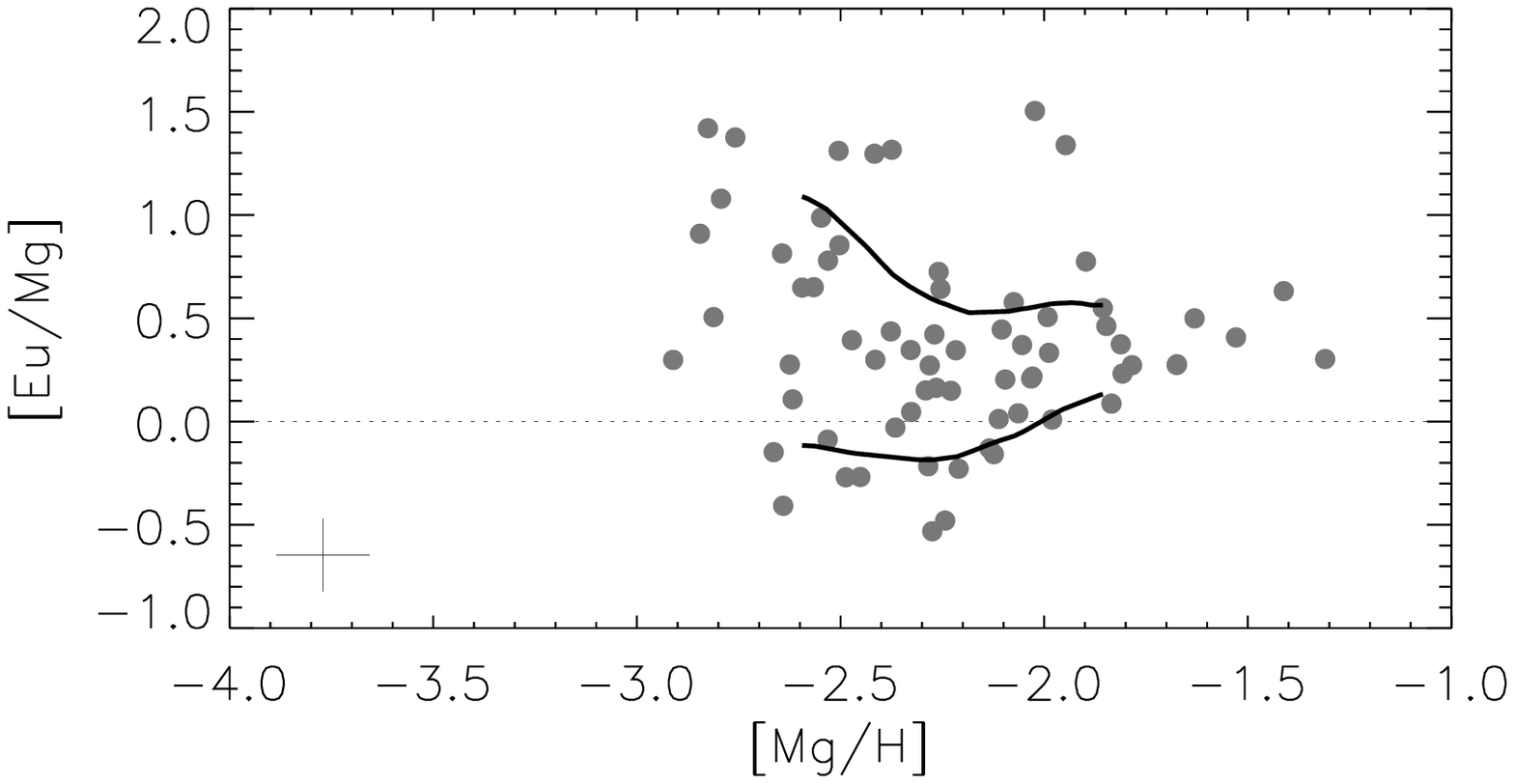}}} \\
\end{tabular}
\end{center}
\caption{Abundances ratios [X/Mg] plotted against [Mg/H] for elements with significant numbers of detections. Full lines show estimated $1\sigma$ scatter.   The average relative error bars are shown in the bottom left.  Note differing scales on the $y$-axes.}
\label{fig:abunds_mg}
\end{figure*}

We now discuss the general behaviour of the abundances, particularly trends and scatter in abundance ratios of interest.  The abundance ratio trends of [X/Fe] with [Fe/H] and [X/Mg] with [Mg/H] have been plotted in Figs.~\ref{fig:abunds} and~\ref{fig:abunds_mg} respectively for elements with a reasonable number of detections.  The estimated $1\sigma$ scatter in the $y$-variable is shown, following the definition for the scatter used by Karlsson \& Gustafsson~(\cite{karlsson05}), such that at a given $x$-coordinate 32\% of the stars lie outside the $1\sigma$ scatter lines, 16\% above and below.  In practice, this has been computed for a given $x$-coordinate by taking the $n$ nearest stars in the $x$-variable, where $n$ is the lesser of $n=50$ or $n_\mathrm{stars}/3$, with $n_\mathrm{stars}$ being the number of stars in the diagram.  The results are smoothed over the average width of bins containing $n$ stars to remove transient behaviour caused by outliers.  In Table~\ref{tab:scatter} we compare the mean measured scatter in [X/Fe] and [X/Mg], across the range of [Fe/H] and [Mg/H] respectively, with the average error estimates.  The minimum and maximum measured scatters are also reported to give an indication of the range of variation with [Fe/H] and [Mg/H].

\begin{table}
\tabcolsep 1mm
\begin{center}
\caption{Comparison of measured scatter in Figs.~\ref{fig:abunds} and~\ref{fig:abunds_mg} with relative errors in the abundance ratios.  For each plot we report the minimum, mean and maximum measured $1\sigma$ scatter $\sigma_\mathrm{meas}$ across the range of [Fe/H] or [Mg/H], which is compared to the average relative error in the abundance ratio $\sigma^\mathrm{rel}$.  The ratio of the mean measured scatter to the estimated error $\langle \sigma_\mathrm{meas} \rangle /\sigma^\mathrm{rel}$ is reported in the last column.}
\label{tab:scatter}
\begin{tabular}{lccccc}
\hline
 & min($\sigma_\mathrm{meas}$) &  $\langle \sigma_\mathrm{meas} \rangle$ & max($\sigma_\mathrm{meas}$) & $\sigma^\mathrm{rel}$ & $\langle \sigma_\mathrm{meas} \rangle /\sigma^\mathrm{rel}$ \\
\hline    
 {   [C/Fe]}   & 0.22   & 0.32  & 0.53   &  0.18   &  1.73  \\  
 {  [Mg/Fe]}   & 0.07   & 0.09  & 0.12   &  0.15   &  0.59  \\  
 {  [Al/Fe]}   & 0.17   & 0.20  & 0.21   &  0.17   &  1.13  \\  
 {  [Ca/Fe]}   & 0.08   & 0.09  & 0.10   &  0.15   &  0.59  \\  
 {  [Sc/Fe]}   & 0.09   & 0.11  & 0.13   &  0.16   &  0.69  \\  
 {  [Ti/Fe]}   & 0.08   & 0.09  & 0.12   &  0.15   &  0.61  \\  
 {   [V/Fe]}   & 0.06   & 0.09  & 0.11   &  0.15   &  0.60  \\  
 {  [Cr/Fe]}   & 0.10   & 0.12  & 0.14   &  0.16   &  0.72  \\  
 {  [Mn/Fe]}   & 0.13   & 0.16  & 0.21   &  0.16   &  1.03  \\  
 {  [Co/Fe]}   & 0.10   & 0.11  & 0.13   &  0.16   &  0.71  \\  
 {  [Ni/Fe]}   & 0.12   & 0.14  & 0.17   &  0.18   &  0.80  \\  
 {  [Zn/Fe]}   & 0.08   & 0.15  & 0.19   &  0.15   &  1.01  \\  
 {  [Sr/Fe]}   & 0.19   & 0.32  & 0.59   &  0.19   &  1.71  \\  
 {   [Y/Fe]}   & 0.21   & 0.25  & 0.32   &  0.17   &  1.47  \\  
 {  [Zr/Fe]}   & 0.13   & 0.19  & 0.23   &  0.16   &  1.18  \\  
 {  [Ba/Fe]}   & 0.37   & 0.50  & 0.69   &  0.18   &  2.77  \\  
 {  [Eu/Fe]}   & 0.21   & 0.37  & 0.62   &  0.16   &  2.34  \\  
 {	   }   &	      &       &        &	 &	  \\  
 {   [C/Mg]}   & 0.24   & 0.35  & 0.51   &  0.18   &  1.89  \\  
 {  [Al/Mg]}   & 0.15   & 0.19  & 0.23   &  0.18   &  1.03  \\  
 {  [Ca/Mg]}   & 0.07   & 0.08  & 0.09   &  0.15   &  0.56  \\  
 {  [Sc/Mg]}   & 0.10   & 0.11  & 0.12   &  0.18   &  0.60  \\  
 {  [Ti/Mg]}   & 0.08   & 0.09  & 0.11   &  0.17   &  0.55  \\  
 {   [V/Mg]}   & 0.12   & 0.13  & 0.14   &  0.16   &  0.79  \\  
 {  [Cr/Mg]}   & 0.12   & 0.15  & 0.19   &  0.18   &  0.84  \\  
 {  [Mn/Mg]}   & 0.17   & 0.21  & 0.30   &  0.17   &  1.26  \\  
 {  [Fe/Mg]}   & 0.08   & 0.10  & 0.14   &  0.15   &  0.63  \\  
 {  [Co/Mg]}   & 0.11   & 0.13  & 0.16   &  0.17   &  0.79  \\  
 {  [Ni/Mg]}   & 0.15   & 0.17  & 0.19   &  0.19   &  0.88  \\  
 {  [Zn/Mg]}   & 0.09   & 0.18  & 0.26   &  0.15   &  1.16  \\  
 {  [Sr/Mg]}   & 0.18   & 0.29  & 0.55   &  0.21   &  1.41  \\  
 {   [Y/Mg]}   & 0.23   & 0.25  & 0.30   &  0.19   &  1.36  \\  
 {  [Zr/Mg]}   & 0.15   & 0.22  & 0.29   &  0.17   &  1.27  \\  
 {  [Ba/Mg]}   & 0.33   & 0.48  & 0.60   &  0.20   &  2.40  \\  
 {  [Eu/Mg]}   & 0.22   & 0.39  & 0.60   &  0.18   &  2.18  \\  
\hline
\end{tabular}
\end{center}
\end{table}

For Mg, Ca, Sc, Ti, Cr, Fe, Co, and Ni, the measured scatters are slightly less than the scatters expected from the typical estimated relative error in the abundance ratios, suggesting that our computed errors overestimate the real relative error.  Bearing in mind the difficulties in accurately calculating the errors, this suggests that the cosmic scatter is small (or even non-existent) at these metallicities.  These results are in line with previous results for smaller samples with smaller uncertainties for some of these elements, such as the studies by Arnone~et~al.~(\cite{arnone05}), Cayrel~et~al.~(\cite{cayrel04}) and Cohen et~al.~(\cite{cohen04}).  It should be noted that in all these elements, apparent outliers may perhaps be explained simply by the errors in the analysis; in a sample of 253 stars outliers due to random errors are expected.  In particular, examining the outliers we noted that they often showed trends in the abundances derived from individual Fe I lines with excitation that were at the edge of the distribution of such slopes for the entire sample (see Fig.~\ref{fig:equilibria_hist}), and occur in elements where particularly temperature sensitive spectral features have been employed, such as Co and Ni.  Thus, we believe that more often than not for these elements, the outliers are more likely due to errors in the analysis, such as $T_\mathrm{eff}$ being in error, than indicative of any real over- or under-abundance of the element in question.   

The results for Al, V, Mn and Zn are less clear.  In the case of Al, the scatters are marginally larger than the typical error bars; however, we note the Al abundances are based on a single line and are very sensitive to the $S/N$ of the spectrum (see Table~\ref{tab:sn_comparison}).  Further, this line is known to be subject to deviations from LTE.   Both these facts may lead to increased scatter in the derived abundances.  In the cases of V and Zn, the number of detections is small and the abundances for these elements are based on quite weak features and are therefore susceptible to overestimation due to unresolved blends.  For Mn, the scatter at low metallicity does marginally exceed the error estimates, with the hint of a ``bump'' of Mn enhanced stars at around $\mathrm{[Fe/H]}\sim -3$.  A similar feature is suggested by the results of Cayrel~et~al.~(\cite{cayrel04}). 

The elements C, Sr, Y, Ba and Eu, and perhaps Zr, show scatter at low metallicities, $\mathrm{[Fe/H]} \la -2.5$ and $\mathrm{[Mg/H]} \la -2.2$, significantly larger than can be explained from the errors in the analysis, implying the scatter is cosmic in origin.  At higher metallicities the scatter among the non-C-rich stars, does not greatly exceed that expected from the errors in the analysis, but perhaps indicates some cosmic scatter.  Note, the scatter in [Ba/Fe] and [Ba/Mg], and to a lesser degree Sr and Y, at higher [Fe/H] and [Mg/H] is affected by C-enhanced stars which have high Ba/Fe and Ba/Mg ratios, see Fig.~\ref{fig:Ba_vs_C}.  The scatter at low metallicity seems, in most cases, to be increasing monotonically with decreasing metallicity.  Similar results have been found, for example, by McWilliam~et~al.~(\cite{mcw95b}), McWilliam~(\cite{mcw98}), Norris~et~al.~(\cite{norris01}), and Burris~et~al.~(\cite{burris00}).  The results of Norris~et~al.~(\cite{norris01}) suggested the existence of a larger scatter in [Sr/Fe] than for [Ba/Fe] at the lowest [Fe/H].  We do not find this result; the scatters are of comparable magnitude.  Note that in contrast to the results for lighter elements discussed above, the cases of under-abundances and over-abundances of these elements with respect to the general trends are sometimes significant.    

In summary, among the stars without strong C enhancement, at about $\mathrm{[Fe/H]} \ga -2.5$ or equivalently $\mathrm{[Mg/H]} \ga -2.2$, our analysis indicates that the cosmic scatter in all abundance ratios is small.  This implies that at around this level of enrichment the Galactic halo was reasonably well mixed.  At lower metallicities C, Sr, Y, Ba and Eu all show evidence for real cosmic scatter, while the results for the remaining elements still admit only a small amount of cosmic scatter within the errors of our analysis.  This dichotomy implies that while the lack of scatter in these elements at $\mathrm{[Fe/H]} \ga -2.5$ might be explained by a well mixed Galactic halo, the small scatter at $\mathrm{[Fe/H]} \la -2.5$ has a different explanation.  One possibilty is suggested by the stochastic models of metal-poor enrichment by Karlsson \& Gustafsson~(\cite{karlsson05}). The small scatter among the most metal-poor stars may be tentatively explained by cosmic selection effects in contributing supernova masses, and a relatively narrow range of masses of gas the newly synthesised elements mix with (so called mixing masses).  Alternatively, as suggested by Arnone et~al.~(\cite{arnone05}), the scatter in C, Sr, Y, Ba and Eu at these metallicities might be explained by the addition of sources of these elements which produce negligible amounts of the other elements such as Mg and Fe, such as low mass supernovae of type II.

\subsection{Carbon}

\begin{figure}
\begin{center}
\resizebox{\hsize}{!}{\rotatebox{0}{\includegraphics{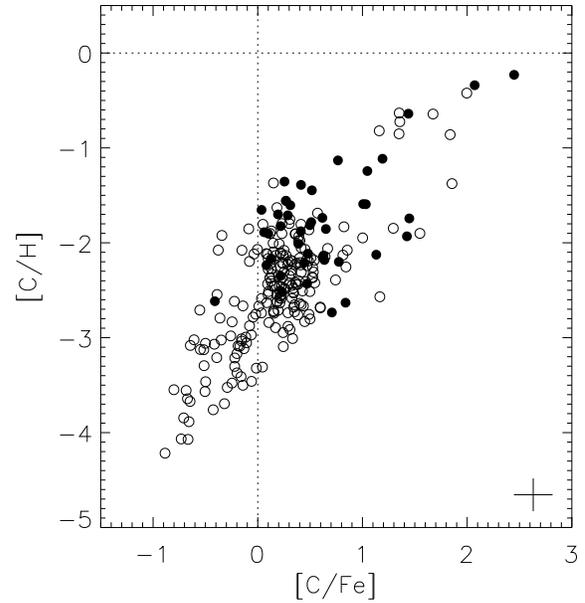}}}
\end{center}
\caption{A plot of [C/H] versus [C/Fe] following Bromm \& Loeb~(\cite{bromm03}).  Giants and red horizontal branch stars are plotted as open circles, while the subgiants and dwarfs are plotted as filled circles.
}
\label{fig:bromm}
\end{figure}

The work of Bromm \& Loeb~(\cite{bromm03}) suggests that halo stars with low C abundances, $\mathrm{[C/H]}\sim-3.5$ according to their model, and low oxygen abundances, are probably true second generation stars.  In Fig.~\ref{fig:bromm} we plot the [C/H] values for our sample, identifying evolved (on the giant branch or red horizontal branch) and unevolved stars (all others).  Though we do not have oxygen abundances, we see the majority of stars lie above the $-3.5$ level.  It would be interesting to obtain O abundances for the stars lying near or below $\mathrm{[C/H]}\sim-3.5$ to identify likely second generation stars among this sample.

\subsection{Heavy Neutron-Capture Elements, Ba--Eu}
\label{subsect:heavyneutron}

Figure~\ref{fig:Ba_Eu}, which we used earlier as a measure of r- vs s-process enrichment, plots [Ba/Eu] against Fe and C abundances.  The plots show a clear separation between two groups in the halo, a separation which correlates with C enrichment.  This distinction was first seen in McWilliam~(\cite{mcw98}), though with fewer stars.   The scatter among the pure r-process stars is consistent with the observational uncertainties and we thus conclude that the cosmic scatter in Ba/Eu among pure r-process halo stars is small.  This implies that the Ba/Eu abundances produced by the r-process in the early Galaxy are universal.  Taking the weighted mean (weights are based on absolute errors), we find for the pure r-process stars analysed in this work $\langle \mathrm{[Ba/Eu]} \rangle = -0.58 \pm 0.03$, where the quoted error is the weighted standard deviation.  The solar system r-process abundance ratio, based on the data from Arlandini et~al.~(\cite{arlandini99}), is $-0.65$.   Our result seems to be in disagreement with that found at low metallicities by Truran et~al.~(\cite{truran02}), who compiled data from a number of studies, and found a large scatter in Ba/Eu.  They noted the difficulties in analysing Ba in metal-poor spectra due to dependence on hyperfine and isotopic structure and microturbulence.  This is particularly problematic when combining data from different studies.  Our data on the other hand are homogeneously analysed, but the Eu abundances are incomplete and biased towards stars with strong r-process enhancement. 

\begin{figure}
\begin{center}
\resizebox{\hsize}{!}{\rotatebox{0}{\includegraphics{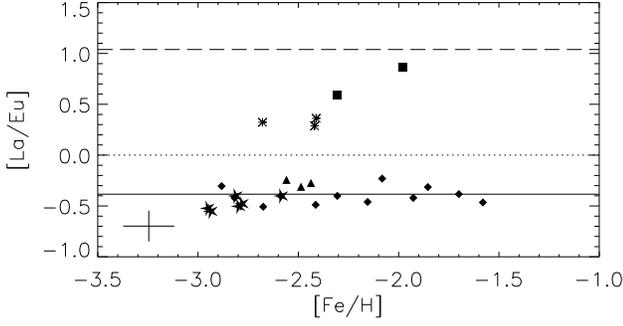}}}
\end{center}
\caption{A plot of [La/Eu] vs [Fe/H].  Symbols and lines have the same meanings as in Fig.~\ref{fig:Ba_Eu}.
}
\label{fig:La_Eu_vs_Fe_H}
\end{figure}

It has been suggested (e.g.\ Burris et~al.~\cite{burris00}, Truran et~al.~\cite{truran02}, Simmerer et~al.~\cite{simmerer04}) that La may be a better alternative as a tracer of the s-process enrichment.  Unfortunately, as the lines available to us are typically weak, La is often undetected in our low $S/N$ spectra according to our detection criteria.  When plotting [La/Eu] vs [Fe/H], Fig.~\ref{fig:La_Eu_vs_Fe_H}, we see no trend or significant scatter.  Among the pure r-process stars $\langle \mathrm{[La/Eu]} \rangle = -0.40 \pm 0.05$.  The solar system r-process abundance ratio is $-0.38$.

As mentioned in Sect.~\ref{sect:new_objects}, the distribution of r-process enhancement in the early Galaxy is of great interest.  This is usually traced by [Eu/Fe], but our data are incomplete due to the inability to reliably detect low Eu abundances in our snapshot spectra.  One may, however, try to reconstruct the r-process enhancement by instead using the Ba abundances, which are practically complete due to the ease of detecting the Ba II resonance line at 4554~\AA.   Following Raiteri~et~al.~(\cite{raiteri99}) and Burris~et~al.~(\cite{burris00}), where possible we attempt to reconstruct the r-process-only contribution to Ba.  We define the ratio $\mathrm{[Ba/Fe]}_\mathrm{r-process} = \log ( N_\mathrm{Ba,r} / N_\mathrm{Fe} )_\star - \log (N_\mathrm{Ba,r} / N_\mathrm{Fe})_\mathrm{\sun}$, where $N_\mathrm{Ba,r}$ indicates the r-process-only Ba abundance.  Note, the zero point of the scale is set at the solar r-process-only Ba abundance, not the total solar Ba abundance.  For the pure r-process stars, $\mathrm{[Ba/Fe]}_\mathrm{r-process} = \mathrm{[Eu/Fe]}$, since the Ba/Eu r-process ratio is fixed at the solar r-process-only ratio and Eu is produced practically entirely by the r-process.  Previous studies indicate that the main s-process does not become important until $\mathrm{[Fe/H]} \ga -2.5$, e.g.\ Burris~et~al.\ and Truran~et~al.~(\cite{truran02}).  Thus we assume that at lower metallicities all Ba in non-C-rich stars is produced by the r-process.  That is, for stars with $\mathrm{[C/Fe]}<0.3$ and $\mathrm{[Fe/H]}<-2.5$, $\mathrm{[Ba/Fe]}_\mathrm{r-process} = \mathrm{[Ba/Fe]} + 0.721$.  Those stars that are not pure r-process stars with higher C enrichment and higher metallicity are disregarded.  

\begin{figure}
\begin{center}
\resizebox{\hsize}{!}{\rotatebox{0}{\includegraphics{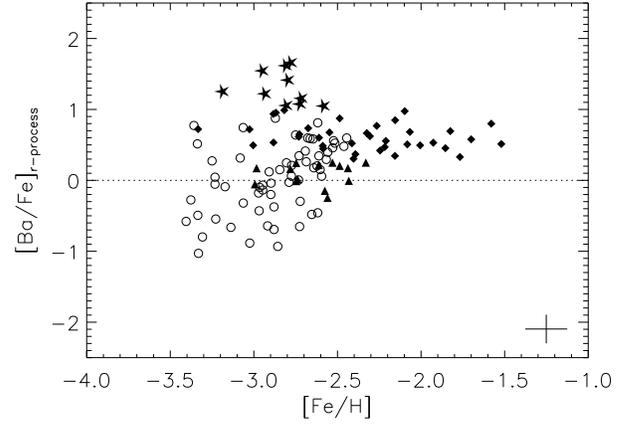}}}
\end{center}
\caption{A plot of $\mathrm{[Ba/Fe]}_\mathrm{r-process}$ vs [Fe/H].  Symbols have the same meanings as in Fig.~\ref{fig:Ba_Eu}.  For the closed symbols (pure r-process stars) $\mathrm{[Ba/Fe]}_\mathrm{r-process} = \mathrm{[Eu/Fe]}$.  Open circles show stars where $\mathrm{[C/Fe]}<0.3$ and $\mathrm{[Fe/H]}<-2.5$, for which we assume $\mathrm{[Ba/Fe]}_\mathrm{r-process} = \mathrm{[Ba/Fe]}+0.721$.  Note the scatter is underestimated at $\mathrm{[Fe/H]} > -2.5$; see text.}
\label{fig:Ba_rprocess}
\end{figure}

The results for $\mathrm{[Ba/Fe]}_\mathrm{r-process}$ are plotted against [Fe/H] in Fig.~\ref{fig:Ba_rprocess}.  Note, since we can only estimate $\mathrm{[Ba/Fe]}_\mathrm{r-process}$ in stars with $\mathrm{[Fe/H]}> -2.5$ if Eu is detected, there is an observational bias here; we will miss stars with low r-process enhancements at $\mathrm{[Fe/H]}> -2.5$.  The results of Burris et~al., which cover this metallicity regime and are much more complete, indicate the scatter in the region  $-2.5 <\mathrm{[Fe/H]}<-1.5$ is larger than seen in our results.  They find stars in this regime with $\mathrm{[Eu/Fe]}$ ranging from as low as $\sim -0.4$ (significantly lower than in our data) to as high as $\sim 0.8$ (similar to our data).  In any case, even accounting for the fact that this bias leads to an underestimation of the scatter at $\mathrm{[Fe/H]}> -2.5$, the scatter at $\mathrm{[Fe/H]}< -2.5$ in our data is significantly larger than seen at higher metallicity in the results of Burris et~al.  If we consider only the upper envelope of the abundance distribution, which should be well defined by our sample, there is an rapid transition from a maximum value of [Eu/Fe] of $\sim 1.0$ to $\sim 1.7$ at $\mathrm{[Fe/H]} \sim -2.5$.  Thus, Fig.~\ref{fig:Ba_rprocess} and the occurrence of r-II stars only at $\mathrm{[Fe/H]}<-2.5$, suggests the r-II stars are extreme cases of a wide range of r-process enrichment in the early, chemically inhomogeneous Galaxy.   Note, the cutoff of $\mathrm{[Eu/Fe]}=1.0$ between r-I and r-II is suggested by the maximum enrichment levels at higher metallicity.

\subsection{Light Neutron-Capture Elements, Sr, Y and Zr}
\label{subsect:lightneutron}

\begin{figure}
\begin{center}
\resizebox{\hsize}{!}{\rotatebox{0}{\includegraphics{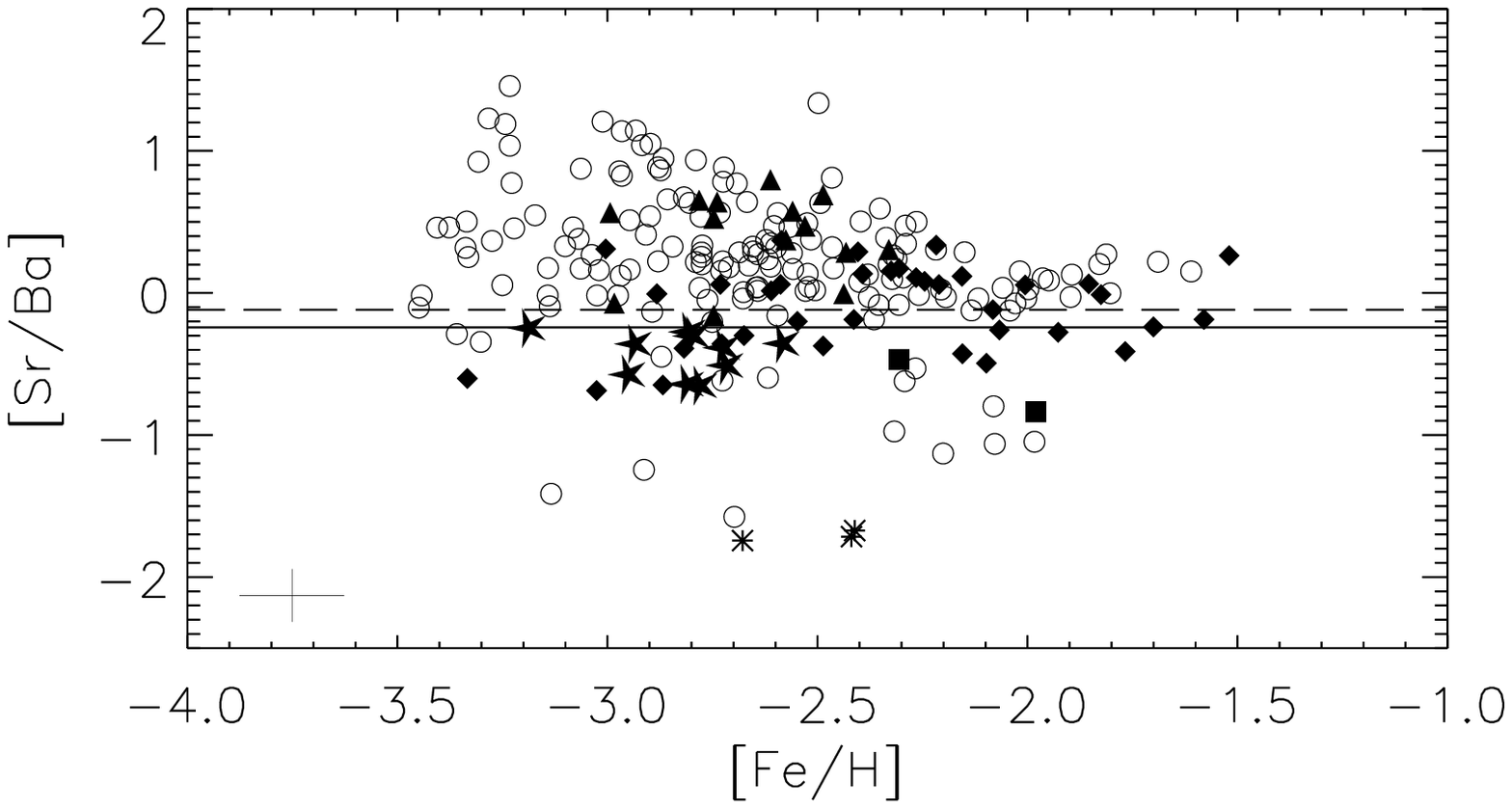}}}
\resizebox{\hsize}{!}{\rotatebox{0}{\includegraphics{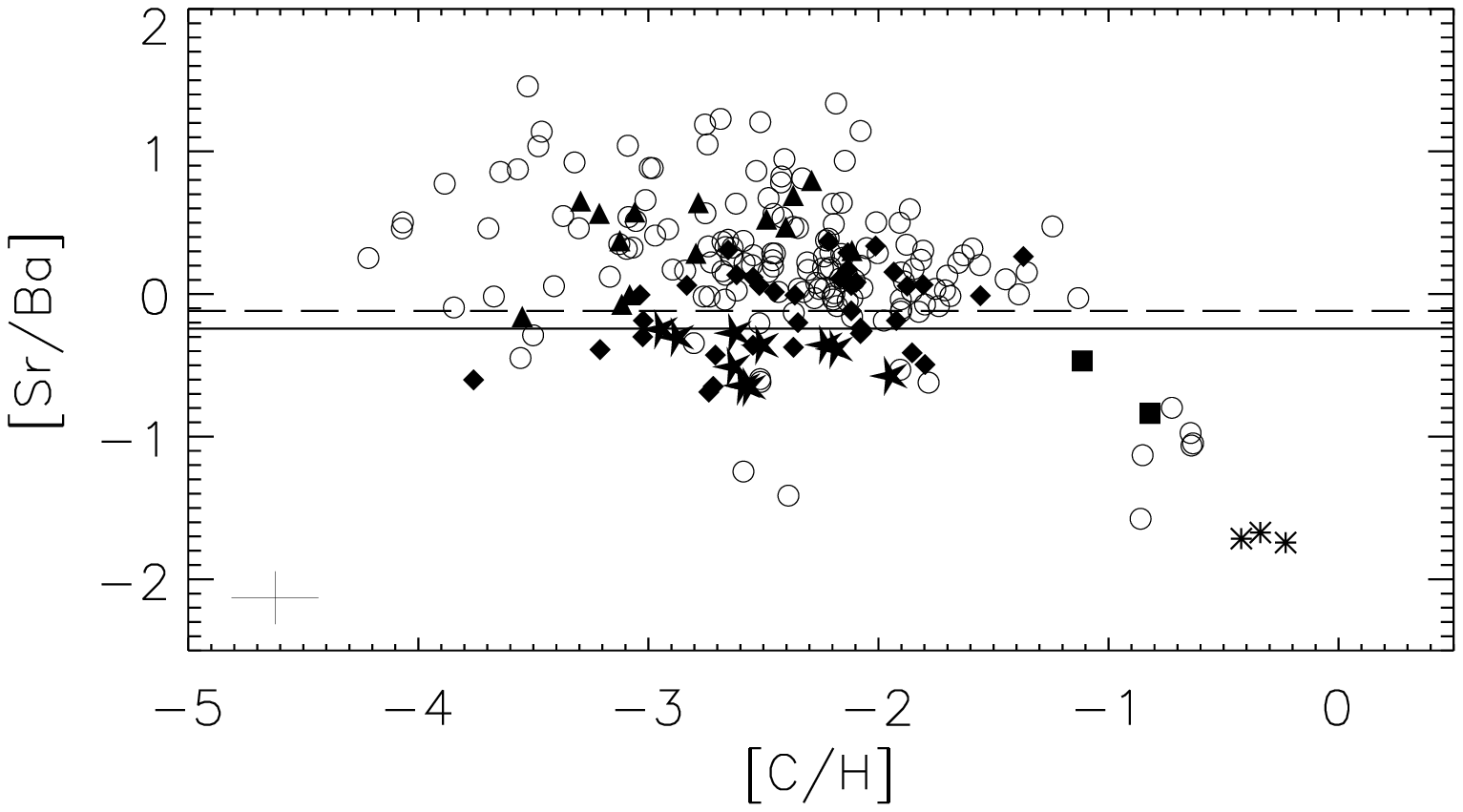}}}
\end{center}
\caption{A plot of [Sr/Ba] vs [Fe/H] (upper panel) and [C/H] (lower panel).  Symbols and lines have the same meanings as in Fig.~\ref{fig:Ba_Eu}, with circles representing stars where Eu is undetected.
}
\label{fig:Sr_Ba}
\end{figure}

\begin{figure}
\begin{center}
\resizebox{\hsize}{!}{\rotatebox{0}{\includegraphics{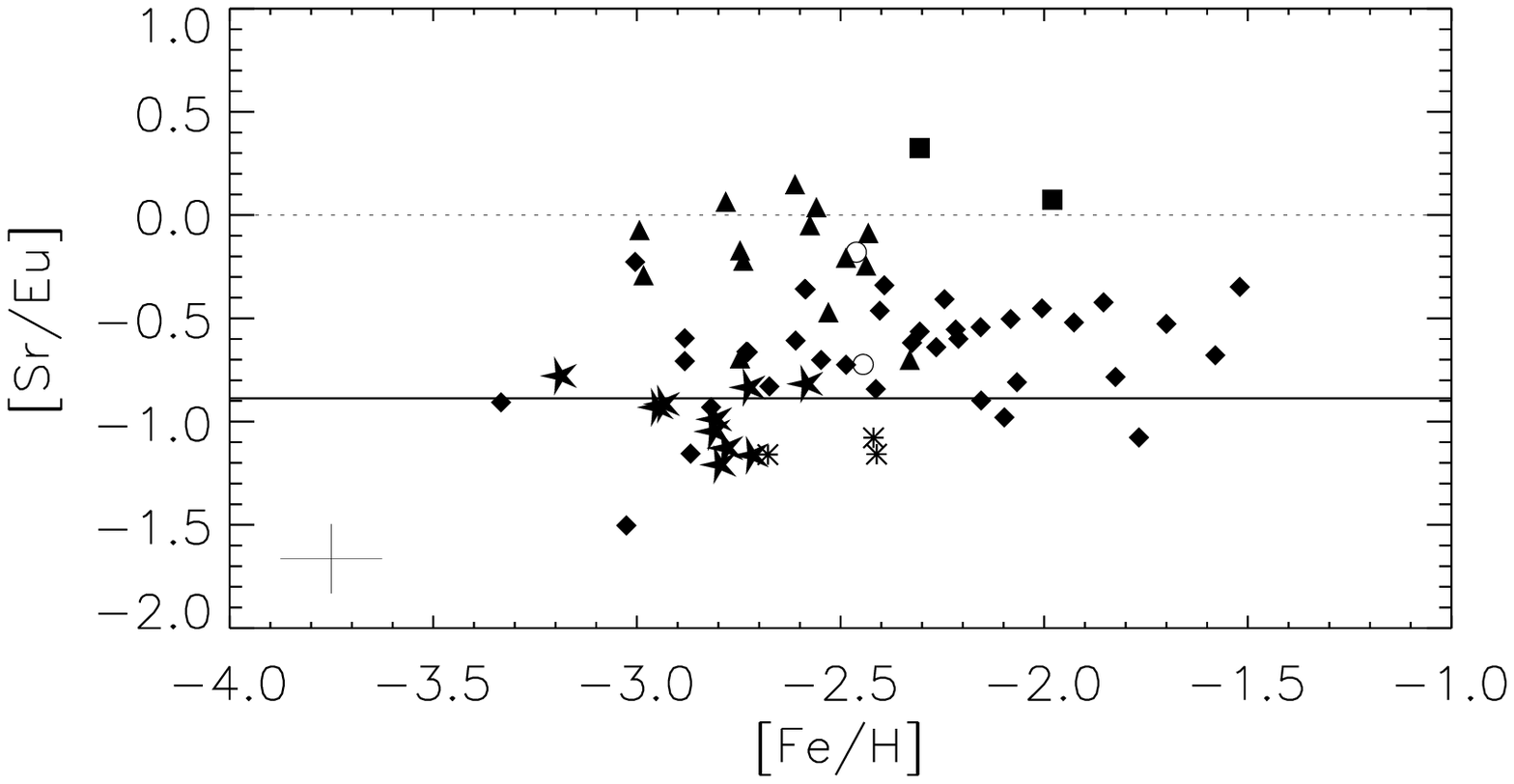}}}
\resizebox{\hsize}{!}{\rotatebox{0}{\includegraphics{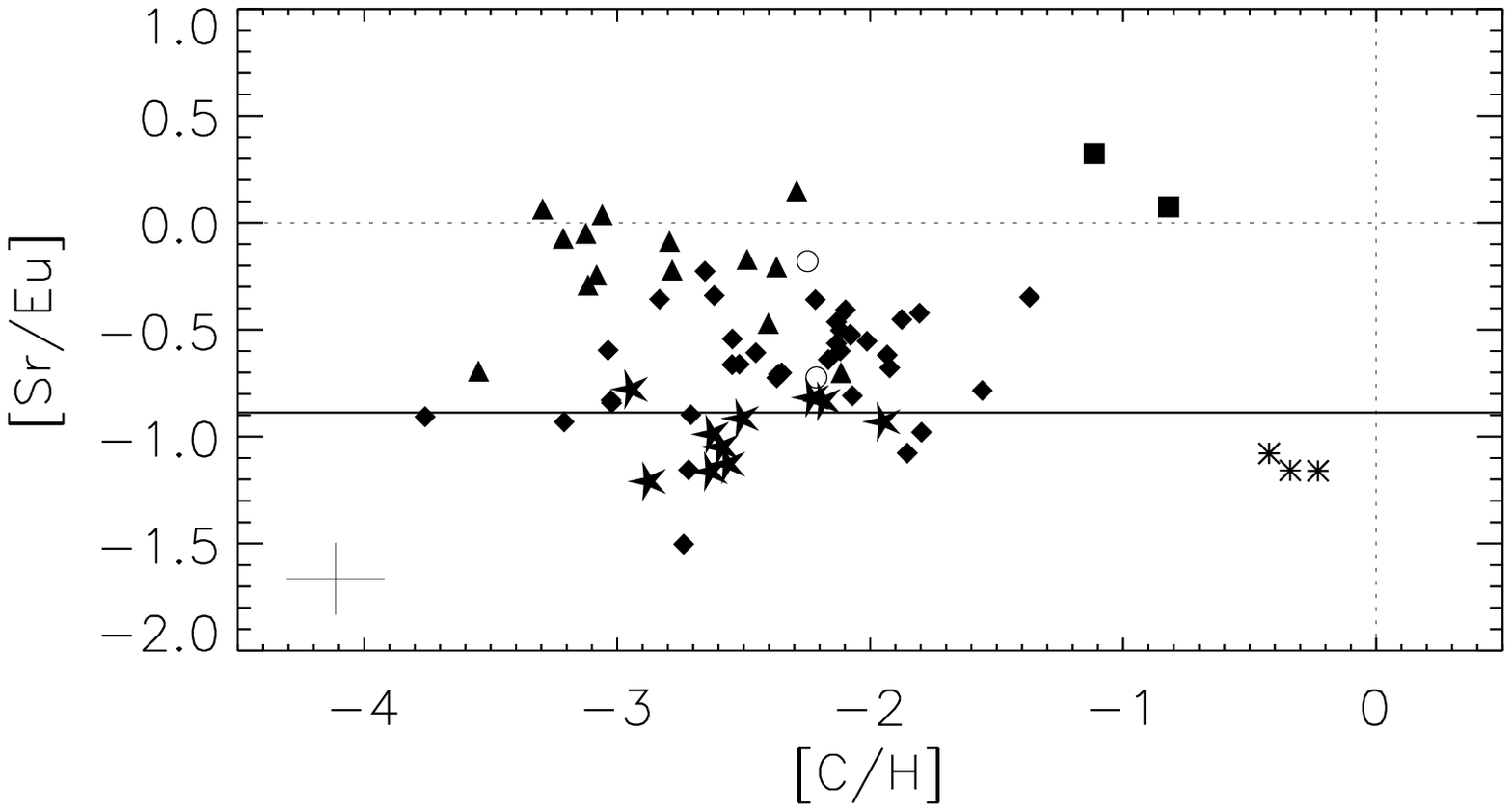}}}
\end{center}
\caption{A plot of [Sr/Eu] vs [Fe/H] (upper panel) and [C/H] (lower panel). Symbols and lines have the same meanings as in Fig.~\ref{fig:Ba_Eu}, with circles representing stars where Ba is undetected.   The solar s-process value is 1.08.
}
\label{fig:Sr_Eu}
\end{figure}

The production of light neutron-capture elements (particularly Sr, Y, Zr) versus heavy neutron-capture elements (such as Ba, Eu), has become a topic of interest due to evidence of production of the former without significant production of the latter, e.g.\  McWilliam ~(\cite{mcw98}), Burris~et~al.~(\cite{burris00}), Truran~et~al.~(\cite{truran02}), Travaglio~et~al.~(\cite{travaglio04}) and Aoki~et~al.~(\cite{aoki05}).  Figures~\ref{fig:Sr_Ba} and~\ref{fig:Sr_Eu} plot [Sr/Ba] and [Sr/Eu] against Fe and C abundances.  Significant scatter is seen in Sr/Ba at low metallicity, as found by McWilliam and Burris et~al.  In both cases, Sr/Ba and Sr/Eu, even among the pure r-process stars, a significant amount of scatter is seen at $\mathrm{[Fe/H]}<-2.5$.   For Sr/Ba, the scatter appears to increase quite uniformly with decreasing [Fe/H], the data showing quite clear upper and lower boundaries, apart from a small number of outliers which are usually C enhanced.  Sr/Eu shows similar tendencies when the s-process rich stars are disregarded, though not as clearly due to the smaller number of stars.   The r-II stars all have similar Sr/Ba and Sr/Eu ratios, which are always among lowest of the non-C-enhanced stars.  For Sr/Ba the weighted mean of the results for r-II stars, is $\langle \mathrm{[Sr/Ba]} \rangle_\mathrm{r-II} = -0.44 \pm 0.08$, and for Sr/Eu $\langle \mathrm{[Sr/Eu]} \rangle_\mathrm{r-II} = -0.98 \pm 0.09$.  The solar system r-process values are $-0.24$ and $-0.89$ respectively.  

\begin{figure}
\begin{center}
\resizebox{70mm}{!}{\rotatebox{0}{\includegraphics{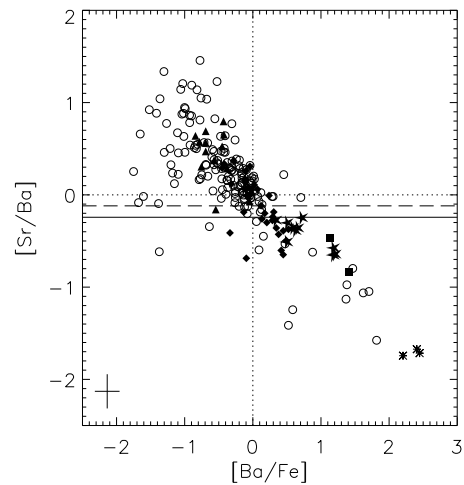}}}
\end{center}
\caption{A plot of [Sr/Ba] vs [Ba/Fe].  Symbols and lines have the same meanings as in Fig.~\ref{fig:Ba_Eu}, with circles representing stars where Eu is undetected.}
\label{fig:Sr_Ba_vs_Ba_Fe}
\end{figure}

\begin{figure}
\begin{center}
\resizebox{70mm}{!}{\rotatebox{0}{\includegraphics{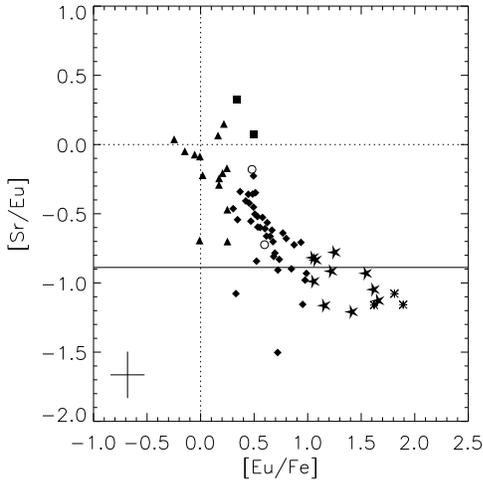}}}
\end{center}
\caption{A plot of [Sr/Eu] vs [Eu/Fe].  Symbols and lines have the same meanings as in Fig.~\ref{fig:Sr_Eu}.}
\label{fig:Sr_Eu_vs_Eu_Fe}
\end{figure}

Truran et~al.~(\cite{truran02}) examined the variation of Sr/Ba with r-process enrichment, as traced by Ba/Fe and Eu/Fe.  Figure~\ref{fig:Sr_Ba_vs_Ba_Fe} shows Sr/Ba with Ba/Fe, and Fig.~\ref{fig:Sr_Eu_vs_Eu_Fe} shows Sr/Eu with Eu/Fe.   We find similar results; however, with our large and homogeneously analysed sample the scatter is well defined, and we see a trend for decreasing scatter in Sr/Ba and Sr/Eu with increasing r-process enrichment.   

\begin{figure}
\begin{center}
\resizebox{\hsize}{!}{\rotatebox{0}{\includegraphics{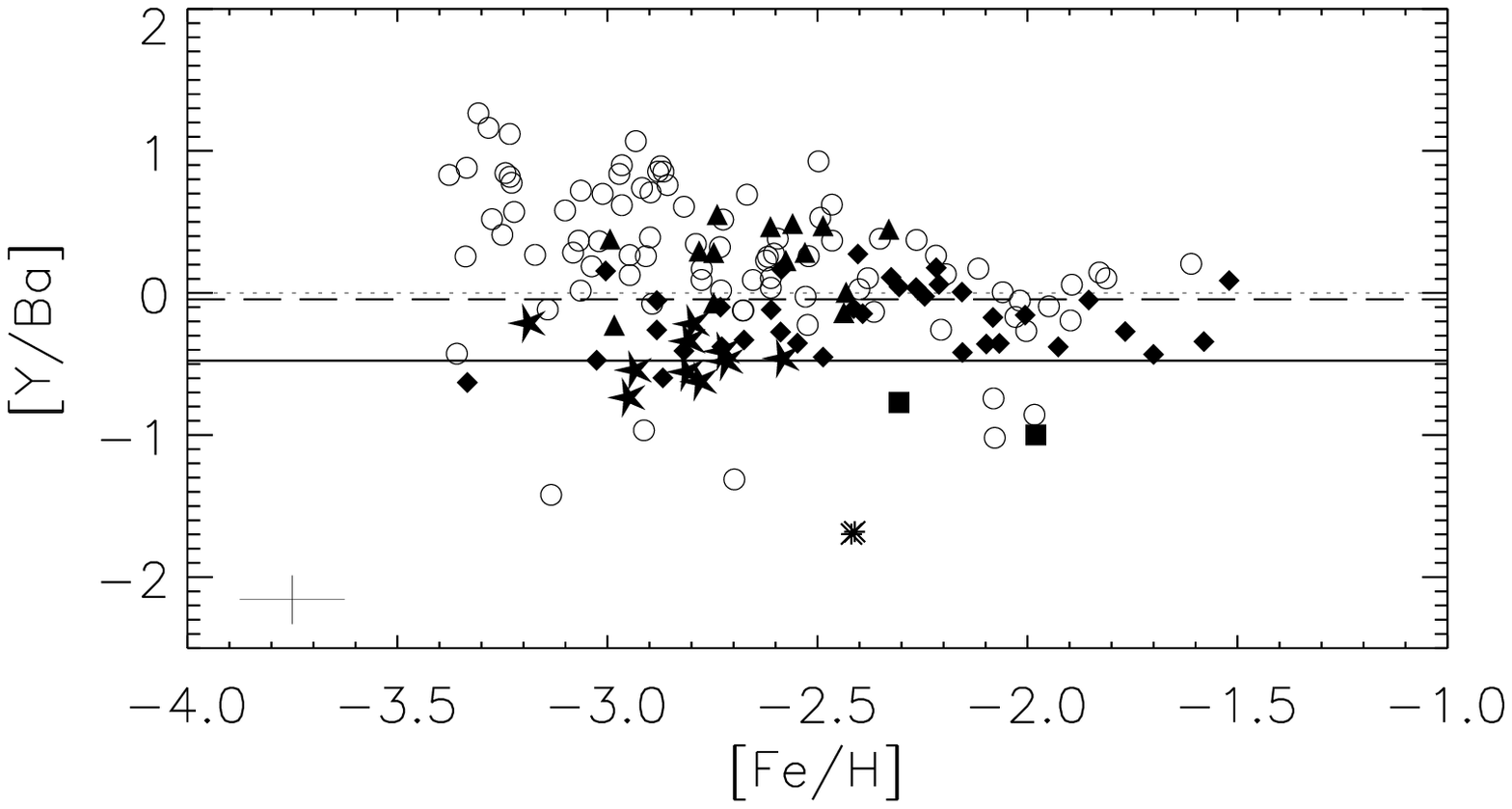}}}
\resizebox{\hsize}{!}{\rotatebox{0}{\includegraphics{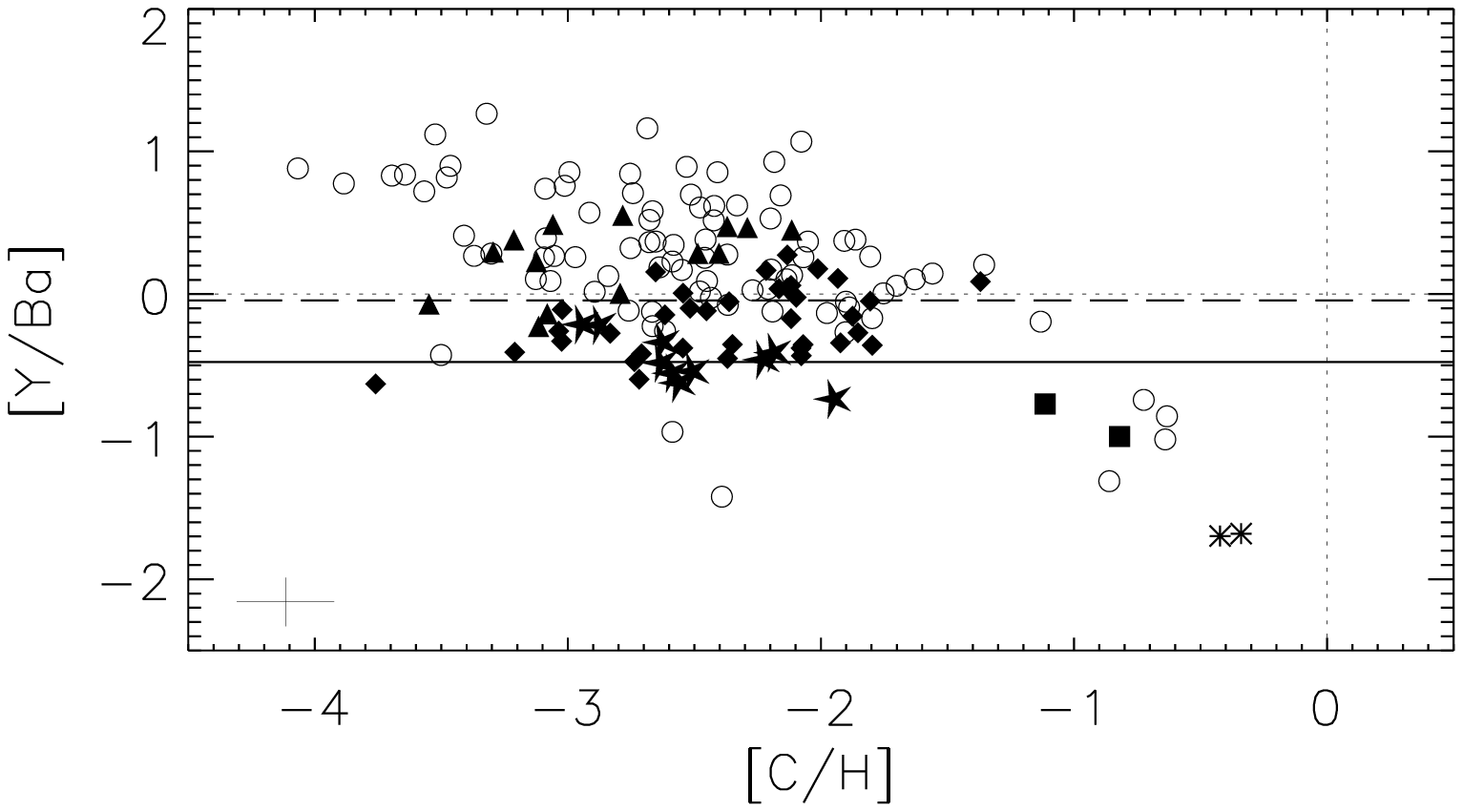}}}
\end{center}
\caption{A plot of [Y/Ba] vs [Fe/H] (upper panel) and [C/H] (lower panel). Symbols and lines have the same meanings as in Fig.~\ref{fig:Ba_Eu}, with circles representing stars where Eu is undetected. 
}
\label{fig:Y_Ba}
\end{figure}

\begin{figure}
\begin{center}
\resizebox{70mm}{!}{\rotatebox{0}{\includegraphics{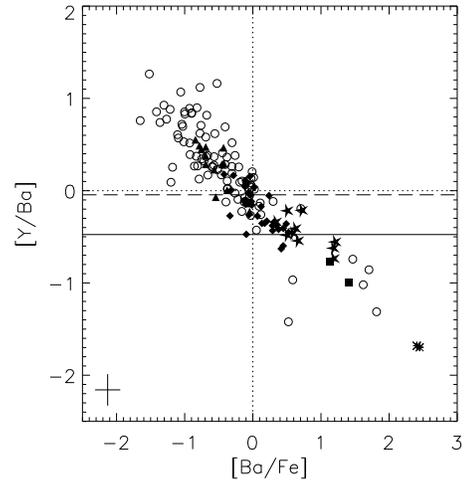}}}
\end{center}
\caption{A plot of [Y/Ba] vs [Ba/Fe].  Symbols and lines have the same meanings as in Fig.~\ref{fig:Y_Ba}.}
\label{fig:Y_Ba_vs_Ba_Fe}
\end{figure}

Similar plots are shown for Y/Ba in Figs.~\ref{fig:Y_Ba} and~\ref{fig:Y_Ba_vs_Ba_Fe}.  The results are similar to those for Sr/Ba, in particular, increasing scatter in Y/Ba with decreasing metallicity and decreasing heavy r-process enrichment, and similar Y/Ba among the r-II stars.  For the r-II stars we find $\langle \mathrm{[Y/Ba]} \rangle_\mathrm{r-II} = -0.47 \pm 0.08$; the solar system r-process value is $\mathrm{[Y/Ba]} = -0.48$.

Figures~\ref{fig:Sr_Ba},~\ref{fig:Sr_Ba_vs_Ba_Fe},~\ref{fig:Y_Ba} and~\ref{fig:Y_Ba_vs_Ba_Fe} identify two stars of interest, two metal-poor stars with $\mathrm{[Sr,Y/Ba]} \la -1$ which are not carbon-rich or overly Ba-rich, which are HE~0305-4520 ($\mathrm{[C/Fe]}\sim +0.3$, $\mathrm{[Sr/Fe]}\sim -0.7$, $\mathrm{[Y/Fe]}\sim -0.4$, $\mathrm{[Ba/Fe]}\sim +0.6$) and HE~2156-3130 ($\mathrm{[C/Fe]}\sim +0.7$, $\mathrm{[Sr/Fe]}\sim -0.9$, $\mathrm{[Y/Fe]}\sim -0.9$, $\mathrm{[Ba/Fe]}\sim +0.5$ ), the latter being the more metal-poor of the two.  These stars perhaps warrant further study.

\begin{figure}
\begin{center}
\resizebox{70mm}{!}{\rotatebox{0}{\includegraphics{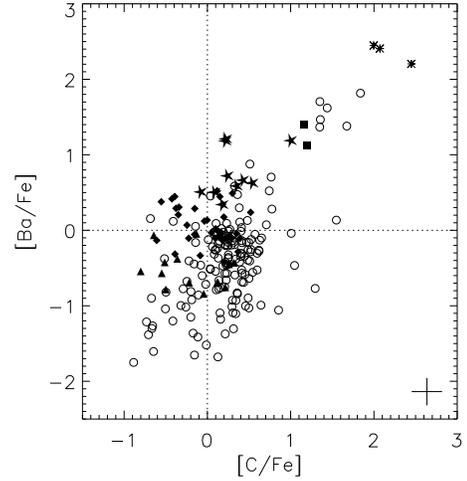}}}
\end{center}
\caption{A plot of [Ba/Fe] vs [C/Fe].  Symbols have the same meanings as in Fig.~\ref{fig:Sr_Ba}.
}
\label{fig:Ba_vs_C}
\end{figure}

We note that Figs.~\ref{fig:Sr_Ba} and~\ref{fig:Sr_Ba_vs_Ba_Fe} show 6 stars with similar Sr/Ba ratios and similar C/H and Fe/H to the stars identified as s-process-rich stars.  These stars also stand out clearly in a plot of [Ba/Fe] vs [C/Fe], see Fig.~\ref{fig:Ba_vs_C}, occupying a similar region of the plot as the s-process-rich stars.  As Eu has not been detected at the 3 sigma level according to our criteria they have not been classified in terms of the likely neutron-capture processes that have contributed to the heavy elements.  It is interesting to look at these stars in more detail.  We find that all these stars appear to be s-process-rich stars based on Eu abundances obtained, though we emphasise the abundances are \emph{not} 3 sigma detections and thus not completely reliable.  HE~1430-1123, the star with the lowest [Sr/Ba] of the six, and thus more closely associated with the s-II stars in Fig.~\ref{fig:Sr_Ba}, appears also to be a s-II star as we determine $\mathrm{[Eu/Fe]}\sim 1.4$.  The other five stars, HE~0231-4016, HE~0430-4404, HE~2150-0825, HE~2227-4044, and HE~2240-0412, which seem more closely associated with the remaining mildly Eu enhanced s-process-rich stars in Fig.~\ref{fig:Sr_Ba}, appear to be similar stars with $\mathrm{[Eu/Fe]}\sim 0.7\leftrightarrow 1.0$.  

Figure~\ref{fig:Ba_vs_C} also identifies a number of mildly carbon-enhanced stars without enhancement of neutron-capture elements, a class of objects identified by Aoki et~al.~(\cite{aoki02a}).  The five objects with $\mathrm{[C/Fe]}>0.8$ and $\mathrm{[Ba/Fe]}<0.2$ are, from largest [C/Fe] to smallest, HE~1351-1049, HE~1300-0641, HE~1330-0354, HE~1300-2201, and HE~1124-2335. 

\section{Conclusions}
\label{sect:conclusions}

We have analysed snapshot spectra of a sample of 253 metal-poor stars, deriving abundances for 22 elements, where detectable, using an automated technique based on SME by Valenti \& Piskunov~(\cite{sme}).  The technique has been shown to give results in agreement with previous work within errors, when one considers differences in temperature scale.  For our particular case (namely resolving power, spectral coverage and line list) the technique has been shown to be quite robust (at around the 0.1~dex level) even for $S/N$ as low as 15.  Our error estimates and comparisons with studies in the literature show the derived elemental abundances to be of moderate precision, relative and absolute errors of order 0.15 and $0.25$~dex respectively.  This work has dealt with the CH weak content of the sample.  The CH strong content will also be examined as part of the HERES programme.  It would be interesting to obtain additional spectra for (the most interesting of) these stars, particularly covering the Eu~II lines at 6437 and 6645~\AA.

Our main goal here has been to identify stars of interest, particularly those enhanced in r-process elements.  We identified 8 new r-II stars, 35 r-I stars and 3 s-process rich stars with strong Eu enhancement. Notably, the r-II stars were found only in a rather narrow metallicity range, $-3.2\la\mathrm{[Fe/H]}\la-2.6$.  These interesting stars should be investigated in more detail, and such work is underway as part of the HERES project large programme.  The spectra obtained will be of much higher quality, in terms of $S/N$, resolving power, and spectral coverage.  The higher quality data, together with a careful manual spectrum analysis, will naturally permit abundances and stellar parameters to be obtained with higher precision and abundances for additional elements to be obtained.  It will also be possible to obtain isotopic ratios for some elements.  Ideally, such analyses should be performed with reference to deviations from the assumptions of LTE and 1D model atmospheres. 

The results presented here provide a database of confirmed metal-poor stars, including a number of new r-process and s-process rich stars, which may be used for selection of stars for further studies.  The results also provide a homogeneous database of moderate precision abundances for comparison with Galactic chemical evolution models.  We stress that such comparisons must consider the limitations of our data set, most importantly the selection effects arising from the significant incompleteness of the data for some elements, but also the precision of the abundances and the assumptions of the modelling such as LTE and the use of 1D model atmospheres.

We investigated trends and scatter in some measured abundance ratios of interest.  Among the stars without strong C enhancement, at about $\mathrm{[Fe/H]} \ga -2.5$ we find that the cosmic scatter in any abundance ratio is small (perhaps even non-existent), implying that at around this level of enrichment the Galactic halo was reasonably well mixed.  At lower metallicities C, Sr, Y, Ba and Eu, and perhaps Zr, show evidence for cosmic scatter, while the results for the Mg, Ca, Sc, Ti, Cr, Fe, Co and Ni still indicate at most small cosmic scatter within the errors of our analysis.  Due to the difficulties in accurately estimating errors, and thus disentangling the observational and modelling uncertainties from the real cosmic scatter, we conclude that to determine the magnitude of the scatter in the cases where it is small will require high precision studies of a large number of stars.

\begin{acknowledgements}

We are indebted to Jeff Valenti and Nikolai Piskunov for making their code SME available to the community, and to Nikolai for help in understanding it.  We are further indebted to the many people who work on the MARCS code for making it available to us.  Bertrand Plez and Alain Jorissen are thanked for providing their CH line list, and permitting us to publish the parts employed.  Bengt Gustafsson is thanked for helpful comments on early versions of the manuscript, and pointing out the importance of correctly treating scattering.  We are grateful to the ESO staff at Paranal and Garching for obtaining the observations and reducing the data respectively.   We also thank T. Sivarani for supplying the routine used for the colour transformations and determination of the effective temperatures.  This work made extensive use of the VALD database and the NIST Atomic Spectra Database.  P.B.\ acknowledges the support of the Swedish Research Council. N.C.\ acknowledges financial support through a Henri Chretien International Research Grant administered by the American Astronomical Society, and from Deutsche Forschungsgemeinschaft under grants CH~214/3-1 and Re~353/44-1.  T.C.B.\ acknowledges partial funding for this work from grants AST 00-98508, AST 00-98549 and AST-04 06784 awarded by the U.S.\ National Science Foundation, and from award PHY 02-16783: Physics Frontiers Center/Joint Institute for Nuclear Astrophysics (JINA).  S.R.\ acknowledges partial financial support from FAPESP and CNPq.         

\end{acknowledgements}

\appendix

\section{Line List}
\label{app:linelist}

The lines and spectral windows used are listed in Tables~\ref{tab:lines_ch1}--\ref{tab:lines_heavy}.  The most important atomic and molecular data, wavelength $\lambda$, excitation potential $\chi$, and $\log gf$, are tabulated.  Line broadening data used may be obtained from the authors on request.  Below we briefly comment on line selection issues and data sources for each element.

First, it is worth commenting on Si which is notably omitted from our analysis.  Our spectra include two Si lines, at 3905 and 4102~\AA.  The $\lambda$4102 line lies in the wing of H$\delta$ and thus presents a considerable challenge for automated analysis.  The $\lambda$3905 line is known to be blended by CH lines (e.g.\ Cayrel~et~al.~\cite{cayrel04}).  Even with known CH lines from the lists of Plez \&\ Jorrisen (see below) included in the synthesis, our Si abundances were found to differ from the literature values (see Sect.~\ref{subsect:snr}), which are usually based on the $\lambda$4102 line, and the difference was seen to correlate with carbon-enhancement.  We have reasonable agreement with the literature for CS~22186-025 which has [C/Fe]$\sim-0.7$, while we have mild disagreement for CS~31082-001 which has [C/Fe]$\sim 0.2$ and strong disagreement for CS~22892-052 which has [C/Fe]$\sim 1$.   This suggests that the $\lambda$3905 line may be affected by further blends of features involving C, and thus we chose to omit Si from our analysis. 

\paragraph{Carbon} Abundances are derived from selected clean regions of CH A--X bands at 4310--4313~\AA\ (G band) and 4362--4367~\AA.  We limited ourselves to these regions to reduce computing times due to the large number of components in these bands. Comparisons with results using larger regions showed no differences to those using the final adopted regions.  We note, however, that we consider it important to use at least two different regions of differing strengths, in particular because the G band region at 4310~\AA\ can become saturated in carbon-enhanced stars.   The observations are not of sufficient quality to determine isotopic abundances, and so we made the assumption that all carbon is in the form of $^{12}$C and there is no $^{13}$C.  The adopted line list given in Tables~\ref{tab:lines_ch1} and~\ref{tab:lines_ch2} was extracted from a list compiled by Plez \&\ Jorissen (private communication), which is described by Hill~et~al.~(\cite{hill02}). 
  
\paragraph{Magnesium} We employed 7 spectral features of Mg I.  The features consist of the generally strong UV triplet near 3835~\AA\ from the $^3$P$^o-^3$D transition, plus some weaker yet cleaner features farther to the red, the strongest of which is generally $\lambda$4703.  Our employed lines are practically the same as those adopted by McWilliam et~al.~(\cite{mcw95b}) and Norris et~al.~(\cite{norris96}) (see also Ryan et~al.~\cite{ryan96}), though we note that there are some significant differences in the adopted $\log gf$ values, particularly for the triplet where the two works adopt values which differ by as much as 0.5 dex. This warrants some discussion. McWilliam et~al. adopted $\log gf$ values for the triplet from the NIST compilation of Wiese et~al.~(\cite{wiese69}), which are based on theoretical calculations of Weiss~(\cite{weiss67}) and re-normalised measurements by Penkin \& Shabanova~(\cite{penkin62}); the compilation can be consulted for details.  We note that the $^3$P$^o-^3$D transition is actually a sextuplet, but due to the practically non-existent fine-structure splitting of the $^3$D level only 3 lines are observed.  Wiese et~al.~(\cite{wiese69}) present $\log gf$ values for each of the 6 components, but McWilliam et~al. seem only to have included the strongest component for each line.  We have also adopted the NIST data, but have added together the appropriate components.  Thus, two of the lines have $\log gf$ values of order 0.1 dex stronger than those used by McWilliam et~al~(\cite{mcw95b}).  Norris et~al. seem not to have been aware of the NIST data, and adopted what amounts to empirical $f$-values guided by the desire to obtain abundances from these lines consistent with those from the other lines, particularly the $\lambda$4703 line, where the theoretical calculations of Froese Fischer~(\cite{froese_fischer75}) are employed (which we also adopt for these lines).  Their adopted values for the triplet differ from our adopted values by as much as 0.6 dex.  Thus, we would expect to find largely discrepant results from the triplet and the remaining lines with our $f$-values.  However, in our sample we always found a good general global fit with a single abundance.  We suspect this apparent discrepancy is due to the fact that we use a profile fitting technique with the line cores removed, while Norris et~al. fit equivalent widths.  The triplet lines are quite strong, having equivalent widths often of order 100~m\AA\ even for stars with [Fe/H]$<-3$, and our fits to the triplet show the observed cores to be typically much deeper than those from the best fit to the wings.  This is a well known problem of LTE analyses with no modelling of chromospheres.  This effect goes in the correct direction, in the sense that Norris et~al. would need to underestimate the $f$-values in order to correct for the overestimation of abundance due to the influence of the line cores.  We note that, since the Norris et~al. $f$-values for the triplet are calibrated to the weaker line abundances, their final abundances should be correct.  

\paragraph{Aluminium} Only one line of Al I was employed, the resonance line $\lambda$3961.  The second resonance line $\lambda$3944 was too blended to be considered reliable. Hyperfine structure was adopted from McWilliam et~al.~(\cite{mcw95b}).

\paragraph{Calcium} The Ca abundance is determined from 9 lines of Ca~I.  The $f$-values are adopted from the NIST critical compilations.

\paragraph{Scandium} The $f$-values are taken from Lawler \& Dakin~(\cite{lawler89}).  Hyperfine structure is taken directly from McWilliam et~al.~(\cite{mcw95b}).  

\paragraph{Titanium} Oscillator strengths for neutral lines are adopted from Blackwell et~al.~(\cite{oxford_ti82a, oxford_ti82b}).  For singly ionised lines, the situation has been markedly improved by the experimental work of Pickering et~al.~(\cite{pickering01}), and their oscillator strengths have been adopted.  For two useful lines where data were not available from this source, we adopted the solar values from Ryabchikova et~al.~(\cite{ryabchikova94}).  We note that for the other Ti II lines we used these two sources agreed within 0.1~dex.

\paragraph{Vanadium} The $\log gf$ data for V I is taken from Doerr et~al.~(\cite{doerr85}), and for V II from Bi\'emont et~al.~(\cite{biemont89v}).  McWilliam et~al.~(\cite{mcw95b}) noted the lack of hyperfine structure data for the transitions used here, and to our knowledge this situation has not changed.  However, as noted by McWilliam et~al. these lines are weak and unsaturated and therefore our abundances should not be greatly affected by this omission.  We note that in the case there was some effect, our abundances would be an upper limit.

\paragraph{Chromium} The $f$-values are taken from the NIST compilation (Martin et~al.~\cite{martin88}).  The values are based on a critical averaging of number of sources, which are detailed in the compilation.

\paragraph{Manganese} We adopted oscillator strengths from Booth et~al.~(\cite{oxford_mn84}).  The hyperfine structure was computed using the data compiled in Lef\`ebvre et~al.~(\cite{lefebvre03}), which for the lines employed here draws on data from Davis et~al.~(\cite{davis71}), and Handrich et~al.~(\cite{handrich69}), as well as their own measurements.  

\paragraph{Iron} A total of 55 lines were used of which 45 were Fe~I lines and 10 Fe~II.  For the neutral lines the oscillator strengths of O'Brian et~al.~(\cite{obrian91}) were adopted.  Oscillator strengths for the Fe~II lines were adopted from Schnabel et~al.~(\cite{schnabel04}) where possible.  Data for some other useful lines were taken from Bi\'emont et~al.~(\cite{biemont91}) and Moity~(\cite{moity83}).  We note that the Fe~II line $f$-values are of lower precision, typically 20\%, than the Fe~I lines, typically 7\%.

\paragraph{Cobalt} Oscillator strengths for the employed lines were taken from Nitz et~al.~(\cite{nitz99}) where possible, otherwise from Cardon et~al.~(\cite{cardon82}).  Hyperfine structure was included using the data of Pickering~(\cite{pickering96}).

\paragraph{Nickel} Four lines of Ni I are employed.  For 3 lines we employed the oscillator strengths from Blackwell et~al.~(\cite{oxford_ni89}); for the remaining line we used the $f$-value from Huber \& Sandeman~(\cite{huber80}).

\paragraph{Zinc} Two neutral lines were employed and the $f$-values taken from Bi\'emont \& Godefroid~(\cite{biemont80}).

\paragraph{Strontium} One line of Sr I and two of Sr II were analysed. Hyperfine and isotopic splitting was accounted for in the $\lambda$4215 line, using the data from Borghs et~al.~(\cite{borghs83}), where solar isotopic ratios have been assumed.  There are unfortunately no data to our knowledge for the upper state of the other resonance line $\lambda$4077.  However, noting that the splitting in the $\lambda$4215 line is dominated by the common lower level $^2S_{1/2}$, we computed hyperfine splitting for the $\lambda$4077 line where we neglected the splitting of the upper level.  The main isotope shifts for the $\lambda$4215 line are interpreted as due to the volume effect which affects the $s$ state most strongly (e.g. Cowan~\cite{cowan}; pg 506). We therefore adopted the same isotopic shifts for $\lambda$4077 as for the $\lambda$4215 line.  We did not find hyperfine structure data for the remaining line.  However, as it is weak, this should not be important.

\paragraph{Yttrium}  The $f$-values for the employed lines of Y II are from Hannaford et~al.~(\cite{hannaford82}).  $^{89}$Y has a small nuclear spin of $I=1/2$, and thus hyperfine splitting is negligible.

\paragraph{Zirconium} Three lines of Zr II are analysed employing the $f$-values from Bi\'emont et~al.~(\cite{biemont81}).

\paragraph{Barium} Our analysis of Ba is based on 2 lines of Ba II, with oscillator strengths from Gallagher~(\cite{gallagher67}) which for common lines are in excellent agreement with Davidson et~al.~(\cite{davidson92}).  Hyperfine structure was included from McWilliam~(\cite{mcw98}), assuming the pure r-process isotopic composition from that paper.  As pointed out by McWilliam~(\cite{mcw98}), if a star would have a significant s-process contribution, then our derived Ba abundance would be a lower limit.  This is because the s-process causes a higher fraction of the Ba to be in even isotopes which have no hyperfine splitting, and thus the stronger lines saturate more quickly, meaning the same line strength would require a higher abundance.

\paragraph{Lanthanum}    The oscillator strengths of Lawler et~al.~(\cite{lawler01la}) are used.  Hyperfine structure is included using the constants from Lawler et~al., which for our lines draws from their own measurements and those of H\"ohle et~al.~(\cite{hohle82}).  

\paragraph{Cerium} We employed 10 lines of Ce II, the same as those employed by Sneden et~al.~(\cite{sneden96}), with the addition of the line $\lambda$3999.  We adopted the same oscillator strengths as used by Sneden et~al~(\cite{sneden96}).  For $\lambda$3999 we adopted the VALD value, which is essentially a solar value averaged from various sources (see Piskunov et~al.~\cite{vald95}).

\paragraph{Neodymium}  We employ 9 lines of Nd II using laboratory $f$-values from Den Hartog et~al.~(\cite{den_hartog03}), which are a significant improvement on those previously available.

\paragraph{Samarium}    Eight lines of Sm II are employed and the $f$-values from Biemont et~al.~(\cite{biemont89sm}) were used.

\paragraph{Europium} Our Eu abundances are based on 4 lines of Eu~II.  Lines at 4435 and 4522 \AA\ were found to be too strongly blended in non-r-process-enhanced stars to be employed.  The line $\lambda$3971 was considered too blended for automated analysis.  Oscillator strengths and hyperfine splitting from Lawler et~al.~(\cite{lawler01}) were employed, assuming solar r-process isotopic fractions (supported by measurements by Sneden~et~al.~\cite{sneden02}). 

\begin{table}
\begin{center}
\caption{Molecular line data for A-X system of the CH molecule near 4310~\AA\ from Plez \&\ Jorissen. The spectral windows employed for this molecular band are defined with respect to an arbitrary $\lambda$ and span from $\lambda+d\lambda_\mathrm{blue}$ to $\lambda+d\lambda_\mathrm{red}$.}
\label{tab:lines_ch1}
\tiny
\begin{tabular}{cccccc}
Species &  $\lambda$ & $\chi$ &  $\log gf $ &   d$\lambda_\mathrm{blue}$ & d$\lambda_\mathrm{red}$ \\
& [\AA] & [eV] & & [m\AA] & [m\AA] \\
\hline
  CH &   4310.150 &          &           &$-$100 &  2300   \\
  CH &   4310.038 &    0.096 &  $-$3.183 &      &      \\
  CH &   4310.090 &    0.096 &  $-$1.412 &      &      \\
  CH &   4310.110 &    0.096 &  $-$1.474 &      &      \\
  CH &   4310.149 &    0.431 &  $-$3.202 &      &      \\
  CH &   4310.162 &    0.096 &  $-$3.025 &      &      \\
  CH &   4310.203 &    0.431 &  $-$1.443 &      &      \\
  CH &   4310.220 &    0.431 &  $-$1.505 &      &      \\
  CH &   4310.272 &    0.431 &  $-$3.039 &      &      \\
  CH &   4310.404 &    0.096 &  $-$3.183 &      &      \\
  CH &   4310.430 &    0.408 &  $-$1.581 &      &      \\
  CH &   4310.458 &    0.096 &  $-$1.412 &      &      \\
  CH &   4310.504 &    0.432 &  $-$3.202 &      &      \\
  CH &   4310.508 &    0.408 &  $-$2.966 &      &      \\
  CH &   4310.556 &    0.432 &  $-$1.443 &      &      \\
  CH &   4310.679 &    0.072 &  $-$1.550 &      &      \\
  CH &   4310.729 &    0.408 &  $-$1.581 &      &      \\
  CH &   4310.757 &    0.072 &  $-$2.952 &      &      \\
  CH &   4310.807 &    0.408 &  $-$2.966 &      &      \\
  CH &   4310.811 &    0.408 &  $-$3.155 &      &      \\
  CH &   4310.889 &    0.408 &  $-$1.508 &      &      \\
  CH &   4310.937 &    0.388 &  $-$1.677 &      &      \\
  CH &   4310.969 &    1.107 &  $-$2.269 &      &      \\
  CH &   4310.991 &    0.072 &  $-$1.550 &      &      \\
  CH &   4311.045 &    0.388 &  $-$2.879 &      &      \\
  CH &   4311.069 &    0.072 &  $-$2.952 &      &      \\
  CH &   4311.075 &    0.408 &  $-$3.155 &      &      \\
  CH &   4311.075 &    0.072 &  $-$3.136 &      &      \\
  CH &   4311.084 &    0.987 &  $-$4.123 &      &      \\
  CH &   4311.145 &    0.987 &  $-$4.123 &      &      \\
  CH &   4311.153 &    0.408 &  $-$1.508 &      &      \\
  CH &   4311.153 &    0.388 &  $-$1.677 &      &      \\
  CH &   4311.153 &    0.072 &  $-$1.477 &      &      \\
  CH &   4311.196 &    1.108 &  $-$3.922 &      &      \\
  CH &   4311.196 &    1.108 &  $-$2.229 &      &      \\
  CH &   4311.261 &    0.388 &  $-$2.879 &      &      \\
  CH &   4311.328 &    0.371 &  $-$1.805 &      &      \\
  CH &   4311.348 &    0.072 &  $-$3.136 &      &      \\
  CH &   4311.394 &    0.388 &  $-$3.105 &      &      \\
  CH &   4311.426 &    0.072 &  $-$1.477 &      &      \\
  CH &   4311.473 &    0.371 &  $-$1.805 &      &      \\
  CH &   4311.476 &    0.371 &  $-$2.771 &      &      \\
  CH &   4311.502 &    0.388 &  $-$1.587 &      &      \\
  CH &   4311.502 &    0.051 &  $-$1.646 &      &      \\
  CH &   4311.545 &    1.108 &  $-$2.269 &      &      \\
  CH &   4311.547 &    0.347 &  $-$2.386 &      &      \\
  CH &   4311.567 &    0.357 &  $-$2.002 &      &      \\
  CH &   4311.580 &    0.388 &  $-$3.105 &      &      \\
  CH &   4311.592 &    0.347 &  $-$2.386 &      &      \\
  CH &   4311.612 &    0.051 &  $-$2.865 &      &      \\
  CH &   4311.618 &    0.662 &  $-$2.054 &      &      \\
  CH &   4311.618 &    0.662 &  $-$2.054 &      &      \\
  CH &   4311.621 &    0.371 &  $-$2.771 &      &      \\
  CH &   4311.655 &    0.357 &  $-$2.002 &      &      \\
  CH &   4311.688 &    0.388 &  $-$1.587 &      &      \\
  CH &   4311.722 &    1.108 &  $-$3.923 &      &      \\
  CH &   4311.727 &    0.051 &  $-$1.646 &      &      \\
  CH &   4311.729 &    1.108 &  $-$2.229 &      &      \\
  CH &   4311.776 &    0.357 &  $-$2.632 &      &      \\
  CH &   4311.837 &    0.051 &  $-$2.866 &      &      \\
  CH &   4311.861 &    0.347 &  $-$2.432 &      &      \\
  CH &   4311.863 &    0.357 &  $-$2.632 &      &      \\
  CH &   4311.897 &    0.371 &  $-$3.052 &      &      \\
  CH &   4311.906 &    0.347 &  $-$2.432 &      &      \\
  CH &   4311.978 &    0.051 &  $-$3.085 &      &      \\
  CH &   4312.017 &    0.371 &  $-$3.052 &      &      \\
  CH &   4312.045 &    0.371 &  $-$1.688 &      &      \\
  CH &   4312.088 &    0.051 &  $-$1.557 &      &      \\
  CH &   4312.153 &    0.033 &  $-$1.773 &      &      \\
  CH &   4312.164 &    0.371 &  $-$1.688 &      &      \\
  CH &   4312.172 &    0.051 &  $-$3.085 &      &      \\
  CH &   4312.280 &    0.051 &  $-$1.557 &      &      \\
  CH &   4312.304 &    0.033 &  $-$1.773 &      &      \\
  CH &   4312.304 &    0.033 &  $-$2.759 &      &      \\
  CH &   4312.317 &    0.358 &  $-$3.007 &      &      \\
  CH &   4312.386 &    0.358 &  $-$3.007 &      &      \\
  CH &   4312.456 &    0.033 &  $-$2.759 &      &      \\
  CH &   4312.527 &    0.358 &  $-$1.831 &      &      \\
  CH &   4312.594 &    0.358 &  $-$1.831 &      &      \\
  CH &   4312.594 &    0.019 &  $-$1.969 &      &      \\
  CH &   4313.620 &          &            &$-$200 &   200   \\
  CH &   4313.377 &    0.020 &   $-$2.985 &      &       \\
  CH &   4313.446 &    0.020 &   $-$2.985 &      &       \\
  CH &   4313.590 &    0.020 &   $-$1.800 &      &       \\
  CH &   4313.660 &    0.020 &   $-$1.800 &      &       \\
  CH &   4313.876 &    0.009 &   $-$3.033 &      &       \\
  CH &   4313.906 &    0.009 &   $-$3.033 &      &       \\
\hline
\end{tabular}
\end{center}
\end{table}

\begin{table}
\begin{center}
\caption{Molecular line data for A-X system of the CH molecule near 4363~\AA\ from Plez \&\ Jorissen. The spectral window employed for this molecular band is defined with respect to two arbitrary wavelengths $\lambda$ and span from $\lambda+d\lambda_\mathrm{blue}$ to $\lambda+d\lambda_\mathrm{red}$. }
\label{tab:lines_ch2}
\tiny
\begin{tabular}{cccrcc}
Species &  $\lambda$ & $\chi$ &  $\log gf $ &   d$\lambda_\mathrm{blue}$ & d$\lambda_\mathrm{red}$ \\
& [\AA] & [eV] & & [m\AA] & [m\AA] \\
\hline
  CH &   4363.300 &          &          & $-$1000 &3700\\
  CH &   4362.021 &    1.247 &   $-$4.054 &      &     \\
  CH &   4362.061 &    1.247 &   $-$1.976 &      &     \\
  CH &   4362.172 &    1.247 &   $-$1.943 &      &     \\
  CH &   4362.202 &    0.777 &   $-$1.982 &      &     \\
  CH &   4362.204 &    1.247 &   $-$4.149 &      &     \\
  CH &   4362.255 &    0.777 &   $-$3.284 &      &     \\
  CH &   4362.531 &    0.777 &   $-$1.917 &      &     \\
  CH &   4362.549 &    0.557 &   $-$1.749 &      &     \\
  CH &   4362.551 &    0.557 &   $-$3.392 &      &     \\
  CH &   4362.697 &    0.777 &   $-$1.982 &      &     \\
  CH &   4362.749 &    0.557 &   $-$1.705 &      &     \\
  CH &   4362.750 &    0.777 &   $-$3.285 &      &     \\
  CH &   4362.985 &    0.777 &   $-$1.917 &      &     \\
  CH &   4363.086 &    0.227 &   $-$3.380 &      &     \\
  CH &   4363.087 &    0.227 &   $-$1.716 &      &     \\
  CH &   4363.162 &    1.248 &   $-$4.055 &      &     \\
  CH &   4363.186 &    1.248 &   $-$1.976 &      &     \\
  CH &   4363.242 &    1.248 &   $-$1.943 &      &     \\
  CH &   4363.276 &    1.248 &   $-$4.149 &      &     \\
  CH &   4363.291 &    0.227 &   $-$1.673 &      &     \\
  CH &   4363.461 &    0.558 &   $-$1.749 &      &     \\
  CH &   4363.463 &    0.558 &   $-$3.392 &      &     \\
  CH &   4363.605 &    0.558 &   $-$1.705 &      &     \\
  CH &   4364.034 &    0.228 &   $-$3.380 &      &     \\
  CH &   4364.036 &    0.228 &   $-$1.716 &      &     \\
  CH &   4364.114 &    1.289 &   $-$4.093 &      &     \\
  CH &   4364.167 &    1.289 &   $-$1.947 &      &     \\
  CH &   4364.181 &    0.228 &   $-$1.673 &      &     \\
  CH &   4364.253 &    1.289 &   $-$1.916 &      &     \\
  CH &   4364.295 &    1.289 &   $-$4.181 &      &     \\
  CH &   4365.416 &    1.290 &   $-$4.093 &      &     \\
  CH &   4365.450 &    1.290 &   $-$1.947 &      &     \\
  CH &   4365.477 &    1.290 &   $-$1.916 &      &     \\
  CH &   4365.522 &    1.290 &   $-$4.181 &      &     \\
  CH &   4365.929 &    0.987 &   $-$3.729 &      &     \\
  CH &   4366.010 &    0.987 &   $-$3.729 &      &     \\
  CH &   4366.230 &    0.987 &   $-$3.828 &      &     \\
  CH &   4366.312 &    0.987 &   $-$3.828 &      &     \\
  CH &   4366.398 &    0.597 &   $-$3.426 &      &     \\
  CH &   4366.407 &    0.597 &   $-$1.699 &      &     \\
  CH &   4366.482 &    0.268 &   $-$3.414 &      &     \\
  CH &   4366.497 &    0.268 &   $-$1.666 &      &     \\
  CH &   4366.520 &    1.333 &   $-$4.129 &      &     \\
  CH &   4366.573 &    0.597 &   $-$1.659 &      &     \\
  CH &   4366.588 &    1.333 &   $-$1.920 &      &     \\
  CH &   4366.647 &    1.333 &   $-$1.892 &      &     \\
  CH &   4366.663 &    0.268 &   $-$1.627 &      &     \\
  CH &   4366.699 &    1.333 &   $-$4.211 &      &     \\
  CH &   4366.897 &    0.805 &   $-$1.903 &      &     \\
  CH &   4366.928 &    0.987 &   $-$3.495 &      &     \\
  CH &   4366.930 &    0.805 &   $-$3.326 &      &     \\
  CH &   4366.991 &    0.987 &   $-$3.495 &      &     \\
  CH &   4367.172 &    0.805 &   $-$1.848 &      &     \\
\hline
\end{tabular}
\end{center}
\end{table}

\begin{table}
\begin{minipage}{8.8cm}
\begin{center}
\caption{Data for atomic lines for species lighter than Fe. The spectral window employed for the line is from $\lambda+d\lambda_\mathrm{blue}$ to $\lambda+d\lambda_\mathrm{red}$.  Lines which are removed in carbon-enhanced stars are marked with asterisks in the wavelength column.  
}
\label{tab:lines_light}
\tiny
\begin{tabular}{crcrccl} 
Species &  $\lambda$ & $\chi$ &  $\log gf $ &   d$\lambda_\mathrm{blue}$ & d$\lambda_\mathrm{red}$ & Refs. \\
& [\AA] & [eV] & & [m\AA] & [m\AA] & \\
\hline
Mg I &   3829.355 &    2.707 &   $-$0.208 &  $-$150 &   100 & WSM69 \\
Mg I &   3832.304 &    2.710 &   $ $0.270 &  $-$300 &   200 & WSM69 \\
Mg I &   3838.292 &    2.715 &   $ $0.490 &  $-$230 &   150 & WSM69 \\
Mg I &  *4057.505 &    4.346 &   $-$0.890 &  $-$120 &   100 & FF75 \\
Mg I &  *4167.271 &    4.346 &   $-$0.710 &  $-$250 &   200 & FF75\\ 
Mg I &   4571.096 &    0.000 &   $-$5.393 &  $-$300 &   300 & WSM69 \\ 
Mg I &   4702.990 &    4.330 &   $-$0.380 &  $-$300 &   300 & FF75 \\ 
&&&&&&\\
Al I &   3961.529 &    0.014 &   $-$0.336 &  $-$200 &   170 & WM80 \\
&&&&&&\\
Ca I &   4226.728 &    0.000 &   $ $0.244 &  $-$150 &   400 & WM80 \\
Ca I &  *4283.011 &    1.886 &   $-$0.224 &  $-$120 &   300 & WM80 \\
Ca I &  *4289.367 &    1.879 &   $-$0.304 &  $-$120 &   170 & WM80 \\
Ca I &   4318.652 &    1.890 &   $-$0.207 &  $-$120 &   120 & WM80 \\
Ca I &  *4425.437 &    1.879 &   $-$0.358 &  $-$300 &   300 & WM80 \\
Ca I &  *4434.957 &    1.886 &   $-$0.005 &  $-$300 &    50 & WM80 \\
Ca I &   4454.780 &    1.898 &   $ $0.258 &  $-$50  &   200 & WM80 \\
Ca I &   4455.887 &    1.899 &   $-$0.526 &  $-$300 &   300 & WM80 \\ 
Ca I &   4578.551 &    2.521 &   $-$0.558 &  $-$200 &   100 & WM80 \\ 
&&&&&&\\
Sc II&   4246.837 &    0.315 &   $ $0.240 &  $-$350 &   250 & LD89 \\
Sc II&  *4314.095 &    0.618 &   $-$0.100 &  $-$250 &    70 & LD89 \\
Sc II&  *4400.399 &    0.605 &   $-$0.540 &  $-$150 &   200 & LD89 \\
Sc II&   4415.563 &    0.595 &   $-$0.670 &  $-$180 &   300 & LD89 \\
Sc II&   4670.417 &    1.357 &   $-$0.580 &  $-$150 &    50 & LD89 \\
&&&&&&\\
Ti I &   3989.759 &    0.021 &   $-$0.140 &  $-$200 &   100 & GBP89, BPSL82  \\
Ti I &   3998.636 &    0.048 &   $ $0.000 &  $-$250 &   150 & GBP89, BPSL82  \\
Ti I &   4533.249 &    0.848 &   $ $0.530 &  $-$150 &   300 & GBP89, BMPS82  \\
Ti I &   4534.776 &    0.836 &   $ $0.340 &  $-$300 &   300 & GBP89, BMPS82  \\
Ti I &   4656.469 &    0.000 &   $-$1.290 &  $-$300 &   300 & GBP89, BPSL82  \\
Ti I &  *4681.909 &    0.048 &   $-$1.020 &  $-$300 &   150 & GBP89, BPSL82  \\
Ti II&  *4337.915 &    1.080 &   $-$0.960 &  $-$150 &   150 & PTP01\\ 
Ti II&  *4394.051 &    1.221 &   $-$1.780 &  $-$100 &   300 & PTP01\\ 
Ti II&  *4395.850 &    1.243 &   $-$1.930 &  $-$200 &   100 & PTP01\\ 
Ti II&  *4417.719 &    1.165 &   $-$1.190 &  $-$200 &   300 & PTP01\\ 
Ti II&   4443.794 &    1.080 &   $-$0.720 &  $-$300 &   250 & PTP01\\ 
Ti II&   4468.507 &    1.131 &   $-$0.600 &  $-$150 &   300 & RHLPS94\\ 
Ti II&  *4470.857 &    1.165 &   $-$2.020 &  $-$250 &   150 & PTP01\\ 
Ti II&   4501.273 &    1.116 &   $-$0.770 &  $-$300 &   300 & PTP01\\ 
Ti II&   4563.761 &    1.221 &   $-$0.690 &  $-$300 &   300 & PTP01\\ 
Ti II&   4571.968 &    1.572 &   $-$0.320 &  $-$300 &   150 & PTP01\\ 
Ti II&   4589.958 &    1.237 &   $-$1.620 &  $-$300 &   300 & RHLPS94\\ 
&&&&&&\\
V  I &  *4379.230 &    0.301 &   $ $0.550 &  $-$150 &   150 & DKKWZ85 \\
V  I &  *4389.976 &    0.275 &   $ $0.270 &  $-$150 &   150 & DKKWZ85 \\
V  II&  *3951.960 &    1.480 &   $-$0.784 &  $-$150 &    60 & BGFML89 \\
&&&&&&\\
Cr I &   4254.332 &    0.000 &   $-$0.114 &  $-$150 &   200 & see MFW88 \\
Cr I &  *4274.796 &    0.000 &   $-$0.231 &  $-$200 &   100 & see MFW88 \\
Cr I &  *4289.716 &    0.000 &   $-$0.361 &  $-$150 &   100 & see MFW88 \\
&&&&&&\\
Mn I &   4030.763 &    0.000 &   $-$0.470 &  $-$150 &   250 & BBPS84 \\
Mn I &   4033.060 &    0.000 &   $-$0.618 &  $-$200 &   150 & BBPS84 \\
Mn I &  *4034.492 &    0.000 &   $-$0.811 &  $-$150 &   300 & BBPS84 \\
Mn I &   4754.040 &    2.282 &   $-$0.086 &  $-$200 &   200 & BBPS84 \\
Mn I &   4823.496 &    2.319 &   $ $0.144 &  $-$100 &   200 & BBPS84 \\
\hline
\end{tabular}
\end{center}
\tiny
References:  BBPS84 = Booth et~al.~(\cite{oxford_mn84}), BGFML89 = Bi\'emont et~al.~(\cite{biemont89v}), BMPS82 = Blackwell et~al.~(\cite{oxford_ti82a}), BPSL82 = Blackwell et~al.~(\cite{oxford_ti82b}),  DKKWZ85 = Doerr et~al.~(\cite{doerr85}), FF75 = Froese Fischer~(\cite{froese_fischer75}), GBP89 = Grevesse et~al.~(\cite{grevesse89}), LD89 = Lawler \& Dakin~(\cite{lawler89}), MFW88 = Martin, Fuhr \& Wiese (\cite{martin88}), 
PTP01 = Pickering et~al.~(\cite{pickering01}), RHLPS94 = Ryabchikova et~al.~(\cite{ryabchikova94}), WM80 = Wiese \& Martin (\cite{wiese80}), WSM = Wiese et~al.~(\cite{wiese69}).
\end{minipage}
\end{table}

\begin{table}
\begin{minipage}{8.8cm}
\begin{center}
\caption{Data for atomic lines of Fe.  The spectral window employed for the line is from $\lambda+d\lambda_\mathrm{blue}$ to $\lambda+d\lambda_\mathrm{red}$.  Lines which are removed in carbon-enhanced stars are marked with asterisks in the wavelength column.}
\label{tab:lines_fe}
\tiny
\begin{tabular}{crcrccl} 
Species &  $\lambda$ & $\chi$ &  $\log gf $ &   d$\lambda_\mathrm{blue}$ & d$\lambda_\mathrm{red}$ & Refs. \\
& [\AA] & [eV] & & [m\AA] & [m\AA] & \\
\hline
Fe  I & *3856.372 &  0.052 & $-$1.286  &  $-$200 &  400 &  OWLWB91 \\
Fe  I & *3859.911 &  0.000 & $-$0.710  &  $-$400 &  350 &  OWLWB91 \\
Fe  I &  3865.523 &  1.011 & $-$0.950  &  $-$250 &  250 &  OWLWB91 \\
Fe  I & *3878.018 &  0.958 & $-$0.896  &  $-$300 &  100 &  OWLWB91 \\
Fe  I &  4005.242 &  1.557 & $-$0.583  &  $-$150 &  150 &  OWLWB91 \\
Fe  I &  4045.812 &  1.485 & $ $0.284  &  $-$100 &  150 &  OWLWB91 \\
Fe  I &  4063.594 &  1.558 & $ $0.062  &  $-$100 &  300 &  OWLWB91 \\
Fe  I &  4071.738 &  1.608 & $-$0.008  &  $-$100 &  300 &  OWLWB91 \\
Fe  I &  4114.445 &  2.831 & $-$1.303  &  $-$250 &  250 &  OWLWB91  \\
Fe  I &  4132.058 &  1.608 & $-$0.675  &  $-$300 &  100 &  OWLWB91  \\
Fe  I &  4143.868 &  1.557 & $-$0.511  &  $-$200 &  300 &  OWLWB91  \\
Fe  I & *4175.636 &  2.845 & $-$0.827  &  $-$150 &  300 &  OWLWB91  \\
Fe  I & *4187.039 &  2.449 & $-$0.514  &  $-$150 &  200 &  OWLWB91  \\  
Fe  I & *4199.095 &  3.047 & $ $0.156  &  $-$200 &  300 &  OWLWB91  \\
Fe  I &  4202.029 &  1.485 & $-$0.689  &  $-$200 &  150 &  OWLWB91  \\
Fe  I &  4222.213 &  2.449 & $-$0.914  &  $-$300 &  200 &  OWLWB91  \\
Fe  I &  4227.426 &  3.332 & $ $0.266  &  $-$150 &  150 &  OWLWB91  \\
Fe  I & *4233.603 &  2.482 & $-$0.579  &  $-$150 &  100 &  OWLWB91  \\
Fe  I & *4250.120 &  2.469 & $-$0.380  &  $-$100 &  300 &  OWLWB91  \\
Fe  I &  4250.787 &  1.557 & $-$0.713  &  $-$300 &  300 &  OWLWB91  \\
Fe  I &  4260.474 &  2.400 & $ $0.077  &  $-$150 &  200 &  OWLWB91  \\
Fe  I & *4271.154 &  2.449 & $-$0.337  &  $-$300 &  100 &  OWLWB91  \\
Fe  I & *4271.761 &  1.485 & $-$0.173  &  $-$150 &  300 &  OWLWB91  \\
Fe  I &  4282.403 &  2.176 & $-$0.779  &  $-$250 &  150 &  OWLWB91  \\
Fe  I & *4325.762 &  1.608 & $ $0.006  &  $-$250 &  100 &  OWLWB91  \\ 
Fe  I & *4375.930 &  0.000 & $-$3.031  &  $-$150 &  300 &  OWLWB91  \\
Fe  I & *4383.545 &  1.485 & $ $0.208  &  $-$300 &  300 &  OWLWB91  \\
Fe  I & *4404.750 &  1.557 & $-$0.147  &  $-$250 &  200 &  OWLWB91  \\
Fe  I & *4415.123 &  1.608 & $-$0.621  &  $-$300 &  200 &  OWLWB91  \\
Fe  I & *4430.614 &  2.223 & $-$1.728  &  $-$250 &  250 &  OWLWB91  \\
Fe  I & *4442.339 &  2.198 & $-$1.228  &  $-$300 &  300 &  OWLWB91  \\
Fe  I &  4443.194 &  2.858 & $-$1.043  &  $-$100 &  250 &  OWLWB91  \\
Fe  I &  4447.717 &  2.223 & $-$1.339  &  $-$300 &  300 &  OWLWB91  \\
Fe  I &  4489.739 &  0.121 & $-$3.899  &  $-$300 &  200 &  OWLWB91  \\
Fe  I &  4494.563 &  2.198 & $-$1.143  &  $-$300 &  300 &  OWLWB91  \\
Fe  I &  4528.614 &  2.176 & $-$0.887  &  $-$100 &  300 &  OWLWB91  \\
Fe  I &  4602.941 &  1.485 & $-$2.208  &  $-$300 &  300 &  OWLWB91  \\
Fe  I & *4736.773 &  3.211 & $-$0.752  &  $-$200 &  200 &  OWLWB91  \\
Fe  I &  4872.137 &  2.882 & $-$0.567  &  $-$300 &  300 &  OWLWB91  \\
Fe  I &  4890.755 &  2.876 & $-$0.394  &  $-$350 &  350 &  OWLWB91  \\
Fe  I &  4891.492 &  2.852 & $-$0.112  &  $-$300 &  300 &  OWLWB91  \\
Fe  I &  4918.994 &  2.845 & $-$0.342  &  $-$350 &  350 &  OWLWB91  \\
Fe  I &  4920.503 &  2.832 & $ $0.068  &  $-$400 &  100 &  OWLWB91  \\
Fe  I &  4938.814 &  2.875 & $-$1.077  &  $-$300 &  200 &  OWLWB91  \\
Fe  I &  4939.687 &  0.859 & $-$3.252  &  $-$250 &  300 &  OWLWB91  \\
Fe  II& *4178.862 &  2.583 & $-$2.443  &  $-$100 &  200 &  SSK04 \\
Fe  II&  4233.172 &  2.583 & $-$1.809  &  $-$150 &  200 &  SSK04 \\
Fe  II&  4416.828 &  2.778 & $-$2.540  &  $-$240 &  300 &  M83 \\
Fe  II&  4508.289 &  2.856 & $-$2.318  &  $-$300 &  300 &  BBKAP91 \\
Fe  II&  4515.343 &  2.844 & $-$2.362  &  $-$100 &  300 &  SSK04 \\
Fe  II&  4520.224 &  2.807 & $-$2.550  &  $-$200 &  300 &  M83  \\
Fe  II& *4541.524 &  2.856 & $-$2.990  &  $-$100 &  200 &  M83 \\
Fe  II& *4555.893 &  2.828 & $-$2.250  &  $-$150 &  100 &  SSK04 \\
Fe  II&  4583.839 &  2.807 & $-$1.740  &  $-$250 &  300 &  SSK04 \\
Fe  II&  4923.927 &  2.891 & $-$1.206  &  $-$400 &  400 &  SSK04 \\
\hline
\end{tabular}
\end{center}
\tiny
References: BBKAP91 = Bi\'emont et~al.~(\cite{biemont91}), M83 = Moity (\cite{moity83}), OWLWB91 = O'Brian et~al.~\cite{obrian91}, SSK04 = Schnabel et~al.~(\cite{schnabel04})
\end{minipage}
\end{table}

\begin{table}
\begin{minipage}{8.8cm}
\begin{center}
\caption{Data for atomic lines of species heavier than Fe.  The spectral window employed for the line is from $\lambda+d\lambda_\mathrm{blue}$ to $\lambda+d\lambda_\mathrm{red}$.  Lines which are removed in carbon-enhanced stars are marked with asterisks in the wavelength column.  }
\label{tab:lines_heavy}
\tiny
\begin{tabular}{crcrccl} 
Species &  $\lambda$ & $\chi$ &  $\log gf $ &   d$\lambda_\mathrm{blue}$ & d$\lambda_\mathrm{red}$ & Refs. \\
& [\AA] & [eV] & & [m\AA] & [m\AA] & \\  
\hline
Co I &  *3842.046 &    0.923 &   $-$0.770 &  $-$200 &   100 & CSSTW82 \\  
Co I &  *3845.461 &    0.923 &   $ $0.010 &  $-$150 &   150 & CSSTW82 \\        
Co I &   3894.073 &    1.049 &   $ $0.090 &  $-$50  &   200 & NKWL99 \\
Co I &   3995.302 &    0.922 &   $-$0.140 &  $-$250 &   250 & NKWL99 \\
Co I &   4118.767 &    1.049 &   $-$0.470 &  $-$100 &   200 & NKWL99 \\
Co I &   4121.311 &    0.922 &   $-$0.300 &  $-$250 &   250 & NKWL99 \\
&&&&&&\\
Ni I &   3775.565 &    0.423 &   $-$1.408 &  $-$200 &   200 &  BBPL89 \\
Ni I &   3783.524 &    0.423 &   $-$1.304 &  $-$70  &   300 &  HS80 \\
Ni I &   3807.138 &    0.423 &   $-$1.220 &  $-$200 &   150 &  BBPL89 \\
Ni I &   3858.292 &    0.423 &   $-$0.951 &  $-$300 &   100 &  BBPL89 \\
&&&&&&\\
Zn I &  *4722.163 &    4.030 &   $-$0.390 &  $-$300 &   300 & BG80 \\
Zn I &   4810.537 &    4.078 &   $-$0.170 &  $-$300 &   300 & BG80 \\
&&&&&&\\
Sr I &  *4607.327 &    0.000 &   $ $0.283 &  $-$300 &   100 & MB87 \\
Sr II&   4077.709 &    0.000 &   $ $0.158 &  $-$150 &   100 & PBL95  \\
Sr II&   4215.519 &    0.000 &   $-$0.155 &  $-$300 &   200 & PBL95 \\
&&&&&&\\
Y  II&   3774.331 &    0.130 &   $ $0.210 &  $-$250 &   150 & HLGBW82 \\
Y  II&  *3788.694 &    0.104 &   $-$0.070 &  $-$150 &   300 & HLGBW82 \\
Y  II&   3818.314 &    0.130 &   $-$0.980 &  $-$300 &   150 & HLGBW82 \\
Y  II&   3950.352 &    0.104 &   $-$0.490 &  $-$200 &   300 & HLGBW82 \\
Y  II&  *4398.013 &    0.130 &   $-$1.000 &  $-$250 &   150 & HLGBW82 \\
Y  II&   4883.684 &    1.084 &   $ $0.070 &  $-$200 &   200 & HLGBW82 \\
&&&&&&\\
Zr II&  *4161.213 &    0.713 &   $-$0.720 &  $-$300 &   130 & BGHL81 \\
Zr II&   4208.985 &    0.713 &   $-$0.460 &  $-$200 &   200 & BGHL81 \\
Zr II&   4317.299 &    0.713 &   $-$1.380 &  $-$300 &   300 & BGHL81 \\
&&&&&&\\
Ba II&   4130.645 &    2.722 &      0.560 &  $-$150 &   300 & G67 \\
Ba II&   4554.000 &    0.000 &   $ $0.163 &  $-$500 &   220 & G67 \\
&&&&&&\\
La II&  *3988.515 &    0.403 &   $ $0.210 &  $-$300 &   300 & LBS01 \\
La II&   3995.745 &    0.173 &   $-$0.060 &  $-$250 &   150 & LBS01 \\
La II&   4086.709 &    0.000 &   $-$0.070 &  $-$200 &   200 & LBS01 \\
La II&   4123.218 &    0.321 &   $ $0.130 &  $-$300 &   180 & LBS01 \\
La II&  *4322.503 &    0.173 &   $-$0.930 &  $-$300 &   150 & LBS01 \\
La II&  *4333.753 &    0.173 &   $-$0.060 &  $-$300 &   100 & LBS01 \\
&&&&&&\\
Ce II&   3999.237 &    0.295 &   $ $0.232 &  $-$100 &   300 &  VALD95 \\
Ce II&   4073.474 &    0.478 &   $ $0.320 &  $-$150 &   150 &  SMPCBA96         \\
Ce II&  *4083.222 &    0.701 &   $ $0.240 &  $-$150 &   200 &  SMPCBA96         \\
Ce II&   4120.827 &    0.320 &   $-$0.240 &  $-$300 &   300 &  SMPCBA96         \\
Ce II&  *4127.364 &    0.684 &   $ $0.240 &  $-$300 &    70 &  SMPCBA96         \\
Ce II&   4222.597 &    0.122 &   $-$0.180 &  $-$150 &   150 &  GS94     \\     
Ce II&  *4418.780 &    0.864 &   $ $0.310 &  $-$200 &   150 &  SMPCBA96         \\
Ce II&   4486.909 &    0.295 &   $-$0.360 &  $-$300 &   300 &  GS94     \\
Ce II&   4562.359 &    0.478 &   $ $0.330 &  $-$300 &   300 &  GS94     \\
Ce II&   4628.161 &    0.516 &   $ $0.260 &  $-$300 &   300 &  GS94     \\
&&&&&&\\
Nd II&  *4018.823 &    0.064 &   $-$0.850 &  $-$300 &  120 &  DLSC03  \\
Nd II&   4021.327 &    0.321 &   $-$0.100 &  $-$200 &  200 &  DLSC03  \\
Nd II&   4061.085 &    0.471 &   $ $0.550 &  $-$300 &  300 &  DLSC03  \\
Nd II&   4069.265 &    0.064 &   $-$0.570 &  $-$200 &  300 &  DLSC03  \\
Nd II&  *4109.448 &    0.321 &   $ $0.350 &  $-$150 &  150 &  DLSC03  \\
Nd II&   4232.374 &    0.064 &   $-$0.470 &  $-$200 &  200 &  DLSC03  \\ 
Nd II&  *4358.161 &    0.321 &   $-$0.160 &  $-$200 &  150 &  DLSC03  \\
Nd II&   4446.384 &    0.205 &   $-$0.350 &  $-$300 &  300 &  DLSC03  \\
Nd II&  *4462.979 &    0.559 &   $ $0.040 &  $-$300 &  300 &  DLSC03  \\
&&&&&&\\
Sm II&   3896.972 &    0.041 &   $-$0.578 &  $-$100 &   200 &  BGHL89\\ 
Sm II&   4068.324 &    0.434 &   $-$0.710 &  $-$150 &   150 &  BGHL89\\
Sm II&  *4318.936 &    0.277 &   $-$0.270 &  $-$150 &   150 &  BGHL89\\
Sm II&   4499.475 &    0.248 &   $-$1.010 &  $-$200 &   200 &  BGHL89\\
Sm II&   4519.630 &    0.544 &   $-$0.432 &  $-$300 &   300 &  BGHL89\\
Sm II&   4537.954 &    0.485 &   $-$0.230 &  $-$300 &   300 &  BGHL89\\
Sm II&   4577.688 &    0.248 &   $-$0.775 &  $-$300 &   300 &  BGHL89\\
Sm II&  *4642.232 &    0.378 &   $-$0.520 &  $-$150 &   150 &  BGHL89\\
&&&&&&\\
Eu II&   3819.670 &    0.000 &   $ $0.510 &  $-$150 &   300 &  LWDS01 \\
Eu II&   3907.110 &    0.207 &   $ $0.170 &  $-$150 &   200 &  LWDS01 \\
Eu II&   4129.720 &    0.000 &   $ $0.220 &  $-$150 &   400 &  LWDS01 \\
Eu II&  *4205.040 &    0.000 &   $ $0.210 &   $-$50 &   200 &  LWDS01 \\
\hline
\end{tabular}
\end{center}
\tiny
References: BBPL89 = Blackwell et~al.~(\cite{oxford_ni89}), BG80 = Biemont \& Godefroid (\cite{biemont80}), BGHL81 = Bi\'emont et~al.~(\cite{biemont81}), CSSTW82 = Cardon et~al.~(\cite{cardon82}), DLSC03 = Den Hartog et~al.~(\cite{den_hartog03}), G67 = Gallagher~(\cite{gallagher67}), GS94 = Gratton \& Sneden (\cite{gratton94}), HLGBW82 = Hannaford et~al.~(\cite{hannaford82}), HS80 = Huber \& Sandeman~(\cite{huber80}), LBS01 = Lawler et~al.~(\cite{lawler01la}), LWDS01 = Lawler et~al.~(\cite{lawler01}),  MB87 = Migdalek \& Baylis~(\cite{migdalek87}), NKWL99 = Nitz et~al.~(\cite{nitz99}),
PBL95 = Pinnington et~al.~(\cite{pinnington95}), VALD95 = see Piskunov et~al.~(\cite{vald95})
\end{minipage}
\end{table}

\section{Error Estimates}
\label{app:errors}

Our method for estimating error propagation is similar to the approach devised by McWilliam et~al.~(\cite{mcw95b}), but modified to suit our abundance analysis method.  The important difference is that our approach fits the spectral features of a given element globally rather than fitting individual lines.  Further, as discussed in Sect.~\ref{subsect:detect_errors}, we will make a distinction between {\em absolute} (the uncertainty in the absolute abundance) and {\em relative} (the uncertainty in the relative abundance between stars) error estimates. First, we develop the formalism quite generally, and later specify the difference between the calculations for these two types of errors.

We consider the propagation of errors in model atmosphere parameters $T_\mathrm{eff}$, $\log g$, $\xi$.  For low metallicity models of the type used here, typical errors in metallicity have negligible effect on the model structure and can be neglected.  Errors due to propagation of uncertainties in $\log gf$, observational error, continuum placement and spectrum modelling uncertainties are also considered.  For compactness, in the following discussion we define the abundance parameter $\varepsilon\equiv\log\epsilon$.  Assuming the behaviour of the considered abundance with small changes in these parameters can be approximated by a first-order Taylor expansion, i.e.\ that the abundance $\varepsilon\equiv\log\epsilon$ varies approximately linearly with changes in parameters on scales of the parameter uncertainties, we obtain (e.g. McWilliam et~al.~\cite{mcw95b}, Taylor~\cite{taylor82}) for the variance in the abundance
\begin{eqnarray}
\sigma_{\varepsilon}^2 & = & 
\left(\frac{\partial\varepsilon}{\partial T}\right)^2 \sigma_T^2  
+ \left(\frac{\partial\varepsilon}{\partial \log g}\right)^2 \sigma_{\log g}^2
+ \left(\frac{\partial\varepsilon}{\partial \xi}\right)^2 \sigma_{\xi}^2 
\nonumber \\
&&
+ \sigma_{\log gf}^2 +  \sigma_{\varepsilon}^2(\mathrm{obs}) +  \sigma_{\varepsilon}^2 (\mathrm{cont}) +  \sigma_{\varepsilon}^2 (\mathrm{model})
\nonumber \\
&&
+ 2 \left\{  \frac{\partial\varepsilon}{\partial T} \frac{\partial\varepsilon}{\partial \log g}\sigma_{T,\log g}   
+ \frac{\partial\varepsilon}{\partial T} \frac{\partial\varepsilon}{\partial \xi} \sigma_{T,\xi}  \right.
\nonumber \\
&&
\left.
+ \frac{\partial\varepsilon}{\partial \log g} \frac{\partial\varepsilon}{\partial \xi} \sigma_{\log g, \xi} \right\} ,
\end{eqnarray}
where $\sigma_i^2$ is the variance in parameter $i$, and  $\sigma_{i,j}$ the covariance of $i$ and $j$.  The variances $\sigma_{\varepsilon}^2(\mathrm{obs})$ and $\sigma_{\varepsilon}^2 (\mathrm{cont})$ represent the variance in the abundance $\varepsilon$ due to observational error and continuum placement uncertainties respectively.  The variance $\sigma_{\varepsilon}^2(\mathrm{model})$ represents the variance in the abundance $\varepsilon$ due to spectrum modelling uncertainties, such as the assumptions of 1D modelling and LTE.  Terms involving covariances of independent parameters have been omitted.  We have used the fact that $\partial\varepsilon / \partial \log gf = 1$.

We adopt $\sigma_{T,\xi}=0$ and $\sigma_{\log g,\xi}=-0.02$ following McWilliam et~al.~(\cite{mcw95b}).  We obtained similar estimates for these quantities from our own numerical experiments for a subsample of stars representative of the complete sample.   Based on these experiments we adopted $\sigma_{T,\log g}=+22$.  Thus, the expression for the total variance becomes
\begin{eqnarray}
\sigma_{\varepsilon}^2 & = & 
\left(\frac{\partial\varepsilon}{\partial T}\right)^2 \sigma_T^2  
+ \left(\frac{\partial\varepsilon}{\partial \log g}\right)^2 \sigma_{\log g}^2
+ \left(\frac{\partial\varepsilon}{\partial \xi}\right)^2 \sigma_{\xi}^2 
\nonumber \\
&&
+ \sigma_{\log gf}^2 +  \sigma_{\varepsilon}^2(\mathrm{obs}) +  \sigma_{\varepsilon}^2 (\mathrm{cont})+  \sigma_{\varepsilon}^2 (\mathrm{model})
\\
&&
+ 2 \left\{  \frac{\partial\varepsilon}{\partial T} \frac{\partial\varepsilon}{\partial \log g}  \sigma_{T, \log g}   
+            \frac{\partial\varepsilon}{\partial \log g} \frac{\partial\varepsilon}{\partial \xi} \sigma_{\log g, \xi}  \right\} . \nonumber
\label{eqn:errors}
\end{eqnarray}
The required partial derivatives are obtained individually for each star and element by direct determination of abundances for shifted stellar parameters (see Sect.~\ref{sect:procedure}).    

Now it simply remains to specify each variance (or standard deviation). The input variances, however, will depend on whether we wish to estimate the absolute error or the relative error.  

\subsection{Relative error estimates} 

The relative error estimates are of interest for comparison of abundances within the sample.   In particular, we wish to estimate the amount of abundance scatter which should be attributed to uncertainties in the data and analysis.  

First, from Sect.~\ref{subsect:photometry}, we adopt $\sigma_T=100$~K.   If the measurement errors in the observed data are normally distributed, the formal random errors in derived spectrum modelling parameters due to this measurement error are given by the SME optimisation routine from the estimated covariance matrix, which is the inverse of the curvature matrix at the solution (see e.g. Press et~al~\cite{numrec}).  These errors represent the propagation of the random observational uncertainties in each point in the spectrum (dominated by photon noise) to the relevant parameter and is computed by the parameter optimisation procedure.  Such contributions are obtained for $\sigma_{\varepsilon}(\mathrm{obs})$, $\sigma_{\log g}$ and $\sigma_{\xi}$; these random components of $\sigma_{\log g}$ and $\sigma_{\xi}$ are usually quite small, around $0.05$~dex and $0.05$ km~s$^{-1}$, while $\sigma_{\varepsilon}(\mathrm{obs})$ varies significantly from element to element.  To this random component of $\sigma_{\log g}$ we add 0.22~dex reflecting the error due to uncertainty in $T_\mathrm{eff}$, and to the random component of $\sigma_{\xi}$ we add 0.1 km~s$^{-1}$ reflecting the error due to uncertainty in $\log g$.  These numbers are based on the numerical experiments for a representative subsample which were mentioned above.   

As mentioned above, errors arise due to uncertainties in the spectrum modelling, for example, assumptions of 1D geometry or LTE.  The contributions of these errors to relative errors in abundances would be expected to cancel if all the stars were identical; however, across our sample, which covers a wide range of stellar atmosphere parameters, we can only expect partial cancellation.  Thus, such uncertainties will lead to a degree of scatter in any quantity derived from the spectrum, arising from differences in modelling errors from star to star.  The relative component of $\sigma_{\varepsilon}(\mathrm{model})$ has been estimated at $0.1$~dex.  This is simply an order-of-magnitude estimate based on indications from 3-D modelling, (e.g. Asplund~\cite{asplund04, asplund05}) and non-LTE spectrum modelling (e.g. Korn et~al.~\cite{korn03}, Asplund~\cite{asplund05}).  However it should be noted this quantity should vary with element and employed spectral feature.  We have neglected any relative error in $\log g$ and $\xi$ which might arise from modelling uncertainties. 

Since the relevant continuum points are located close to lines employed in the analysis, they will have very similar error bars to the points in the line, and thus the sensitivity of the abundance to continuum placement errors will be essentially identical to that for the points in the line.  Thus we expect that the error due to uncertainty in the continuum placement can be approximated by $\sigma_{\varepsilon}(\mathrm{cont})\approx\sigma_{\varepsilon}(\mathrm{obs})/\sqrt{m}$, where the $1/\sqrt{m}$ factor accounts for the fact that we typically have $m$ times more independent pixels to define the continuum than the line.  Based on inspection of spectra we adopt $m=5$ for all elements except carbon, where we adopt $m=1$ since it is determined from wide molecular bands.  The estimates for $m$ approximately account for possible correlations between pixels introduced by very weak lines.  The uncertainty in $\log gf$ will not contribute to relative errors, and thus we adopt $\sigma_{\log gf}=0$. 

\subsection{Absolute error estimates}

While relative errors are generally of most interest, it is also important to have some estimate of the absolute error in our obtained abundances.  The calculation of absolute errors follows that of the relative errors with a few changes which we now list.  

The main differences arise in the modelling uncertainties, as there is no cancellation as in the relative error case.  To $\sigma_{\log g}$ and $\sigma_{\xi}$ we add an additional $0.1$~dex and $0.1$~km~s$^{-1}$ respectively to account for modelling uncertainties.  We estimate $\sigma_{\varepsilon}(\mathrm{model})$ at $0.15$~dex.  Further, the errors $\sigma_{\log gf}$ must be included in the absolute errors.  Since our procedure globally fits the spectral lines of a given element, the value for $\sigma_{\log gf}$ should be representative of the typical error for the chosen spectral lines.  In Table~\ref{tab:loggf_unc} we provide an estimated average uncertainty for each element with reference to the original literature, which are adopted for $\sigma_{\log gf}$.

\begin{table}
\begin{center}
\caption{Assigned average values of $\sigma_{\log gf}$ for each element.}
\label{tab:loggf_unc}
\begin{tabular}{cccc}
\hline
Element &  $\sigma_{\log gf}$ & Element &  $\sigma_{\log gf}$ \\
\hline
C & 0.10  & Ni & 0.03 \\
Mg & 0.07 & Zn & 0.10 \\
Al & 0.11 & Sr & 0.10 \\
Ca & 0.11 & Y  & 0.03 \\
Sc & 0.04 & Zr & 0.03 \\
Ti & 0.05 & Ba & 0.03 \\
V  & 0.05 & La & 0.03 \\
Cr & 0.05 & Ce & 0.10 \\
Mn & 0.06 & Nd & 0.03 \\
Fe & 0.03 & Sm & 0.05 \\
Co & 0.10 & Eu & 0.03 \\
\hline
\end{tabular}
\end{center}
\end{table} 

\subsection{Errors in abundance ratios}

We are often interested in elemental abundance ratios.  Following McWilliam~et~al.~(\cite{mcw95b}), if the abundances $\varepsilon_A$ and $\varepsilon_B$ are expressed in logarithms such that $\varepsilon_{A/B}=\varepsilon_A-\varepsilon_B$, the variance in the abundance ratio $\varepsilon_{A/B}$ is given by
\begin{equation}
\sigma_{A/B}^2 =  \sigma_{A}^2 + \sigma_{B}^2 - 2 \sigma_{A,B}
\label{eqn:var_ratio}
\end{equation} 
where $\sigma_{A}^2$ and  $\sigma_{B}^2$ are the variance in $\varepsilon_A$ and $\varepsilon_B$ respectively and $\sigma_{A,B}$ is the covariance of $\varepsilon_A$ and $\varepsilon_B$ given by
\begin{eqnarray}
\sigma_{A,B} & = & 
\frac{\partial\varepsilon_A}{\partial T} \frac{\partial\varepsilon_B}{\partial T} \sigma_T^2  
+ \frac{\partial\varepsilon_A}{\partial \log g} \frac{\partial\varepsilon_B}{\partial \log g} \sigma_{\log g}^2
+ \frac{\partial\varepsilon_A}{\partial \xi}\frac{\partial\varepsilon_B}{\partial \xi} \sigma_{\xi}^2 
\nonumber \\
&&
+ \left\{  \frac{\partial\varepsilon_A}{\partial T}\frac{\partial\varepsilon_B}{\partial \log g} + \frac{\partial\varepsilon_A}{\partial \log g}\frac{\partial\varepsilon_B}{\partial T}\right\} \sigma_{T, \log g}
\nonumber \\
&& 
+ \left\{  \frac{\partial\varepsilon_A}{\partial \xi}\frac{\partial\varepsilon_B}{\partial \log g} + \frac{\partial\varepsilon_A}{\partial \log g}\frac{\partial\varepsilon_B}{\partial \xi}\right\} \sigma_{\log g, \xi}.
\end{eqnarray}
Thus, as pointed out by McWilliam et~al.~(\cite{mcw95b}), there may be partial cancellation of errors if the abundances of elements $A$ and $B$ have similar sensitivity to atmospheric parameters, or partial compounding of errors if the elements have contrary sensitivity.  Note, abundance ratio error estimates may be computed in both the absolute and relative senses.


\begin{thebibliography}{}

\bibitem[2002]{allende02} 
Allende Prieto C., Lambert D.L., Asplund M., 2002, \apj\ 573, L137

\bibitem[1999]{alonso99} 
Alonso A., Arribas S., Martinez-Roger C. 1999, \aaps\ 139, 335

\bibitem[1991]{anstee91} 
Anstee S.D., O'Mara B.J., 1991, \mnras\ 253, 549

\bibitem[2005]{aoki05} 
Aoki W., Honda S., Beers T.C., et~al., 2005, in press

\bibitem[2002a]{aoki02a} 
Aoki W., Norris J.E., Ryan S.G., Beers T.C., Ando H., 2002, \apj\ 567, 1182

\bibitem[2002b]{aoki02b} 
Aoki W., Ryan S.G., Norris J.E., et~al., 2002, \apj\ 580, 1149

\bibitem[1999]{arce99} 
Arce H.G., Goodman A.A. 1999, \apj\ 512, L135

\bibitem[1999]{arlandini99} 
Arlandini C., K\"appeler F., Wisshak K., et~al., 1999, \apj\ 525, 886

\bibitem[2005]{arnone05} 
Arnone E., Ryan S.G., Argast D., Norris J.E., Beers T.C., 2005, \aap\ 430, 507

\bibitem[1987]{arribas87} 
Arribas S. Martinez-Roger C.M., 1987, \aaps\ 70, 303

\bibitem[1997]{marcs_asp} 
Asplund M., Gustafsson B., Kiselman D., Eriksson K., 1997, \aap\ 318, 521

\bibitem[2004]{asplund04} 
Asplund M., 2004, Mem.\ S.\ A.\ It.\ 75, 300

\bibitem[2005]{asplund05} 
Asplund M., 2005, \araa\ in press

\bibitem[2000]{barklem00} 
Barklem P.S., Piskunov N., O'Mara B.J., 2000, \aaps\ 142, 467

\bibitem[1997]{baumueller97}
Baum\"uller D., Gehren T., 1997, \aap\ 325, 1088

\bibitem[1990]{beers90} 
Beers T.C., Flynn K., Gebhardt K., 1990, \aj\ 100, 32

\bibitem[2005]{beers05} 
Beers T.C., Christlieb N., 2005,  \araa\ in press

\bibitem[1999]{beers99} 
Beers T.C., Rossi S., Norris J.E., Ryan S.G., Shefler T., 1999, \aj\ 117, 981

\bibitem[2002]{beers02} 
Beers T.C., Drilling J.S., Rossi S., et~al., 2002, \aj\ 124, 931

\bibitem[1991]{biemont91} 
Bi\'emont E., Baudoux M., Kurucz R.L., Ansbacher W., Pinnington E.H., 1991, \aap\ 249, 539

\bibitem[1980]{biemont80} 
Bi\'emont E., Godefroid M., 1980, \aap\ 84, 361

\bibitem[1989]{biemont89v} 
Bi\'emont E., Grevesse N., Faires L.M., Marsden G., Lawler J.E., 1989, \aap\ 209, 391

\bibitem[1981]{biemont81} 
Bi\'emont E., Grevesse N., Hannaford P., Lowe R.M., 1981, \apj\ 248, 867

\bibitem[1989]{biemont89sm} 
Bi\'emont E., Grevesse N., Hannaford P., Lowe R.M., 1989, \aap\ 222, 307

\bibitem[1989]{oxford_ni89} 
Blackwell D.E., Booth A.J., Petford A.D., Laming J.M., 1989, \mnras\ 236, 235

\bibitem[1982a]{oxford_ti82a} 
Blackwell D.E., Menon S.L., Petford A.D., Shallis M.J., 1982a, \mnras\ 201, 611

\bibitem[1982b]{oxford_ti82b} 
Blackwell D.E., Petford A.D., Shallis M.J., Leggett S., 1982b, \mnras\ 199, 21

\bibitem[1984]{oxford_mn84} 
Booth A.J., Blackwell D.E., Petford A.D., Shallis M.J., 1984, \mnras\ 208, 147

\bibitem[1983]{borghs83} 
Borghs G., De Bisschop P., Van Hove M., Silverans R.E., 1983, Hyperfine Interactions 15, 177

\bibitem[2003]{bromm03} 
Bromm V., Loeb A., 2003, \nat\ 425, 812

\bibitem[2000]{burris00}
Burris D.L., Pilachowski C.A., Armandroff T.E., et~al. 2000, \apj\ 544, 302

\bibitem[1982]{burstein82}
Burstein D., Heiles C., 1982, \aj\ 87, 1165

\bibitem[1999]{busso99}
Busso M., Gallino R., Wasserburg G.J., 1999, \araa\ 37, 239 

\bibitem[1971]{cannon71}
Cannon C.J., 1971, Pub.\ Astron.\ Soc.\ Aust.\ 2, 42

\bibitem[1982]{cardon82}
Cardon B.L., Smith P.L., Scalo J.M., Testerman L., Whaling W., 1982, \apj\ 260, 395

\bibitem[2004]{cayrel04}
Cayrel R., Depagne E., Spite M., et~al. 2004, \aap\ 416, 1117

\bibitem[2004]{heres1}
Christlieb N., Beers T.C., Barklem P.S., et~al., 2004, \aap\ 428, 1027 (Paper I)

\bibitem[2003]{cohen03}
Cohen J.G., Christlieb N., Qian Y.-Z., Wasserburg G.J., 2003, \apj\ 588, 1082

\bibitem[2004]{cohen04}
Cohen J.G., Christlieb N., McWilliam A., et~al. 2004, \apj\ 612, 1107

\bibitem[1981]{cowan}
Cowan R.D., 1981, The Theory of Atomic Structure and Spectra, University of California Press, Berkeley

\bibitem[2003]{cutri03}
Cutri, R.M., Skrutskie M.F., van Dyk S., et~al. 2003, VizieR Online Data Catalog, 2246

\bibitem[1992]{davidson92}
Davidson M.D., Snoek L.C., Volten H., D\"onszelmann A., 1992, \aap\ 255, 457

\bibitem[1971]{davis71}
Davis S.J., Wright J.J., Balling L.C., 1971, \pra\ 3, 1220

\bibitem[2003]{den_hartog03}
Den Hartog E.A., Lawler J.E., Sneden C., Cowan J.J., 2003, \apjs\ 148, 543

\bibitem[1985]{doerr85}
Doerr A., Kock M., Kwiatkowski M., Werner K., Zimmermann P., 1985, \jqsrt\ 33, 55

\bibitem[1975]{froese_fischer75}
Froese Fischer C., 1975, Canadian J.\ Phys.\, 53, 184

\bibitem[1988]{fuhr88} 
Fuhr J.R., Martin G.A., Wiese W.L., 1988, J.\ Phys.\ Chem.\ Ref.\ Dat.\ 17, Suppl.\ 4

\bibitem[2000]{fulbright00} 
Fulbright J.P., 2000, \aj\ 120, 1841

\bibitem[2003]{fulbright03} 
Fulbright J.P., Johnson J.A., 2003, \apj\ 595, 1154

\bibitem[1967]{gallagher67}
Gallagher A., 1967, Phys.\ Rev.\ 157, 24

\bibitem[1994]{gratton94}
Gratton R.G., Sneden C., 1994, \aap 287, 927 

\bibitem[1989]{grevesse89} 
Grevesse N., Blackwell D.E., Petford A.D., 1989, \aap\ 208, 157

\bibitem[1998]{grevesse98} 
Grevesse N., \&\ Sauval A.J. 1998, in Solar composition and its evolution -- from core to corona, ed. C. Fr\"olich, M.C.E. Huber, \&\ S.K. Solanki (Dordrecht: Kluwer), 161

\bibitem[1982]{griffin82} 
Griffin R., Gustafsson B., Vieira T., Griffin R., 1982, \mnras\ 198, 637 

\bibitem[1975]{marcs_gust} 
Gustafsson B., Bell R.A., Eriksson K., Nordlund \AA., 1975, \aap\ 42, 407

\bibitem[1969]{handrich69} 
Handrich E., Steudel A., Walther H., 1969, Phys.\ Lett.\ 29A, 486

\bibitem[1982]{hannaford82} 
Hannaford P., Lowe R.M., Grevesse N., Bi\'emont E., Whaling W., 1982, \apj\ 261, 736

\bibitem[2000]{hill00} 
Hill V., Barbuy B., Spite M., et~al. 2000, \aap\ 353, 557

\bibitem[2002]{hill02} 
Hill V., Plez B., Cayrel R., et~al. 2002, \aap\ 387, 560

\bibitem[2004]{honda04} 
Honda S., Aoki W., Kajino T., et~al. 2004, \apj\ 607, 474

\bibitem[1980]{huber80} 
Huber M.C.E, Sandeman R.J., 1980, \aap\ 86, 95

\bibitem[1982]{hohle82} 
H\"ohle C., H\"uhnermann H., Wagner H., 1982, Z.\ Phys.\ A 304, 279

\bibitem[2004]{johnson04}
Johnson J., Bolte M., 2004, \apj\ 605, 462

\bibitem[2001]{karlsson01}
Karlsson T., Gustafsson B., 2001, \aap\ 379, 461

\bibitem[2005]{karlsson05}
Karlsson T., Gustafsson B., 2005, \aap\ in press

\bibitem[2003]{korn03}
Korn A.J., Shi J., Gehren T., 2003, \aap\ 407, 691

\bibitem[1999]{vald} 
Kupka F., Piskunov N., Ryabchikova T.A., Stempels H.C., Weiss W.W., 1999, \aaps\ 138, 119

\bibitem[1995]{kurucz}
Kurucz R.L., 1995, CDROMs, Cambridge, SAO 

\bibitem[1989]{lawler89} 
Lawler J.E., Dakin J.T., 1989, J.\ Opt.\ Soc.\ Am.\ B 6, 1457

\bibitem[2001]{lawler01la} 
Lawler J.E., Bonvallet G., Sneden C., 2001, \apj\ 556, 472

\bibitem[2001]{lawler01} 
Lawler J.E., Wickliffe M.E., Den Hartog E.A., Sneden C., 2001, \apj\ 563, 1075

\bibitem[2003]{lefebvre03} 
Lef\`ebvre P.H., Garnir H.P., Bi\'emont E., 2003, \aap\ 404, 1153

\bibitem[1963]{marquardt63} 
Marquardt D.W., 1963, J.\ Soc.\ Ind.\ Appl.\ Math.\ 11, 431

\bibitem[1988]{martin88} 
Martin G.A., Fuhr J.R., Wiese W.L., 1988, J.\ Phys.\ Chem.\ Ref.\ Dat.\ 17, Suppl.\ 3

\bibitem[1995a]{mcw95a} 
McWilliam A., Preston G.W., Sneden C., Shectman S., 1995a, \aj\ 109, 2736

\bibitem[1995b]{mcw95b} 
McWilliam A., Preston G.W., Sneden C., Searle L., 1995b, \aj\ 109, 2757

\bibitem[1998]{mcw98} 
McWilliam A., 1998, \apj\ 115, 1640

\bibitem[1987]{migdalek87} 
Migdalek J., Baylis W.E., 1987, Can.\ J.\ Phys.\ 65, 1612

\bibitem[1978]{mihalas}
Mihalas D., 1978, Stellar Atmospheres. Freeman \&\ Co., San Francisco

\bibitem[1983]{moity83} 
Moity J., 1983, \aaps\ 52, 37

\bibitem[2002]{nilsson02} 
Nilsson H., Zhang Z., Lundberg H., Johansson S., Nordstr\"om B., 2002, \aap\ 382, 368

\bibitem[1999]{nitz99} 
Nitz D.E., Kunau A.E., Wilson K.L., Lentz L.R., 2002, \apjs\ 122, 557

\bibitem[1996]{norris96} 
Norris J.E., Ryan S.G., Beers T.C., 1996, \apjs\ 107, 391

\bibitem[2001]{norris01} 
Norris J.E., Ryan S.G., Beers T.C., 2001, \apj\ 561, 1034

\bibitem[1991]{obrian91} 
O'Brian T.R., Wickliffe M.E., Lawler J.E., Whaling W., Brault J.W., 1991, J.\ Opt.\ Soc.\ Am.\ B 8, 1185

\bibitem[1962]{penkin62}
Penkin N.P., Shabanova L.N., 1962, Optics and Spectroscopy (U.S.S.R.), 12, 1 

\bibitem[1996]{pickering96}
Pickering J.C., 1996, \apjs\ 107, 811

\bibitem[2001]{pickering01}
Pickering J.C., Thorne A.P., Perez R., 2001, \apjs\ 132, 403

\bibitem[1996]{pilachowski96}
Pilachowski C.A., Sneden C., Kraft R., 1996, \aj\ 111, 1689

\bibitem[1995]{pinnington95}
Pinnington E.H., Berends R.W., Lumsden M., 1995, J.\ Phys.\ B  28, 2095 

\bibitem[1995]{vald95} 
Piskunov N., Kupka F., Ryabchikova T.A., Weiss W.W., Jeffrey C.S., 1995, \aaps\ 112, 525

\bibitem[1992]{numrec} 
Press W.H., Flannery B.P., Teukolsky S.A., Flannery B.P., 1992, Numerical Recipes.  Cambridge University Press, Cambridge

\bibitem[1999]{raiteri99} 
Raiteri C.M., Villata M., Gallino R., Busso M., Cravanzola A., 1999, \apj\ 518, L91

\bibitem[1994]{ryabchikova94} 
Ryabchikova T.A., Hill G.M., Landstreet J.D., Piskunov N., Sigut T.A., 1994, \mnras\ 267, 697 

\bibitem[1996]{ryan96} 
Ryan S.G., Norris J.E., Beers T.C., 1996, \apj\ 471, 254

\bibitem[1998]{schlegel98} 
Schlegel D.J., Finkbeiner D.P., Davis M., 1998, \apj\ 500, 525

\bibitem[2004]{schnabel04} 
Schnabel R., Schultz-Johanning M., Kock M., 2004, \aap\ 414, 1169

\bibitem[2004]{simmerer04}
Simmerer J., Sneden C., Cowan J.J., et~al. 2004, \apj\ 617, 1091

\bibitem[2004]{sivarani04}
Sivarani T., Bonifacio P., Molaro P., et~al. 2004, \aap\ 413, 1073

\bibitem[1996]{sneden96} 
Sneden C., McWilliam A., Preston G.W., et~al. 1996, \apj\ 467, 819

\bibitem[2000]{sneden00} 
Sneden C., Cowan J.J., Ivans I.I., et~al. 2000, \apj\ 533, L139

\bibitem[2002]{sneden02} 
Sneden C., Cowan J.J., Lawler J.E., et~al. 2002, \apjl\ 566, L25

\bibitem[2003]{sneden03} 
Sneden C., Cowan J.J., Lawler J.E., et~al., 2003, \apj\ 591, 936

\bibitem[2001]{snider01} 
Snider, S., Allende Prieto, C., von Hippel, T., et~al., 2001, \apj\ 562, 528

\bibitem[1982]{taylor82} 
Taylor J.R., {\em An Introduction to Error Analysis}, Oxford University Press, 1982, chapter 9

\bibitem[2004]{travaglio04} 
Travaglio C., Gallino R., Arnone E., et~al., 2004, \apj\ 601, 864

\bibitem[2002]{truran02} 
Truran J.W., Cowan J.J., Pilachowski C.A., Sneden C., et~al., 2002, \pasp\ 114, 1293

\bibitem[1955]{unsold55} 
Uns\"old A., 1955, Physik der Stern Atmosp\"aren, Zweite Auflage. Springer-Verlag

\bibitem[1996]{sme} 
Valenti J.A., Piskunov N., 1996, \aaps\ 118, 595

\bibitem[1998]{valenti98} 
Valenti J.A., Piskunov N., Johns-Krull, C.M., 1998, \apj\ 498, 851

\bibitem[1967]{weiss67} 
Weiss A.W., 1967, J.\ Chem.\ Phys.\ 47, 3573

\bibitem[1980]{wiese80} 
Wiese W.L., Martin G.A., 1980, Wavelengths and Transition Probabilities for Atoms and Atomic Ions, NSRDS-NBS No.~68, Part II.

\bibitem[1969]{wiese69} 
Wiese W.L., Smith M.W., Miles B.M., 1969, Atomic Transition Probabilities, Vol.~2, Sodium through Calcium, NSRDS-NBS 22

\bibitem[2000]{wisotzki00} 
Wisotzki L., Christlieb N., Bade N., et~al., 2000, \aap\ 358, 77

\end{thebibliography}
\end{document}